\documentclass[11pt]{article}
\newcommand{\pmb}[1]{{\setbox0=\hbox{#1}%
  \kern-.025em\copy0\kern-\wd0
  \kern.05em\copy0\kern-\wd0
  \kern-.025em\raise.0433em\box0 }}
\newcommand{\fg}{\boldsymbol}   
\usepackage{citesort}
\usepackage{amssymb,amsfonts,mathrsfs,color}
\usepackage{latexsym}
\usepackage{amsmath}
\usepackage{comment}
\usepackage{epstopdf}
\usepackage[pdftex]{graphicx}
\usepackage{graphicx}
\usepackage[titletoc,title]{appendix}
\numberwithin{equation}{section}
\DeclareGraphicsExtensions{.pdf,.jpg,.eps}
\usepackage{epsf}
\usepackage{subfigure}
\usepackage{calc}

\newlength{\spacer}
\newsavebox{\mybox}

\newcommand{\bey}{\begin{eqnarray}}
\newcommand{\eey}{\end{eqnarray}}

\newcommand{\bec}{\begin{center}}
\newcommand{\eec}{\end{center}}

\def\tanh{\operatorname{tanh}}

\newcommand{\drop}[1]{}

\usepackage[lmargin=2.0cm,rmargin=2.0cm,tmargin=2.0cm,bmargin=2.0cm]{geometry}
\begin{document}
\pagestyle{myheadings}
\setcounter{tocdepth}{2}
\baselineskip19pt
\belowdisplayskip11pt
\belowdisplayshortskip11pt
\renewcommand{\thefootnote}{\fnsymbol{footnote}}
\bec
{\Large Grain boundary-induced martensitic transformations: A phase-field study of nucleation, size-effect, triple junction-effect, microstructures, and compatibility at the nanoscale}
\eec
\bec
{\large Anup Basak \footnote{abasak@iittp.ac.in (corresponding author)}}\\
{\it
 Department of Mechanical Engineering,\\
 Indian Institute of Technology Tirupati,  \\
 Tirupati, Yerpedu, A.P. 517619, India.\\

}
\eec
An original thermodynamically consistent large strains-based multiphase phase-field (PF) approach of Ginzburg-Landau type is developed for studying the grain boundary (GB)-induced martensitic transformations (MTs) in polycrystalline materials at the nanoscale considering the structural stresses within the interfaces. 
In this general PF approach, $N$ independent order parameters are used for describing the austenite ($\sf A$)$\leftrightarrow$martensite ($\sf M$) transformations and $N(>1)$ martensitic variants, and another $M$ independent order parameters are considered for describing $M(>1)$ grains in the polycrystalline samples. The change in the GB energy due to its structural rearrangement during MTs is considered using  variable energy for the GB(s) as a function of the order parameter related to the $\sf A\leftrightarrow\sf M$ transformation. 
A rich plot for the temperatures of transformations between the $\sf A$, premartensite, and $\sf M$ in a bicrystal with a symmetric planar tilt GB are plotted for the varying austenitic GB width.
The strong effects of the parameters, including the austenitic GB width, change in GB energy due to MTs,  GB misorientation, applied strains, and sample size on heterogeneous nucleation of the phases and the subsequent complex martensitic microstructures evolution are explored in various bicrystals with symmetric or asymmetric planar or circular tilt GBs during the forward and reverse transformations.  The triple junction (TJ) energy and the energy and width of the adjacent GBs are also shown to strongly influence the nucleation and microstructures using the tricrystals having three symmetric planar tilt GBs meeting at $120^\circ$ dihedral angles. The compatibility of the microstructures across the GBs is studied. The elastic and structural stresses across the GBs and TJ regions are plotted, which is essential for understanding the role of GBs and TJs in materials failure. 
The plausible reasons for the nonintuitive behaviour of the GBs on martensite nucleation observed in experiments and atomistic simulations are explained.

\vspace{2mm}\noindent {\bf Keywords:} Grain boundary-induced martensitic transformations;  Triple junction; Phase-field approach; Heterogeneous nucleation;  Size-effect.

 \section{Introduction}
 \label{introd} 
 \noindent{\bf Grain boundary-induced martensitic transformations:}
Austenite ($\sf A$)$\leftrightarrow$martensite ($\sf M$) phase transformations play a central role in various industrially important materials, including shape memory alloys (SMAs) which can recover large strains and exhibit superelasticity \cite{Bha04,Ball-James-87,Porter-Easterling,RoytburdSl-01};
dual phase steels where a proper mixture of ferrite and martensite yields suitable mechanical properties \cite{TasanIJP-14}; ferroic materials which are used for actuation and control \cite{Wadhawan}; various ceramics and minerals \cite{Wadhawan}, etc. 
 In materially uniform bodies, the  heterogeneities, including the dislocations (\cite{OlsonCohen-1981,Mishin-09}, Chapter 6 of \cite{Porter-Easterling}), accumulated plastic strains and shear bands \cite{OlsonCohen-1981,OlsonCohen-1975,Rabbe-JMPS-11,Kundin-JMPS-15,Paranjape-IJP-2012,Paranjape-IJP-2012,Levitas-IJP-2021}, crack tips \cite{Hangen-1999},  free surface \cite{Mishin-09,Gall-05}, anti-phase boundaries \cite{Mishin-09}, or any other regions within the materials with concentrated stresses \cite{Kyriakides-2022IJP,Kyriakides-2014IJSS}, etc., are usually the  preferred sites for nucleation of $\sf M$ in a single grain or polycrystalline samples of materials capable of undergoing martensitic transformations (MTs). The reduction in energies during MTs at  other heterogeneous sites such as the matrix-precipitate interfaces  \cite{Wang-2015-Acta}, external surfaces of the samples \cite{Gall-05,Lovely-Chandrasekaran-1983,Umantsev-2017,Gall2001,Kyriakides-2014IJSS}, and heterophase interfaces \cite{Schryvers-2004,Waitz-2004-Acta} also promotes the nucleation of $\sf M$. 
On the other hand, point defects such as vacancies, interstitials, and ani-site defects may suppress the nucleation of $\sf M$ by significantly reducing the transformation temperatures \cite{Tehrani-2015,Yang-2022}.  

The experimental studies have shown that  the grain boundaries (GBs) and triple junctions (TJs) are  additional heterogeneities favourable for the $\sf M$ nucleation in the polycrystalline samples of such phase-changing materials \cite{Butler-1982,Ueda-2001Acta,Ueda-2004ISIJ,Tsuzaki-1995,Kajiwara-1986,Song-14,Schuh-13,Arlt-90,Mana-2020Mater,Bauer-16,Liu-2014MaterDesg}, which was also confirmed through atomistic simulations \cite{Zhang-2018,Dmitriev-2018,Meiser-16,Lazarev-2008,Uehara-09}. In particular, the parameters such as the grain orientation distribution (see, e.g. \cite{Shu-Bhattacharya-98,Bha04,Uehara-09,Bauer-16},  and the references therein), average grain size \cite{LeeLee-2005,Kajiwara-1986,SunAPL-2013}, type of the GBs (tilt or twist) \cite{Song-14,Ueda-2001Acta,Dmitriev-2018},  GB energy change during MTs \cite{Butler-1982},  stress concentration within the GBs during transformations \cite{Bauer-16,Schuh-13,Schuh-12-1}, degree of reduction in symmetry from $\sf A$ to $\sf M$ \cite{Shu-Bhattacharya-98,Bhattacharya-Kohn-96}, and direction of loadings \cite{Gall2001} have been observed to play an important role on nucleation and evolution of very rich martensitic microstructures. The MTs in nanocrystalline materials, which contain a large volume fraction of the GBs and TJs, are greatly influenced by these heterogeneous sites    \cite{FischerJMPS-2007,SunJMPS-2018}. 
The  temperatures of $\sf M$ nucleation from the GBs  are usually significantly higher than the  homogeneous transformation temperature of $\sf M$ \cite{Ueda-2001Acta,Ueda-2004ISIJ}.  The start and finish temperatures for both the forward ($\sf A\to \sf M$) and the reverse ($\sf M\to \sf A$) transformations from the tilt GBs are much higher than that for the twist GBs \cite{Song-14,Ueda-2001Acta,Zhang-2018,Dmitriev-2018}. The reason for this was attributed to  the local stress fields associated with the intrinsic dislocations in the low angle tilt GBs \cite{Song-14,Ueda-2001Acta}. 
A reduction in the energy of the GB region after $\sf M$ nucleation was clearly observed in atomistic simulations \cite{Qin-2018}.  While the atomistic simulations in \cite{Zhang-2018} showed that the GBs are preferred over the external surface for $\sf M$ nucleation, an opposite result is, however, reported in \cite{Clapp-2004}.
 Furthermore, the experiments \cite{Kajiwara-1986} and recent atomistic studies \cite{Qin-2018} showed that not all the tilt GBs could promote MTs.  The heterogeneous nucleation of $\sf M$ from the GBs in polycrystals is clearly a complex phenomenon 
 and is still far from being well-understood. The TJs have also been observed to be potential $\sf M$ nucleation sites \cite{Schuh-13}. The stresses were found to be concentrated mainly within the GBs and TJ regions causing nucleation of cracks in the brittle SMAs \cite{Schuh-13,Schuh-12-1,Bauer-16,SchuhAPL-09}.  Because of this, great efforts are being made to minimize the GBs and TJs in designing the phase-changing materials \cite{Schuh-13,Schuh-12-1,Bauer-16,SchuhAPL-09}. Because of the additional system parameters influencing the MTs in polycrystals, a significant difference in the shape memory effects in polycrystalline SMAs is observed from its single crystal counterpart (see, e.g.  \cite{Shu-Bhattacharya-98} and the references therein). 

\begin{figure}[t!]
\centering
\hspace{-8mm}
\subfigure[]{
  \includegraphics[width=2.1in, height=1.9in] {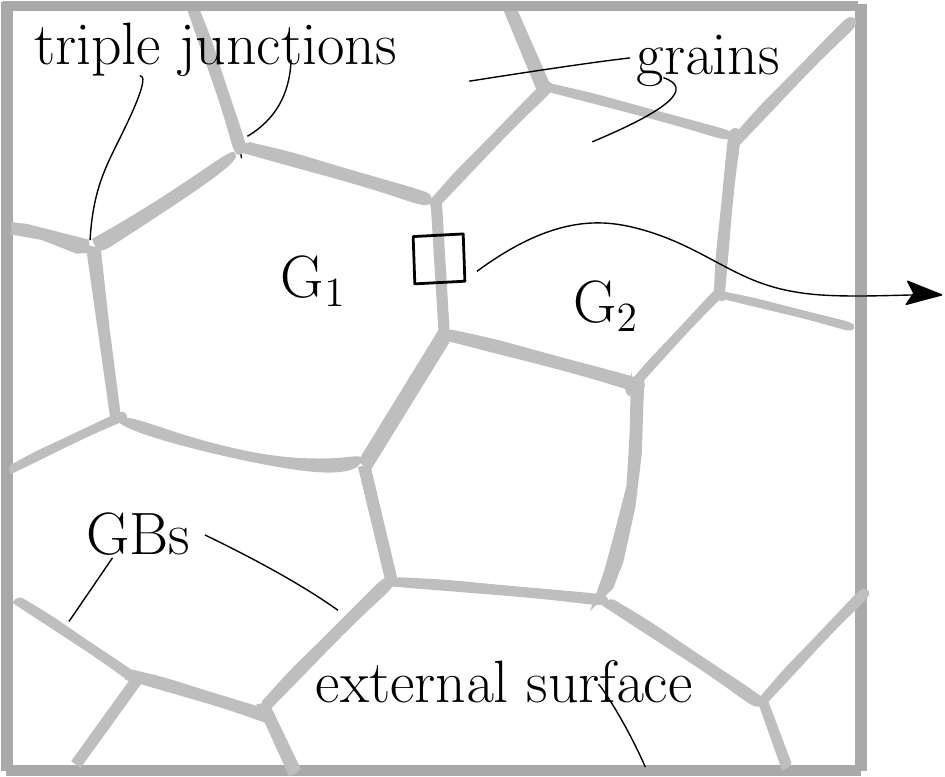}
	}
\hspace{2mm}
\subfigure[]{
  \includegraphics[width=4.2in, height=2.1in] {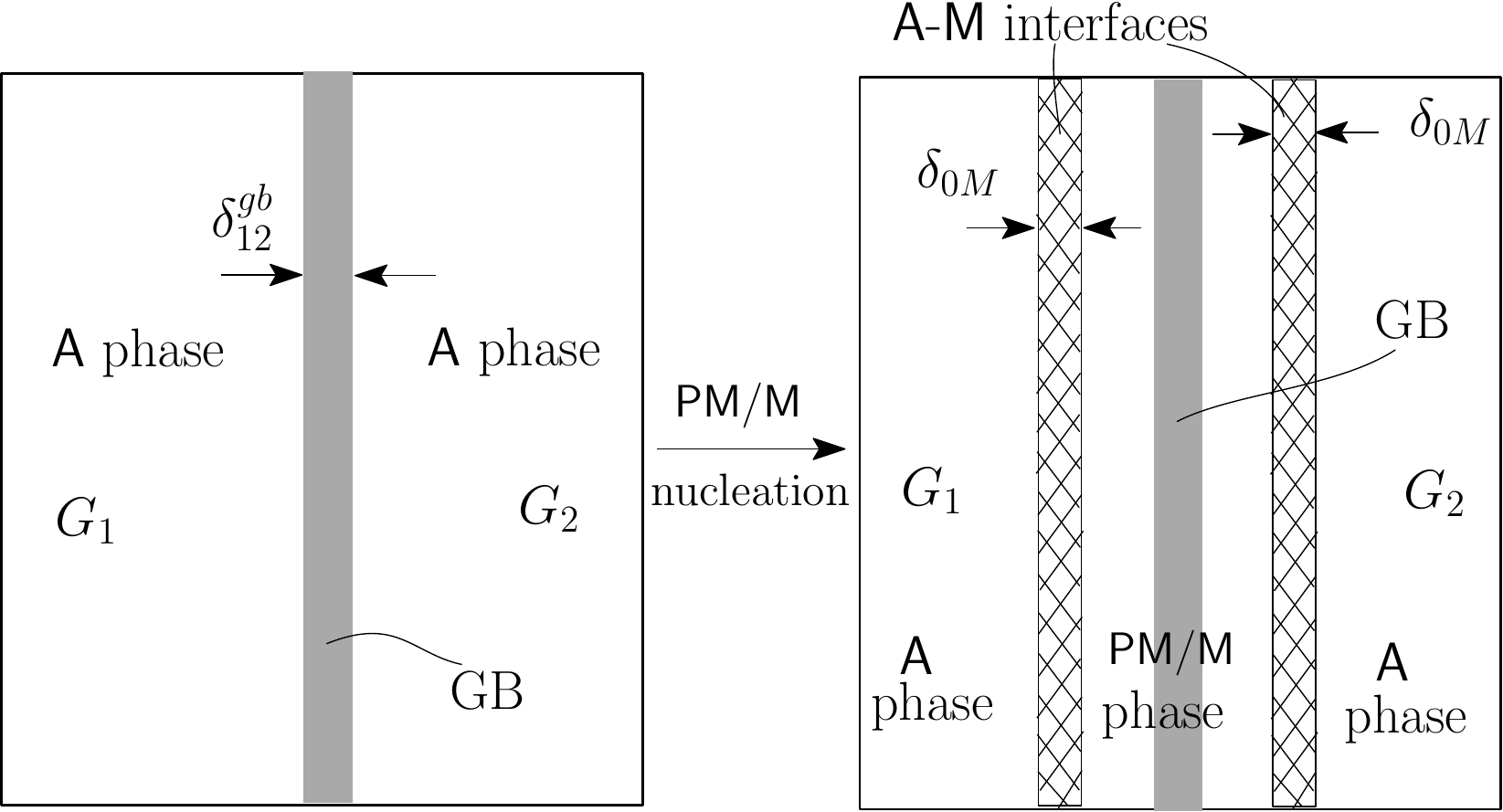}
	}
\caption{(a) Schematic of a stress-free austenitic polycrystal. (b) An exaggerated view of the part (we call it a bicrystal) is highlighted in (a) and  contains a part of the planar GB between grains G$_1$ and $G_2$. The rightmost figure shows the bicrystal after a $\sf PM$/$\sf M$ layer appeared on the GB and the neighbouring bulk. $\delta_{12}^{gb}$- austenitic GB width between grains $G_1$ and $G_2$; 
 $\delta_{0M}$- width of $\sf A$-$\sf M$ interface.}
\label{GBschemat}
\end{figure}

Although multiple reasons have been observed behind the GBs to promote $\sf M$ nucleation, as discussed above,
a complete understanding of the GB-induced MTs is still missing. 
The general understanding of any GB-induced phase transformation is that a thin layer of a new phase (e.g. a precipitate, a disordered phase like premelt/melt, an ordered phase like $\sf M$, etc.) nucleates within a GB region (maybe locally)  at an energetically favourable thermomechanical condition, which is driven by a reduction in the total energy of the transformed region (see, e.g. Sec. 6.3.2.2 of \cite{Sutton-Balluffi},  \cite{Lipowsky1986PRL,Mishin2009-I,Basak-2021-PCCP}, and the references therein for explanation about GB-induced premelting/melting).
We explain the GB-induced MTs here through a schematic shown in Fig. \ref{GBschemat}(b), where a stress-free austenitic bicrystal is shown with a planar GB of  width $\delta_{12}^{gb}$. Here we see the bicrystal as a small region between the grains $G_1$ and $G_2$ from a large polycrystal shown in Fig. \ref{GBschemat}(a). The suffix $12$ in $\delta_{12}^{gb}$ is now obvious.  At a thermo-mechanical condition within the  stability regime of $\sf A$ phase, a layer of premartensite (denoted by $\sf PM$), or $\sf M$ phase (say, a single variant) of thickness $w$ nucleates within the austenitic GB and grains as shown in Fig. \ref{GBschemat}(b) if
\begin{equation}
\gamma_{12}^M +2\gamma_{0M} +(\psi^\theta_M+\Delta\psi^e-\psi^\theta_0)w<\gamma_{12}^A,
   \label{constraintNuclm}
 \end{equation}
  where $\sf PM$ is an intermediate state between $\sf A$ and $\sf M$ (see, e.g. Chapter 6 of \cite{Porter-Easterling} and \cite{Lovely-Chandrasekaran-1983}),  $\gamma_{12}^A$ and $\gamma_{12}^M$ are the energies of the GB  when they are in $\sf A$ and $\sf M$ phases, respectively, $\gamma_{0M}$ is the $\sf A$-$\sf M$ interface energy,  $\psi^\theta_0(\theta)$ and $\psi^\theta_M(\theta)$ are the thermal energies of $\sf A$ and $\sf M$ phases, respectively, $\theta>0$ is the absolute temperature, and $\Delta\psi^e$ is the change in average elastic energy  of the transformed region. We have taken twice $\gamma_{0M}$ in inequality \eqref{constraintNuclm} as there are two $\sf A$-$\sf M$ interfaces in the transformed region. The criterion \eqref{constraintNuclm} should be seen from the micron scale and when the GB and $\sf A$-$\sf M$ widths, which vary between a few angstroms to a few nanometres, are negligible compared to $w$. Note that we have assumed a single variant appears in the transformed region in the relation \eqref{constraintNuclm}; the interfacial energies between the variants should be taken into account therein if multiple variants are considered. The above criterion in the nanoscale, when the interfacial widths and the transformed region thickness $w$ are comparable, can be expressed as (see  \cite{Basak-2021-PCCP,RappazMMTA2003} for the nanoscale criterion of GB-induced premelting/melting)
 \begin{eqnarray}
\gamma_{12}^M +2\gamma_{0M} +(\gamma_{12}^A-\gamma_{12}^M -2\gamma_{0M} )\exp \left(-\frac{w}{\delta}\right)+(\psi^\theta_M+\Delta\psi^e-\psi^\theta_0)w &<& \gamma_{12}^A \nonumber\\
\Rightarrow\,\,  \Delta \gamma_{12}^{gb}+2\gamma_{0M} -(\Delta\gamma_{12}^{gb}+2\gamma_{0M} )\exp \left(-\frac{w}{\delta}\right)+(\psi^\theta_M+\Delta\psi^e-\psi^\theta_0)w &<& 0,
   \label{constraintNucl}
 \end{eqnarray}
 where $\delta$ is the width of the interfaces (both GB width and $\sf A$-$\sf M$ interface width are assumed to be identical in inequality \eqref{constraintNucl}), and $\Delta \gamma_{12}^{gb}=\gamma_{12}^M-\gamma_{12}^A$ is the change in GB energy due to MT. When $w\gg \delta$, the inequality \eqref{constraintNucl} coincides with the criterion given by inequality \eqref{constraintNuclm} for the micron scale. Based on the nanoscale criterion  \eqref{constraintNucl}, we identify that the following two GB parameters which determine the thermo-mechanical condition for nucleation of $\sf M$ from the GBs in polycrystals: (i) the change in GB energy $\Delta \gamma_{12}^{gb}$ due to MTs, and (ii) austenitic GB width. 
 
 Note that in the inequalities \eqref{constraintNuclm} and \eqref{constraintNucl}, we have considered different energies for the GB in $\sf A$ and $\sf M$ phases are based on the atomistic results of \cite{Qin-2018} on NiTi alloy bicrystals, which is arising because of the structural changes of the GBs due to the transformations \cite{Qin-2018}. Although no experimental studies demonstrating the change in GB energies due to MTs are there in the literature to the best of the author's knowledge, the atomistic simulations in \cite{Qin-2018} clearly showed a decrease in the GB energy during the $\sf A\to\sf M$ PT. Another exciting and counter-intuitive result reported in \cite{Qin-2018} is that a $\Sigma 9$ austenitic GB promoting $\sf M$ nucleation has much lower energy ($0.22$ times) than another  $\Sigma 9$ GB which does not promote nucleation. Furthermore, the drop in the average energy in the former sample due to $\sf A\to\sf M$ transformation is higher than that of the later one at the respective transformation temperatures (see Table-1, Figs. 6a and 8a of \cite{Qin-2018}). We also note that the $\Sigma 9$ GB promoting nucleation has a  smaller width than the other  (see Table-1 and Fig. 2 of \cite{Qin-2018}). It is important to mention here that the size (width) of the surface/interface and its change in energy due to the PT have been observed to strongly influence the nucleation of the phases during the solid$\leftrightarrow$melt transformations induced by the external surface \cite{Levitas-Samani-PRB-14,Basak-Levitas-2018-nanovoid} or the GBs \cite{Mishin2009-I,RappazMMTA2003,Basak-2021-PCCP}, and also the MTs induced by the external surfaces \cite{levitas-javanbakht-PRL-10,levitas-javanbakht-PRL-11} or the matrix-precipitate interfaces \cite{BasakLevitas2020Acta-2004}. It is thus of utmost importance to investigate the role of the similar parameters, i.e. the austenitic GB width and the change in GB energy due to MTs for the  understanding of the MTs in polycrystalline samples, which is yet to be done. 

 \noindent{\bf Phase-field studies of heterogeneous nucleation:}
   The phase-field (PF) approaches of Ginzburg-Landau (similar to Allen-Cahn approach) provide an appropriate framework for studying various  structural transformations  in materials, including the MTs by homogeneous or heterogeneous nucleation and the subsequent growth of the phases (\cite{Umantsevbook,levitas-javanbakht-PRL-10,levitas-javanbakht-PRL-11,Basak-Levitas-2018JMPS,Basak-Levitas-2019CMAME,Tuma-Stupkiewicz-16,Tuma-Stupkiewicz-Petryk-16,Stupkiewicz-Petryk-22,Clayton-knap-11,Finel-2012,Schryvers-12,Artemev-08,Artemev-08_1,Artemev-Wang-Khachaturyan-2000-Acta,Khachaturyana-2002,Landis-19,Levitasetal-PRB-15,Levitas-2013IJP,Levitasetal-PRB-twin-13,Chen-20,Chen-Material_sc,Chen-02,BasakLevitasCMT2022,Levitas-14a,Steinbach-09,PF_review,Clayton2021,ClaytonCMT2014,levitas-javanbakht-JMPS-14a,levitas-javanbakht-JMPS-14} and the references therein). The surface-effect on the PTs within the Ginzburg-Landau theory was first introduced by Lipowsky \cite{Lipowsky84-PRL}  to the best of the author's knowledge, and an analytical treatment for a semi-infinite domain was given. The analytical treatment for a finite-sized sample was presented in \cite{Levitas2006-EPL}, where variable energy for the heterogeneity as a quadratic function of the order parameter was considered, which is, however, oversimplified. A more realistic and non-contradictory variable surface energy, which is quartic in the order parameter, was introduced in \cite{levitas-javanbakht-PRL-10}. The model was used to study the surface-induced multivariant MTs in  \cite{levitas-javanbakht-PRL-11}. 
%
%
%
%
%
%
%
%
%
%
%
%
%
%
%
%
%
%
%
%
%
%
%
%
%
   Several PF approaches have been further developed and successfully used for studying surface-induced melting from the flat surfaces \cite{Tang16} and  curved surfaces of nanoparticles \cite{Levitas-Samani-PRB-14}  and  nanovoids \cite{Basak-Levitas-2018-nanovoid}; GB and TJ-induced solid$\leftrightarrow$melt transformations (e.g. \cite{Mishin2009-I,Basak-2021-PCCP} and the references therein); surface-induced paraelectric-ferroelectric PT  \cite{Umantsev-2017}; matrix-precipitate interface induced MTs \cite{BasakLevitas2020Acta-2004} in single grains; see \cite{Levitas_scripta_2018} for a review. 
     The  studies in \cite{Levitas-Samani-PRB-14,Basak-Levitas-2018-nanovoid,Basak-2021-PCCP,levitas-javanbakht-PRL-11,BasakLevitas2020Acta-2004}  have shown that the width and the change in energy of the corresponding surface/interface during the PTs strongly influence the respective transformation temperatures, which may vary with the characteristic size of the heterogeneities in a complex and nonintuitive way. The TJ energy also plays a vital role in premelt/melt nucleation, and kinetics during the forward and reverse PTs  \cite{Basak-2021-PCCP}.

 The small strains-based multiphase phase-field approaches have been developed to study  MTs in polycrystalline  materials in \cite{Steinbach-2019,Cho-Levitas-12-IJSS,Artemev-Jin-Khachaturyan-2002-Acta,Xua-20021-IJMS,Xua-2022-EJMS,Fan-2022-CMS,Xi-20021-AMech,Cisse-2020-AM,Mamivanda-2014-IJP,Paliwal-12-1,Chen-14,Khachaturyan-01,Nestler-2018,Yamanaka2010,Cui2007,Sutou2005,Sun-2018-CMS,Yeddu-2018-CMS,Zhu-2020-CMS}. The only PF model used for polycrystalline samples by Malik et al. \cite{Malik-2013-Acta,Malik-2013-MSMSE} considers a geometric nonlinear total strain, which was additively decomposed into inelastic and elastic parts. Such a strain decomposition is well-known to have limitations and is applicable only in some special cases (see  Sec. 8.2 of \cite{Lublinerbook} and the references therein).  Both concentration based \cite{Steinbach-2019,Cho-Levitas-12-IJSS,Artemev-Jin-Khachaturyan-2002-Acta,Xua-20021-IJMS,Fan-2022-CMS,Xi-20021-AMech,Cisse-2020-AM,Mamivanda-2014-IJP,Paliwal-12-1,Chen-14,Khachaturyan-01,Nestler-2018,Yamanaka2010,Cui2007,Sutou2005,Sun-2018-CMS,Yeddu-2018-CMS,Zhu-2020-CMS} and transformation strains \cite{Cho-Levitas-12-IJSS,Artemev-Jin-Khachaturyan-2002-Acta,Chen-14,Khachaturyan-01} based order parameters have been used for describing the $\sf A$ and the $N$ variants ${\sf M}_1\,{\sf M}_2,\ldots,{\sf M}_{N}$. In \cite{Chen-14}, the order parameters were indexed in a way to take care of both the phase changes and the variation of the grains. However, the remaining papers for polycrystals cited here did not introduce any order parameter for describing the grains. A single constraint, which is the summation of all the order parameters (volume fraction) is equal to the unity, was used in  \cite{Steinbach-2019,Nestler-2018,Fan-2022-CMS}. Such a single constraint cannot assure that each of the MTs can be systematically described by a single order parameter only, which is necessary for identifying the relations between the system parameters and the measurable material parameters; see \cite{Basak-Levitas-2018JMPS}. The other papers \cite{Artemev-Jin-Khachaturyan-2002-Acta,Xua-20021-IJMS,Xua-2022-EJMS,Xi-20021-AMech,Cisse-2020-AM,Mamivanda-2014-IJP,Paliwal-12-1,Chen-14,Khachaturyan-01,Yamanaka2010,Cui2007,Sutou2005,Sun-2018-CMS,Yeddu-2018-CMS,Zhu-2020-CMS,Malik-2013-Acta,Malik-2013-MSMSE} also did not consider all necessary constraints which allow controlling all the transformation paths and calibration of all the system  parameters; see \cite{Basak-Levitas-2018JMPS} for a detailed discussion.
 Various strategies were adopted to initiate the MTs in those papers as follows. 
While the random noise in the order parameters \cite{Artemev-Jin-Khachaturyan-2002-Acta,Khachaturyan-01,Nestler-2018,Fan-2022-CMS,Lookman2004,Mamivanda-2014-IJP,Paliwal-12-1,Chen-14} or in the displacements \cite{Ahluwalia2015,Xu-2016} was considered to trigger nucleation of $\sf M$ from the GBs, artificial nuclei were placed on the GBs in \cite{Yamanaka2010,Xua-2022-EJMS} for that purpose. In the PF model of \cite{Chen-14}, a relaxation parameter for the misfit strain between $\sf A$ and $\sf M$ variants within the GB  is considered to promote nucleation. Nucleation was also observed due to the stress concentration within dislocation pileups near the GBs and shear bands (see, e.g. \cite{Levitas-IJP-2021} and the references therein) and due to the geometric constraint within the samples in  \cite{Xua-20021-IJMS}. In \cite{Steinbach-2019,Cho-Levitas-12-IJSS,Malik-2013-MSMSE,Malik-2013-Acta,Xi-20021-AMech,Yeddu-2018-CMS}, where the primary objective was to study the effect of resistance provided by the GBs on the growth of the variant plates, artificial nuclei were placed inside the grains to initiate MTs. These studies have also shown that with a reduction in the average grain size in the polycrystals, the barrier energy for the nucleation of $\sf M$ increases, drastically affecting the subsequent transformations. The GB texture also was shown to greatly influence the nucleation and growth of the variants \cite{Chen-14,Yamanaka2010,Artemev-Jin-Khachaturyan-2002-Acta,Cui2007,Sutou2005}. 
  
It is to be mentioned that none of the existing papers on MTs in polycrystalline solids \cite{Steinbach-2019,Cho-Levitas-12-IJSS,Levitas-IJP-2021,Malik-2013-Acta,Artemev-Jin-Khachaturyan-2002-Acta,Xua-20021-IJMS,Fan-2022-CMS,Xi-20021-AMech,Cisse-2020-AM,Mamivanda-2014-IJP,Paliwal-12-1,PF_review,Chen-14,Khachaturyan-01,Nestler-2018,Yamanaka2010,Cui2007,Sutou2005,Malik-2013-MSMSE,Sun-2018-CMS,Yeddu-2018-CMS,Zhu-2020-CMS} studied the role of the GB width on the temperature of nucleation of the phases and the subsequent microstructure evolution.  The change in GB energy due to MTs, an important material parameter discussed above, is also yet to be incorporated to develop a robust model. Furthermore, the role of the TJ energy on GB-induced MTs is  not yet studied. The PF models in \cite{Steinbach-2019,Cho-Levitas-12-IJSS,Levitas-IJP-2021,Artemev-Jin-Khachaturyan-2002-Acta,Xua-20021-IJMS,Fan-2022-CMS,Xi-20021-AMech,Cisse-2020-AM,Mamivanda-2014-IJP,Paliwal-12-1,Chen-14,Khachaturyan-01,Nestler-2018,Yamanaka2010,Cui2007,Sutou2005,Sun-2018-CMS,Yeddu-2018-CMS,Zhu-2020-CMS} considered small strains, whereas the MTs usually involve large strains, and hence a large strains-based general PF model for MTs in polycrystals is yet to be developed. Recall that the only large strains-based model in \cite{Malik-2013-Acta,Malik-2013-MSMSE} has issues with the kinematic model, as pointed out above.  The structural stresses \cite{Levitas-14a,Basak-Levitas-2018JMPS}, which play an important role in MTs \cite{levitas-javanbakht-PRL-10,levitas-javanbakht-PRL-11} and are still missing in all the existing models for polycrystals cited here, must be considered for completeness.

  \noindent{\bf Contributions of the paper:} The contributions of the paper are as follows:

 \noindent {\bf (I)} A thermodynamically consistent original and non-contradictory multiphase PF approach for studying GB-induced $\sf A\leftrightarrow\sf M$ transformations  at the nanoscale is developed for polycrystalline solids capable of undergoing MTs with $M(>1)$ grains and considering $N(>1)$ martensitic variants. The present model considers the following aspects which are missing in the existing ones for MTs in polycrystals  \cite{Steinbach-2019,Cho-Levitas-12-IJSS,Levitas-IJP-2021,Malik-2013-Acta,Artemev-Jin-Khachaturyan-2002-Acta,Xua-20021-IJMS,Fan-2022-CMS,Xi-20021-AMech,Cisse-2020-AM,Mamivanda-2014-IJP,Paliwal-12-1,PF_review,Chen-14,Khachaturyan-01,Nestler-2018,Yamanaka2010,Cui2007,Sutou2005,Malik-2013-MSMSE,Sun-2018-CMS,Yeddu-2018-CMS,Zhu-2020-CMS}:
 
 (a) Variable  energy of the GB region as a function for the order parameter related to ${\sf A}\leftrightarrow{\sf M}$ transformation is considered, which incorporates the change in GB energy due to the structural transformation between $\sf A$ and $\sf M$ phases within the free energy of the system.
 
 (b) The order parameters related to the variants and those related to the GBs are constrained to particular hypersurfaces in the respective order parameter spaces, which is essential to control the transformation paths between the variants and between the grains and also calibration of all the system parameters \cite{Basak-Levitas-2018JMPS}. 
 
 (c) Our model considers large strains, and hence a multiplicative decomposition for the total deformation gradient into elastic and transformational parts is used.
 
 (d) The structural part of the total interfacial stress tensor is considered for all the interfaces. 
 
 (e) An excess energy for the TJ regions connecting the GBs is considered  which is necessary to control the energy and size of the TJ regions.
 
 (f) A non-contradictory and new kinetic model for the coupled Ginzburg-Landau equations related to the MTs and GBs are derived.

 
 \vspace{2mm}
  
 \noindent {\bf (II)} The numerical results are presented for 
 (a) bicrystals with a planar symmetric or asymmetric tilt GB, and  a bicrystal with a circular symmetric tilt GB; and  
 (b) tricrystals consisting of three planar symmetric tilt GBs meeting the interior of the sample at $120^\circ$ dihedral angles. 
 The results for both forward and reverse transformations are presented. The strong effect of the width of the GBs in stress-free $\sf A$ polycrystals on $\sf M$ nucleation, temperature hysteresis, and microstructure evolution is studied. A plot for the temperatures of transformations between $\sf A$, $\sf PM$, and $\sf M$ within the GBs for varying austenitic GB width is presented for a bicrystal with a planar symmetric GB. The role of  change in the GB energy due to MTs, energy of the TJ, GB misorientation angle, applied strain, and the sample size is studied. The compatibility of the microstructures across the GBs is studied. The plausible reasons for various nonintuitive effects of the GBs on MTs observed in the experiments, and atomistic studies are discussed.
 
 \vspace{2mm} 
 \noindent {\bf (III)} The distribution of the elastic and structural stresses within the bicrystals and tricrystals are studied, which is essential for failure analysis and design of materials.

The paper is organized as follows. The multiphase PF model is developed in Sec. \ref{system_eqns}; the material parameters are listed in Sec. \ref{MaterParam}; the numerical results are presented in Sec. \ref{simulations};  we conclude the paper in Sec. \ref{consls}.
 \vspace{5mm}

 \noindent{\em Notation:} We denote the inner product and multiplication between two arbitrary second-order tensors as
 ${\fg A}:{\fg D}=A_{ab} D_{ba}$ and $({\fg A}\cdot{\fg D})_{ab}=A_{ac} D_{cb}$, respectively, where the repeated
indices denote Einstein's summation, and $A_{ab}$ and $D_{ab}$ are the components of the tensors
in a right-handed Cartesian basis $\{\fg e_1,\fg e_2,\fg e_3\}$. The Euclidean norm of $\fg A$ is defined as $|\fg A|=\sqrt{{\fg A}:{\fg A}^T}$; ${\fg I}$ is the second order identity tensor and its, say, $ab^{th}$ component is denoted by the Kronecker delta $\delta_{ab}$; $\fg A^{-1}$, $\fg A^T$, $det\,\fg A$, and $tr\,\fg A$ denote inversion, transpose, determinant, and trace of $\fg A$, respectively. The dyadic
product between two arbitrary vectors $\fg a$ and $\fg b$ is denoted as $\fg a\otimes\fg b$. We denote the reference, stress-free intermediate, and current configurations of a body by $\Omega_0$, $\Omega_t$, and $\Omega$, respectively. The volumes from the reference and current configurations are denoted by $V_0$ and $V$, and their boundaries by $S_0$ and $S$, respectively.
The symbols $\nabla_0$ and $\nabla$ represent the gradient operators in  $\Omega_0$ and $\Omega$, respectively; $\nabla_0^2 := \nabla_0 \cdot \nabla_0 $ and $\nabla^2 := \nabla \cdot \nabla$ are the Laplacian operators in $\Omega_0$ and $\Omega$, respectively. The symbol $:=$ stands for equality by definition.

 \section{System of coupled phase field and mechanics equations}
 \label{system_eqns}
In this section, we present our novel multiphase PF approach to GB-induced MTs in polycrystals. The coupled mechanics and PF equations are derived, followed by an introduction to the order parameters.

 \subsection{ Order parameters}
 \label{order_params}
We introduce $N+1$  order parameters $\eta_0,\,\eta_1, \eta_2,\ldots,\eta_{N}$ related to the $\sf A$ and $\sf M$ phases, where 
 $\eta_0$ describes ${\sf A}\leftrightarrow{\sf M}$ PTs, and $\eta_1, \eta_2,\ldots,\eta_{N}$ describe $N$ martensitic variants. 
We consider $0\leq \eta_0\leq 1$ and  assign $\eta_0=0$ in $\sf A$ and $\eta_0=1$ in $\sf M$. For the order parameters relatd to the variants, we consider $0\leq \eta_i\leq 1$, and assign $\eta_i=1$ in ${\sf M}_i$ and $\eta_i=0$ in ${\sf M}_j$ for all $i,j=1,2,\ldots,N$ and $j\neq i$. We accept a particle as $\sf A$,  $\sf PM$, and $\sf M$ if $\eta_0$ therein lies within $0\leq \eta_0<0.05$, $0.05\leq \eta_0\leq 0.95$, and $0.95< \eta_0\leq 1$, respectively. The order parameters $\eta_1, \eta_2,\ldots,\eta_{N}$ are taken to be constrained to lie on the plane \cite{Basak-Levitas-2018JMPS}
\begin{equation}
\sum_{i=1}^{N} \eta_i-1=0,
   \label{constraint11}
 \end{equation}
 which is necessary for controlling the transformation paths between the variants; see \cite{Basak-Levitas-2018JMPS} for details. We introduce another $M$ order parameters $\xi_1,\,\xi_1,\,\ldots,\xi_{M}$,  which continuously vary between $0$ and $1$, and describe $M$  grains  in a polycrystal  such that $\xi_I=1$   in $G_I$  and $\xi_I=0$   in $G_J$ for  all   $I,J=1,2,\ldots,M$ and $J\neq I$. The denote the GB between grains $G_I$ and $G_J$ by $\Gamma_{IJ}$. We assume these order parameters also satisfy the following constraint for controlling the transformation path between the grains (see, e.g. \cite{Basak-Levitas-2018JMPS,Basak-2021-PCCP}):
\begin{equation}
\sum_{I=1}^{M} \xi_I-1=0.
   \label{constraint}
 \end{equation}
 The set of the order parameters related to the MTs is denoted as $\tilde{\eta}=\{\eta_0, \eta_1, \ldots , \eta_i,\ldots, \eta_N\}$, and
a subset for the $\sf M$ variants $\tilde{\eta}_M=\{\eta_1, \ldots , \eta_i,\ldots, \eta_N\}$ is introduced following \cite{Basak-Levitas-2018JMPS}.
Similarly, for the order parameters related to the grains, we introduce the set 
$\tilde{\xi}=\{\xi_1, \xi_2, \ldots , \xi_I,\ldots, \xi_M\}$. The set of the gradient of the order parameters in $\Omega$ are introduced as $\tilde{\eta}^\nabla=\{\nabla\eta_0,\nabla\eta_1,\ldots,\nabla\eta_i,\ldots, \nabla\eta_N\}$ and $\tilde{\xi}^\nabla=\{\nabla\xi_1, \nabla\xi_2, \ldots , \nabla\xi_I,\ldots, \nabla\xi_M\}$.

\subsection{Kinematics}
\label{kinmetcss}
The position vector of a particle in the deformed configuration $\Omega$ is given by $\fg r(\fg r_0,t)=\fg r_0+\fg u(\fg r_0,t)$, where $\fg r_0$, $\fg u$, and $t$ denote the particle's position vector in $\Omega_0$, the displacement vector, and time instance, respectively. We consider the following multiplicative decomposition for the total deformation gradient $\fg F:=\nabla_0\fg r$ \cite{Levitas-14a}:
  \begin{equation}
{\boldsymbol F} = \fg F_e\cdot\fg F_t=\fg V_e\cdot\fg R\cdot\fg U_t,
 \label{multdecom}
\end{equation}
where we have used the decompositions $\fg F_e=\fg V_e\cdot\fg R_e$ and $\fg F_t=\fg R_t\cdot \fg U_t$,  the subscripts $e$ and $t$ stand for the elastic and transformational parts, respectively, $\fg U_t$ is the symmetric right transformation stretch tensor, $\fg V_e$ is the symmetric left elastic stretch tensor, $\fg R_e$ and $\fg R_t$ are the elastic and transformational rotations, respectively, and $\fg R=\fg R_e\cdot\fg R_t$ is the resultant lattice rotation.  We designate $J=det\,\fg F $, $J_t=det\,\fg F_t=det\,\fg U_t $, and $J_e=det\,\fg F_e=det\,\fg V_e $. Hence $J=J_t J_e$ by Eq. (\ref{multdecom}). We will use the Lagrangian total and elastic strains defined as (see, e.g. Chapter 2 of \cite{Lubardabook})
\begin{equation}
 {\boldsymbol E} = \frac{1}{2}(\fg F^T\cdot\fg F-{\boldsymbol I}), \quad\text{and}\quad {\boldsymbol E}_e = \frac{1}{2}(\fg F_e^T\cdot\fg F_e-{\boldsymbol I}),
   \label{lagstrainsCG}
 \end{equation}
 and the spatial total and elastic strain tensors defined as (see e.g. Chapter 2 of \cite{Lubardabook})
 \begin{equation}
 {\boldsymbol b} = \frac{1}{2}(\fg F\cdot\fg F^T-{\boldsymbol I}), \quad\text{and}\quad {\boldsymbol b}_e = \frac{1}{2}(\fg F_e\cdot\fg F_e^T-{\boldsymbol I}).
   \label{strainsCG}
 \end{equation}

\subsubsection{Transformation stretch tensor}
\label{trsnstrtet}
Within each grain, we assume the transformation stretch tensor as a linear combination of the Bain tensor multiplied with the nonlinear interpolation functions related to $\eta_0$ and $\eta_i$ ($i=1,\ldots,N$) \cite{Basak-Levitas-2018JMPS}
\begin{equation}
 {\boldsymbol U}_t^{(I)} = {\boldsymbol F}_t^{(I)} =\fg I+ \varphi(a_\varepsilon,\eta_0)\sum_{i=1}^{N}(\fg U_{ ti}^{(I)}-\fg I)\,\phi_i \quad \text{for all } I=1,\ldots, M,
 \label{utildeIIp}
\end{equation}
where $\fg U_{ti}^{(I)}$ and $\fg\varepsilon_{ti}^{(I)}:=\fg U_{ti}^{(I)}-\fg I$ are the Bain stretch tensor and Bain strain tensor, respectively, for
$\mathsf M_i$ in the grain $G_I$ obtained using the relation
\begin{equation}
 {\boldsymbol U}_{ti}^{(I)} = \fg Q^{(I)} \cdot{\boldsymbol U}_{ti}\cdot \fg Q^{(I)T} \quad \text{for all } i=1,\ldots, N, \text{ and for all } I=1,\ldots, M.
 \label{utildeII}
\end{equation}
In Eq. \eqref{utildeIIp}, ${\boldsymbol U}_{ti}$ is the Bain stretch tensor for ${\sf M}_i$ with respect to a reference orthonormal basis attached to a $\sf A$ unit cell, and $ \fg Q^{(I)}$ is the orientation of the austenitic grain $G_I$ with respect to that reference basis. The interpolation functions $\varphi(a,\eta_0)$ and $\phi_i(\eta_i)$ are taken as  \cite{Basak-Levitas-2018JMPS}
\begin{equation}
\varphi(a, \eta_0) =a\eta_0^2+(4-2a)\eta_0^3 +(a-3)\eta_0^4,
\quad\text{and} \quad \phi_i = \eta_i^2(3-2\eta_i)=\varphi(3, \eta_i), \quad \text{for all }i=1,\ldots, N,
\label{interpolations}
\end{equation}
which satisfy the necessary conditions derived from the thermodynamic equilibrium of each of the homogeneous phases in each grain \cite{Basak-Levitas-2018JMPS} as follows
\begin{eqnarray}
&& \varphi(a, 0)=0,\quad \varphi(a, 1)=1, \quad \text{and}\quad \frac{\partial\varphi(a, 0)}{\partial\eta_0}=\frac{\partial\varphi(a, 1)}{\partial\eta_0}=0; \nonumber\\
&&\phi_i(0)=0,\quad \phi_i(1)=1, \quad \text{and}\quad \frac{\partial\phi_i(0)}{\partial\eta_i}=\frac{\partial\phi_i(1)}{\partial\eta_i}=0 , \quad \text{for all }i=1,\ldots, N.
\label{thermodyn_eqls}
\end{eqnarray}
The parameter $a$ in Eqs. \eqref{interpolations}$_1$ and \eqref{thermodyn_eqls}$_1$ is a constant lying in the range $0\leq a\leq 6$ \cite{Levitas-14a}.  We consider the following rule for the the transformation stretch tensor $\fg U_t$ introduced in Eq. \eqref{multdecom} for the polycrystal:
 \begin{equation}
 {\boldsymbol U}_t = {\boldsymbol F}_t = \fg I+ \sum_{I=1}^M  ({\boldsymbol U}_t^{(I)}-\fg I) \phi_I^\xi(\xi_I)=
 \fg I+\varphi(a_\varepsilon,\eta_0)\sum_{I=1}^M  \sum_{i=1}^{N}\fg \varepsilon_{ ti}^{(I)}\,\phi_i(\eta_i) \phi_I^\xi(\xi_I), \quad\text{where}
 \label{utilde}
\end{equation}
\begin{equation}
 \phi_I^\xi(\xi_I) = \xi_I^2(3-2\xi_I), 
\label{interpolationsxi}
\end{equation}
and it satisfies the following conditions in the corresponding grain
\begin{equation}
\phi_I^\xi(0)=0,\quad\phi_I^\xi(1)=1, \quad \text{and}\quad \frac{\partial\phi_I^\xi(0)}{\partial\xi_I}=\frac{\partial\phi_I^\xi(1)}{\partial\xi_I}=0 , \quad \text{for all }I=1,\ldots, M.
\label{interpolationsxidd}
\end{equation}

 \subsection{Free energy}
 The Helmholtz free energy per unit mass of the polycrystalline sample is considered as 
\begin{eqnarray}
\psi(\fg F,\fg E_e,\tilde\eta,\tilde\xi,\theta,\tilde\eta^\nabla,\tilde\xi^\nabla) = \frac{J_t}{\rho_0}\psi_e(\fg E_e,\tilde\eta, \tilde\xi,\theta)
+J\breve{\psi}^{\theta}(\tilde\eta,\theta) +\tilde{\psi}^\theta(\tilde\eta,\theta)+{\psi}^p(\tilde\eta)  + J\psi^\nabla(\tilde\eta^\nabla) +J\psi_\xi(\tilde\xi,\tilde\xi^\nabla,\eta_0),
\label{MF0}
\end{eqnarray}
 where $\psi_e$ is the strain energy per unit volume of $\Omega_t$ and it is considered here to have the standard quadratic form in $\fg E_e$,
 $\rho_0$ is the density of material in $\Omega_0$;
  $\breve\psi^\theta$ is the specific energy barrier related to ${\sf A}\leftrightarrow{\sf M}$ and ${\sf M}_i\leftrightarrow{\sf M}_j$
 transformations, $\tilde{\psi}^\theta $ is the specific thermal energy, $\psi^\nabla$ is the specific gradient energy accounting for the interfacial energies between $\sf A$-$\sf M$ and between the variants, $\psi^p$ is the specific energy for penalizing variant-variant transformation path deviations from the specified paths, and the $\psi^\xi$  is the total of the specific barrier energy between the grains and the energies of the GBs and TJs. The first five terms on the right-hand side of Eq. \eqref{MF0} were earlier used in \cite{Basak-Levitas-2018JMPS} for modeling MTs in single grains. The additional last term in Eq. \eqref{MF0} accounts for the excess free energy of the heterogeneities, including the GB and TJ regions  and the interactions between these heterogeneities and the MTs in the polycrystal. The expressions for all the energies are as follows  \cite{Levitas-14a,Basak-Levitas-2018JMPS}:
 \begin{eqnarray}
 \psi_e &=& \frac{1}{2} (\mathbb C: \fg E_e): \fg E_e,  \, \text{where }{\boldsymbol{\mathbb C}}(\tilde\eta,\tilde\xi) = \sum_{I=1}^M  \left[(1-\varphi(a,\eta_0)) {\boldsymbol{\mathbb C}}_0^{(I)}
 + \varphi(a,\eta_0)\sum_{i=1}^N \phi_i (\eta_i)  {\boldsymbol{\mathbb C}}_i^{(I)}\right] \phi_I^\xi(\xi_I);\nonumber\\ 
\breve\psi^\theta &=& [A_{0M}+(a_\theta-3)\Delta\psi^\theta(\theta)]\eta_0^2(1-\eta_0)^2+\varphi(a_b,\eta_0)\sum_{i=1}^{N-1}\sum_{j=i+1}^{N}{A}_{ij}^\eta\eta_i^2\eta_j^2 ; \nonumber\\
 \tilde{\psi}^\theta  &=&  {\psi^\theta}_0(\theta)  + \eta_0^2(3-2\eta_0) \Delta \psi^{\theta} (\theta); \nonumber\\
\psi^p &=& \sum_{i=1}^{N-1}\sum_{j=i+1}^N  K_{ij}^\eta( \eta_i+\eta_j-1)^2\eta_i^2\eta_j^2; \nonumber\\
 \psi^\nabla &=& \frac{1}{2\rho_0} \left[\beta_{0M}|\nabla\eta_0|^2- \tilde\varphi(a_\beta,a_0,\eta_0) \sum_{i=1}^{N-1}\sum_{j=i+1}^{N}\beta_{ij}^\eta\nabla\eta_i\cdot\nabla\eta_j \right], \quad \text{where } \beta_{ij}^\eta=\beta_{ji}^\eta, \,\,\text{and}\nonumber\\
 \tilde\varphi(a_\beta,a_c,\eta_0) &=& \varphi(a_\beta,\eta_0)+a_c = a_\beta\eta_0^2+(4-2a_\beta)\eta_0^3 +(a_\beta-3)\eta_0^4+a_c; \nonumber\\
  \psi^\xi &=& \psi^\xi_{loc} + \psi^\xi_{\nabla},  \qquad\text{where} \quad
 \psi^\xi_\nabla = -\sum_{I=1}^{M-1}\sum_{J=I+1}^{M}  \frac{\beta_{IJ}^{gb}(\eta_0)}{2\rho_0} \nabla\xi_I\cdot\nabla\xi_J , \quad \beta_{IJ}^{gb}=\beta_{JI}^{gb} \nonumber\\
 \psi^\xi_{loc}&=& \sum_{I=1}^{M-1}\sum_{J=I+1}^{M} \left[ A_{IJ}^{gb}(\eta_0)\xi_I^2\xi_J^2 +  K_{IJ}^{gb}( \xi_I+\xi_J-1)^2\xi_I^2\xi_J^2 \right]+ \sum_{H=1}^{M-2}\sum_{I=H+1}^{M-1} \sum_{J=I+1}^{M} K_{HIJ}\xi_H^2\xi_I^2\xi_J^2.
  \nonumber\\
\label{MF2}
\end{eqnarray}
The material parameters introduced in the energies (Eq. \eqref{MF2}) will now be defined. The fourth order tensor $\mathbb C$ is the elasticity tensor of the polycrystalline sample;  $\mathbb C_0^{(I)}$ and $\mathbb C_i^{(I)}$ are the elasticity tensors of $\sf A$ phase and the variant ${\sf M}_i$, respectively,  in grain $G_I$, and they should be obtained using e.g. (Chapter 1 of \cite{Jog-2015})
\begin{equation}
(\mathbb C_l^{(I)})_{abcd}= Q^{(I)}_{ae} Q^{(I)}_{bf}Q^{(I)}_{cg}Q^{(I)}_{dh}(\mathbb C_l)_{efgh} \qquad \text{for } l=0,1,\ldots,N,
\label{propertstiff}
\end{equation}
where $\mathbb C_l$ is the elasticity tensor of the $\sf A$ ($l=0$) and $\sf M_i$ ($l=i=1,\ldots,N$) with respect to the reference basis; 
$A_{0M}$ and ${A}_{ij}^\eta$ are the coefficients for barrier energies between ${\sf A}$-${\sf M}$ and  ${\sf M}_i$-${\sf M}_j$, respectively;  $A_{IJ}^{gb}$ is the barrier energy coefficient between the grains $G_I$ and $G_J$. Assuming the specific heat at constant pressure to be identical for both $\sf A$ and $\sf M$, we consider the thermal energy difference between these two phases as
$\Delta \psi^{\theta}=\psi^\theta_M-\psi^\theta_0=-\Delta s\,(\theta-\theta_e)$, where $\Delta s$ is the change in entropy due to ${\sf A}$ to ${\sf M}$ transformation;  $\theta_e$ is the thermodynamic equilibrium temperature between $\sf A$ and $\sf M$. The coefficient $K_{ij}^\eta$ is a positive constant  which determines the deviation of the path for ${\sf M}_i\leftrightarrow {\sf M}_i$ transformations from the lines $\eta_i+\eta_j=1$,  $\eta_i=0$, and  $\eta_j=0$ for all $i,j=1,\ldots, N$ and $i\neq j$ (see \cite{Basak-Levitas-2018JMPS} for a detailed explanation); $K_{IJ}^{gb}$ has a similar meaning as $K_{ij}^\eta$ has but is related to the transformations between the grains $G_I$ and $G_J$. The constant coefficient $K_{HIJ}$ is used for penalizing the TJs between the grains $G_H$, $G_I$, and $G_J$, i.e. it allows  control of the energy and size of the TJs region (also see \cite{Garcke1999,Basak-Levitas-2018JMPS,Basak-2021-PCCP}). The gradient energies considered in Eq. \eqref{MF2} are similar to that of \cite{Steinbach-96}; see \cite{Toth_et_al-2015} for a comparative study for various gradient energies.
The parameters $\beta_{0M}$, $\beta_{ij}^\eta$, and $\beta_{IJ}^{gb}(\eta_0)$ are the gradient energy coefficients for the ${\sf A}$-${\sf M}$, ${\sf M}_i$-${\sf M}_j$, and $\Gamma_{IJ}$ interfaces, respectively.  The constant parameters $a_\theta, a_b,a_\beta$ lie within the range $0\leq a_\theta, a_b,a_\beta \leq 6$ \cite{Levitas-14a,Basak-Levitas-2018JMPS}, and the constant $a_c$ in Eq. \eqref{MF2}$_{5,6}$ is taken as  $0<a_c\ll 1$; see \cite{Basak-Levitas-2018JMPS} for introducing $a_c$.
The energies $\breve{\psi}^{\theta}$, $ \psi^\nabla$, and $ \psi^\xi$ are multiplied with $J=det\,\fg F$, and the gradients of all the order parameters are taken in the current configuration $\Omega$ in Eq. (\ref{MF0})  to obtain the desired expression for the structural stresses given by Eqs. \eqref{hl4fjgffg},  and \eqref{hl5f34} (see \cite{Levitas-14a,Basak-Levitas-2018JMPS} for further discussion on the structural stresses). The material parameters, which are dependent on the grain orientations, at any point in the sample are determined using the following interpolation
\begin{equation}
B(\tilde\eta,\tilde\xi,\theta,{\boldsymbol F})=\sum_{I=1}^M\left[ (1-\varphi(a,\eta_0))B_{0}^{(I)}+\varphi(a,\eta_0)\sum_{i=1}^{N}\phi_iB_i^{(I)}\right]\phi_I^\xi(\xi_I),
\label{properties}
\end{equation}
where $B_0^{(I)}$ and $B_i^{(I)}$ are the material properties of $\mathsf A$ and $\mathsf M_i$, respectively, in $G_I$, and $\varphi(a,\eta_0)$, $\phi_i(\eta_i)$, and $\phi_I^\xi(\xi_I)$ are given by  Eqs. \eqref{interpolations} and \eqref{interpolationsxi}. 

\subsubsection{GB energy}
\label{GBbsseng}
The energy of the GBs $\Gamma_{IJ}$ appeared in Eqs. \eqref{MF2}$_{7,8,9}$ is assumed to be isotropic, i.e. it is a function of GB misorientations and independent of the GB inclination (see, e.g. Chapter 1 of \cite{Sutton-Balluffi}), and it is taken as   
 \begin{equation}
 \gamma_{IJ}^{gb}(\Theta_{IJ},\eta_0) = \gamma_{IJ}^A(\Theta_{IJ}) + \Delta\gamma_{IJ}^{gb}(\Theta_{IJ})\, \varphi(a,\eta_0),
\label{propertiesf}
\end{equation}
where $\Theta_{IJ}$ is the list of three misorientation angles associated with $\Gamma_{IJ}$,
$\Delta\gamma_{IJ}^{gb}=\gamma_{IJ}^M-\gamma_{IJ}^A$ is the change in GB energy due to MT, $\varphi(a,\eta_0)$ is given by Eq. \eqref{interpolations}$_1$, and the parameters $\gamma_{IJ}^A$ and $\gamma_{IJ}^B$ are energies of GB $\Gamma_{IJ}$ in $\sf A$ and $\sf M$ phases, respectively. From Eq. \eqref{propertiesf}, $ \gamma_{IJ}^{gb} = \gamma_{IJ}^A$ when $\eta_0=0$ (in $\sf A$), and $ \gamma_{IJ}^{gb} = \gamma_{IJ}^M$ when $\eta_0=1$ (in $\sf M$). The energy of the heterogeneities similar to Eq. \eqref{propertiesf} was  earlier used to study the surface-induced solid$\leftrightarrow$melt PTs in nanoparticles  \cite{Levitas-Samani-PRB-14} and  nanovoids \cite{Basak-Levitas-2018-nanovoid}, surface-induced MTs \cite{levitas-javanbakht-PRL-11}), matrix-precipitate interface-induced MTs \cite{BasakLevitas2020Acta-2004}, and GB-induced solid$\leftrightarrow$melt PTs in polycrystals \cite{Basak-2021-PCCP}.


\subsection{Dissipation rate, coupled mechanics  and Ginzburg-Landau equations}
\label{dississipat}
We start with the following local form of the dissipation inequality, whose detailed derivation is given in Appendix \ref{psidot}:
\begin{equation}
\rho_0{\mathcal D}= \fg P_d : \dot{\fg F}^T+X_0\dot\eta_0+\sum_{i=1}^NX_i^\eta\dot\eta_i +\sum_{I=1}^M X_I^\xi\dot\xi_I  \geq 0 \quad\text{in }\Omega_0,
\label{dissiIneq1}
\end{equation}
where ${\mathcal D}$ is the total power dissipation per unit mass, $\fg P_d$ is the dissipative or viscous first Piola-Kirchhoff stress tensor (see  Eq. \eqref{ptotaal}), the overdot denotes material time derivative, and $X_l^\eta$ and $X_I^\xi$ are the conjugate thermodynamic forces (see Eq. \eqref{dissiIneq2}) for the rates of the order parameters $\dot\eta_l$ (for $l=0,\ldots,N$) and $\dot\xi_l$ (for $I=1,\ldots,M$), respectively. The exact expressions for $\fg P_d$, $X_l^\eta$, and $X_I^\xi$ are given in Secs. \ref{stressessubsec}  and \ref{GLsubsec}.
We simplify the present theory by assuming that the power dissipation due to the dissipative stress, the evolution of the phases, and evolution of the grains are independent, and hence we decouple the inequality \eqref{dissiIneq1} into the following three, which respect the original inequality \eqref{dissiIneq1}:   
\begin{equation}
\rho_0{\mathcal D}_d=\fg P_d:\dot{\fg F}^T\geq 0;  \qquad  \rho_0{\mathcal D}_\eta= X^\eta_0\dot\eta_0+\sum_{i=1}^N X_i^\eta\dot\eta_i \geq 0;
\qquad\text{and}\qquad \rho_0{\mathcal D}_\xi= \sum_{I=1}^M X_I^\xi \dot\xi_I \geq 0; 
\quad\text{in }\Omega_0.
\label{dissiIneqdec}
\end{equation}
\subsubsection{ Stresses and mechanical equilibrium}
\label{stressessubsec} 
The dissipative first Piola-Kirchhoff stress tensor $ {\boldsymbol P}_d$ in the inequality \eqref{dissiIneqdec}$_1$ is defined as
\begin{equation}
 {\boldsymbol P}_d:= \fg P-{\boldsymbol P}_e-\fg P_{st},
 \label{ptotaal}
\end{equation}
where $\fg P$ is the total Piola-Kirchhoff stress tensor, and $\fg P_e$ and $\fg P_{st}$ are the elastic and structural parts of $\fg P$, respectively, and they are given by
\begin{eqnarray}
   {\boldsymbol P}_e &=&  J_t\frac{\partial\psi_e(\fg F_e)}{\partial\fg F_e}\cdot\fg F_t^{-T}, \,\,\text{and}\,\,
  {\boldsymbol P}_{st}=  {\boldsymbol P}_{st}^\eta+{\boldsymbol P}_{st}^\xi, \quad\text{where} \nonumber\\
  {\boldsymbol P}_{st}^\eta &=& J\rho_0(\breve{\psi}^{\theta}+\psi^\nabla){\fg F}^{-T}
 -J\rho_0 \sum_{l=0}^N \left( \nabla\eta_l \otimes\frac{\partial\psi^\nabla}{\partial\nabla\eta_l}\right)\cdot\fg F^{-T};  \nonumber\\
  {\boldsymbol P}_{st}^\xi &=&J\rho_0\psi^\xi{\fg F}^{-T}
 -J\rho_0  \sum_{I=1}^M  \left(\nabla\xi_I \otimes\frac{\partial\psi^\xi_\nabla}{\partial\nabla\xi_I}\right)\cdot\fg F^{-T}.
 \label{hl4fjgffg}
\end{eqnarray}
Considering the dissipative stress to be negligible in the present theory (also see \cite{Basak-Levitas-2018JMPS}), we express  the total Piola-Kirchhoff stress tensor using Eq. \eqref{ptotaal} as
\begin{equation}
   \fg P = {\boldsymbol P}_e+ {\boldsymbol P}_{st}.
 \label{hl4fjgffg1stpk}
\end{equation}
Recalling that the density of material in $\Omega$ is given by $\rho=J^{-1}\rho_0$, and  the relation between the first Piola-Kirchhoff and Cauchy stress tensors is (see Chapter 3 of \cite{Jog-2015})
\begin{equation}
{\boldsymbol\sigma}= J^{-1}  {\boldsymbol P}\cdot {\boldsymbol F}^T,
 \label{PKCauStresses}
\end{equation}
 we get the total, elastic, and structural Cauchy stresses as (recall that $\fg\sigma_d$ is neglected)
\begin{eqnarray}
&& \fg\sigma=\fg\sigma_e+\fg\sigma_{st}; \qquad   \fg\sigma_e=J^{-1}_e \frac{\partial\psi_e(\fg F_e)}{\partial {\boldsymbol F}_e}\cdot{\boldsymbol F}_e^T
=J_t\,{\boldsymbol F}_e\cdot\frac{\partial \psi_e(\fg E_e)}{\partial{\boldsymbol  E}_e }\cdot{\boldsymbol F}_e^T;  \qquad
 \fg\sigma_{st}=  \fg\sigma_{st}^\eta+\fg\sigma_{st}^\xi, \quad\text{where} \nonumber\\
 && \fg\sigma_{st}^\eta=J\rho(\breve{\psi}^{\theta}+\psi^\nabla)\fg I-J\rho\ \sum_{l=0}^N \nabla\eta_i\otimes
 \frac{\partial\psi^\nabla}{\partial\nabla\eta_i}, \qquad  \fg\sigma_{st}^\xi =J\rho\psi^\xi{\fg I}
 -J\rho  \sum_{I=1}^M  \nabla\xi_I \otimes\frac{\partial\psi^\xi_\nabla}{\partial\nabla\xi_I}.
\label{hl5f34}
\end{eqnarray}
Using Eqs. \eqref{MF0} and \eqref{MF2} in Eqs. \eqref{hl5f34}$_{1,2,3,4,5}$ the structural stresses can be finally expressed as
\begin{eqnarray}
  {\boldsymbol P}_{st}^\eta &=& J\rho_0(\breve{\psi}^{\theta}+\psi^\nabla){\fg F}^{-T} -J \beta_{0M}\left(\nabla\eta_0 \otimes\nabla\eta_0\right)\cdot\fg F^{-T}
 +\frac{J\tilde\varphi(a_\beta,a_c,\eta_0)}{2} \sum_{i=1}^{N} \sum_{j=1,\neq i}^N \beta^\eta_{ij} \left( \nabla\eta_i \otimes \nabla\eta_j\right)\cdot\fg F^{-T}; \nonumber \\
 \\
  {\boldsymbol P}_{st}^\xi &=&J\rho_0\psi^\xi{\fg F}^{-T}
  +\frac{J}{2} \sum_{H=1}^{M} \sum_{I=1,\neq H}^M  \beta^{gb}_{HI}(\eta_0)\left( \nabla\eta_H \otimes\nabla\eta_I\right)\cdot\fg F^{-T}; 
\label{hl5f34fg}
\end{eqnarray}
\begin{eqnarray}
   \fg\sigma_{st}^\eta &=&J \rho(\breve{\psi}^{\theta}+\psi^\nabla)\fg I -\beta_{0M}\nabla\eta_0 \otimes\nabla\eta_0
 +\frac{\tilde\varphi(a_\beta,a_c,\eta_0)}{2} \sum_{i=1}^{N} \sum_{j=1,\neq i}^N  \beta^\eta_{ij} \nabla\eta_i \otimes\nabla\eta_j;  \,\,\text{and}\nonumber\\
  \fg\sigma_{st}^\xi &=&J\rho\psi^\xi\fg I
  + \sum_{H=1}^{M} \sum_{I=1,\neq H}^M \frac{ \beta^{gb}_{HI}(\eta_0)}{2} \nabla\eta_H \otimes\nabla\eta_I.
\label{hl5f34fge}
\end{eqnarray}
For isotropic elastic response, the Cauchy elastic stress tensor can alternatively be expressed as \cite{Levitas-14a}
\begin{equation}
{\boldsymbol\sigma}_e= J_e^{-1} \fg V_e^2\cdot \frac{\partial\psi_e(\fg b_e)}{\partial \fg b_e}.
 \label{elasticCauStresses}
\end{equation}
The total stresses satisfy the mechanical equilibrium equations, which are obtained using the balance of linear momentum, when the inertia and body forces are neglected,  as \cite{Levitas-14a}
\begin{equation}
\nabla_0\cdot\fg P=\fg 0 \quad\text{in }\Omega_0, \quad\text{or equivalently,}
\quad \nabla\cdot\fg\sigma=\fg 0 \quad\text{in }\Omega.
 \label{lin_momentum}
 \end{equation}
The boundary conditions (BCs) on the displacement boundary $S_{0u}\subset S_0$ and on the traction boundary $S_{0t}\subset S_0$ are  given by 
\begin{equation}
\fg u=\fg u^{sp} \quad \text{on } S_{0u}, \quad \text{and} \quad \fg p^{sp}=\fg P\cdot\fg n_0 \quad \text{on } S_{0t},
 \label{lin_momentumtt}
 \end{equation}
  where $ S_0=S_{0u} \cup S_{0t}$, $\fg n_0$ is the outward unit normal to $S_0$, and $\fg p^{sp}$ is the traction specified.
 The exact displacement and traction BCs for the numerical calculations will be specified while presenting the results in Sec. \ref{simulations}.
 
 \subsubsection{Thermodynamic forces and Ginzburg-Landau equations}
 \label{GLsubsec}
\noindent{\em Thermodynamic forces:}  The  thermodynamic forces $X_0^\eta$, $X_i^\eta$, and $X_I^\xi$ appearing in inequalities \eqref{dissiIneqdec}$_{2,3}$ are given by
\begin{eqnarray}
X_0^\eta &=& \left(\fg P_e^T\cdot\fg F_e-J_t\psi_e\fg F_t^{-1}\right):\frac{\partial\fg F_t}{\partial\eta_0}-
J_t\left.\frac{\partial\psi_e}{\partial\eta_0}\right|_{\fg F_e}-
\rho_0J\frac{\partial (\breve\psi^\theta+\psi^\nabla+\psi^\xi)}{\partial\eta_0}-
\rho_0 \frac{\partial(\tilde\psi^\theta+\psi^p)}{\partial\eta_0}+ \nonumber\\
&& \nabla_0\cdot\left(\rho_0 J\fg F^{-1}\cdot\frac{\partial\psi^\nabla}{\partial \nabla\eta_0}\right); \nonumber\\
X_i^\eta &=& \left(\fg P_e^T\cdot\fg F_e-J_t\psi_e\fg F_t^{-1}\right):\frac{\partial\fg F_t}{\partial\eta_i}-
J_t\left.\frac{\partial\psi_e}{\partial\eta_i}\right|_{\fg F_e}-
\rho_0J\frac{\partial (\breve\psi^\theta+\psi^\nabla)}{\partial\eta_i}-
\rho_0 \frac{\partial(\tilde\psi^\theta+\psi^p)}{\partial\eta_i}
+ \nonumber\\
&& \nabla_0\cdot\left(\rho_0 J\fg F^{-1}\cdot \frac{\partial\psi^\nabla}{\partial \nabla\eta_i}\right) \quad \text{for } i=1,2, \ldots, N; \nonumber\\
X_I^\xi &=& \left(\fg P_e^T\cdot\fg F_e-J_t\psi_e\fg F_t^{-1}\right):\frac{\partial\fg F_t}{\partial\xi_I}-
J_t\left.\frac{\partial\psi_e}{\partial\xi_I}\right|_{\fg F_e}-
\rho_0J\frac{\partial \psi_{loc}^\xi}{\partial\xi_I}
+\nabla_0\cdot\left(\rho_0 J\fg F^{-1}\cdot \frac{\partial\psi^\xi_\nabla}{\partial \nabla\xi_I}\right), \, \text{for } I=1, \ldots, M.
\label{dissiIneq2}
\end{eqnarray}
The thermodynamic forces given in Eq. \eqref{dissiIneq2}$_{1,2,3}$ can also be expressed as functions of the Cauchy stresses and the gradients in the current configuration as 
\begin{eqnarray}
X_0^\eta &=& \left(J\fg F^{-1}\cdot\fg \sigma_e\cdot\fg F-J_t\psi_e\fg I\right):\fg F_t^{-1}\cdot\frac{\partial\fg F_t}{\partial\eta_0}-
J_t\left.\frac{\partial\psi_e}{\partial\eta_0}\right|_{\fg F_e}-
\rho_0J\frac{\partial (\breve\psi^\theta+\psi^\nabla+\psi^\xi)}{\partial\eta_0}-
\rho_0 \frac{\partial(\tilde\psi^\theta+\psi^p)}{\partial\eta_0}+\nonumber\\
&& J\nabla\cdot\left( \rho_0\frac{\partial\psi^\nabla}{\partial \nabla\eta_0}\right); \nonumber\\
X_i^\eta &=& \left(J\fg F^{-1}\cdot\fg \sigma_e\cdot\fg F-J_t\psi_e\fg I\right):\fg F_t^{-1}\cdot\frac{\partial\fg F_t}{\partial\eta_i}-
J_t\left.\frac{\partial\psi_e}{\partial\eta_i}\right|_{\fg F_e}-
\rho_0J\frac{\partial (\breve\psi^\theta+\psi^\nabla)}{\partial\eta_i}-
\rho_0 \frac{\partial(\tilde\psi^\theta+\psi^p)}{\partial\eta_i}+  \nonumber\\
&& J\nabla\cdot\left( \rho_0\frac{\partial\psi^\nabla}{\partial \nabla\eta_i}\right), \quad\text{for } i=1, \ldots, N; \nonumber\\
X_I^\xi &=& \left(J\fg F^{-1}\cdot\fg \sigma_e\cdot\fg F-J_t\psi_e\fg I\right):\fg F_t^{-1}\cdot\frac{\partial\fg F_t}{\partial\xi_I}-
J_t\left.\frac{\partial\psi_e}{\partial\xi_I}\right|_{\fg F_e}-
\rho_0J\frac{\partial \psi_{loc}^\xi}{\partial\xi_I}
+J\nabla\cdot\left(\rho_0 \frac{\partial\psi^\xi_\nabla}{\partial \nabla\xi_I}\right), \quad \text{for } I=1, \ldots, M,
\label{dissiIneq2c}
\end{eqnarray}
where Eq. \eqref{PKCauStresses}, and the relations $\frac{\partial\psi^\nabla}{\partial \nabla_0\eta_l}= \fg F^{-1}\cdot\frac{\partial\psi^\nabla}{\partial \nabla\eta_l}$ \cite{Levitas-14a},  and $\displaystyle\nabla_0\cdot(J\fg F^{-T})=\fg 0$ (see Sec. 2.2.2 of \cite{Jog-2015} for the proof) have been used.  The thermodynamic forces given in Eqs. \eqref{dissiIneq2} and  \eqref{dissiIneq2c} can also be compactly expressed  as 
\begin{eqnarray}
X_l^\eta &=&  -\rho_0\frac{\partial\psi}{\partial\eta_l} +\nabla_0\cdot\left(\rho_0 J
\frac{\partial\psi^\nabla}{\partial \nabla_0\eta_l}\right) = -J \rho\frac{\partial\psi}{\partial\eta_l} + J\nabla\cdot\left(\rho J
\frac{\partial\psi^\nabla}{\partial \nabla\eta_l}\right) \quad\text{for } l=0,1,2, \ldots, N; \nonumber\\
X_I^\xi &=&  -\rho_0\frac{\partial\psi}{\partial\xi_I} +\nabla_0\cdot\left(\rho_0 J
\frac{\partial\psi^\xi_\nabla}{\partial \nabla_0\xi_I}\right) = -J \rho\frac{\partial\psi}{\partial\xi_I} + J\nabla\cdot\left(\rho J
\frac{\partial\psi^\xi_\nabla}{\partial \nabla\xi_I}\right)  \quad\text{for } I=1,2, \ldots, M.
\label{dissiIneq2com}
\end{eqnarray}
%

\vspace{2mm} 
\noindent{\em Ginzburg-Landau equations:} In deriving the Ginzburg-Landau equations, we further simplify our theory by decoupling the inequality \eqref{dissiIneqdec}$_2$  into the following sufficient conditions:
\begin{equation}
\rho_0 D_{\eta 0} =  X^\eta_0\dot\eta_0 \geq 0, \qquad \text{ and}\qquad \rho_0 {\mathcal D}_{\eta i}=\sum_{i=1}^N X_i^\eta\dot\eta_i \geq 0.
 \label{dissiineqfurth}
 \end{equation}
 Using the inequality \eqref{dissiineqfurth}$_1$, we derive
\begin{equation}
\dot\eta_0 = L_{0M} X_0^\eta,
 \label{dissiineq}
\end{equation}
where $ L_{0M}\geq 0$ is the kinetic coefficient for $\sf A\leftrightarrow\sf M$ transformations. Following \cite{Basak-Levitas-2018JMPS,BasakLevitasCMT2022} we introduce the rates of the order parameters as
\begin{eqnarray}
\dot\eta_i &=& \sum_{j=1}^N\dot\eta_{ij}, \qquad\text{where }\dot\eta_{ij}=-\dot\eta_{ji} \text{ and } \dot\eta_{ii}=0 \quad \text{(no sum) for all } i,j=1,\ldots,N; \nonumber\\
\dot\xi_I &=& \sum_{J=1}^M\dot\xi_{IJ}, \qquad\text{where }\dot\xi_{IJ}=-\dot\xi_{JI} \text{ and } \dot\xi_{II}=0 \quad \text{(no sum) for all } I,J=1,\ldots,M
 \label{vari_toj}
\end{eqnarray}
for driving the kinetic equations for the order parameters $\eta_i$ ($i=1,\ldots,N$) and $\xi_I$ ($I=1,\ldots,M$). 
 Using Eq. \eqref{vari_toj} into the dissipation inequalities \eqref{dissiineqfurth}$_2$ and \eqref{dissiIneqdec}$_3$ we can show that the corresponding dissipations can be rewritten as (see \cite{Basak-Levitas-2018JMPS} for the proof)
\begin{eqnarray}
  \rho_0 {\mathcal D}_{\eta i}  = \sum_{j=1}^{N-1}\sum_{i=j+1}^N X_{ij}^\eta\dot\eta_{ij} \geq 0, \quad \text{and}\quad 
\rho_0 {\mathcal D}_{\xi}  = \sum_{J=1}^{M-1}\sum_{I=J+1}^M X_{IJ}^\xi\dot\xi_{IJ} \geq 0,
 \label{dissiinedd1}
\end{eqnarray}
where $X_{ij}^\eta= X_{i}^\eta-X_{j}^\eta$ and $X_{IJ}^\xi= X_{I}^\xi-X_{J}^\xi$. 
Using the inequality \eqref{dissiinedd1}$_1$ we derive the kinetic equations for the evolution of the variants as
\begin{eqnarray}
\dot\eta_{ij} = L_{ij}^\eta(X_i^\eta-X_j^\eta) ,
\label{kiness}
\end{eqnarray}
where $L_{ij}^\eta \geq 0$ is the kinetic coefficient for ${\sf M}_i\leftrightarrow{\sf M}_j$ transformations, and $L_{ij}^\eta =L_{ji}^\eta$ by the Onsager's reciprocity theorem \cite{Onsager-1931}. Following \cite{Idesman-Levitas-JMPS-05} we consider the coefficient  as   
\begin{eqnarray}
    L_{ij}^\eta  \left\{\begin{array}{@{}lr@{}}
        \neq 0 & \text{if }(X_i^\eta-X_j^\eta)\geq 0 \quad \text{and} \quad \{0\leq \eta_i<1 \,\,\&\,\, 0<\eta_j\leq 1\}\\
                 \neq 0               & \text{if }(X_i^\eta-X_j^\eta)\leq 0 \quad \text{and} \quad \{0< \eta_i\leq 1 \,\,\&\,\, 0\leq \eta_j <1\} \\
        =0 & \text{if }(X_i^\eta-X_j^\eta)\geq 0 \quad \text{and} \quad \{\eta_i=1 \,\,\text{or}\,\, \eta_j=0\}\\
       =0 & \text{if }(X_i^\eta-X_j^\eta)\leq 0 \quad \text{and} \quad \{ \eta_i=0 \,\,\text{or}\,\,  \eta_j=1\};
        \end{array}\right.
\label{kinesss}
\end{eqnarray}
 also see \cite{BasakLevitasCMT2022}. Substituting Eq. \eqref{kiness} into Eq. \eqref{vari_toj}$_1$, the Ginzburg-Landau equations for all $N$ order parameters $\eta_1,\ldots,\eta_N$ are obtained as
\begin{eqnarray}
 \dot\eta_i = \sum_{j=1, j\neq i}^NL_{ij}^\eta(X_i^\eta-X_j^\eta) \qquad \text{for } i=1,2,\ldots, N.
\label{kinessg}
\end{eqnarray}
The kinetic coefficient, for example, $L_{ij}^\eta$ in Eq. \eqref{kinessg}, is considered as a piecewise constant function of the driving force $X_i^\eta-X_j^\eta$ given by Eq. \eqref{kinesss} rather than considering it as a constant similar to that of \cite{Basak-Levitas-2018JMPS}. We now show that considering $L_{ij}^\eta$ as constant leads to contradictions. Without the loss of generality, let us assume a martensitic region ($\eta_0=1$) where three variants (denoted by $\sf M_1$, $\sf M_2$, and $\sf M_3$) are evolving. The order parameters for the variants are  $\eta_1$, $\eta_2$, and $\eta_3$, which are constrained by $\eta_1+\eta_2+\eta_3=1$ as per  Eq. \eqref{constraint11}. We can treat $\eta_1$, and $\eta_2$ as two independent order parameters, and the corresponding Ginzburg-Landau equations using from Eq. \eqref{kinessg} are written as
\begin{eqnarray}
\dot\eta_1 =L_{12}^\eta(X_1^\eta-X_2^\eta)+L_{13}^\eta(X_1^\eta-X_3^\eta), \quad \text{and} \quad \dot\eta_2 =L_{12}^\eta(X_2^\eta-X_1^\eta)+L_{23}^\eta(X_2^\eta-X_3^\eta),
\label{kinessgp}
\end{eqnarray}
and $\dot\eta_3=-\dot\eta_1-\dot\eta_2$. We now consider a subregion within the $\sf M$ region where only the variants  $\sf M_1$ and $\sf M_2$ are coexisting, and $\sf M_3$ is absent. The order parameters $\eta_1$ and $\eta_2$ must be determined using the Ginzburg-Landau equation $\dot\eta_1 =L_{12}^\eta(X_1^\eta-X_2^\eta)$ and the constraint $\eta_1+\eta_2=1$, which are, however, possible to achieve from Eqs. \eqref{kinessgp}$_1$ and \eqref{kinessgp}$_2$ if and only if $L_{13}^\eta=L_{23}^\eta=0$ in that subregion. Such conditions clearly cannot be achieved with a constant $L_{ij}^\eta$, but when $L_{ij}^\eta$ is a piecewise function, for example, given by Eq. \eqref{kinesss}. The last two conditions in Eq. \eqref{kinesss} say that it cannot be transformed to any other variant when, say, $\sf M_i$ is absent. 

Similarly, using the dissipation inequality \eqref{dissiIneqdec}$_3$, we derive the Ginzburg-Landau equations for $\xi_1,\ldots,\xi_M$ as
\begin{eqnarray}
 \dot\xi_I = \sum_{J=1, J\neq I}^M L_{IJ}^\xi (X_I^\xi-X_J^\xi)  \qquad \text{for } I=1,2,\ldots, M,
\label{kinessgxi}
\end{eqnarray} 
where the kinetic coefficients $L_{IJ}^\xi=L_{JI}^\xi\geq 0$ for transformations between the grains $G_I$ and $G_J$ is taken similar to that for $L_{ij}^\eta$ given by Eq. \eqref{kinesss}, where obviously we read $X_i^\eta$ and $X_j^\eta$  as $X_I^\xi$ and $X_J^\xi$, respectively.

\vspace{2mm} 
\noindent{\em Boundary conditions:} Assuming that the external surface energy of the sample (say, $V_0$)  remains unchanged during the MTs, we consider the following Neumann BC for $\eta_0,\eta_1,\ldots,\eta_N$ expresses on the both undeformed and deformed surfaces of the body (see \cite{Levitas-14a} for derivation):
\begin{eqnarray}
 \rho_0\frac{\partial\psi}{\partial\nabla_0\eta_l}\cdot \fg n_0 &=&0 \quad\text{on }S_0 \quad \text{for }l=0,1,\ldots, N;\nonumber\\
 \rho\frac{\partial\psi}{\partial\nabla\eta_l}\cdot \fg n &=&0 \quad\text{on }S\quad \text{for }l=0,1,\ldots, N;
  \label{gleqsbc}
 \end{eqnarray}
 where $\fg n$ is the unit outward normal to the surface in the deformed body. Similarly, assuming the surface energy does not change during the evolution of the GBs, the Neumann BC for $\xi_1,\ldots,\xi_M$ is expressed as   
 \begin{eqnarray}
 \rho_0\frac{\partial\psi}{\partial\nabla_0\xi_I}\cdot \fg n_0 &=&0 \quad\text{on }S_0 \quad \text{for }I=1,\ldots, M;\nonumber\\
 \rho\frac{\partial\psi}{\partial\nabla\xi_I}\cdot \fg n &=&0 \quad\text{on }S \quad \text{for }I=1,\ldots, M.
  \label{gleqsbcxi}
 \end{eqnarray}
 The initial conditions (ICs) on the order parameters for numerical calculations will be specified while discussing the results in Sec. \ref{simulations}.

The complete system of coupled mechanics and PF equations derived in this section is collected and enlisted in Appendix \ref{Listequns2d}. 

\section{Material parameters identification}
\label{MaterParam}  
We now calibrate the material parameters involved in the kinematic model for the transformation stretch,  free energy, GB energy, Ginzburg-Landau equations (see Eqs. \eqref{utilde}, \eqref{MF2}, \eqref{propertiesf}, \eqref{dissiineq}, \eqref{kinesss}, and \eqref{kinessgxi}) using the atomistic simulation and experimental data from the literature. Following \cite{Levitas-14a,Basak-Levitas-2018JMPS}, we simplify the Ginzburg-Landau equation  \eqref{dissiineq} for $\sf A\leftrightarrow\sf M$ transformations in a single grain (say, $\xi_1=1$ and $\xi_I=0$ for $I=2,\ldots,M$) for a system with $\sf A$ and a single $\sf M$ variant and a planar interface between them as (neglecting mechanics)
\begin{eqnarray}
\dot\eta_0= L_{0M} \left[  -\rho_0\Delta\psi^\theta(6\eta_0-6\eta_0^2)-\rho_0A_{0M} (2\eta_0-6\eta_0^2+4\eta_0^3)+
 \beta_{0M}\frac{\partial^2\eta_0}{\partial r_{02}^2}\right], 
 \label{kineqsbx1sol0}
\end{eqnarray}
where the parameters $A_{0M}$,  $\beta_{0M}$, and $L_{0M}$ are assumed to be constants. The solution to Eq. \eqref{kineqsbx1sol0}  is \cite{Steinbach-09,Levitasetal-PRB-15} 
\begin{eqnarray}
 \eta_0(r_{02},t) &=& 0.5+0.5\tanh [3(r_{02}-r_{0c}-c_{0M}  t)/\delta_{0M}],  \quad \text{where}  \nonumber\\
  \delta_{0M} &=& \sqrt{\frac{18\beta_{0M}}{\rho_0A_{0M}}},  \quad \gamma_{0M}=\frac{\beta_{0M}}{\delta_{0M}}, \quad\text{and}\quad  c_{0M} =L_{0M}\delta_{0M}\rho_0\Delta\psi^\theta(\theta),
 \label{analyticeta1}
\end{eqnarray}
are the width, energy, and speed, respectively, of the $\sf A$-$\sf M$  interface, and $r_{0c}$ is the position of the centre of the interface where $\eta_0=0.5$. Note that the model parameters $A_{0M}$,  $\beta_{0M}$, and $L_{0M}$ can be determined using the directly measurable quantities (either experimentally or by atomistic simulations) $\delta_{0M}$, $\gamma_{0M}$, $c_{0M}$, and $\Delta\psi^\theta(\theta)$ from Eqs. \eqref{analyticeta1}$_{2,3,4}$. 

In a similar manner, we can calibrate the parameters related to the ${\sf M}_i\leftrightarrow{\sf M}_j$ transformations (for all $i,j=1,\ldots, N; \, i\neq j$) and $G_I\leftrightarrow G_J$ transformations (for all  $I,J=1,\ldots, M; \, I\neq J$). We consider the equations for ${\sf M}_i\leftrightarrow{\sf M}_j$ transformations in a single grain, and $G_I$-$G_J$ transformations in a stress-free austenitic bicrystal ($\eta_0=0$), which are given by (obtained using Eqs. \eqref{bicrystgovt}$_2$ and \eqref{bicrystgovt}$_3$, respectively, and neglecting mechanics)
\begin{eqnarray}
\dot{\eta}_i &=& L_{ij}^\eta\left[-\rho_0A_{ij}^\eta(2\eta_i-6\eta_i^2+4\eta_i^3)+ \beta_{ij}^\eta\frac{\partial^2\eta_i}{\partial r_{02}^2}\right], \,\, \text{and}\nonumber\\
\dot{\xi}_I &=& L_{IJ}^{gb}\left[-\rho_0A_{IJ}^{gb}(0)(2\xi_I-6\xi_I^2+4\xi_I^3)+ \beta_{IJ}^{gb}(0)\frac{\partial^2\xi_I}{\partial r_{02}^2}\right],
 \label{kineqsbx1sol}
\end{eqnarray}
respectively, where there is no sum on $i,\,j,\,I,\,J$, and we have used the constrains $\eta_i+\eta_j=1$ and $\xi_I+\xi_J=1$. The ${\sf M}_i$-${\sf M}_j$ interface and GB $\Gamma_{IJ}$ in Eqs. \eqref{kineqsbx1sol}$_1$ and \eqref{kineqsbx1sol}$_2$ are considered as planar. The solution for Eqs. \eqref{kineqsbx1sol}$_1$ and  \eqref{kineqsbx1sol}$_2$ are \cite{Steinbach-09,Levitasetal-PRB-15}
\begin{eqnarray}
\eta_i(r_{02},t) = 0.5+0.5\tanh [3(r_{02}-r_{0c})/\delta_{ij}^\eta], \quad \text{and}\quad  \xi_I(r_{02},t) = 0.5+0.5\tanh [3(r_{02}-r_{0c})/\delta_{IJ}^{gb}],  
 \label{analyticeta}
\end{eqnarray}
respectively, where $r_{0c}$ is the location of the middle of the respective interfaces, and the width, energy, and speed of these respective interfaces are
\begin{eqnarray}
 \delta_{ij}^\eta = \sqrt{\frac{18\beta_{ij}^\eta}{\rho_0A_{ij}^\eta}}, \quad \gamma_{ij}^\eta=\frac{\beta_{ij}^\eta}{\delta_{ij}^\eta},\quad c_{ij}^\eta =0;\quad \text{and}\quad 
\delta_{IJ}^{gb}=\sqrt{\frac{18\beta_{IJ}^{gb}(0)}{\rho_0A_{IJ}^{gb}(0)}},\quad\gamma_{IJ}^{gb}(0)=\frac{\beta_{IJ}^{gb}(0)}{\delta_{IJ}^{gb}},\quad c_{IJ}^{gb} =0. 
 \label{analyticeta2}
\end{eqnarray}

For the numerical calculations presented in Sec. \ref{simulations}, we consider the material constants for Ni-rich NiAl alloy, which exhibit cubic ($\mathsf A$) to tetragonal ($\sf M$) transformations  \cite{Bha04,Mishin-09}. The corresponding Bain tensors for variants $\sf M_1$, $\sf M_2$, and $\sf M_3$, in the standard $\{\fg c_1,\fg c_2,\fg c_3\}$ basis attached to a cubic $\sf A$ unit cell in a reference grain,  are (see Chapter 4 of \cite{Bha04})
\begin{equation}
\fg U_{t1}=diag( \chi, \alpha, \alpha), \quad\fg U_{t2}=diag(\alpha,\chi,\alpha), \quad\text{and}\quad\fg U_{t3}=diag(\alpha,\alpha,\chi),
 \label{trans_strn} 
\end{equation}
respectively, where $\alpha = 0.92$, and $\chi=1.22$ for NiAl alloy \cite{Mishin-09}, and the argument of $diag(\cdot)$ shows the diagonal elements of the tensor.  For all the simulations,  we will consider $a=a_b=a_\beta=a_\theta=3$, $a_c=10^{-3}$, $\delta_{0M}=1.5$ nm, $\gamma_{0M}=0.2$ N/m, $\delta_{12}^\eta= 0.75$ nm, and $\gamma_{12}^\eta=0.03$ N/m (see, e.g. \cite{Tuma-Stupkiewicz-Petryk-16,Basak-Levitas-2018JMPS} and the references therein for typical values of the widths and energies). Hence using Eqs. \eqref{analyticeta1}$_{2,3,4}$ and \eqref{analyticeta2} we have $\rho_0A_{0M} =2400$ MJ/m$^3$, $\rho_0 A_{12}^\eta=720$ MJ/m$^3$, $\beta_{0M}= 3\times 10^{-10}$ N, and $\beta_{12}^\eta=2.25\times 10^{-11}$ N.  A typical austenitic GB energy $\gamma^A_{IJ}=0.9$ N/m \cite{Mishin-2005Acta} is considered for all the calculations, whereas different values of $\Delta\gamma_{IJ}^{gb}$ (and hence different $\gamma^M_{IJ}$) are considered, and they would be specified for the simulations at hand in Sec. \ref{simulations}; $I,J=1,2,3$.  In this paper, the results for the austenitic GB width $\delta_{IJ}^{gb}$ between $0.5$ nm and $1.7$ nm are presented; see, e.g. \cite{Hagege1982} and Chapter 7 of \cite{German2014} for typical values. Based on the atomistic simulation results of \cite{Mishin-09}, we take the critical temperatures for ${\sf A}\to{\sf M}$ and ${\sf M}\to{\sf A}$ transformations for NiAl alloy as $\theta^c_{{\sf A}\to{\sf M}}= 240$ K and $\theta^c_{{\sf M}\to{\sf A}}= 490$ K, respectively. The thermodynamic equilibrium temperature between $\sf A$ and $\sf M$ is considered as $\theta_e=0.5(\theta^c_{{\sf A}\to{\sf M}}+\theta^c_{{\sf M}\to{\sf A}})$; see, e.g. \cite{Levitas-Roy-Acta-16} and the references therein. Thus, $\theta_e=365$ K for NiAl alloy. In \cite{Levitas-Roy-Acta-16}, the critical temperatures are shown to be related to the parameters $\rho_0A_{0M}$, $\theta_e$, and $\rho_0\Delta s$ by $\theta^c_{{\sf A}\to{\sf M}}= \theta_e+\rho_0A_{0M}/(3\rho_0\Delta s)$  and $\theta^c_{{\sf M}\to{\sf A}}= \theta_e-\rho_0A_{0M}/(3\rho_0\Delta s)$, respectively, using which we get $\rho_0\Delta s = -6.4$ MJ/(m$^3$K). We assume the isotropic elastic response of the phases, and identical Lam\'{e} constants ($\lambda,\,\mu$) are assumed for $\sf A$, $\sf M_1$, and $\sf M_2$:  $\lambda=74.6 $ GPa, and $\mu=72$ GPa \cite{Basak-Levitas-2018JMPS}. The kinetic coefficients of the phases are taken as $L_{0M}=L_{12}^\eta=2600$ (Pa-s)$^{-1}$ \cite{Basak-Levitas-2018JMPS}.

\begin{figure}[t!]
\centering
  \includegraphics[width=4.0in, height=1.1in] {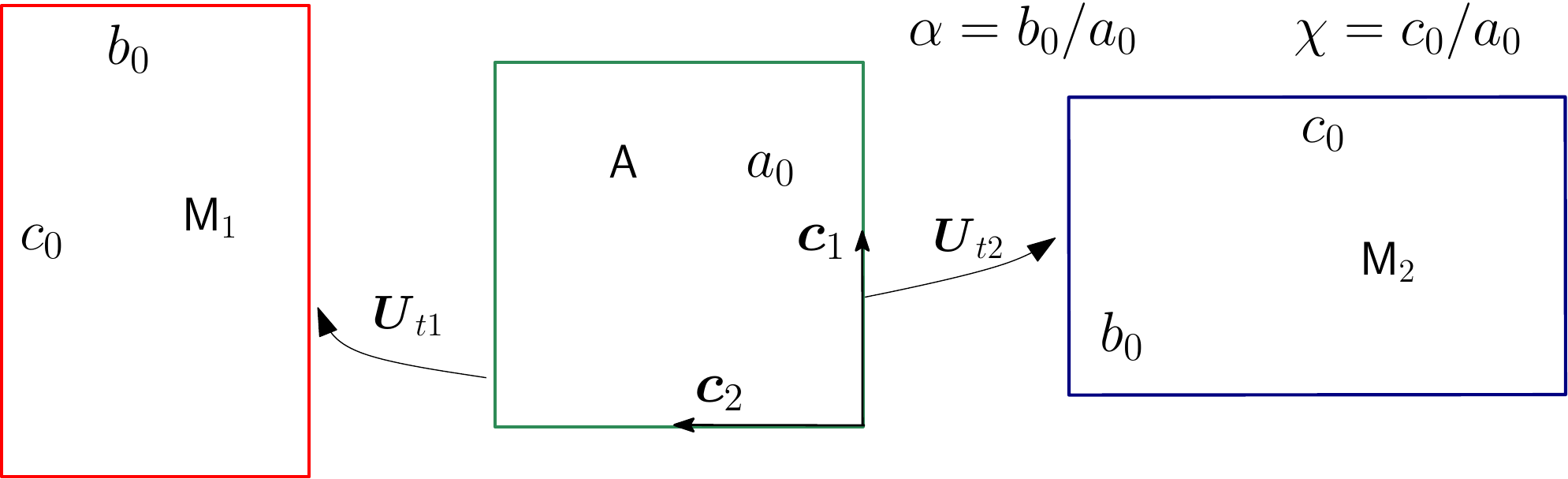}
\caption{Schematic of the unit cells of cubic austenite and  tetragonal martensitic variants ${\sf M}_1$ and ${\sf M}_2$ in the $\fg c_1$-$\fg c_2$ plane, where  $\{\fg c_1, \fg c_2,\fg c_3\}$ is an orthonormal basis attached to the cubic $\sf A$ unit cell. } 
\label{unit_cells}
\end{figure}

\section{Numerical results and discussions}
\label{simulations}
We now present the numerical results. Assuming the plane stress condition and isotropic elastic response, the results for the 2D bicrystals and tricrystals with tilt GBs are presented in Secs. \ref{bicrystls} and \ref{tricryslt}, respectively, considering the PTs between $\sf A$ and two variants. We consider $\sf M_1$ and $\sf M_2$ as those two variants without a loss of generality. The schematic of the unit cells is shown in Fig. \ref{unit_cells} in the $\fg c_1$-$\fg c_2$ plane, where $\fg c_3$-axis is parallel to the outward normal to that plane. The governing equations for both the bicrystals and tricrystals, obtained using the general system of equations derived in Sec. \ref{system_eqns}, are enlisted in Appendix \ref{planestresseqs}. Figures \ref{xiplots}(a) and (b) show two typical undeformed austenitic bicrystals ($V_0$) with a planar GB and a circular GB, respectively. Figure \ref{xiplots}(c) shows a typical undeformed austenitic tricrystal ($V_0$)  with three planar GBs with identical energy and width, which are meeting in the middle of the sample at $120^\circ$ dihedral angles according to the equilibrium condition at the TJ (see, e.g. Sec. 3.3.3. of \cite{Porter-Easterling}). The exact  sample size and the GB parameters, including $\gamma_{IJ}^M$, $\delta^{gb}_{IJ}$, $\rho_0K_{IJ}^{gb}$, and $\rho_0K_{HIJ}$ are specified while presenting the simulation results. The initial stress-free austenitic bicrystals  are simply obtained by using the analytical solution for $\xi_1$ given by Eq. \eqref{analyticeta}$_2$. The Eqs. \eqref{bicrystgovt}$_3$, \eqref{bicrystgovt}$_4$, and \eqref{bicrystgovt}$_5$ have been solved, neglecting mechanics until the stationary distribution of the order parameters $\xi_1$, $\xi_2$, and $\xi_3$ are obtained to get all the stress-free austenitic tricrystals (see e.g. Fig. \ref{xiplots}(c)), where we have used the Neumann BC given by Eq. \eqref{neumbcs1}$_3$ on all the surfaces, and $L^\xi_{12}=L^\xi_{13}=L^\xi_{23}=100$ (Pa-s)$^{-1}$ when the coefficients are non-zero; see the discussion of Sec. \ref{GLsubsec}. The nonlinear finite element procedure based on the algorithm presented in \cite{Basak-Levitas-2019CMAME} has been developed, and it is implemented using the open-source package deal.ii \cite{Bangerth-16}. The quadratic quadrilateral elements are used for all the simulations. The mesh sizes in all the samples are so chosen that at least four grid points are present across all the interfaces to ensure the mesh-independent results. The time derivatives in the Ginzburg-Landau equations \eqref{bicrystgov}$_{1,2}$ for the bicrystals and \eqref{bicrystgovt}$_{1,2}$ for the tricrystals are discretized using the Euler's backward difference scheme of order one, and a time step of size $\Delta t=10^{-14}$ s is used for all the calculations. While solving the equilibrium equations given by Eq. \eqref{eql_eqs} and the PF equations for the bicrystals \eqref{bicrystgov}$_{1,2}$ or tricrystals \eqref{bicrystgovt}$_{1,2}$, we have applied the following BCs during both the forward and reverse MTs:  (i) the external surfaces are free of traction in their tangential directions; (ii) all the external surfaces are supported by rollers, i.e. the normal displacement at each boundary is zero, except for the example of Sec. \ref{sussub4}, and it will be specified there itself; (iii) the homogeneous Neumann BCs for $\eta_0$ and $\eta_1$ given by Eqs. \eqref{neumbcs1}$_1$ and \eqref{neumbcs1}$_2$ are applied on the surfaces. The same time discretization scheme and time step size mentioned above are used for solving Eqs. \eqref{bicrystgovt}$_{3,4,5}$ for obtaining the stress-free tricrystals. The orientation of the basis $\{\fg c_1,\fg c_2\}$ for each austenitic grain is shown in Figs. \ref{xiplots}(a,b,c). A fixed basis $\{\fg e_1,\fg e_2,\fg e_3\}$ is also considered, where $\fg e_3$ being parallel to the $\fg c_3$-axis. The Bain tensors  $\fg U_{ti}^{(I)}$ ($i=1,2$) for grain $G_I$, introduced in Eq. \eqref{utilde}, is obtained using  Eq. \eqref{trans_strn}  in Eq. \eqref{utildeII}, considering
\begin{equation}
[Q^{(I)}] = \begin{bmatrix}
   \cos\vartheta_I &  \sin\vartheta_I & 0\\
    -\sin\vartheta_I  &  \cos\vartheta_I & 0 \\
     0 &  0 &  1\\
\end{bmatrix},
 \label{trans_rule} 
\end{equation}
where $\vartheta_I$ is the angle of the $\fg c_1$-axis from the $\fg e_1$-axis for $G_I$  (see Fig. \ref{xiplots}), and it is considered positive if anticlockwise.  In the 2D samples, each of the GBs $\Gamma_{IJ}$  has only one misorientation angle, i.e. the list $\Theta_{IJ}$ introduced in Eq. \eqref{propertiesf} has one element, which we define as 
\begin{equation}
\vartheta_{IJ}=\vartheta_I-\vartheta_J.
 \label{misortnn} 
\end{equation}
\begin{figure}[t!]
\centering
  \includegraphics[width=6.3in, height=2.8in] {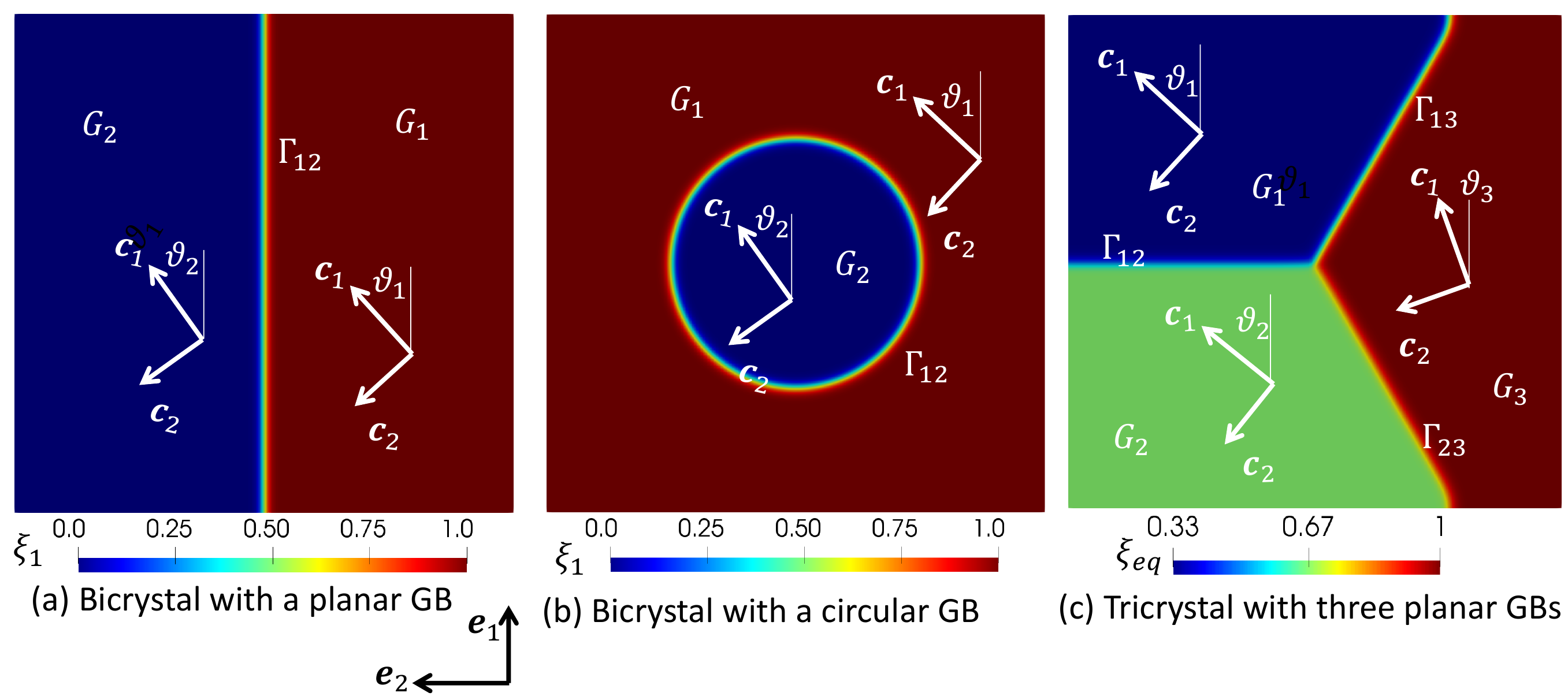}
\caption{Typical stress-free austenitic samples ($V_0$) of $30$ nm$\times$30 nm size:  bicrystals with (a) a planar tilt GB and  (b) a circular tilt GB of $15$ nm diameter; (iii) tricrystal with planar tilt GBs meeting at $120^\circ$ dihedral angles at the TJ. We used $\gamma_{12}^A=0.9$ N/m and $\delta_{IJ}^{gb}=1$ nm. For the tricrystal, $\rho_0K_{IJ}^{gb}=\rho_0 K_{123}=800$ GPa. Orientation of the basis $\{\fg c_1,\fg c_2\}$ for each grain is shown; $\{\fg e_1,\fg e_2\}$ is a fixed reference frame.
 }
\label{xiplots}
\end{figure}

\subsection{Results for bicrystals}
\label{bicrystls}
The results for the bicrystals are presented here. The typical austenitic initial samples are shown in Fig.  \ref{xiplots}(a) and (b). The effect of the GB width $\delta_{12}^{gb}$, the difference in GB energy $\Delta\gamma_{12}^{gb}$,  misorientation $\vartheta_{12}$, applied strains, sample size, and GB curvature are studied for samples with a symmetric tilt GB during the forward and reverse  transformations. A bicrystal with an asymmetric planar tilt GB is also considered. The role of compatibility of the Bain strains across the GB is studied.

\begin{figure}[t!]
\centering
  \includegraphics[width=3.0in, height=2.5in] {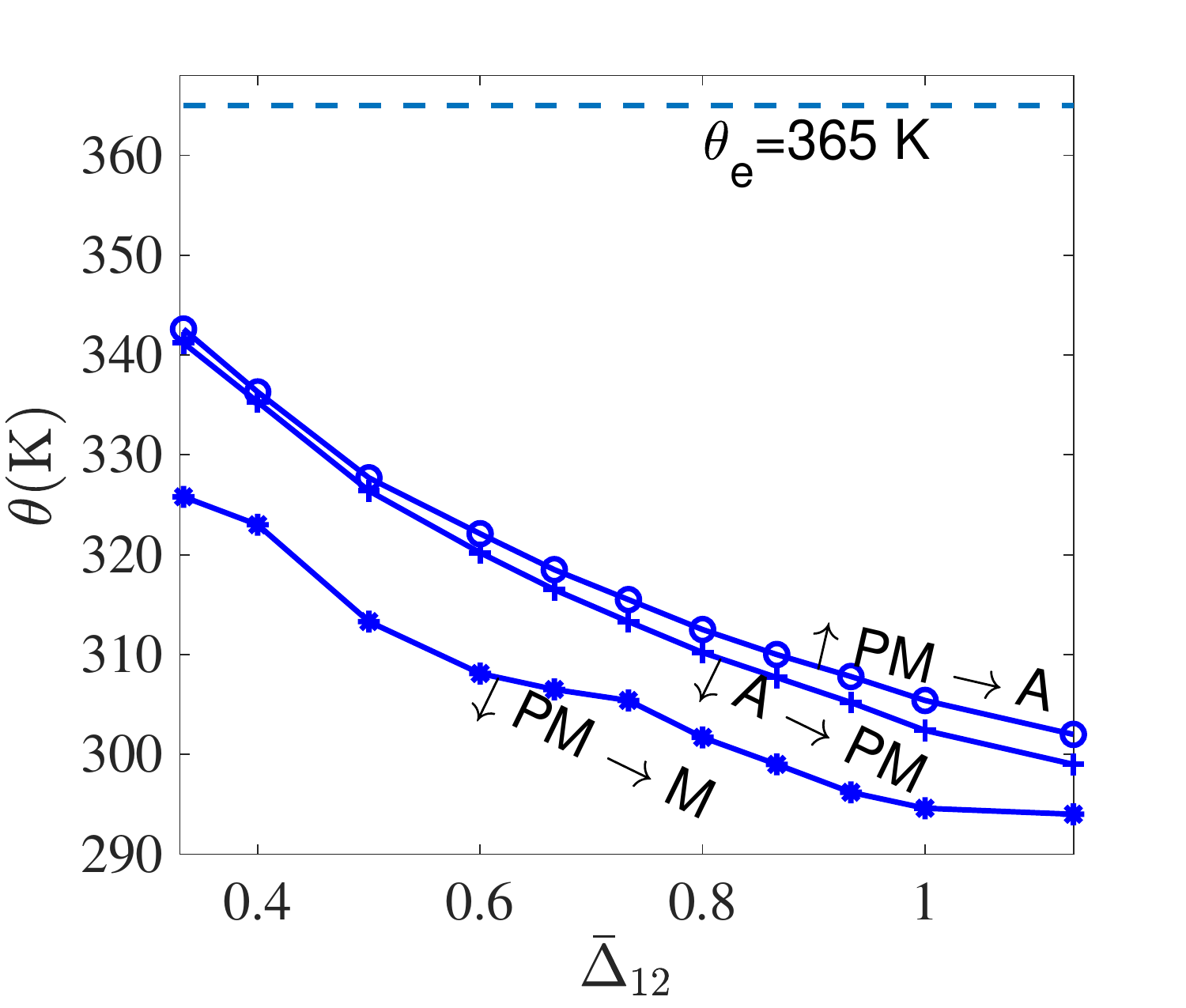}
\caption{Transformation temperature $\theta$ versus a nondimensional  width ratio $\bar\Delta_{12}=\delta_{12}^{gb}/\delta_{0M}$ in a $30$ nm $\times$ 30 nm bicrystal with a planar symmetric tilt GB. We used $\gamma_{12}^A=0.9$ N/m, $\Delta\gamma_{12}^{gb}=-0.5$ N/m, $\vartheta_{12}=40^\circ$, and no external strain.}
\label{phasedia}
\end{figure}

\subsubsection{ GB size (width)-effect}
\label{sussub1}
\begin{figure}[t!]
\centering
\hspace{-8mm}
\subfigure[]{
  \includegraphics[width=3.2in, height=2.7in] {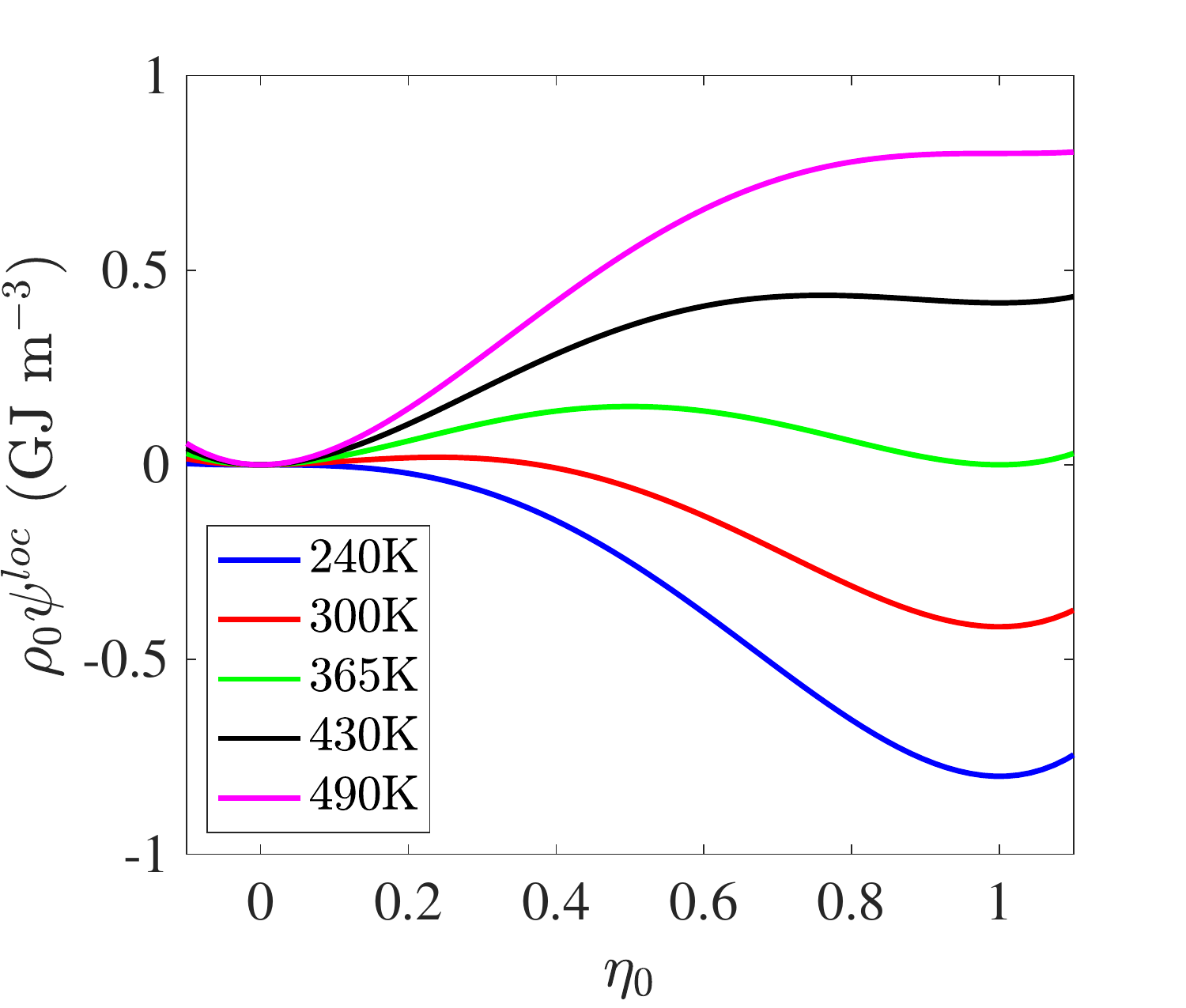}
	}
\hspace{-1mm}
    \subfigure[]
   {
    \includegraphics[width=3.2in, height=2.7in] {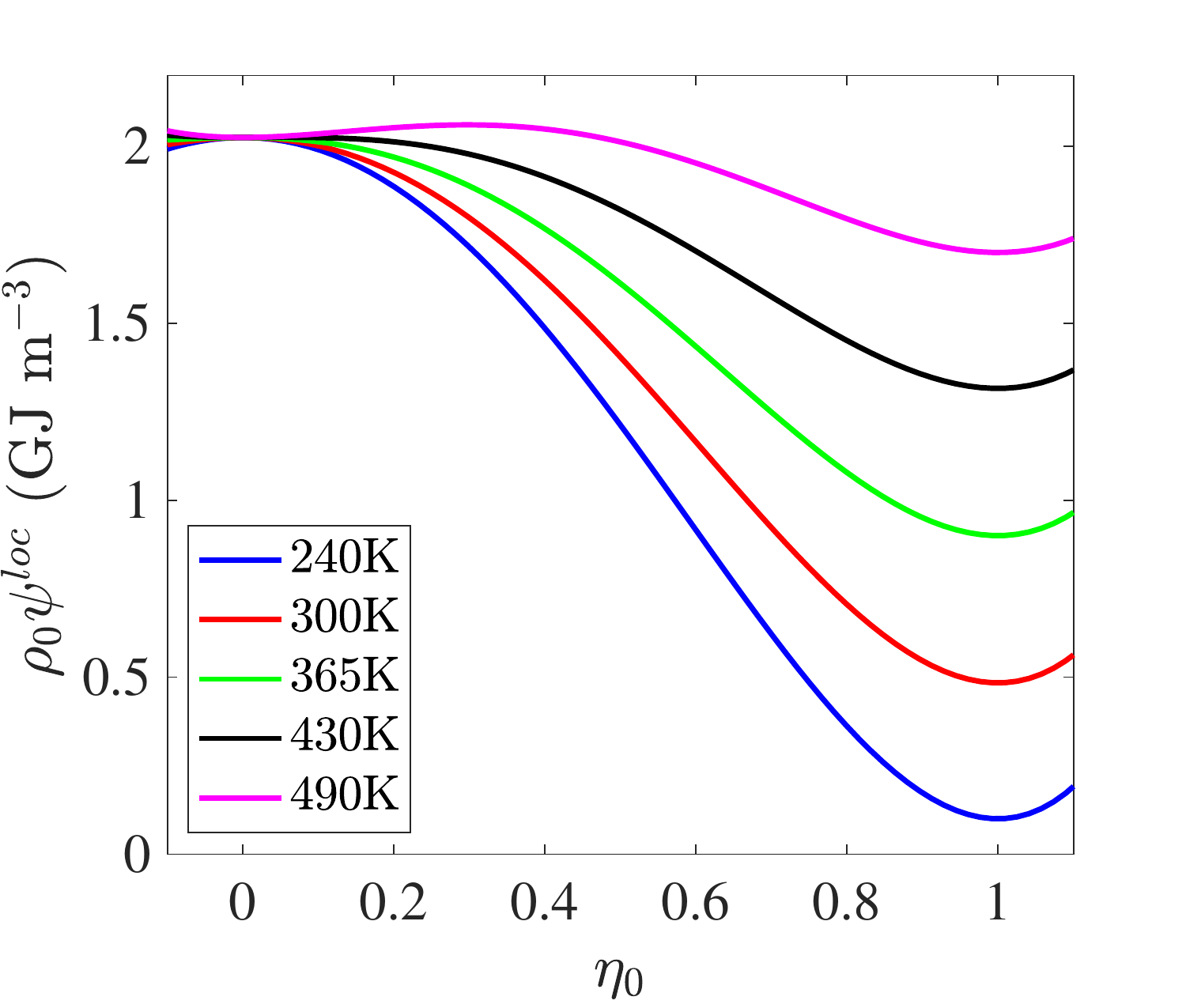}
}
\caption{Free energy per unit volume (see Eq. \eqref{MF0}) variation with $\eta_0$ neglecting mechanics at  $\theta^c_{A\to M}\leq \theta\leq \theta^c_{M\to A}$ in a (a) homogeneous grain, and (b) in the middle of a long planar GB of a bicrystal, where $\xi_1=0.5$. We used  $\gamma_{12}^A=0.9$ N/m, $\Delta\gamma_{12}^{gb} = -0.5$ N/m, and $\delta_{12}^{gb}=1$ nm.}
\label{fig:sub-firstse}
\end{figure}
We consider an austenitic bicrystal of $30$ nm $\times$ $30$ nm size with a planar symmetric tilt GB, where $\vartheta_1=-\vartheta_2=20^\circ$, and hence $\vartheta_{12}=40^\circ$ by Eq. \eqref{misortnn}  (also see Fig.  \ref{xiplots}(a)). We vary  $\delta_{12}^{gb}$ between $0.5$ nm and $1.7$ nm, while  $\Delta\gamma_{12}^{gb}$ is fixed to $-0.5$ N/m. The GB energy in the $\sf M$ phase is thus  $\gamma_{gb}^M = 0.4$ N/m. 

\noindent{\bf Transformation temperature variation with $\bar\Delta_{12}$:}

The variation of the temperatures for transformations between $\sf A$, $\sf PM$, and $\sf M$ within the GB region for varying non-dimensional GB width $\bar\Delta_{12}:=\delta_{12}^{gb}/\delta_{0M}$ is shown in Fig. \ref{phasedia}, where the value for $\delta_{0M}$ is fixed and is given in Sec. \ref{MaterParam}. The curve indicated by  $\sf A\to \sf PM$ represents the temperature (denoted by $\theta_{\sf A\to \sf PM}$) at which a $\sf PM$ layer appears in a jump-like manner in the GB region with  $0.05\leq max(\eta_0)\leq 0.95$. However,  $0.1$ K above the corresponding $\theta_{\sf A\to \sf PM}$, the entire sample was austenitic. For obtaining $\theta_{\sf A\to \sf PM}$, the ICs on $\eta_0$ and $\eta_1$ were taken as $0\leq\eta_0(\fg r_0,t=0)\leq 0.02$ and $0\leq\eta_1(\fg r_0,t=0)\leq 0.1$, both distributed randomly all over the sample. The initial sample was hence in the $\sf A$ phase (see Sec. \ref{order_params}). As the sample temperature is further decreased from $\theta_{\sf A\to \sf PM}$, the $\sf PM$ layer thickness increases, and at the temperatures corresponding to the curve indicated by $\sf PM\to \sf M$ and denoted by $\theta_{\sf PM\to \sf M}$, the GB region and the adjacent part of the grains transform to complete $\sf M$ with $0.95< max(\eta_0)\leq 1$  in a jump-like manner. However, that region was $\sf PM$ only above $0.1$ K of $\theta_{\sf PM\to \sf M}$. We see from Fig. \ref{phasedia} that all the $\theta_{\sf A\to \sf PM}$ and $\theta_{\sf PM\to \sf M}$ lie within the temperature regime of stable $\sf A$, and they are significantly higher than the critical temperature $\theta_{\sf A\to \sf M}^c=240$ K. To explain the reason, we note that the local part of the free energy per unit volume of the bicrystal within the GB region is given by
\begin{equation}
\rho_0\psi^{loc}(\eta_0,\theta) = \rho_0A_{0M}\eta_0^2(1-\eta_0)^2- \rho_0\Delta s\, (\theta-\theta_e) \eta_0^2(3-2\eta_0) +\frac{36}{\delta_{12}^{gb}}\left[\gamma^A_{12}+\Delta\gamma_{12}^{gb}\eta_0^2(3-2\eta_0)\right] \xi_1^2(1-\xi_1)^2,
\label{energy1var}
\end{equation}
which we have obtained from Eqs. \eqref{MF0} and \eqref{MF2} by neglecting mechanics and considering the transformations between $\sf A$ and a single $\sf M$ variant only. In Eq. \eqref{energy1var}, we have also used the relation
$$A_{12}^{gb}(\eta_0) \xi_1^2(1-\xi_1)^2 + 0.5 \beta_{12}^{gb}(\eta_0)\left( \frac{\partial\xi_1}{\partial r_{02}}\right)^2 =\frac{36\gamma_{12}^{gb}(\eta_0)}{\delta_{12}^{gb}}\xi_1^2(1-\xi_1)^2,$$
which can be easily proved using the analytical solution for $\xi_1(r_{02})$ given by Eq. \eqref{analyticeta}$_2$, and Eqs. \eqref{analyticeta2}$_{4,5}$. Note that we have also used $\partial\xi_1/\partial r_{01}=0$. The last term on the right-hand side of Eq. \eqref{energy1var} vanishes within the homogeneous grains where $\xi_1=0$ or $1$. We  show the local energy $\rho_0\psi^{loc}$ plots within the particles of a homogeneous grain and in the middle line of the planar GB where $\xi_1=0.5$ and the energy is given by Eq. \eqref{energy1var} is the maximum in Figs. \ref{fig:sub-firstse}(a) and \ref{fig:sub-firstse}(b), respectively, for the temperatures between $\theta^c_{A\to M}= 240$ K and $\theta^c_{M\to A}=490$ K. We have used $\gamma_{12}^A=0.9$ N/m, $\Delta\gamma_{12}^{gb} = -0.5$ N/m, and $\delta_{12}^{gb}=1$ nm and all other material parameters are listed in Sec. \ref{MaterParam}. From Fig. \ref{fig:sub-firstse}(b), it is obvious that the barrier energy between $\sf A$ ($\eta_0=0$) and $\sf M$ ($\eta_0=1$) on the $\xi_1=0.5$ line vanishes at $\theta$ much above the $\theta^c_{A\to M}$, at which the  barrier energy vanishes in the homogeneous grains (see Fig. \ref{fig:sub-firstse}(a)). Hence the instability temperatures at which $\sf PM$ or $\sf M$ nucleates in the GB region lie within the stability regime of $\sf A$, and it is determined by the values of $\delta_{IJ}^{gb}$ and $\Delta\gamma_{IJ}^{gb}$. The actual scenario, however, is more complex due to the interaction of the thermal problem with the mechanics, which is the case, for example, in Fig. \ref{phasedia} also. 

We now determine the temperature for the reverse ($ \sf PM\to \sf A$) transformation, which is denoted as $\theta_{\sf PM\to \sf A}$ and indicated by  $\sf PM\to \sf A$ curve in Fig. \ref{phasedia}. For that, we have used the sample obtained at $\theta_{\sf A\to \sf PM}$ as the initial sample and increased its temperature. As $\theta$ increases, the $\sf PM$ layer thickness decreases along with a decrease in the value of  $max(\eta_0)$. However, at $\theta=\theta_{\sf A\to \sf PM}$, the $\sf PM$ layer on the GB region abruptly disappears, and the original austenitic bicrystal is recovered. Below $0.1$ K of $\theta_{\sf PM\to \sf A}$, the  $\sf PM$  layer was, however, present. 
We see from Fig. \ref{phasedia} that the difference between $\theta_{\sf A\to \sf PM}$ and $\theta_{\sf PM\to \sf A}$ increases as $\delta_{12}^{gb}$ (and so $\bar\Delta_{12}$) increases, implying a monotonic increase in the temperature hysteresis, which is measured as the difference in the corresponding temperatures, i.e. $\theta_{\sf PM\to \sf A}-\theta_{\sf A\to \sf PM}$ (see, e.g. \cite{Basak-Levitas-2018-nanovoid,Levitas-Samani-PRB-14} in case of surface-induced melting). The hysteresis is caused by a jump-like discontinuous  evolution of the order parameter $\eta_0$ during these two transformations. The hysteresis decreases as the diffused GB approaches a sharp one, i.e. $\bar\Delta_{12}\to 0$. The difference in the temperatures of transformations between $\theta_{\sf A\to \sf PM}$ and $\theta_{\sf PM\to \sf M}$ is going down as the GB width increases, similar to other interface-induced PTs \cite{levitas-javanbakht-PRL-10,levitas-javanbakht-PRL-11,BasakLevitas2020Acta-2004,Levitas-Samani-PRB-14,Basak-Levitas-2018-nanovoid,Basak-2021-PCCP}. The transformation temperatures $\theta_{\sf A\to \sf PM}$ is $59$ K to $101.2$ K higher, and $\theta_{\sf PM\to \sf M}$ is $54$ K to $85.8$ K higher than the critical temperature $\theta_{\sf A\to \sf M}^c=240$ K over the entire range of $\bar\Delta_{12}$ shown in Fig. \ref{phasedia}. 

Although we did not find any relevant experimental or atomistic data for GB-induced MTs in Ni-rich NiAl alloy in the literature, we found the following limited data for a ferrous alloy and the NiTi alloy, which we would use for comparing our numerical results qualitatively. The  martensitic start temperature was observed to be $66$ K higher  in bicrystal with a planar tilt GB  than that of the single crystal of a ferrous alloy \cite{Tsuzaki-1995}. Ueda et al. considered a Fe-Ni alloy and studied MTs in two different bicrystals with identical (crystallographically) tilt GBs in \cite{Ueda-2001Acta} and \cite{Ueda-2004ISIJ},  where both the samples were obtained using an identical procedure. The martensitic start temperatures for these two bicrystals were reported as $234$ K \cite{Ueda-2001Acta} and $210$ K \cite{Ueda-2004ISIJ}, i.e. there is a difference of $24$ K. Since the GBs in these two bicrystals are crystallographically identical, their energies are also the same. 
However, a plausible reason for that nucleation temperature difference could be the difference in GB widths. 
Since the GB widths were not reported in these two references, we cannot confirm that.  The nucleation temperatures for those bicrystals (with tilt GB) were reported as $54$ K \cite{Ueda-2001Acta} and $68$ K  \cite{Ueda-2004ISIJ} higher than that of the single crystals. The atomistic simulations for NiTi alloy in  \cite{Qin-2018} showed that a tilt GB promoting MTs nucleates $\sf M$ at $16$ K higher than its single crystal counterpart. 
 The temperature ranges in our simulations mentioned in the previous paragraph are close to the ranges reported in these experimental and atomistic studies. Furthermore, the nucleation of the product phase ($\sf PM$ or $\sf M$) on the GB in a jump-like manner was also observed in experiments \cite{Ueda-2001Acta,Ueda-2004ISIJ}, which has been referred to as `the burst transformation' therein. The experiments and atomistic simulations in polycrystalline solids have shown that not all the tilt GBs promote $\sf M$ nucleation \cite{Tsuzaki-1995,Kajiwara-1986,Qin-2018}, the reason for which was not clearly mentioned. Therefore, it would be interesting to conduct systematic experiments and molecular simulations to confirm the effects of GB width on the temperatures of transformations between $\sf A$, $\sf PM$, and $\sf M$ to compare the present PF results shown in Fig. \ref{phasedia}. 

Note that very rich transformation temperature plots for MTs with the variation of the width of the corresponding external surface \cite{levitas-javanbakht-PRL-10,levitas-javanbakht-PRL-11} or a matrix-precipitate interface \cite{BasakLevitas2020Acta-2004} were developed, which were shown to agree with the experiments. Similar plots were also developed for the transformations between solid, melt, and premelt induced by external surface in nanoparticles \cite{Levitas-Samani-PRB-14} and nanovoids \cite{Basak-Levitas-2018-nanovoid}, and also induced by GBs in polycrystals \cite{Basak-2021-PCCP}, which explain the different temperature ranges for premelt nucleation to complete melting reported in various experimental and atomistic studies.


\begin{figure}[t!]
\centering
\hspace{-8mm}
\subfigure[$\eta_0$ plots during forward MT]{
  \includegraphics[width=3.0in, height=3.0in] {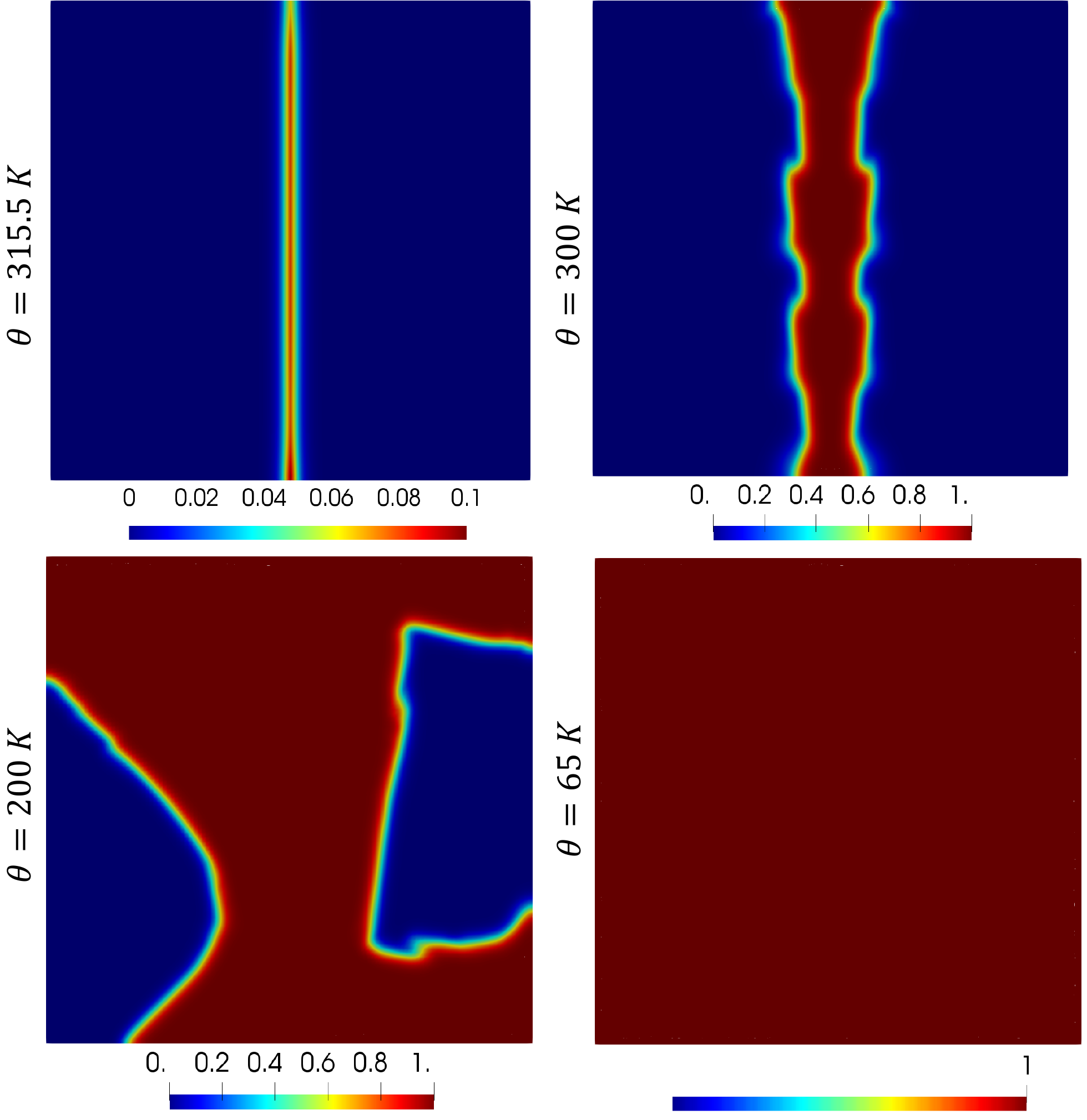}
	}
\hspace{-1mm}
    \subfigure[$\eta_{eq}$ plots during forward MT]
   {
    \includegraphics[width=3.0in, height=3.0in] {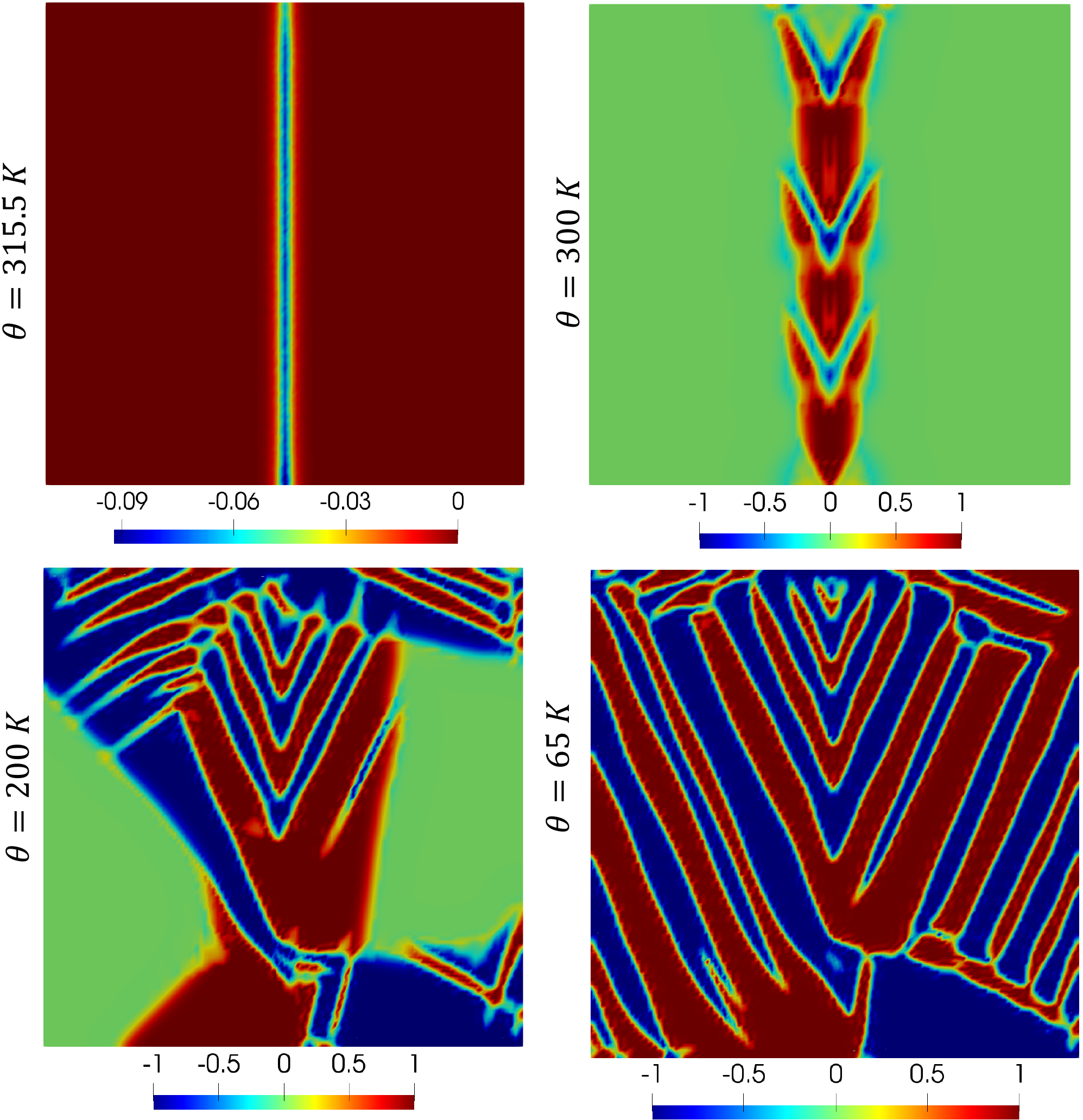}
}
\subfigure[$\eta_0$ plots during reverse MT]{
  \includegraphics[width=3.0in, height=3.0in] {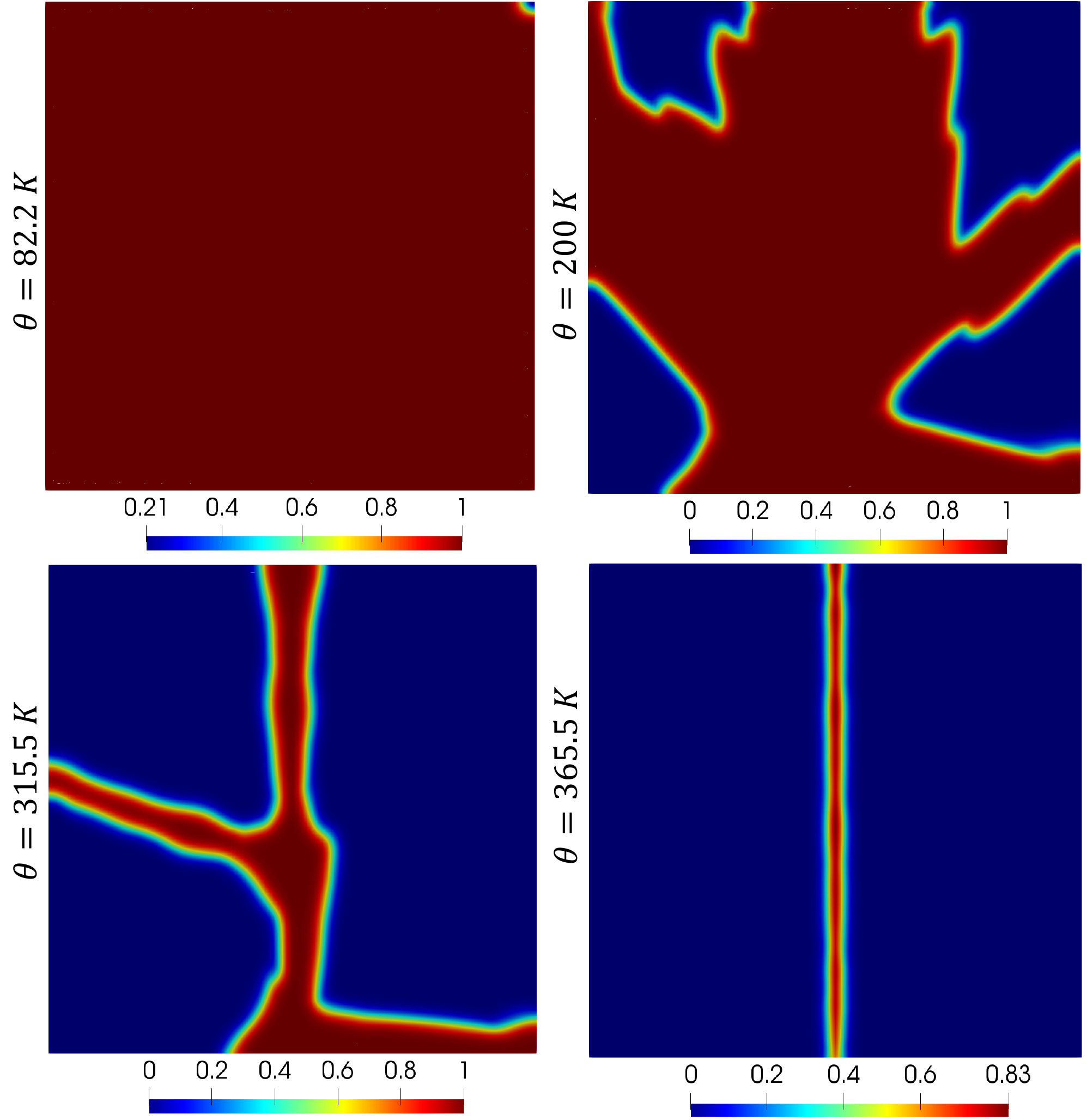}
	}
\hspace{-1mm}
    \subfigure[$\eta_{eq}$ plots during reverse MT]
   {
    \includegraphics[width=3.0in, height=3.0in] {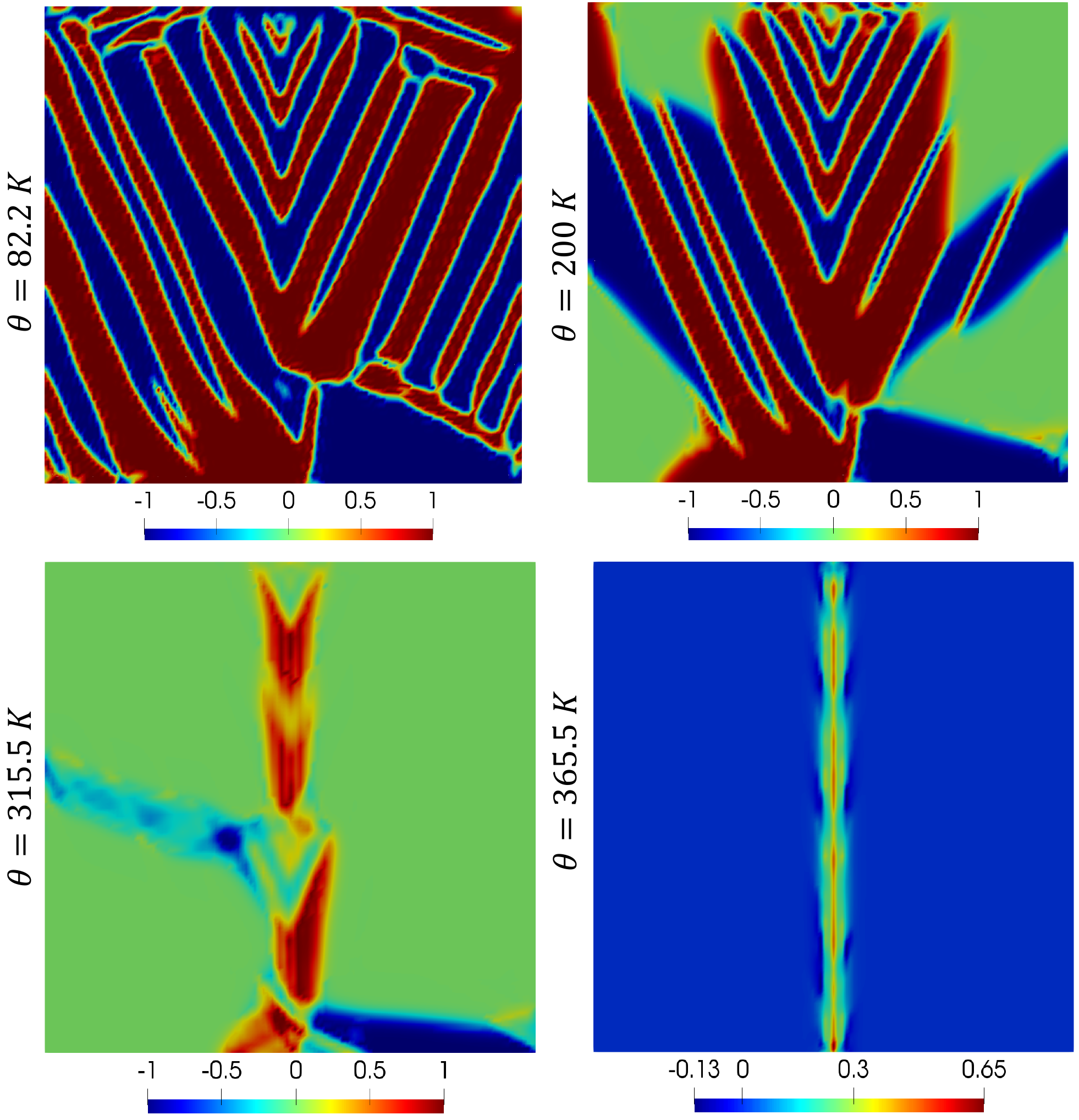}
}
\caption{Microstructure evolution during (a,b) forward, and (c,d) reverse MTs in a $30$ nm $\times$ $30$ nm bicrystal having a symmetric planar GB with $\theta$ variation. We used $\Delta\gamma_{12}^{gb} = -0.5$ N/m, $\delta_{12}^{gb}=1$ nm, and $\vartheta_{12}=40^\circ$. No strain applied.}
\label{fig:sub-firsts}
\end{figure}

\begin{figure}[t!]
\centering
\hspace{-8mm}
\subfigure{
  \includegraphics[width=5.7in, height=3.6in] {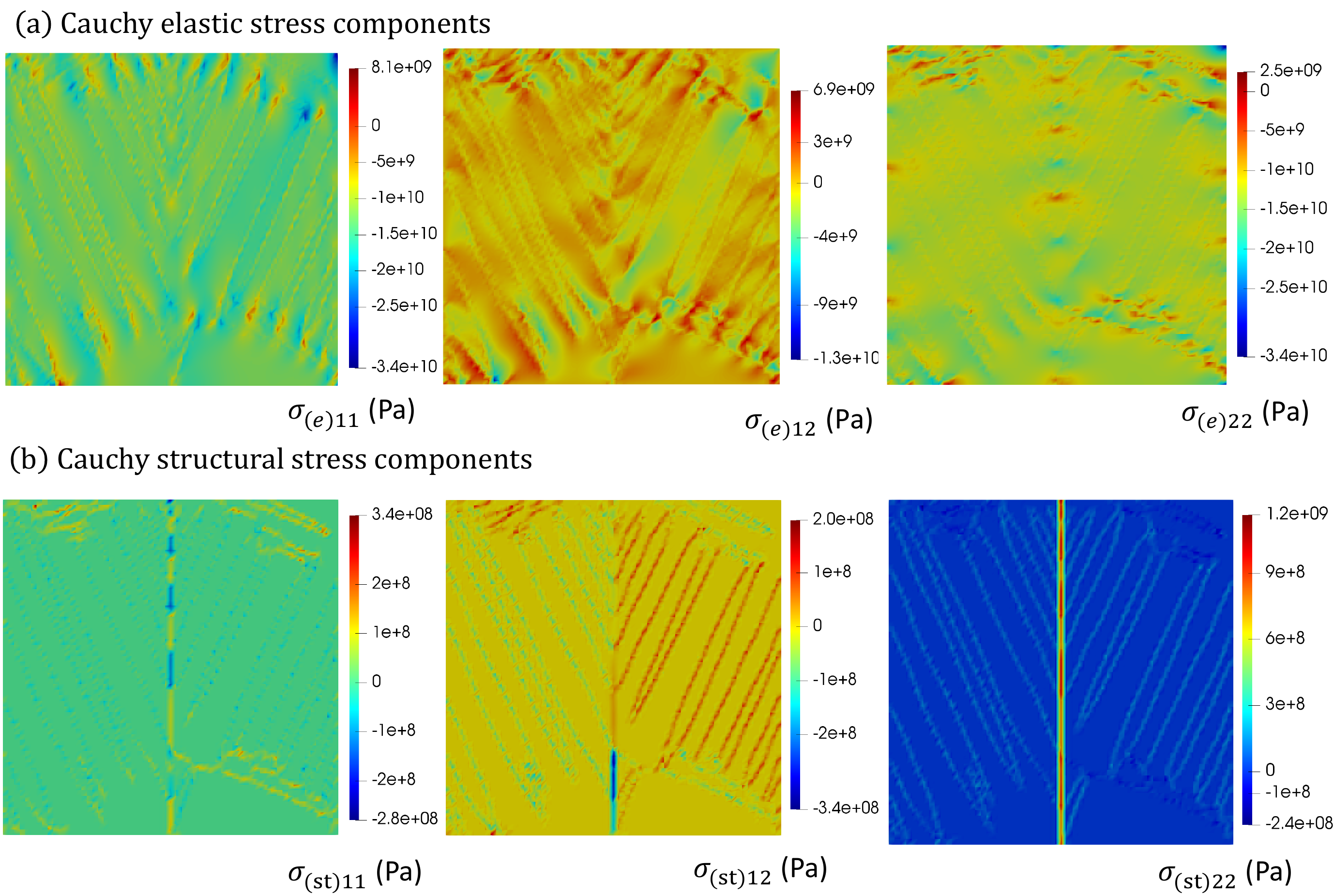}
	}
\hspace{-1mm}
    \subfigure
   {
    \includegraphics[width=4.0in, height=1.6in] {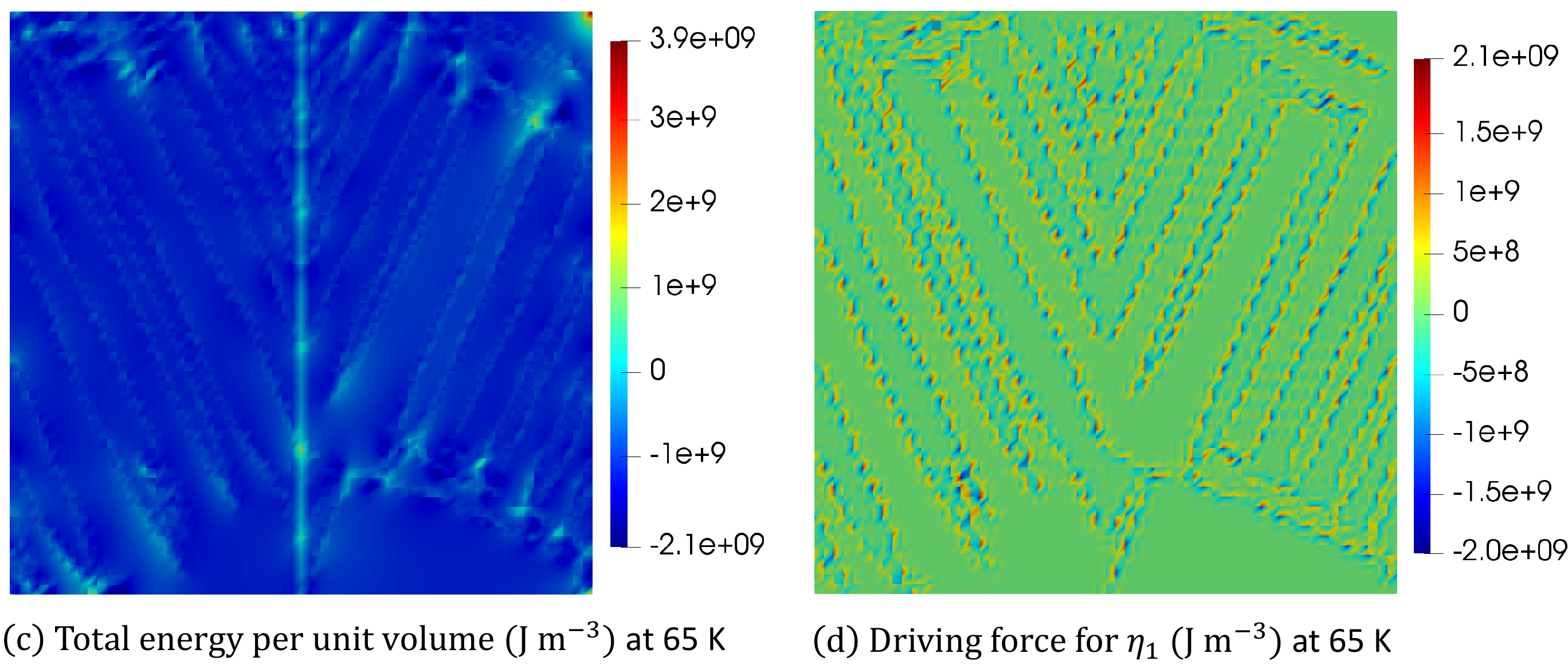}
}
\caption{Plots for the (a) Cauchy elastic stresses $\sigma_{(e)ij}$, (b) Cauchy structural stresses $\sigma_{(st)ij}$, (c) $\rho_0\psi$ - total energy per unit volume of $\Omega_0$, and (d) driving force $X_{12}^\eta$ at  $\theta=65$ K within the martensitic sample shown in Fig. \ref{fig:sub-firsts}(a,b) obtained by forward MT.}
\label{fig:sub-firstsstress}
\end{figure}

\begin{figure}[t!]
\centering
\hspace{-8mm}
\subfigure[$\eta_0$ plots during forward MT]{
  \includegraphics[width=3.0in, height=3.0in] {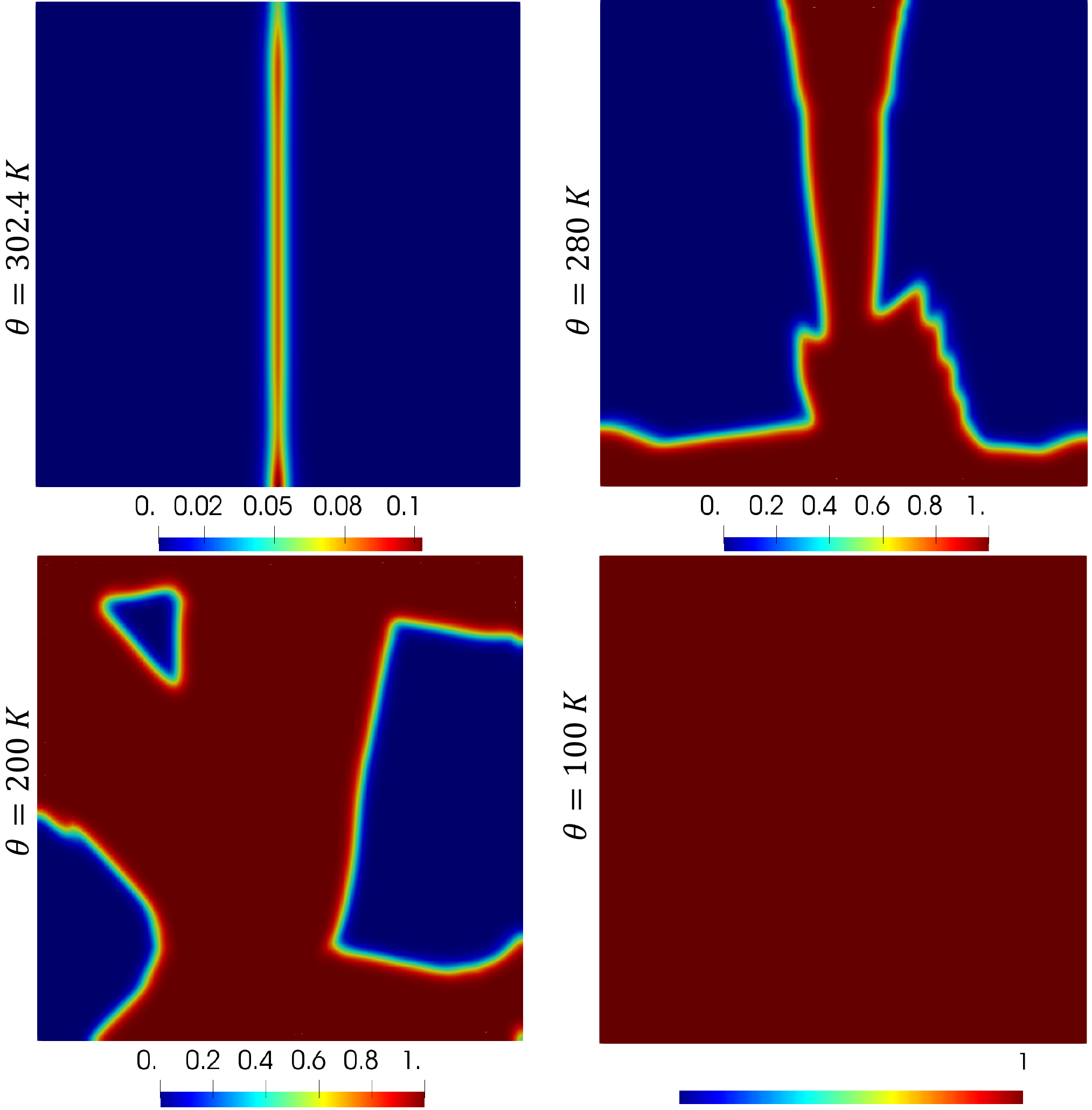}
	}
\hspace{-1mm}
    \subfigure[$\eta_{eq}$ plots during forward MT]
   {
    \includegraphics[width=3.0in, height=3.0in] {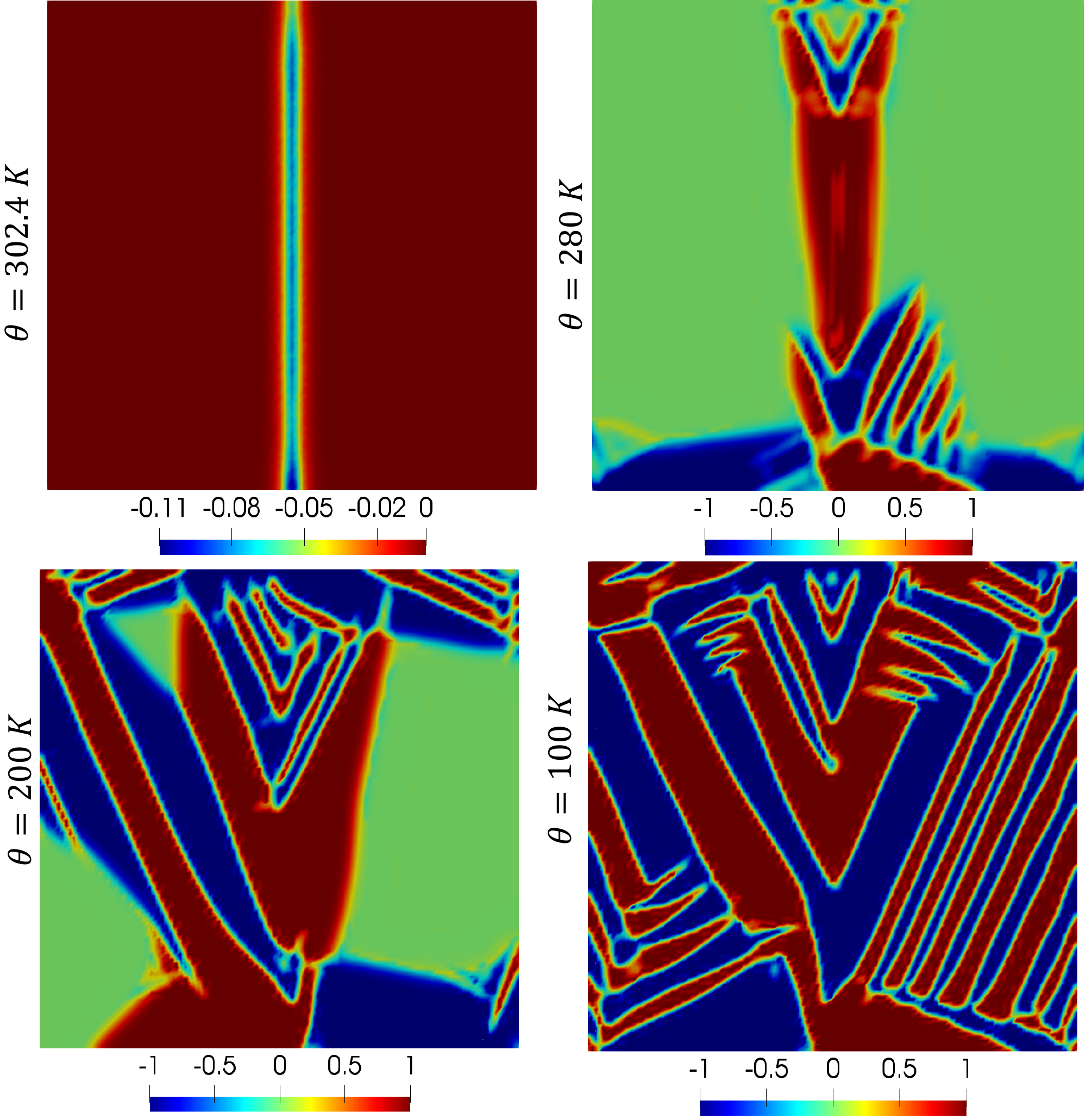}
}
\subfigure[$\eta_0$ plots during reverse MT]{
  \includegraphics[width=3.0in, height=3.0in] {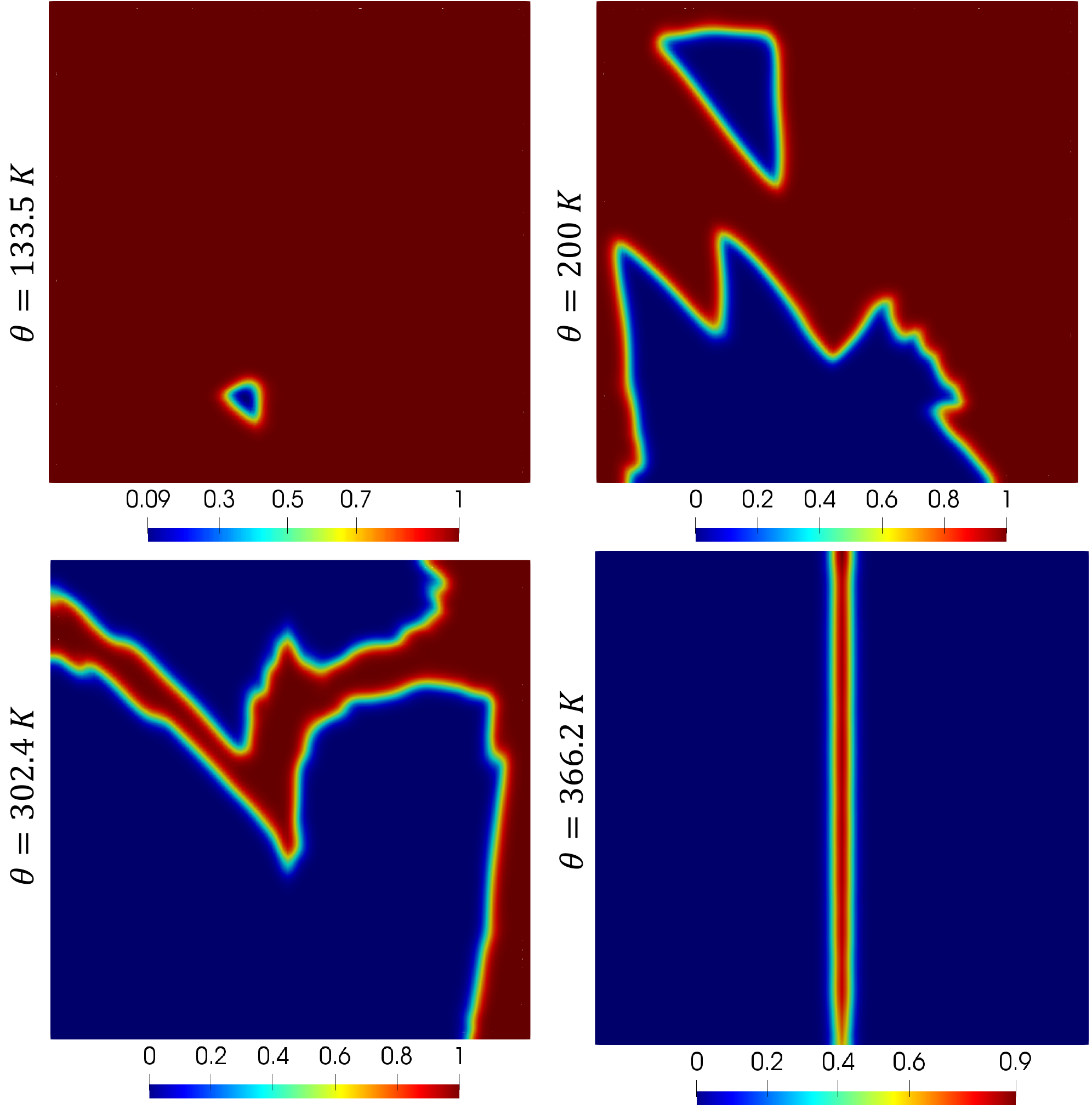}
	}
\hspace{-1mm}
    \subfigure[$\eta_{eq}$ plots during reverse MT]
   {
    \includegraphics[width=3.0in, height=3.0in] {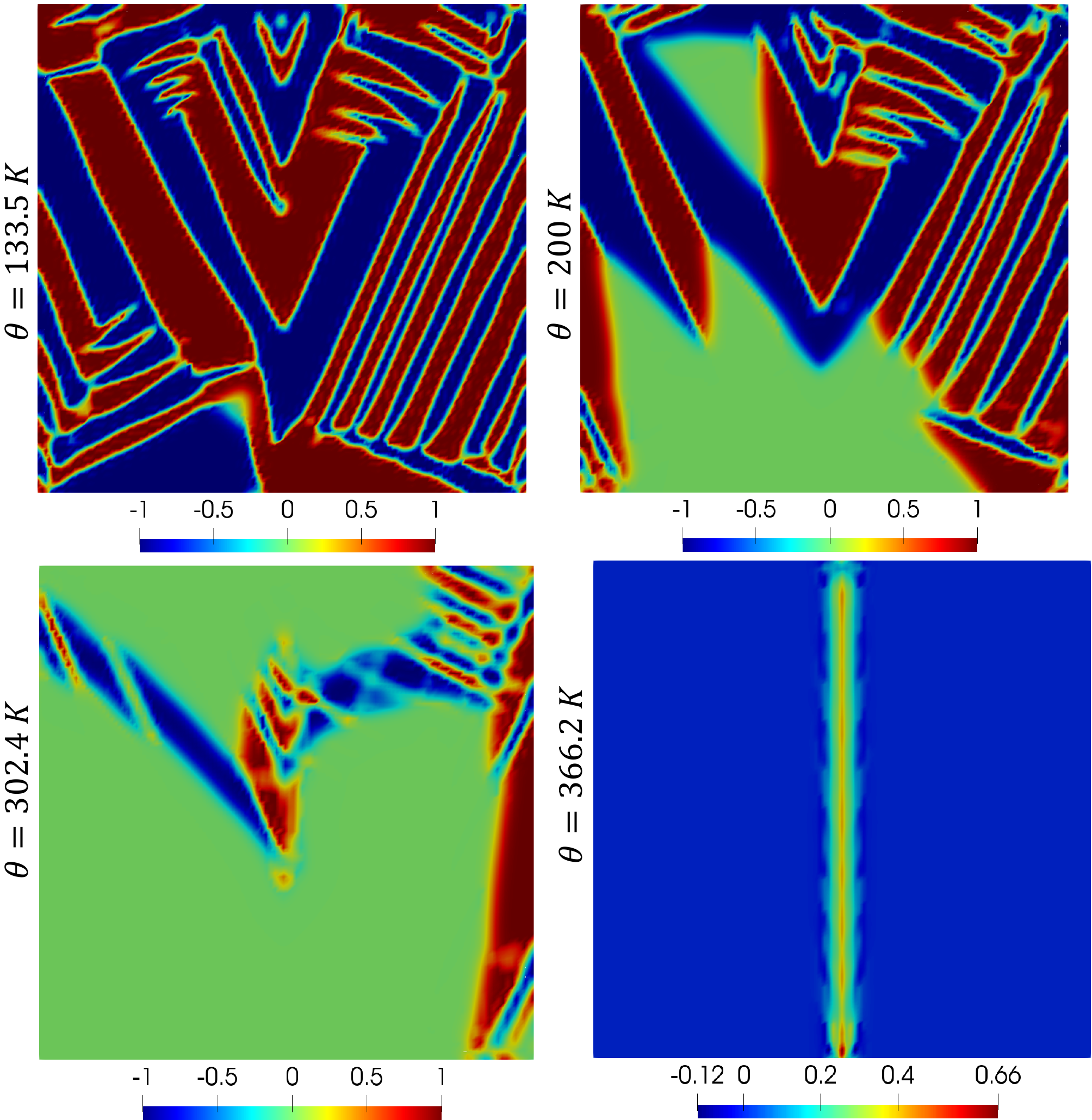}
}
\caption{Microstructure evolution during (a,b) forward, and (c,d) reverse MTs in a $30$ nm $\times$ $30$ nm bicrystal having a symmetric planar GB with $\theta$ variation. We  used $\Delta\gamma_{12}^{gb} = -0.5$ N/m, $\delta_{12}^{gb}=1.5$ nm, and $\vartheta_{12}=40^\circ$. No strain applied.}
\label{fig:sub-firstsdr}
\end{figure}

\begin{figure}[t!]
\centering
\hspace{-8mm}
\subfigure{
  \includegraphics[width=5.7in, height=3.6in] {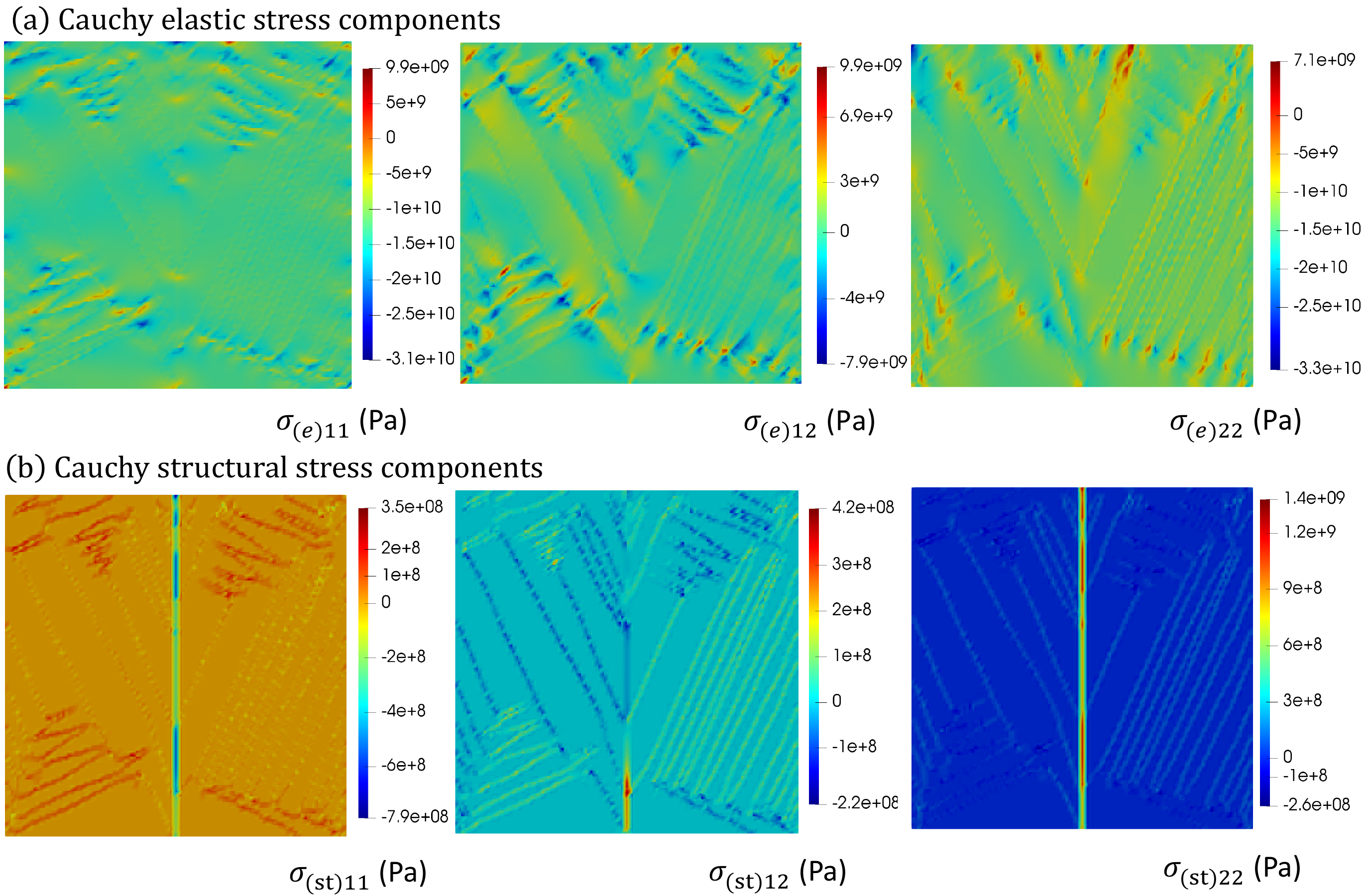}
	}
\hspace{-1mm}
    \subfigure
   {
    \includegraphics[width=4.0in, height=1.6in] {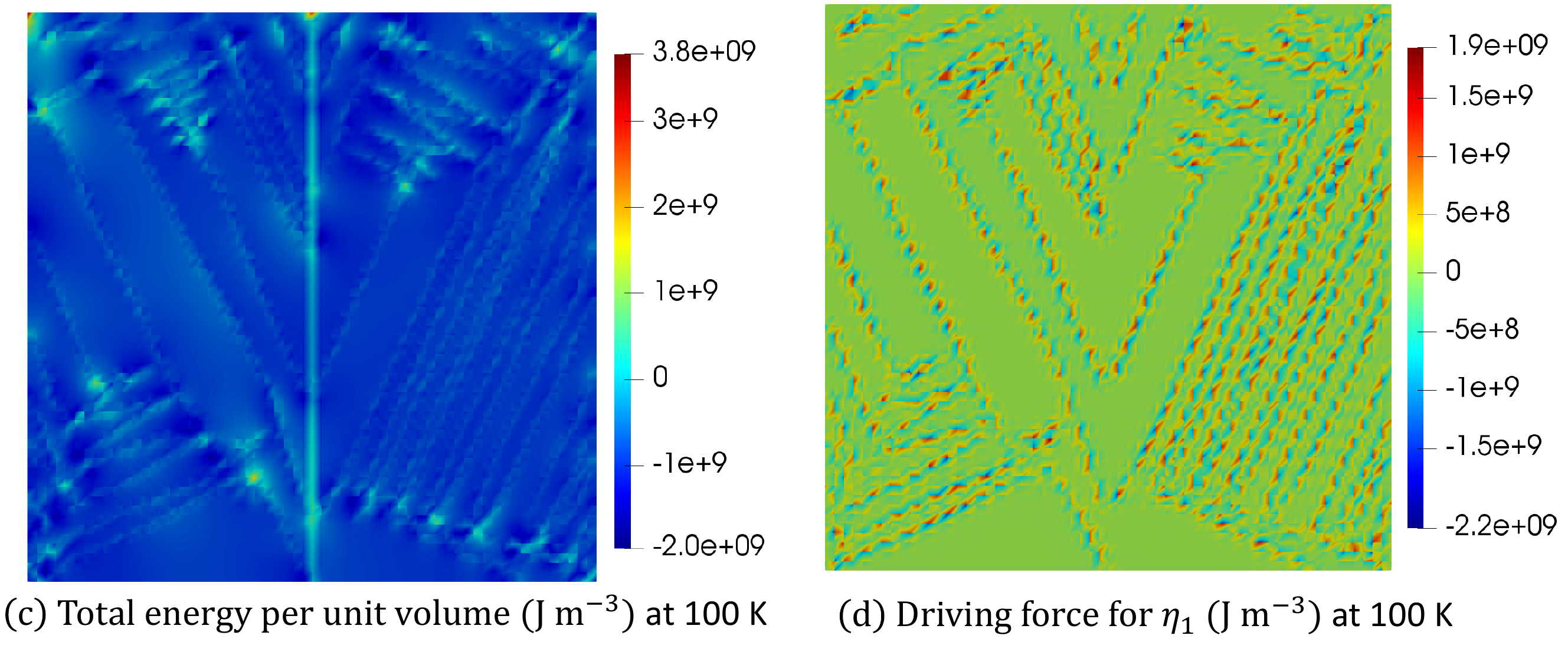}
}
\caption{Plots for the (a) Cauchy elastic stresses $\sigma_{(e)ij}$, (b) Cauchy structural stresses $\sigma_{(st)ij}$, (c) $\rho_0\psi$ - total energy per unit volume of $\Omega_0$,  and (d) driving force $X_{12}^\eta$ at $\theta=100$ K within the martensitic sample shown in Fig. \ref{fig:sub-firstsdr}(a,b) obtained by forward MT.}
\label{fig:sub-firstsStr2}
\end{figure}

\noindent{\bf Microstructures evolution:}

The microstructure evolution at different $\theta$ during the forward and reverse transformations is shown in  Figs. \ref{fig:sub-firsts} and  \ref{fig:sub-firstsdr} for $\delta_{12}^{gb}=1$ nm ($\bar\Delta_{12}=0.667$) and $\delta_{12}^{gb}=1.5$ nm $(\bar\Delta_{12}=1$), respectively, while all other  parameters are identical to that used for plotting Fig. \ref{phasedia}.   We have introduced an equivalent order parameter defined as $\eta_{eq}=2\eta_0(\eta_1-0.5)$ for plotting the evolution of the overall austenite-martensite microstructures in Figs. \ref{fig:sub-firsts}(b,d) and  \ref{fig:sub-firstsdr}(b,d). We note that $\eta_{eq}=0$ in $\sf A$, and it appears as green, $\eta_{eq}=1$ in ${\sf M}_1$, and it appears as dark red, and $\eta_{eq}=-1$ in ${\sf M}_2$, and it appears as dark blue in the plots. 
The ICs for $\eta_0$ and $\eta_1$ would be indicated separately for each case. The subfigures (a) and (b) of Figs. \ref{fig:sub-firsts} and \ref{fig:sub-firstsdr} show the evolution of $\eta_0$ and $\eta_{eq}$, respectively, during the forward MTs when $\theta$ is decreased in steps. The subfigures (c) and (d) show the the evolution of $\eta_0$ and $\eta_{eq}$, respectively, during the reverse MTs when $\theta$ is increased.

\noindent{\bf  Forward MT.} \hspace{3mm} The $\sf PM$  nucleates in the GB regions in a jump-like manner at $\theta_n^f=315.5$ K and $\theta_n^f= 302.4$ K, as shown in Figs. \ref{fig:sub-firsts}(a,b) and  \ref{fig:sub-firstsdr}(a,b), when $\delta_{12}^{gb}=1$ nm (or $\bar\Delta_{12}=0.667$) and $1.5$ nm (or $\bar\Delta_{12}=1$), respectively. The reason for introducing a new symbol $\theta_n^f$ for denoting the nucleation temperature during the forward MT (instead of simply using $\theta_{\sf A\to \sf PM}$ introduced above) is that the austenitic GB might,  depending on the material parameters, directly nucleate $\sf M$ instead of transforming via $\sf PM$. These two temperature values were also reported on the $\sf A\to\sf PM$ curve for the respective $\bar\Delta_{12}$ in  Fig. \ref{phasedia}. In both the layers of $\sf PM$ nucleated in the GB region, $max(\eta_0)=0.1$, and the $\eta_{eq}$ ranges are  $-0.09\leq \eta_{eq}\leq 0$ and $-0.11\leq \eta_{eq}\leq 0$, respectively. Noticing that only a single variant was observed to nucleate in atomistic studies \cite{Dmitriev-2018} (also see \cite{Chen-14}), we have considered an IC on $\eta_1$, which is biased towards one of the variants, say, $\sf M_2$; this resulted in the nucleation of a layer of pre-$\sf M_2$. On choosing an unbiased initial $\eta_1$, say, $\eta_1=0.5$ in the whole sample, we see that the $\sf PM$ nucleates at $\theta_n^f=310.2$ K for the sample with $\delta_{12}^{gb}=1$ nm, which is just $5.3$ K lower than that with the biased IC described above. Such a $\sf PM$ consists of both pre-$\sf M_1$ and pre-$\sf M_2$, and there are clear interfaces between them, which penalizes some energy resulting in the said drop in $\theta_n^f$. However, this paper will consider the IC $0\leq \eta_1(\fg r_0,0)\leq 0.1$ distributed randomly for all the simulations related to the $\sf A\to\sf PM$/$\sf M$ transformations.  Our results clearly show that the higher the GB width, the lower the $\theta_n^f$ (also see Fig. \ref{phasedia} for other values of $\delta_{12}^{gb}$). The samples shown in Figs. \ref{fig:sub-firsts}(a,b) and  \ref{fig:sub-firstsdr}(a,b) however remained in $\sf A$ phase with $\eta_0=0$ everywhere at $315.6$ K and $302.5$ K, respectively. The jump-like nucleation of $\sf PM$ or $\sf M$, also called `the burst transformation', has been observed experimentally on the GBs within the temperature regime of $\sf A$ stability  \cite{Ueda-2001Acta,Ueda-2004ISIJ}.

As  the temperatures of the samples are further reduced from the respective $\theta_n^f$, the respective single-variant pre-$\sf M_2$ layers grow and transform to complete $\sf M$ consisting of two different-scaled and mutually orthogonal plates of two variants in each of the grains, which is driven by a reduction in the strain energy. The twins in these two grains are disoriented by approximately equal to the misorientation angle $40^\circ$. The post-nucleation microstructures in Fig. \ref{fig:sub-firsts}(a)-(b) are shown at $\theta=300$ K, $200$ K, and $65$ K, where at  $65$ K, the entire bicrystal is martensitic. In Fig. \ref{fig:sub-firstsdr}(a)-(b), the post-nucleation microstructures are shown at $\theta=280$ K, $200$ K, and $100$ K, where the sample is entirely in $\sf M$ at $100$ K. At $\theta=300$ K and $\theta=280$ K in the individual cases, we see that the GB region has predominantly transformed to $\sf M_1$ and a much smaller volume fraction of $\sf M_2$.
 By noticing the orientation of the transformation strains (see Figs. \ref{unit_cells}, \ref{xiplots}), we see by the appearance of a large volume fraction of $\sf M_1$, the system tries to maximize the transformation strain. 
 However, as the temperature is further reduced, the volume fraction of $\sf M_2$ also increases, and both the variants evolve with apparently the same volume fraction. In both the samples, two mutually orthogonal twins have developed, where the twin boundaries  far from the GB region are making approximately $45^\circ$ angles with the $\fg c_1$ and $\fg c_2$-axes in the respective grains, which is in accordance with the crystallographic solutions for twins in large single grains undergoing cubic$\to$tetragonal MTs; see, e.g. \cite{Ball-James-87} and Chapter 5 of  \cite{Bha04}. The variant plates meeting in the GB region from both the grains are coherent, which we can explain using the fact that the variants $\sf M_1$ and $\sf M_2$ are compatible about a symmetric tilt GB (see Appendix \ref{compatibilityBiTri} for the proof). The plates about the GB towards the middle of the samples have a larger thickness than the other variant plates. Similar self-accommodating twins arrangements evolution (shown in Fig. \ref{fig:sub-firsts}(a)-(b) and Fig. \ref{fig:sub-firstsdr}(a)-(b)) about the tilt GBs  have also been observed in experiments  \cite{Ueda-2001Acta,Ueda-2004ISIJ} and  atomistic simulations \cite{Qin-2018}.  Finer twins within the thick $\sf M$ plates in the sample shown in Fig. \ref{fig:sub-firstsdr}(a)-(b) at $100$ K are observed, which is driven by a reduction in the elastic energy of the system. Near the top and bottom boundaries of the samples,  irregularities in microstructures are observed due to the boundary effect. Such random nucleation of twin plates away from the GB region was observed in experiments also \cite{Ueda-2001Acta,Ueda-2004ISIJ}. Although the GBs are symmetric in both samples, due to the biased IC on $\eta_1$, we get  asymmetries in the microstructures about the GBs. However, with an unbiased IC,  say, $\eta_1=0.5$ all over the samples, we verified that fully symmetric microstructure evolves in the two grains about the GB at any $\theta\leq\theta_n^f$.

 The components of the Cauchy elastic and structural stresses in the fully $\sf M$ samples are shown in Figs. \ref{fig:sub-firstsstress}(a) and \ref{fig:sub-firstsstress}(b), respectively, when $\delta_{12}^{gb}=1$ nm, and in Figs. \ref{fig:sub-firstsStr2}(a) and \ref{fig:sub-firstsStr2}(b), respectively, they are shown for $\delta_{12}^{gb}=1.5$ nm. The pockets of high stresses are mainly concentrated across the variant-variant interfaces and the GB regions in both the samples, similar to that observed in experiments; see, e.g. \cite{Schuh-13}. The elastic stresses are much higher than the structural stresses, and in some places, the difference may be of two orders of magnitude.  A detailed study with the analytical solution for the elastic and structural stresses across the variant-variant interfaces is given in  \cite{Basak-Levitas-2017Acta}, and the reason for such a large difference in these two stresses for the present kinematic model (Eq. \eqref{utilde}) can be seen therein and also in \cite{Basak-Levitas-2018JMPS,Basak-Levitas-2019CMAME}. In the realistic solids, such high stresses would relax by nucleation of defects such as dislocations, and cracks, as observed in, say, e.g. \cite{Schuh-13}. However, our model does not account for such coupled phenomena. 
 The total free energy  per unit volume given by Eq. \eqref{MF0}, and the driving force $X_{12}^\eta$ for variant-variant transformation given by Eq. \eqref{kineqs20b}  are shown across the fully $\sf M$ samples in Figs. \ref{fig:sub-firstsstress}(c) and \ref{fig:sub-firstsstress}(d), respectively, when $\delta_{12}^{gb}=1$ nm, and in Figs. \ref{fig:sub-firstsStr2}(c) and \ref{fig:sub-firstsStr2}(d), respectively, when $\delta_{12}^{gb}=1.5$ nm. Since $X_{0M}=0$ in  fully $\sf M$ samples, we do not show its plot.
Although the large values of the driving force $X_{12}^\eta$ are concentrated  across the interfaces, its average over the area of the respective samples is vanishing (note the sign changes), and hence the microstructure does not evolve further. We note that the highest energy densities are concentrated at the right upper corner of the sample with $\delta^{gb}_{12}=1$ nm (see Fig. \ref{fig:sub-firstsstress}(c)) and in a region near the GB and close to the bottom surface of the left grain for the sample with $\delta^{gb}_{12}=1.5$ nm (see Fig. \ref{fig:sub-firstsStr2}(c)). This information we will use it in the next paragraph to study the nucleation of $\sf PM$ or $\sf A$ during the reverse MT.


%
%

%
%
%
 \noindent{\bf Reverse MT.}\hspace{3mm} We now study the reverse transformations, i.e. $\sf M\to \sf PM$ and $\sf A$ transformation, from the fully martensitic samples. It is to be clarified that in Fig. \ref{phasedia}, the reverse transformation of the $\sf PM$/$\sf M$ layer nucleated at $\theta^f_n$ in the GB region and its small neighbourhood only was studied.
 However, we consider the fully martensitic samples obtained at $65$ K and $100$ K  shown in Figs. \ref{fig:sub-firsts}(a,b) and \ref{fig:sub-firstsdr}(a,b) as the reference bodies ($V_0$) when $\delta_{12}^{gb}=1$ nm and $1.5$ nm, respectively.
We assume an IC for $\eta_0(\fg r_0,t=0)=0.98$ distributed uniformly in both the samples, whereas the same $\eta_1$ distribution in the complete $\sf M$ sample obtained during the forward transformation is used. The same IC for $\eta_0$ and the similar IC for $\eta_1$ (i.e. as it is in the $\sf M$  sample) for studying the reverse MTs in fully martensitic samples are used for all the examples related to the bicrystals and tricrystals.
The microstructures evolution during the reverse transformations for these two samples is shown in Figs.  \ref{fig:sub-firsts}(c,d) and \ref{fig:sub-firstsdr}(c,d). The temperatures at which the $\sf PM$, i.e. incomplete $\sf A$ nucleated are $\theta_n^r=82.2$ K and $\theta_n^r=133.5$ K, respectively, which are obviously in the regime of stability of $\sf M$. In  Fig.  \ref{fig:sub-firsts}(c,d), the $\sf PM$ nucleated at the right upper corner of the sample where the total energy density was the highest in the fully $\sf M$ sample,  as shown in  \ref{fig:sub-firstsstress}(c) (see the above paragraph). When $\delta_{12}^{gb}=1.5$ nm, the $\sf PM$ nucleated at a region in the lower part of the left grain, which is a region with high total energy density, as shown in Fig. \ref{fig:sub-firstsStr2}(c). Such high energetic regions provide the  driving force required for nucleation of $\sf PM$ and $\sf A$ on heating of the martensitic samples.
Post $\sf PM$ nucleation, the temperature of both the samples is further increased, and the microstructures at different $\theta$ are shown.
 A significant portion of both the samples is transformed to $\sf A$ at $200$ K. The third of the plots in Figs.  \ref{fig:sub-firsts}(c,d) and \ref{fig:sub-firstsdr}(c,d) depict the microstructures at the respective $\theta_n^f$, the temperature at which the $\sf PM$/$\sf M$ nucleated during the forward MT. The  microstructures of Fig. \ref{fig:sub-firsts}(a,b) and that of Fig. \ref{fig:sub-firsts}(c,d) at $\theta_n^f=315.5$ K are significantly different. At that $\theta$, the GB region contains a thin layer of pre-$\sf M_2$ during the forward MT. In contrast, the GB region contains a thick M layer containing both variants, and a significant volume fraction of M is present in the interior of the grains during the reverse MT.  A similar conclusion is  drawn for the other sample as well. The last plots in Figs.  \ref{fig:sub-firsts}(c,d) and \ref{fig:sub-firstsdr}(c,d) show the microstructures with thin layers of $\sf PM$ containing both pre-$\sf M_1$ and pre-$\sf M_2$ only on the GB regions at temperatures indicated therein. The samples fully transform to $\sf A$ (i.e. $\eta_0=0$) at $0.1$ K above the indicated temperatures. The experimental study of \cite{Ueda-2004ISIJ}  in Fe-Ni bicrystal with a tilt GB also showed that the reverse transformation starts at high energetic regions within the grains, usually far from the GB. It was also reported that the variant plates far from the GBs transform to A first on heating the sample further, and the twins about the GB finally disappear to transform the sample back to austenite. Our simulation results are hence in good agreement with the experimental observations of \cite{Ueda-2004ISIJ}.

\begin{figure}[t!]
\centering
\hspace{-8mm}
\subfigure[$\eta_0$ plots during forward MT]{
  \includegraphics[width=3.0in, height=3.0in] {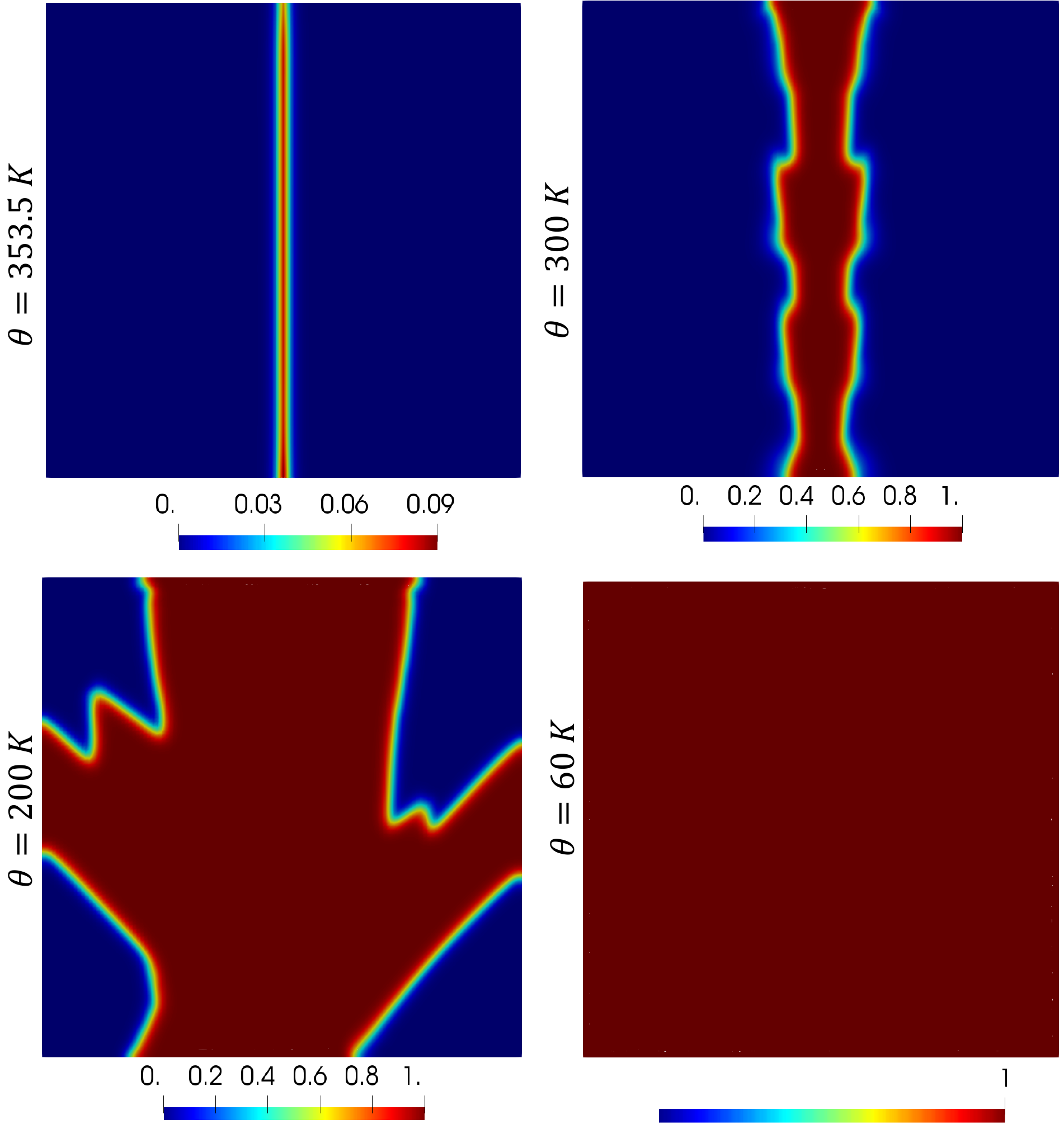}
	}\hspace{1mm}
    \subfigure[$\eta_{eq}$ plots during forward MT]{
    \includegraphics[width=3.0in, height=3.0in] {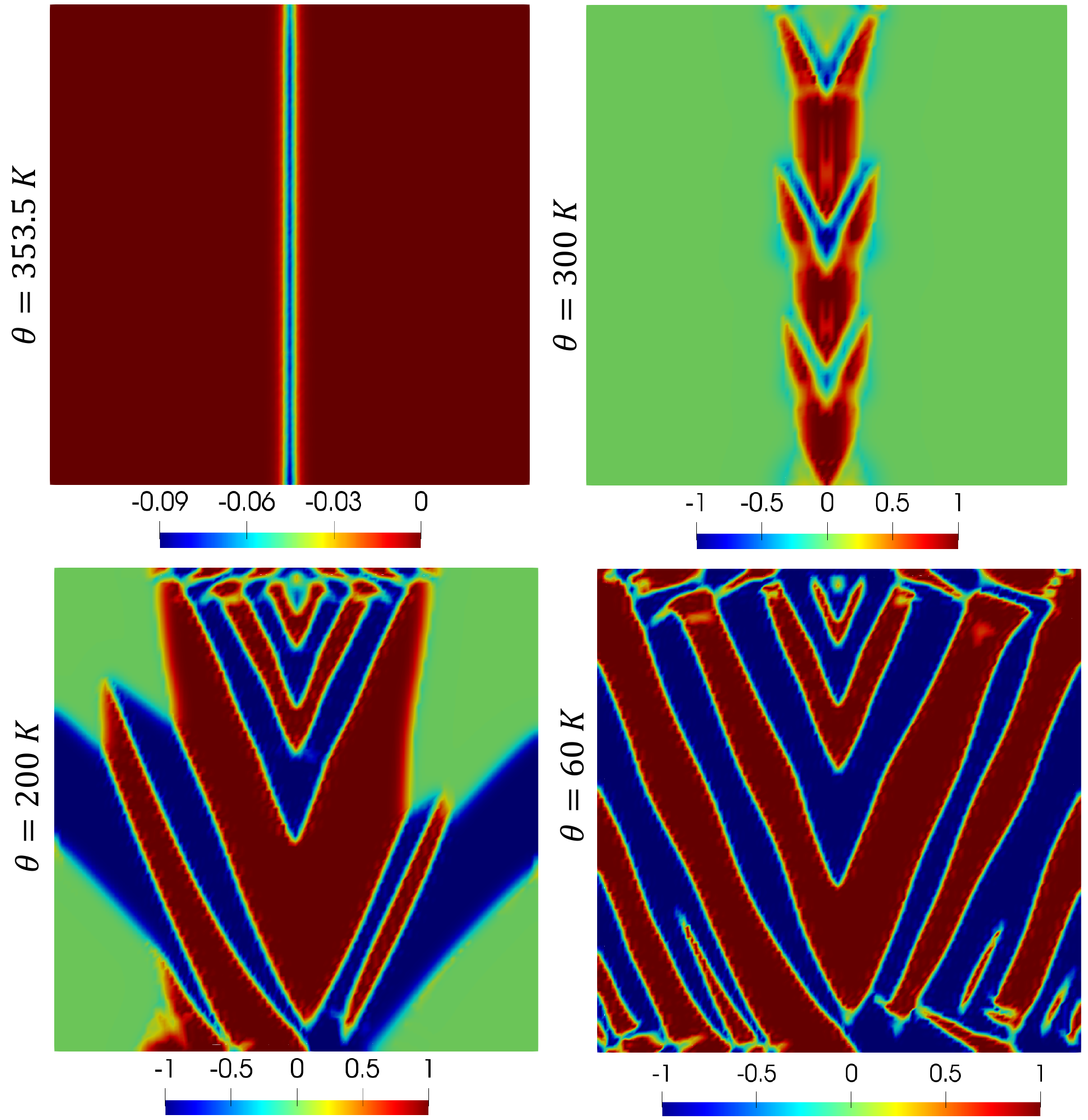}
}
\subfigure[$\eta_0$ plots during reverse MT]{
  \includegraphics[width=3.0in, height=3.0in] {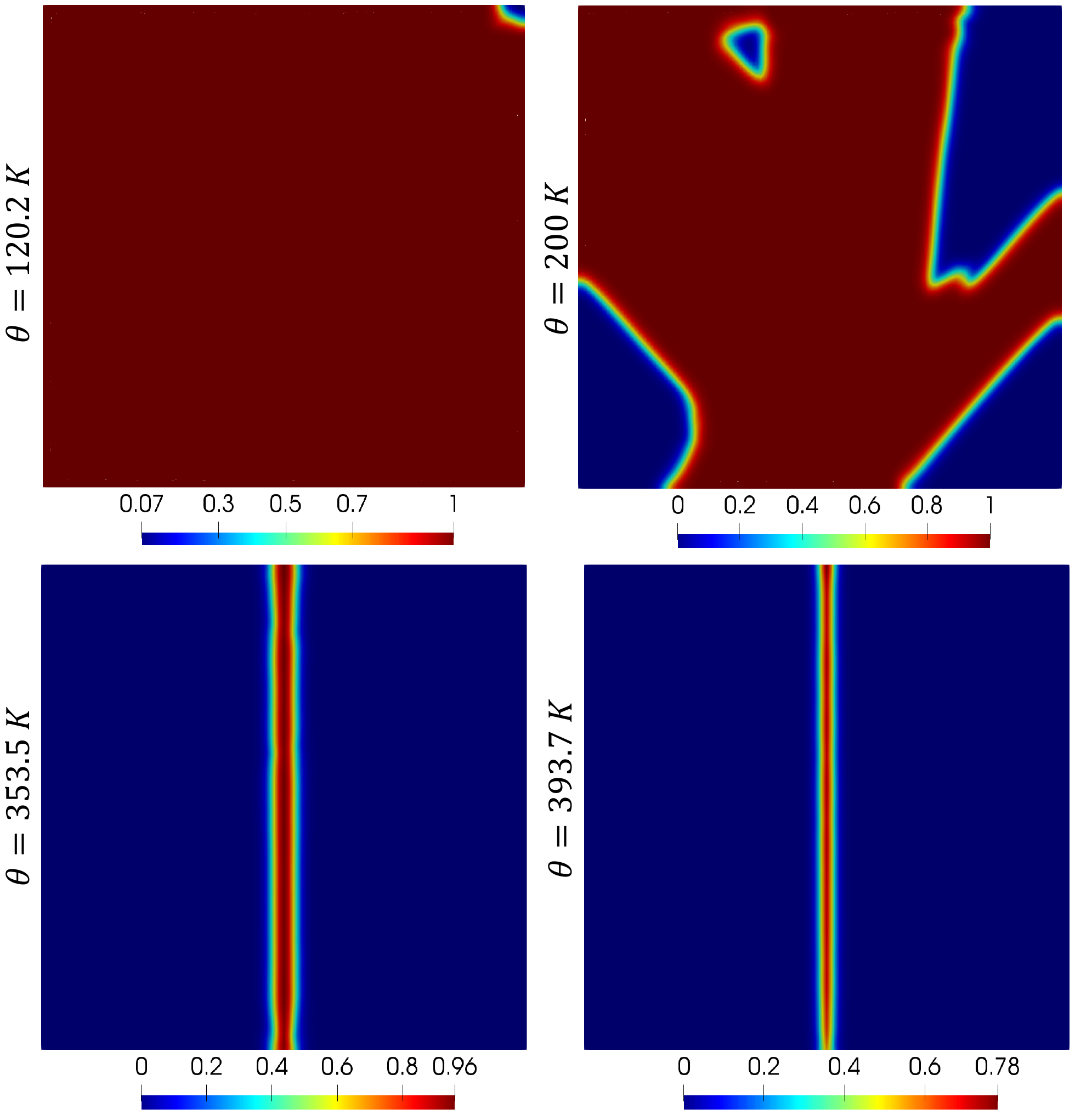}
	}\hspace{1mm}
    \subfigure[$\eta_{eq}$ plots during reverse MT]{
    \includegraphics[width=3.0in, height=3.0in] {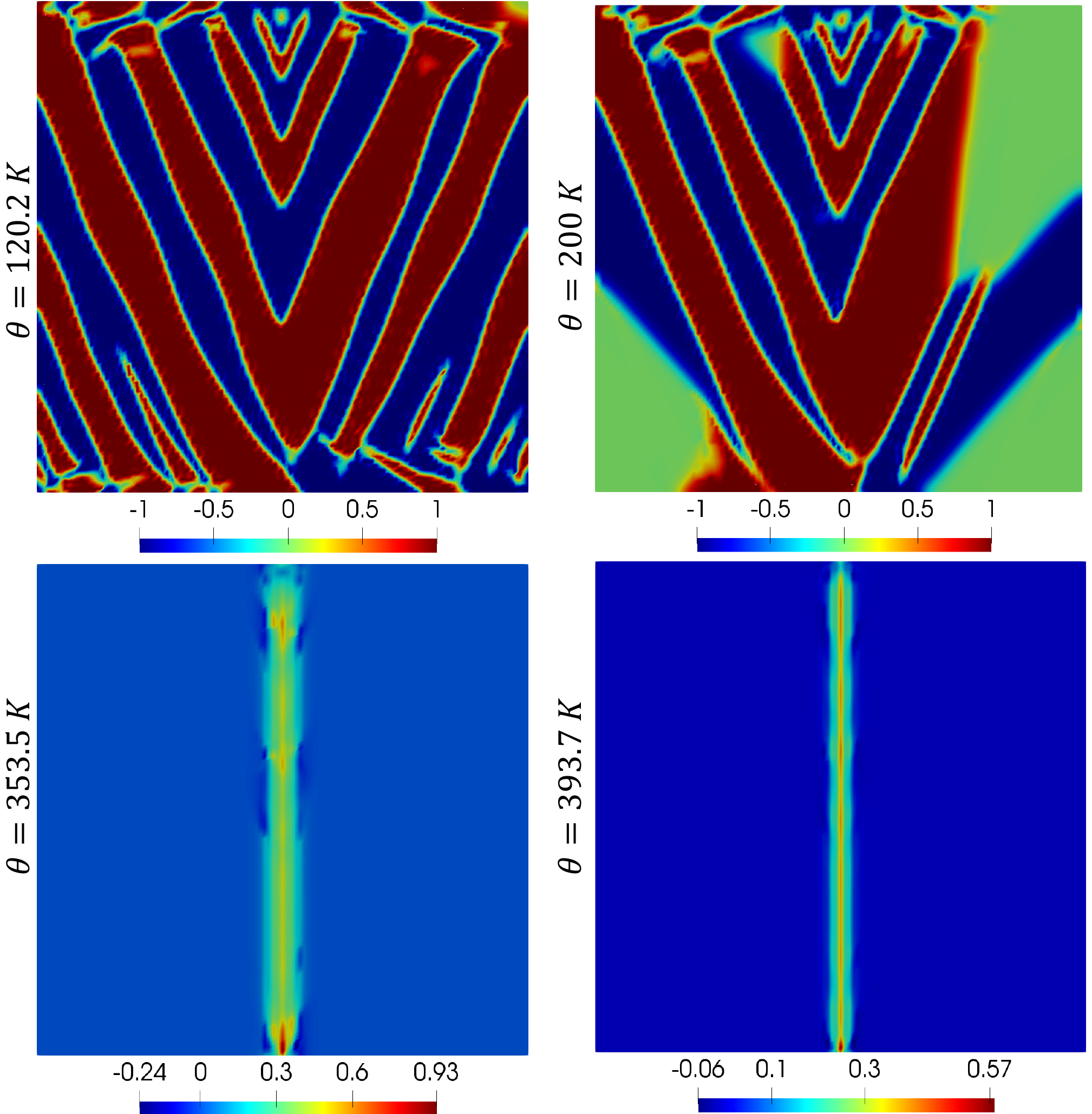}
}
\caption{Microstructure evolution during (a,b) forward, and (c,d) reverse MTs in a $30$ nm $\times$ $30$ nm bicrystal having a symmetric planar GB with $\theta$ variation. We took $\Delta\gamma_{12}^{gb} = -0.7$ N/m,  $\delta_{12}^{gb}=1$ nm, and $\vartheta_{12}=40^\circ$. No strain applied.}
\label{BivaryGbenergy}
\end{figure}

%

\subsubsection{Effect of change in GB energy during MTs} 
\label{sussub2}
We now study the effect of change in GB energy $\Delta\gamma_{12}^{gb}$ considering the same austenitic bicrystal  with a symmetric planar tilt GB considered in Sec \ref{sussub1}  for $\Delta\gamma_{12}^{gb} = -0.7$ N/m, while all other parameters are identical, i.e.  $\delta_{12}^{gb}=1$ nm, $\vartheta_1=-\vartheta_2=20^\circ$ ($\vartheta_{12}=40^\circ$), and  no strains applied. The results for  $\Delta\gamma_{12}^{gb} = -0.5$ N/m and $\Delta\gamma_{12}^{gb} = -0.7$ N/m are finally compared. The martensitic GB energies in these two cases are obviously $\gamma_{gb}^M = 0.4$ N/m and $\gamma_{gb}^M = 0.2$ N/m, respectively.  

\noindent{\bf Forward MT.} \hspace{3mm}  Figures  \ref{BivaryGbenergy}(a,b) show the evolution of the microstructures with a variation of $\theta$ when  $\Delta\gamma_{12}^{gb} = -0.7$ N/m during the forward MT. Thin layers of pre-$\sf M_2$, similar to the sample for $\Delta\gamma_{12}^{gb} = -0.5$ N/m, appear jump-like at $\theta_n=353.5$ K on the GB. The reason for pre-$\sf M_2$ is already explained in Sec \ref{sussub1}.
 However, the samples were fully austenitic at $353.6$ K. A larger reduction in the GB energy due to $\sf A\to\sf M$ transformations increases the temperature for nucleation of $\sf PM/\sf M$. The sample temperature  is further reduced, and the subsequent stationary microstructures at $\theta=300$ K, $200$ K, and $60$ K are shown in Fig.  \ref{BivaryGbenergy}(a,b). The microstructures at $\theta=300$ K are similar in both  samples. More than half of both the samples are transformed to $\sf M$ phase consisting of mostly the self-accommodating twins about the GB at $200$ K. At $\theta=60$ K, the entire sample is martensitic, consisting of both the variants with almost equal volume fraction, when $\Delta\gamma_{12}^{gb}=-0.7$ N/m. The overall microstructures in both the $\sf M$ samples are qualitatively similar, except that the average thickness of the variant plates for the case with $\Delta\gamma_{12}^{gb}=-0.5$ N/m is smaller than the case when $\Delta\gamma_{12}^{gb}=-0.7$ N/m. The elastic and structural stresses in the fully martensitic sample are concentrated mainly across the GB and the variant-variant interfaces in case of $\Delta\gamma_{12}^{gb}=-0.7$ N/m, which is similar to the other sample and hence not shown here.
 

\noindent{\bf Reverse MT.} \hspace{3mm} We have studied the reverse MT using the sample obtained at $60$ K shown in Fig.  \ref{BivaryGbenergy}(a,b) when  $\Delta\gamma_{12}^{gb} = -0.7$ N/m, whereas the result for $\Delta\gamma_{12}^{gb} = -0.5$ N/m was already discussed in Sec. \ref{sussub1}.  As the temperature of the sample is increased from $60$ K, a small region of $\sf PM$ appears in a jump-like manner at the upper right corner at $\theta_n^r=120.2$ K, shown in Fig.  \ref{BivaryGbenergy}(c,d), where the total energy density was the maximum in the $\sf M$ sample at $60$ K. A significant portion of both the grains transform to $\sf A$ at $200$ K. At $\theta=\theta_n^f=353.5$ K, only a small volume fraction of $\sf M$ remains in the GB region, which is of a slightly higher volume fraction than that obtained during the forward MT at the same temperature. On further increase in $\theta$, the $\sf M$ layer gets thinner and eventually transforms to $\sf PM$. The $\sf PM$ layer on the GB region at $393.7$ K is shown in Fig.  \ref{BivaryGbenergy}(c,d), which disappears jump-like at $393.8$ K to transform the sample into austenite.  Thus we see that a higher $\Delta\gamma^{gb}_{12}$ significantly increases the austenite-finish temperature in the bicrystals (compare with Fig. \ref{fig:sub-firsts}(c,d)). 

We cannot compare our results for the effect of $\Delta\gamma_{12}^{gb}$ on the transformation temperatures and the microstructures with any atomistic or experimental result due to the scarcity of relevant data in the literature. However, the overall transformation kinetics during the forward and reverse MTs are qualitatively in agreement with the experimental and atomistic results \cite{Ueda-2001Acta,Ueda-2004ISIJ,Qin-2018}.

\begin{figure}[t!]
\centering
\hspace{-8mm}
\subfigure[$\eta_0$ plots during forward MT]{
  \includegraphics[width=3.0in, height=3.00in] {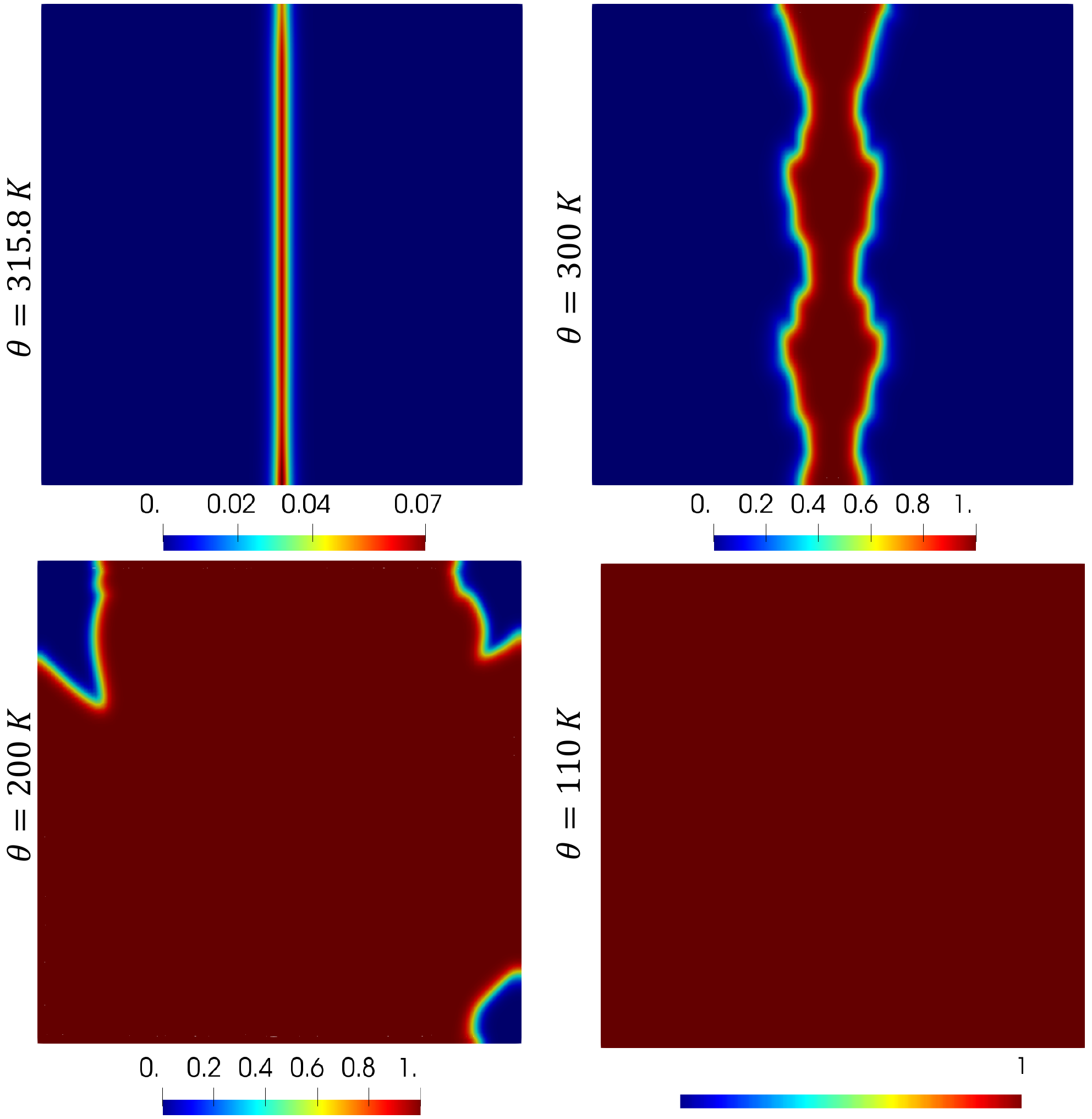}
	}\hspace{1mm}
    \subfigure[$\eta_{eq}$ plots during forward MT]{
    \includegraphics[width=3.0in, height=3.0in] {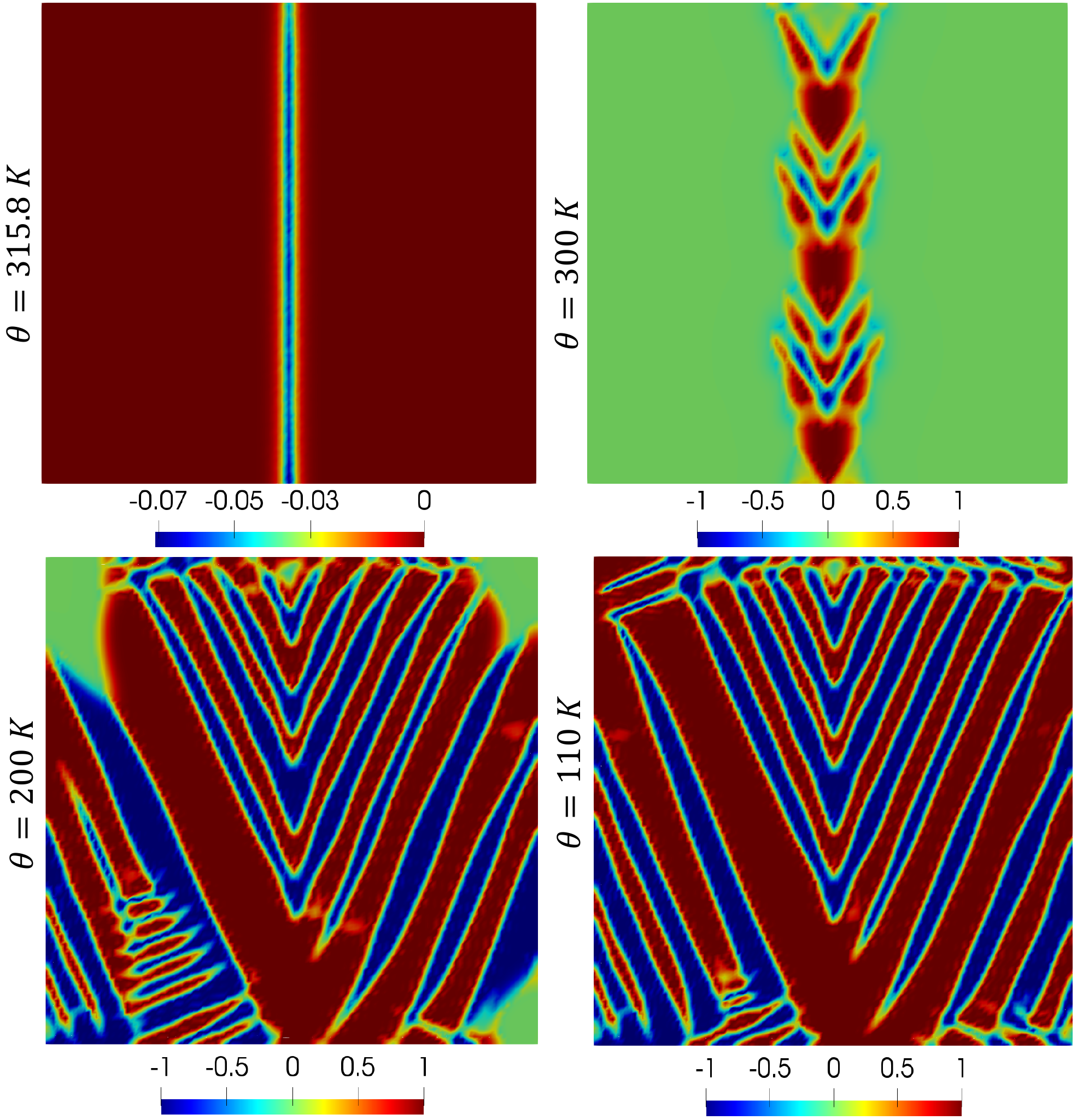}
}
\subfigure[$\eta_0$ plots during reverse MT]{
  \includegraphics[width=3.0in, height=3.0in] {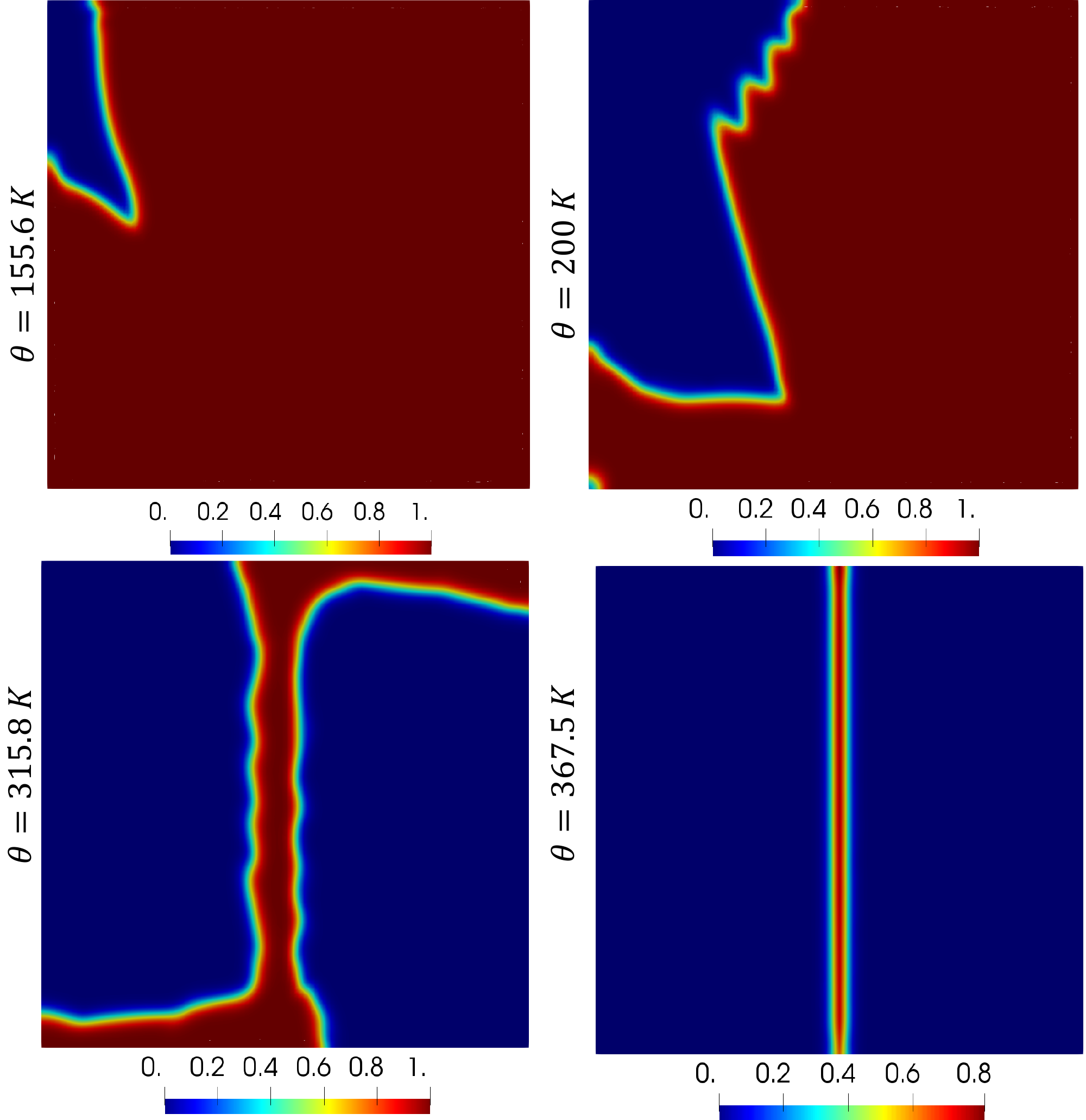}
	}\hspace{1mm}
    \subfigure[$\eta_{eq}$ plots during reverse MT]{
    \includegraphics[width=3.0in, height=3.0in] {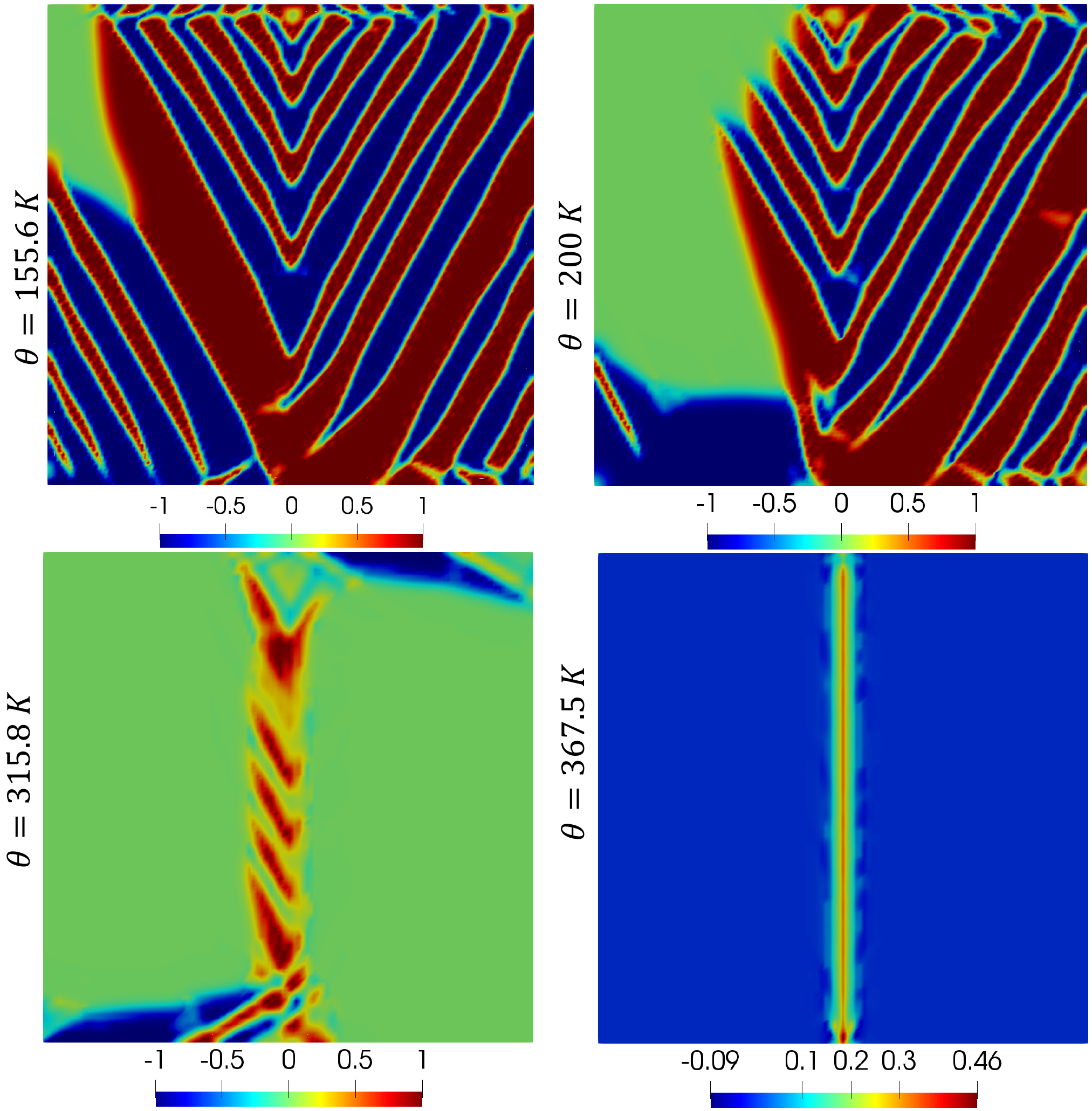}
}
\caption{Microstructure evolution during (a,b) forward, and (c,d) reverse MTs in a $30$ nm $\times$ $30$ nm bicrystal having a symmetric planar GB with $\theta$ variation. We considered $\Delta\gamma_{12}^{gb} = -0.5$ N/m,  $\delta_{12}^{gb}=1$ nm, and $\vartheta_{12}=30^\circ$. No strain applied.}
\label{BivaryGbmisor}
\end{figure}

\subsubsection{Effect of GB misorientation} 
\label{sussub3}
We now consider another  $30$ nm $\times$ $30$ nm sample with a symmetric tilt planar GB where the grain orientations are $\vartheta_1=-\vartheta_2=15^\circ$, and hence the misorientation is $\vartheta_{12}=30^\circ$. The other parameters are the same as that considered for the case shown in Fig. \ref{fig:sub-firsts}:  $\delta_{12}^{gb}=1$ nm,  $\Delta\gamma_{12}^{gb} = -0.5$ N/m, and no strain is applied. The result for this sample is shown in Fig. \ref{BivaryGbmisor}, which is compared with that of Fig. \ref{fig:sub-firsts}. The differences in these two cases are only in the misorientation angle and the orientation of the Bain strains ($40^\circ$ used in Fig. \ref{fig:sub-firsts}). Although the GB misorientations are different for these two samples, the GB energies $\gamma_{gb}^A$ and $\gamma_{gb}^M$ are considered to be identical. We can justify this by
noticing that isotropic GBs with different misorientations can have the same energy; see  \cite{Mishin-2005Acta} for the examples in NiAl alloy.

\noindent{\bf Forward MT.} \hspace{3mm} 
Comparing the Figs. \ref{fig:sub-firsts} and \ref{BivaryGbmisor}, we see that a $10^\circ$ reduction in the misorientation, keeping all other parameters the same, increases the nucleation temperature just by  $0.3$ K. The  thin $\sf PM$ layers appeared with $max(\eta_0)=0.1$ and $max(\eta_0)=0.07$ in the samples having $\vartheta_{12} =40^\circ$ and $30^\circ$, respectively. At $300$ K, no significant difference in the microstructures is observed. At $200$ K, the volume fraction of the $\sf M$ phase is, however, significantly larger when $\vartheta_{12}=30^\circ$ than in the other sample. The sample with $\vartheta_{12}=30^\circ$  is completely martensitic at $\theta=110$ K, while the sample having $\vartheta_{12}=40^\circ$ had to be cooled down to $60$ K to transform it to fully $\sf M$ (compare between Figs.  \ref{fig:sub-firsts} and \ref{BivaryGbmisor}). The martensitic plates are finer for the sample with smaller misorientation.

\noindent{\bf Reverse MT.}
The results for the reverse MT for $\vartheta_{12}=30^\circ$ are shown in Fig. \ref{BivaryGbmisor}(c,d). The $\sf A$ phase nucleated barrierlessly at $\theta_n^r=155.6$ K at the top-left  corner of the left grain, as shown in Fig. \ref{BivaryGbmisor}(c,d). The right grain, however, remains completely martensitic at that $\theta$. The $\sf A$-$\sf M$ phase boundary then propagates with the left grain on further increase in $\theta$ to increase the volume fraction of $\sf A$. The right grain remains in the $\sf M$ phase even at $\theta=200$ K, as shown in Fig. \ref{BivaryGbmisor}(c,d). The $\sf M$ phase is still observed in the GB region and at the top and bottom surfaces of the sample at $\theta=\theta_n^f=315.8$ K, having a larger volume fraction of $\sf M$ than that obtained during the forward MT (see Fig. \ref{BivaryGbmisor}). The entire sample transforms to $\sf A$ barrierlessly at $\theta=367.6$ K, whereas this happened at $\theta=365.6$ K for the other sample with $\vartheta_{12}=40^\circ$.  The thin layer of $\sf PM$ in the GB region observed at $367.5$ K and with $max(\eta_0)=0.8$ is shown in \ref{BivaryGbmisor}(c,d). 

We thus conclude that a $5^\circ$ difference in the orientation of the transformation strains, keeping all other parameters identical, changes the nucleation of $\sf M/\sf PM$ and the $\sf A$-finish temperatures marginally. However, the microstructure evolution significantly differs.

\begin{figure}[t!]
\centering
\hspace{-8mm}
\subfigure[$\eta_0$ plots during forward MT]{
  \includegraphics[width=3.0in, height=3.00in] {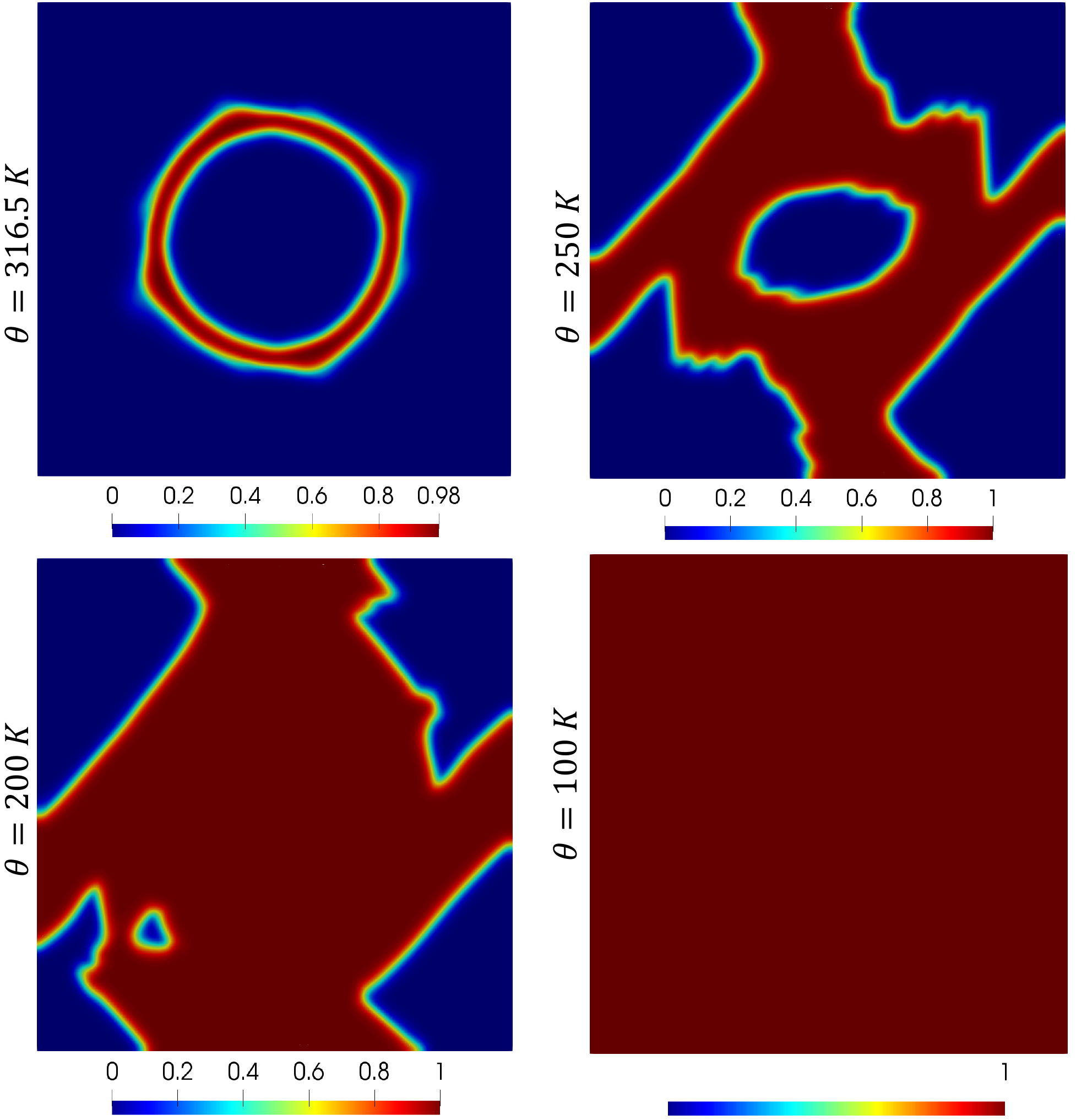}
	}\hspace{1mm}
    \subfigure[$\eta_{eq}$ plots during forward MT]{
    \includegraphics[width=3.0in, height=3.00in] {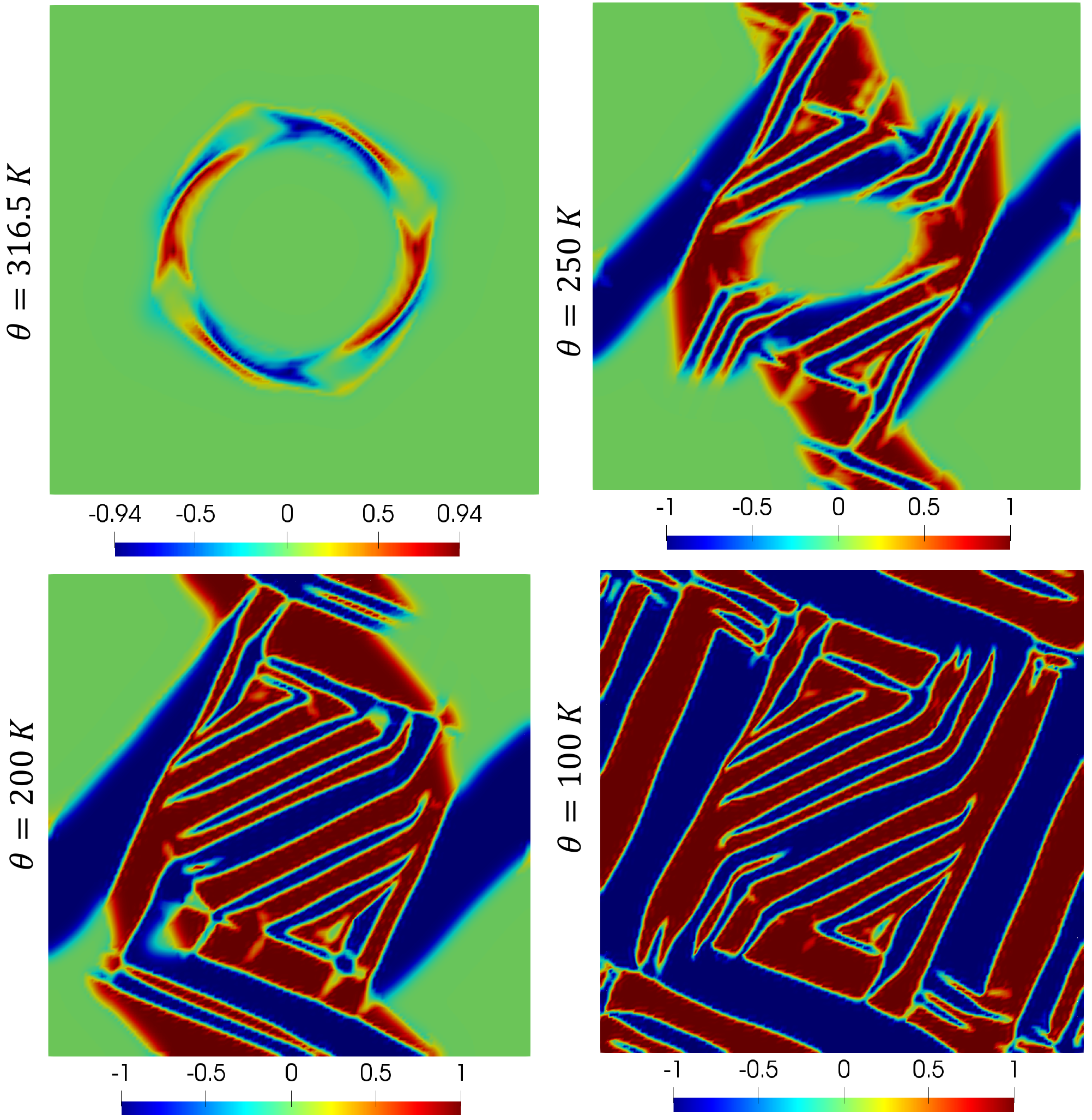}
}
\subfigure[$\eta_0$ plots during reverse MT]{
  \includegraphics[width=3.0in, height=3.00in] {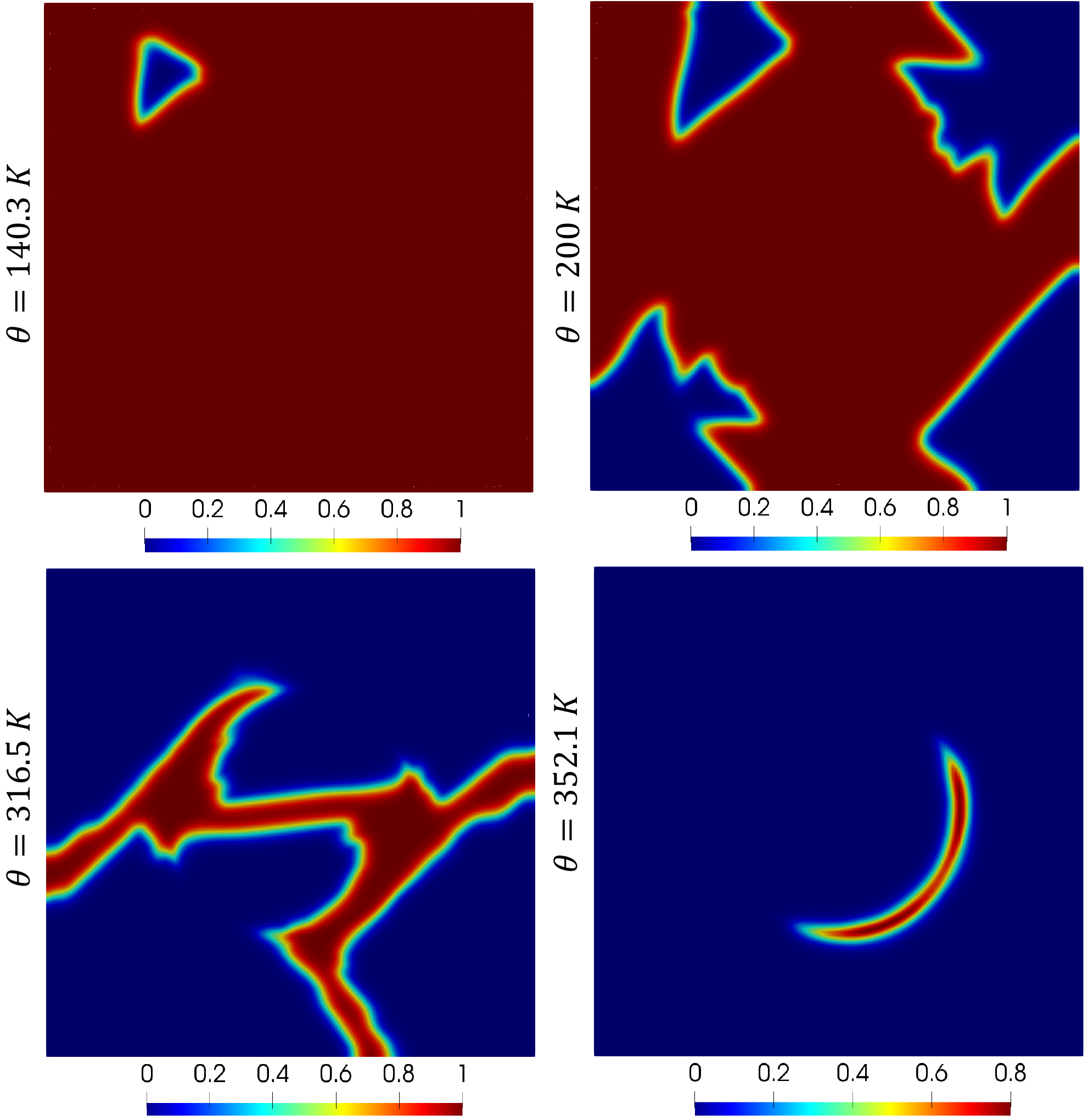}
}
    \subfigure[$\eta_{eq}$ plots during reverse MT]{
    \includegraphics[width=3.0in, height=3.00in] {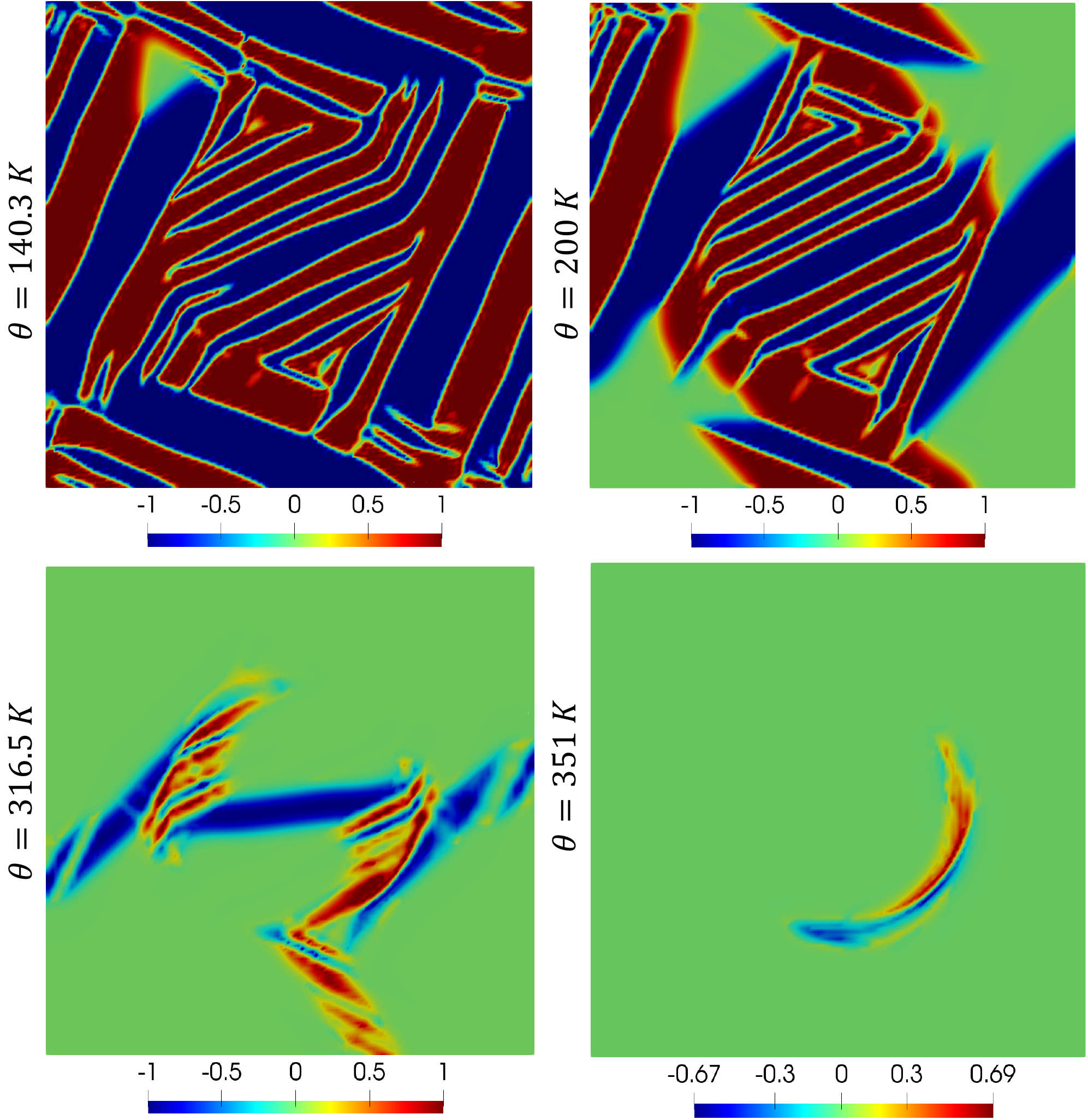}
}
\caption{Microstructure evolution during (a,b) forward, and (c,d) reverse MTs in a $30$ nm $\times$ $30$ nm bicrystal having a circular GB of $15$ nm radius with $\theta$ variation. We used $\Delta\gamma_{12}^{gb} = -0.5$ N/m,  $\delta_{12}^{gb}=1$ nm, and $\vartheta_{12}=40^\circ$. No strain applied.}
\label{BivaryGbCurvature}
\end{figure}

\subsubsection{Effect of GB curvature}
\label{sussub5}
 We now consider a stress-free $\sf A$ bicrystal with a circular symmetric tilt GB of $15$ nm radius embedded inside the $30$ nm $\times$  $30$ nm sample as shown in Fig. \ref{xiplots}(b) to study the role of GB curvature. We take $\vartheta_1=-\vartheta_2=20^\circ$, i.e. $\vartheta_{12}=40^\circ$, and all other parameters are identical to the bicrystal with a planar GB considered in Fig. \ref{fig:sub-firsts}: $\Delta\gamma_{12}^{gb} = -0.5$ N/m,  $\delta_{12}^{gb}=1$ nm,  and no strain is applied. The microstructure evolution during the forward and reverse MTs are shown in Fig. \ref{BivaryGbCurvature}.
 
\noindent{\bf Forward MT.} \hspace{3mm} The $15$ nm radius of curvature increases the nucleation temperature $\theta_n^f$ only by $1$ K (compare between Figs. \ref{BivaryGbCurvature}(a,b) and \ref{fig:sub-firsts}(a,b)). However, the nucleated layer in the curved GB region is fully martensitic, containing both the variants with equal volume fractions in contrast to the sample with a planar GB where it was a pre-$\sf M_2$ layer only (see Fig. \ref{fig:sub-firsts}). As $\theta$ is further reduced, the martensitic plates grow from the GB regions into both grains. The embedded grain, however, transforms much faster than the outer grain due to the additional driving force contributed by the $\sf A$-$\sf M$ interface curvature, and we see that at  $200$ K itself, it is fully martensitic, whereas the outer grain is partially transformed (see Fig. \ref{BivaryGbCurvature}(a,b). On the other hand, both the grains were partially transformed in the sample with planar GB at $200$ K. On the further decrease in $\theta$, a fully martensitic sample is seen at $100$ K.  In the outer grain, two sets of mutually orthogonal arrangements of $\sf M_1$ and $\sf M_2$ plates are observed away from the GB region at $65^\circ$ and $25^\circ$ angles with respect to the $\fg e_1$-axis, which agrees  with the crystallographic theory in an isolated large grain.  The average plate thickness of the embedded grain is smaller than that of the outer grain. A continuous change in orientation of the GB region results in a very complex microstructure there. The GB curvature was shown to promote MTs with nucleation of multiple variants in the PF study of Heo and Chen \cite{Chen-14}, which is similar to our results. 

\noindent{\bf Reverse MT.} \hspace{3mm} A small region of $\sf A$ nucleates near the top surface of the outer grain at $\theta_n^r=140.3$ K on heating the martensitic sample. On further increase in $\theta$, a significant volume fraction of the outer grain transforms to $\sf A$ at $200$ K, while the inner grain is still completely martensitic. As the temperature is raised to $\theta=\theta_n^f=316.5$ K, almost half of the GB region transforms to $\sf A$, and only a small volume fraction of the inner and outer grains are still martensitic. At $351$ K, two thin layers of $\sf PM$ still remain in the GB region, which  vanish in a jump-like manner at $351.1$ K.

Although the circular GB raises the martensitic nucleation temperature $\theta_n^f$ by $1$ K only compared to the planar GB, the austenite finish temperature is decreased significantly, i.e. by $14.5$ K.


\begin{figure}[t!]
\centering
\hspace{-8mm}
\subfigure[$\eta_0$ plots during forward MT]{
  \includegraphics[width=3.2in, height=3.20in] {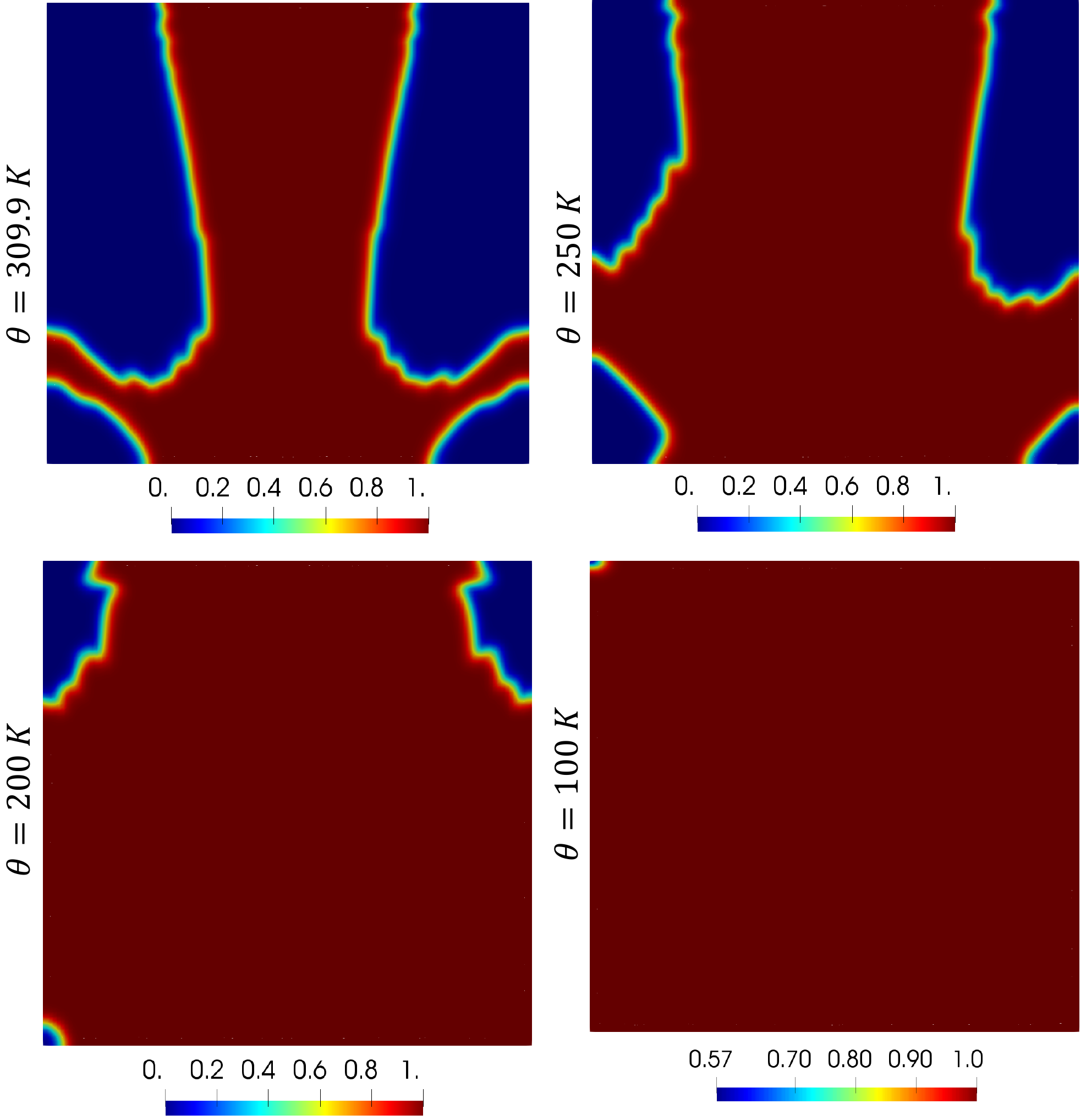}
	}\hspace{1mm}
    \subfigure[$\eta_{eq}$ plots during forward MT]{
    \includegraphics[width=3.2in, height=3.20in] {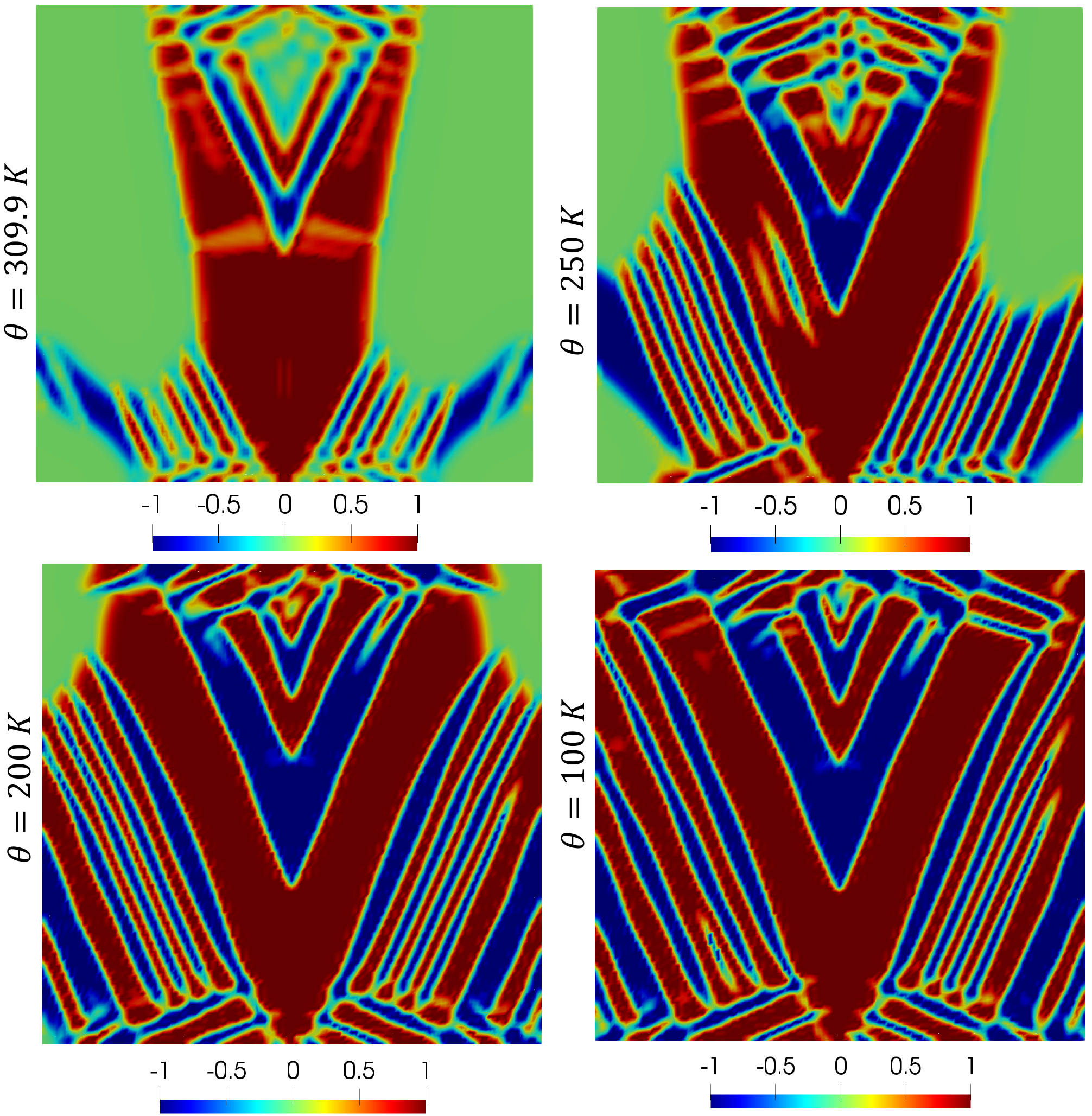}
}
\caption{ Microstructure evolution during (a,b) forward MT in a $30$ nm $\times$ $30$ nm bicrystal having a symmetric planar GB with $\theta$ variation. We used $\Delta\gamma_{12}^{gb} = -0.5$ N/m, $\delta_{12}^{gb}=1$ nm, $\vartheta_{12}=40^\circ$, and applied $\varepsilon_1=0.04$.}
\label{sBivarystrain}
\end{figure}
\subsubsection{Effect of applied strain} 
\label{sussub4}
We consider the same initial austenitic bicrystal with a planar symmetric tilt GB having $\delta_{12}^{gb}=1$ nm considered in Fig. \ref{fig:sub-firsts} (see Sec. \ref{sussub1}), which is now subjected to a normal strain $\varepsilon_1=0.04$ on the right vertical surface, having the outward unit normal parallel to $-\fg e_2$-axis. All other parameters are identical, i.e. $\Delta\gamma_{12}^{gb} = -0.5$ N/m,  and  $\vartheta_{12}=-40^\circ$. Recall that in Fig. \ref{fig:sub-firsts}, no external strain was applied.  We have only shown the results for the forward MT in the present case. We have applied $\varepsilon_1=0.04$ on the mentioned surface at $t=0$ and hold it fixed for all $t>0$, while the normal displacement to all other surfaces is zero for all the time. All the surfaces are free of the tangential component of the traction.
A complete $\sf M$ phase nucleates at $\theta_n^f=309.9$ K in a jump-like manner on the GB and in its neighbourhood, as shown in Fig. \ref{sBivarystrain}, where the $\theta_n^f$ is $5.6$ K lower than the case when no strain was applied. In the present case, the volume fraction of nucleated $\sf M$ is much larger than when $\varepsilon_1=0$  (compare between Figs. \ref{sBivarystrain} and \ref{fig:sub-firsts}). 
A significant drop in the nucleation temperature of $\sf M$ from the tilt GB was also observed in the experiment with a Fe-Ni bicrystal under tensile strain in \cite{Ueda-2001Acta}. A larger volume fraction of variant plates was observed around the GB therein when the strain was applied than in the case with no external strain \cite{Ueda-2001Acta}.  Under the applied strain, the volume fraction of the $\sf M$ phase increases faster for the decrease in $\theta$, as shown in Fig. \ref{fig:sub-firsts}. Thick $\sf M_1$ and $\sf M_2$ plates have grown near the GB region, whereas the plates away from the GB are much thinner. The sample is transformed to $\sf M$ except for a tiny region in the left corner of the top surface of the sample at $100$ K. The simulation results under applied strain are qualitatively in agreement with the experimental observations  described above.

\begin{figure}[t!]
\centering
\hspace{-8mm}
\subfigure[$\eta_0$ plots during forward MT]{
  \includegraphics[width=3.0in, height=3.00in] {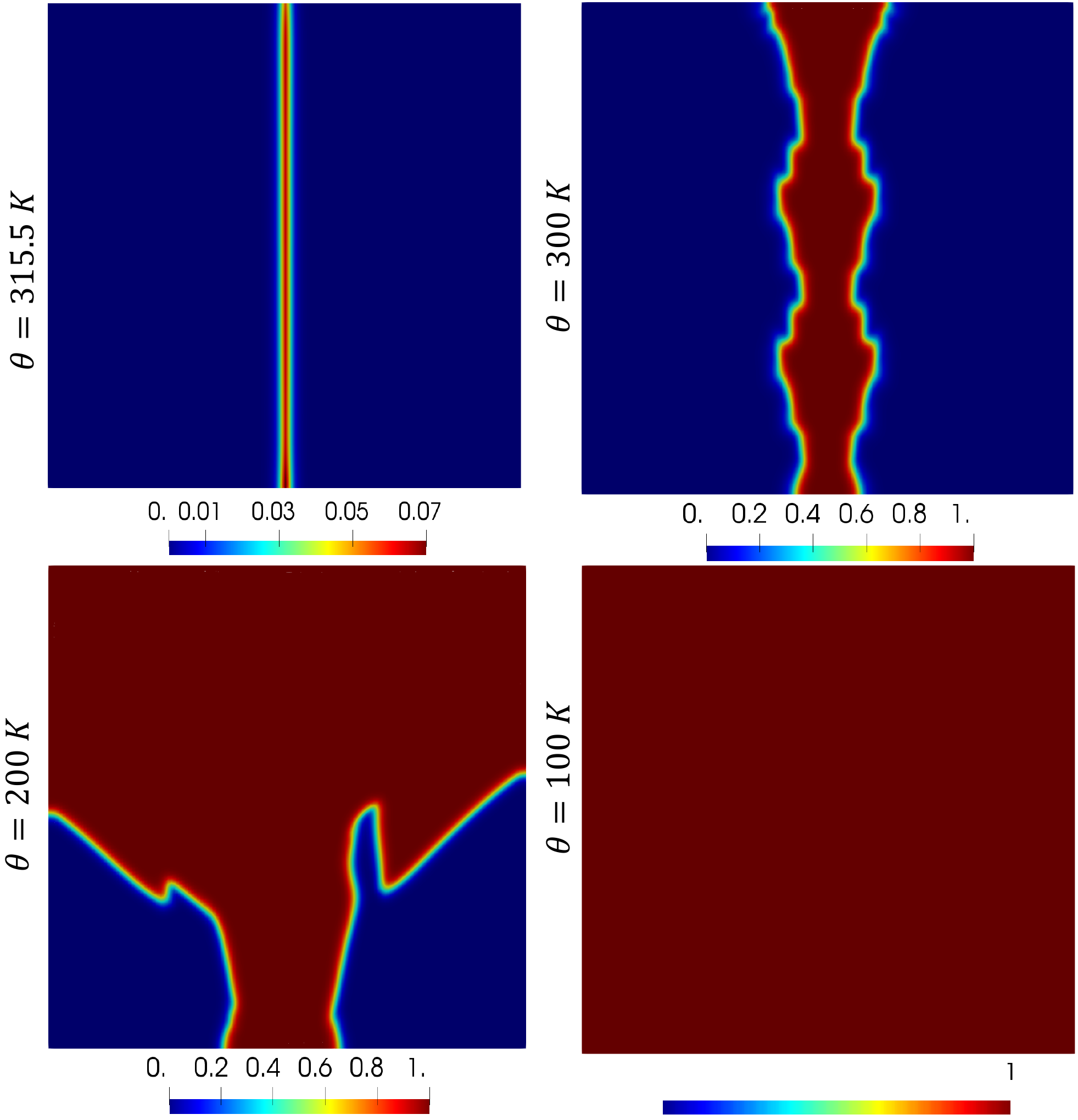}
	}\hspace{1mm}
    \subfigure[$\eta_{eq}$ plots during forward MT]{
    \includegraphics[width=3.0in, height=3.00in] {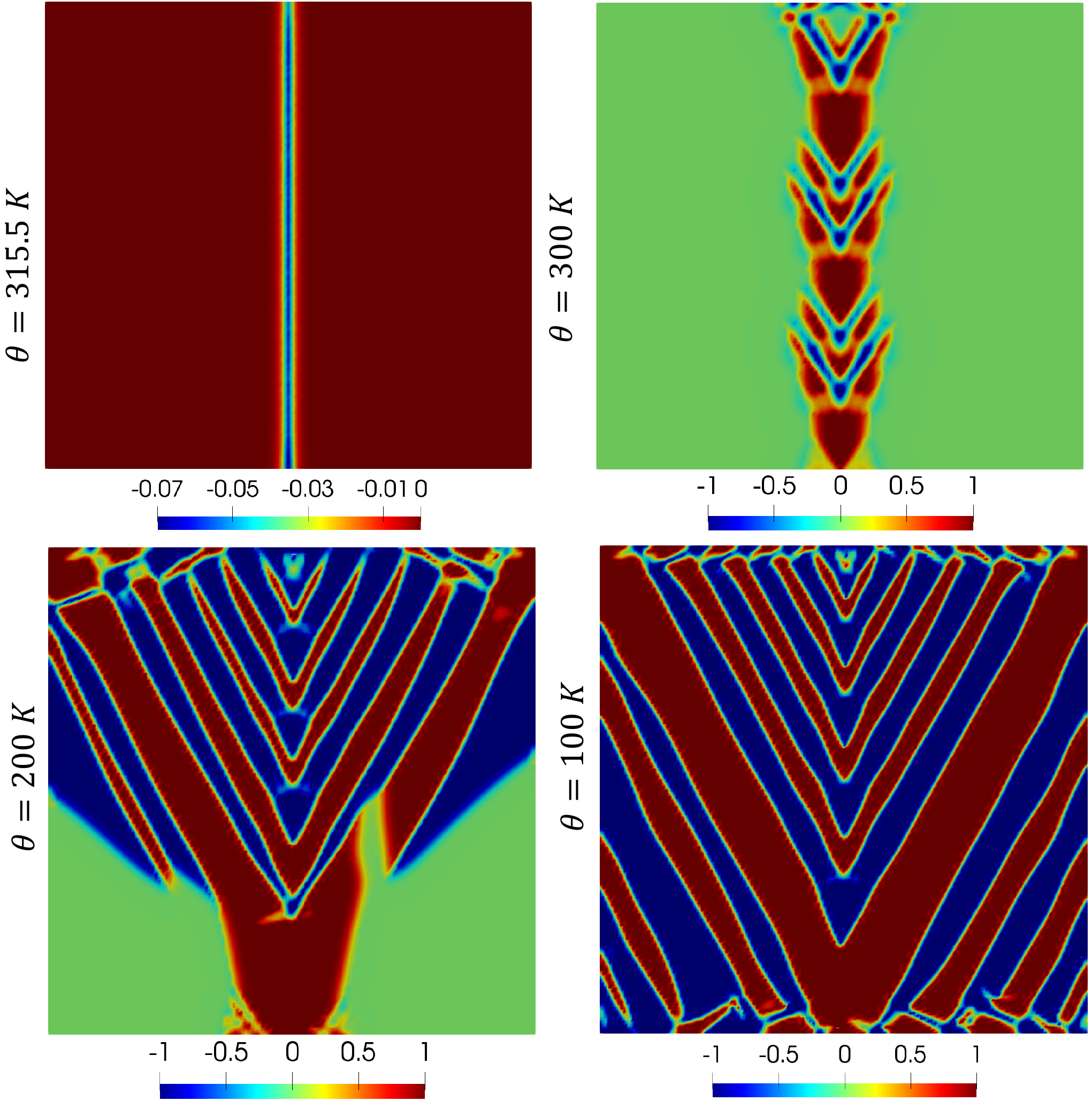}
}
\caption{Microstructure evolution during (a,b) forward MT in a $40$ nm $\times$ $40$ nm bicrystal having a symmetric planar GB with $\theta$ variation. We consider $\Delta\gamma_{12}^{gb} = -0.5$ N/m, $\delta_{12}^{gb}=1$ nm, and $\vartheta_{12}=40^\circ$. No strain applied.}
\label{BivarySamplesize}
\end{figure}
\subsubsection{ Effect of sample  size}
\label{sussub41}
We now consider a larger bicrystal of $40$ nm $\times$ $40$ nm size with a symmetric planar tilt GB (similar to Fig. \ref{BivarySamplesize}). The microstructure evolution during the forward MT is presented in Fig. \ref{BivarySamplesize}. All the ICs and other parameters are identical to the case shown in Fig. \ref{fig:sub-firsts}: $\Delta\gamma_{12}^{gb} = -0.5$ N/m, $\delta_{12}^{gb}=1$ nm,   $\vartheta_1=-\vartheta_2=20^\circ$, i.e. $\vartheta_{12}=40^\circ$, and  no strain is applied.   The nucleation temperature $\theta_n^f$  remains unchanged, but $\max(\eta_0)$ is slightly less in this larger sample. However, the martensitic microstructures evolution in these two samples significantly differ  on decreasing $\theta$. The sample is completely martensitic at $100$ K, and the variant plates are more regularly arranged in this larger sample than in the smaller sample shown in Fig. \ref{fig:sub-firsts}. 

\begin{figure}[t!]
\centering
\hspace{-8mm}
\subfigure[$\eta_0$ plots during forward MT]{
  \includegraphics[width=3.0in, height=3.00in] {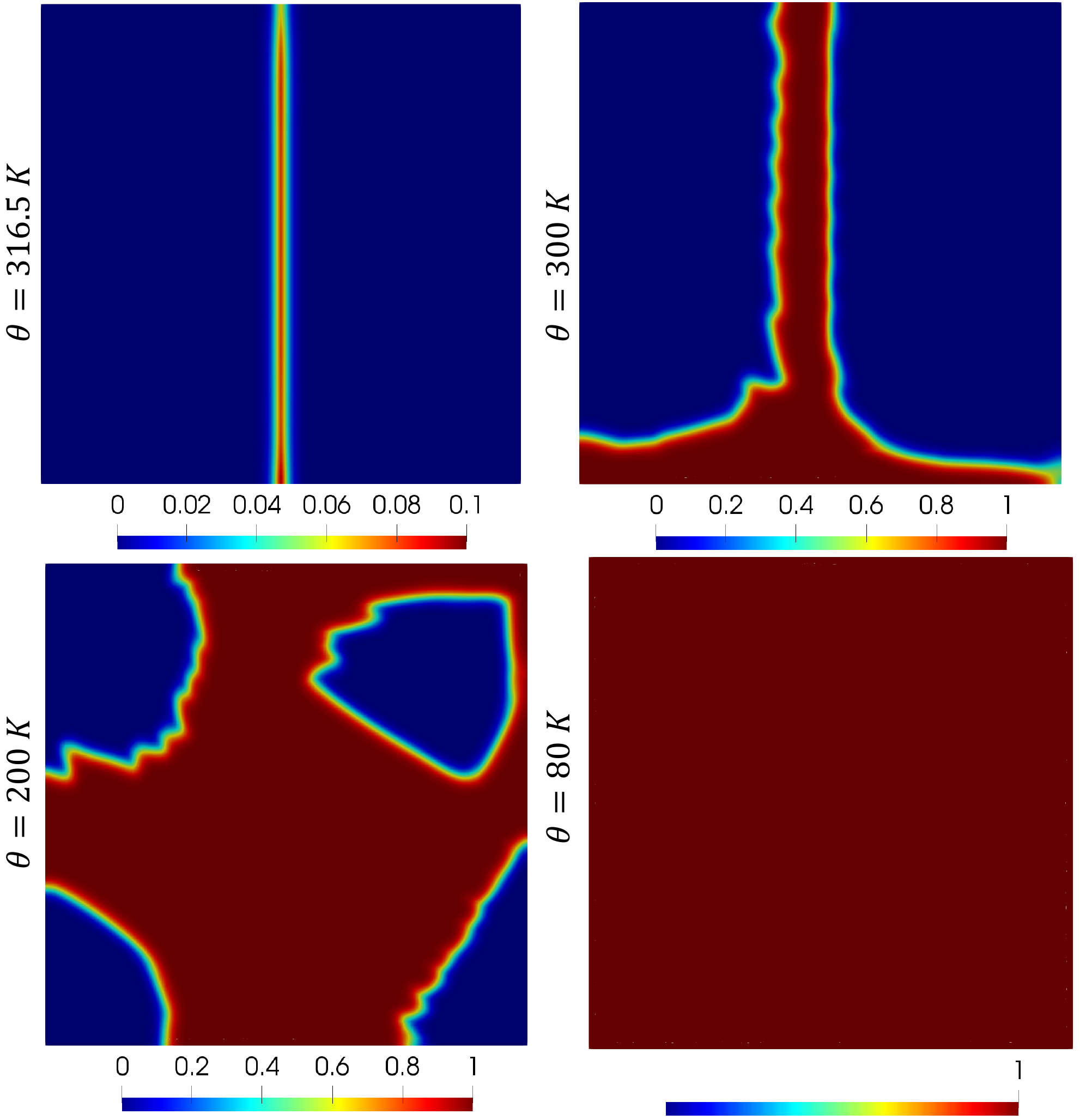}
	}\hspace{1mm}
    \subfigure[$\eta_{eq}$ plots during forward MT]{
    \includegraphics[width=3.0in, height=3.00in] {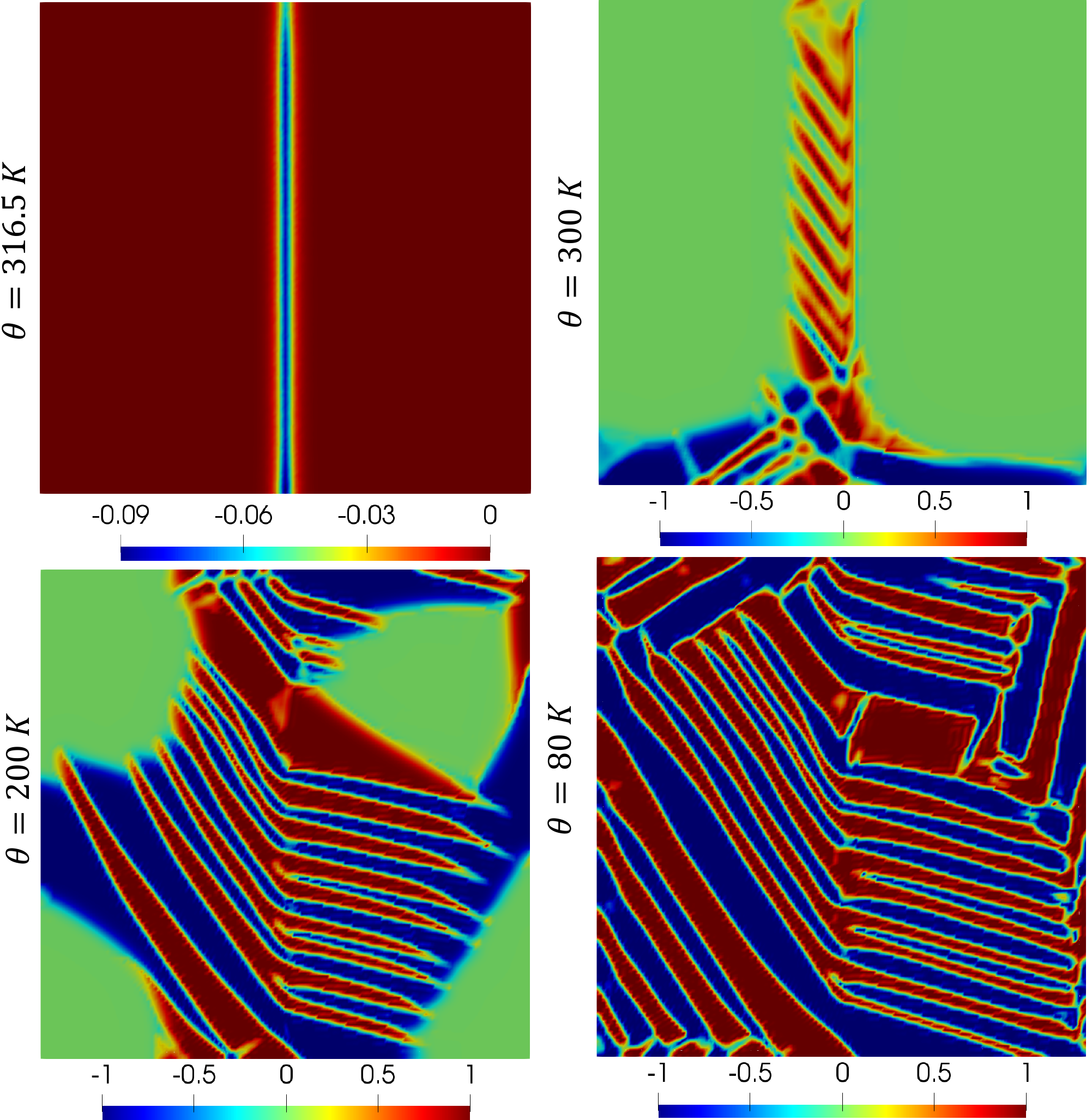}
}
\subfigure[$\eta_0$ plots during reverse MT]{
  \includegraphics[width=3.0in, height=3.00in] {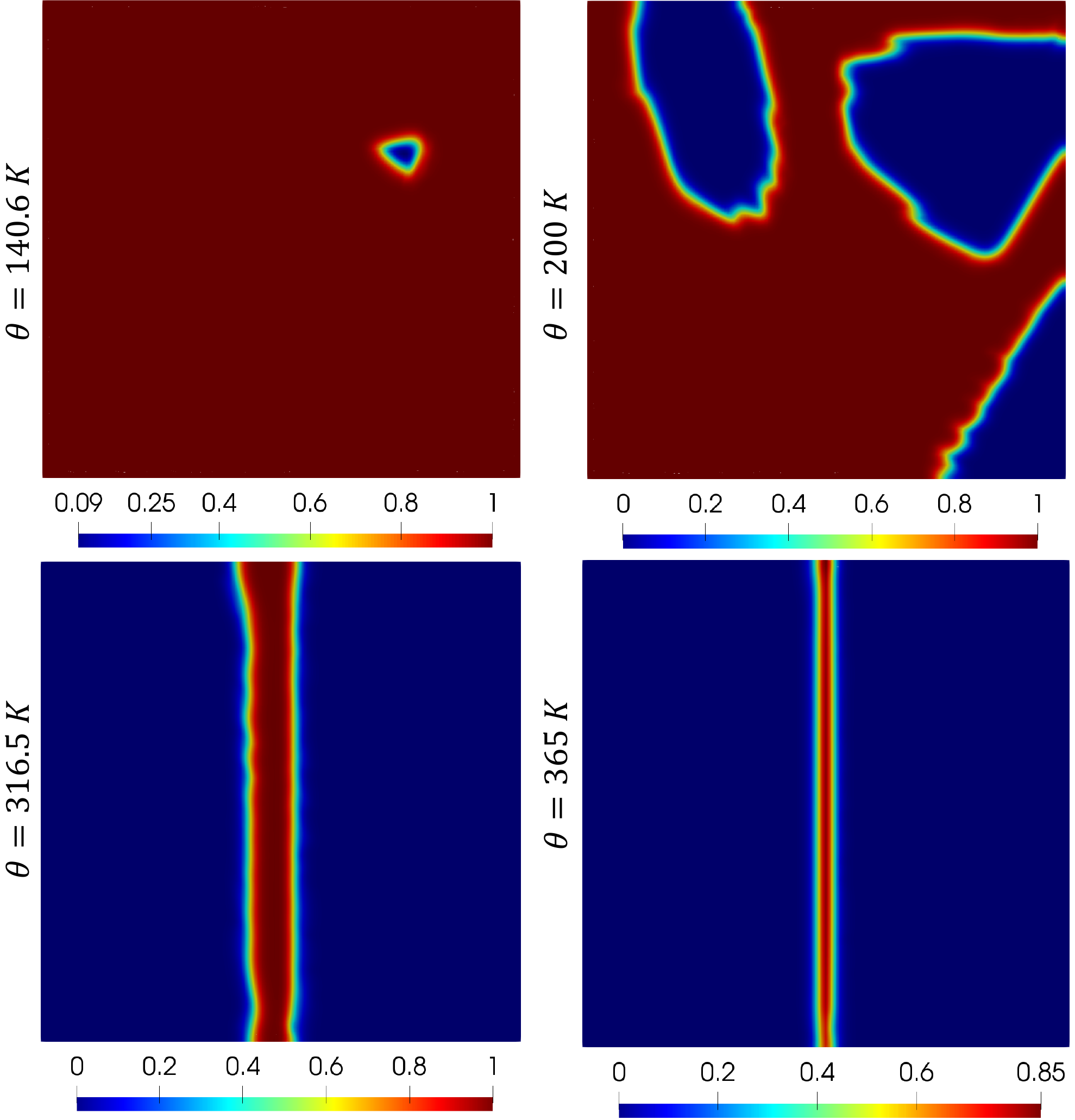}
	}\hspace{1mm}
    \subfigure[$\eta_{eq}$ plots during reverse MT]{
    \includegraphics[width=3.0in, height=3.00in] {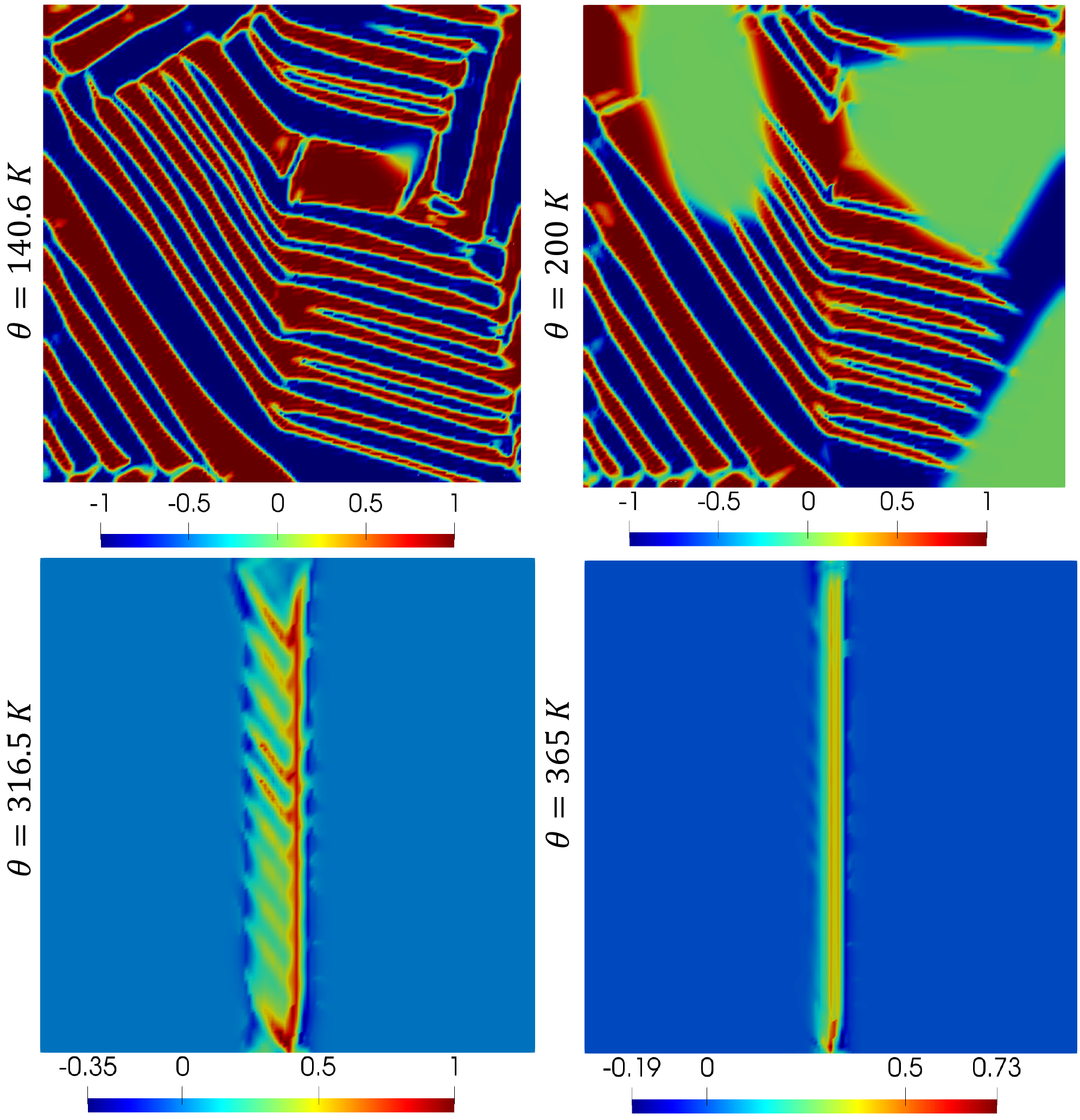}
}
\caption{Microstructure evolution during (a,b) forward, and (c,d) reverse MTs in a $30$ nm $\times$ $30$ nm bicrystal having a asymmetric planar GB with $\theta$ variation. We consider $\Delta\gamma_{12}^{gb} = -0.5$ N/m, $\gamma_{gb}^A = 0.9$ N/m, $\delta_{12}^{gb}=1$ nm, $\vartheta_1=30^\circ$, and $\vartheta_2=-10^\circ$. No strain applied.}
\label{BiAsymmetricGB}
\end{figure}

\begin{figure}[t!]
\centering
\hspace{-8mm}
{
  \includegraphics[width=4.9in, height=3.2in] {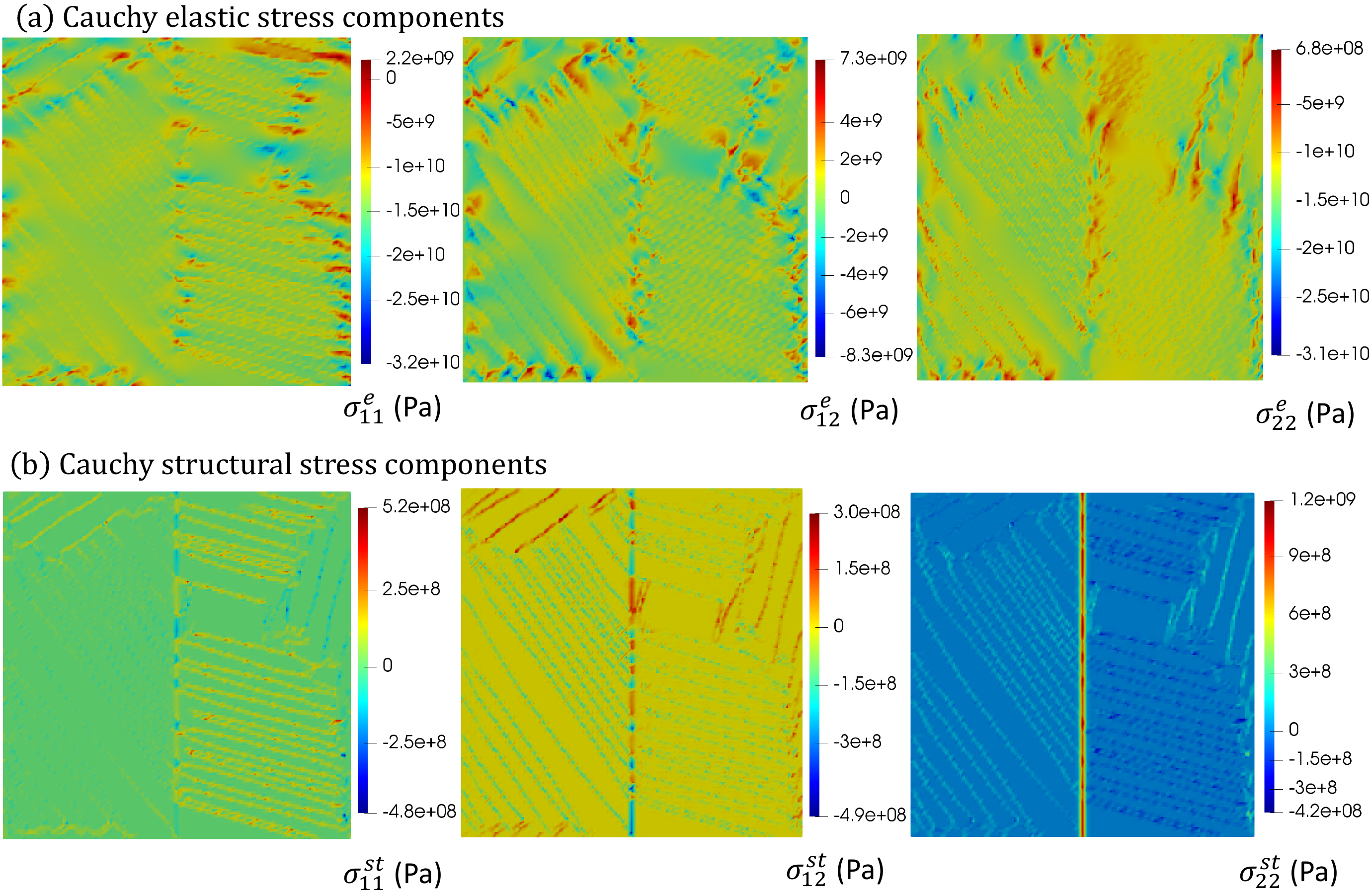}
	}
\caption{Plots for the (a) Cauchy elastic stresses $\sigma_{(e)ij}$, and (b) Cauchy structural stresses $\sigma_{(st)ij}$ at $\theta=80$ K within the martensitic sample shown in Fig. \ref{BiAsymmetricGB}(a,b) obtained by forward MT.}
\label{BiAsymGBStres}
\end{figure}

\subsubsection{ Results for asymmetric GB}
\label{sussub41as}
We now consider a stress-free austenitic bicrystal with an asymmetric planar tilt GB, where the grain orientations are  $\vartheta_1=30^\circ$ and $\vartheta_2=-10^\circ$ (see Fig. \ref{xiplots}(a)). The GB misorientation is thus $\vartheta_{12}=40^\circ$. All other parameters are identical to that considered for Fig. \ref{fig:sub-firsts}: $\delta_{12}^{gb}=1$ nm, $\Delta\gamma_{12}^{gb}=-0.5$ N/m, and no strain is applied. The stationary microstructures obtained during the forward and reverse MTs are shown in Fig. \ref{BiAsymmetricGB}. 

\noindent{\bf Forward MT.} \hspace{3mm} As discussed in Appendix \ref{compatibilityBiTri}, neither of the variants $\sf M_1$ and  $\sf M_2$ is incompatible across the asymmetric GB of this sample as $|\vartheta_1|\neq |\vartheta_2|$. We note that the nucleation temperature $\theta_n^f$ is $1$ K higher than the sample with a symmetric GB (compared with Fig. \ref{fig:sub-firsts}), where the misorientation is identical.  Only incomplete $\sf M_1$ variant plates are developed in $G_1$ about the GB at $300$ K, and the bottom part of it is transformed. The grain $G_2$, on the other hand, is still primarily austenitic except for a region near the bottom surface. On the further decrease in $\theta$, the variant-variant plates grow mainly from the GB region at $200$ K, and also few plates grow from other parts of the grains. The sample is fully martensitic at $80$ K. The orientation of the variant-variant interfaces within each grain, which are away from the GB, agrees with the crystallographic solutions (\cite{Ball-James-87} and Chapter 5 of \cite{Bha04}). The coherency of the $\sf M$ plates of each variant from the two grains is clearly absent in the asymmetric GB (see Fig. \ref{BiAsymmetricGB}(b)), which is due to the reason that the variants are not compatible across such a GB. However, the variant plates were coherently structures about the symmetric planar GB shown in Fig. \ref{fig:sub-firsts} (also in Figs. \ref{fig:sub-firstsdr}, \ref{BivaryGbenergy}, \ref{BivaryGbmisor}, \ref{sBivarystrain}, \ref{BivarySamplesize}). 

The elastic and structural Cauchy stresses within the fully martensitic sample are shown in Fig. \ref{BiAsymGBStres}(a) and (b), respectively. Compared to the martensitic bicrystal with a symmetric GB, more high amplitude elastic and structural stress pockets are observed within the GB region of the martensitic bicrystal with an asymmetric GB (compare between Figs. \ref{fig:sub-firstsstress} and \ref{BiAsymGBStres}).

\noindent{\bf Reverse MT.} \hspace{3mm} On heating the fully $\sf M$ sample from $80$ K, a $\sf PM$ region nucleates in $G_2$ at $140.6$ K, as shown in Fig. \ref{BiAsymmetricGB}(c,d). The microstructure evolution with a further increase in $\theta$ is also shown. The entire sample becomes austenitic at $365.1$ K. Note that the bicrystal with the symmetric GB transformed to $\sf A$ at $365.6$ K (see Fig. \ref{fig:sub-firsts}(c,d)).

The incompatibility of the variants at the GB marginally influences the $\sf M$ nucleation and $\sf A$ finish temperatures. The microstructures, however, significantly differ, especially within the GB region.

\begin{figure}[t!]
\centering
\hspace{-8mm}
\subfigure[$\eta_0$ plots during forward MT]{
  \includegraphics[width=2.8in, height=2.8in] {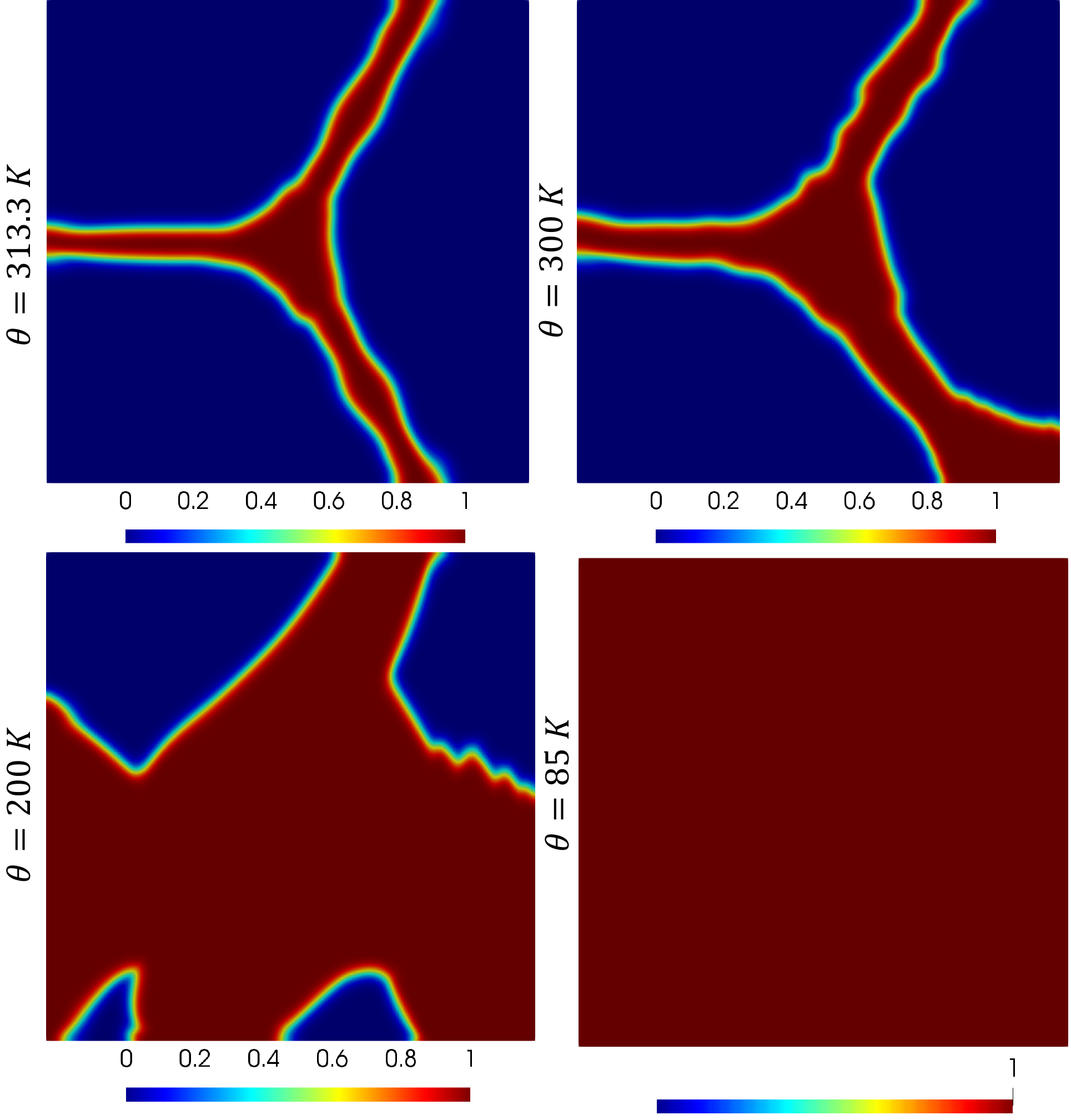}
	}\hspace{-1mm}
    \subfigure[$\eta_{eq}$ plots during forward MT]{
    \includegraphics[width=2.8in, height=2.8in] {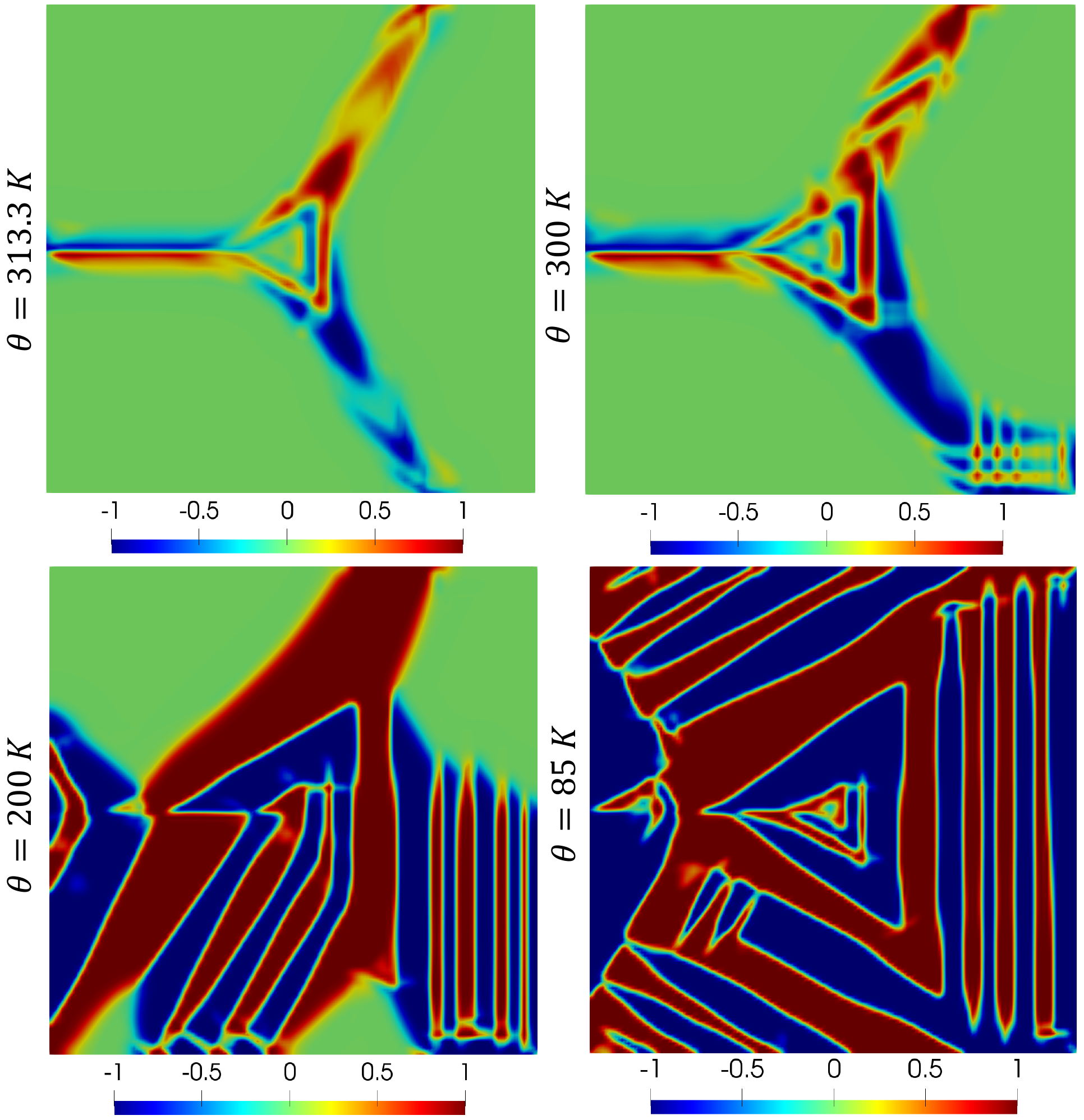}
}
\subfigure[$\eta_0$ plots during reverse MT]{
  \includegraphics[width=2.8in, height=2.8in] {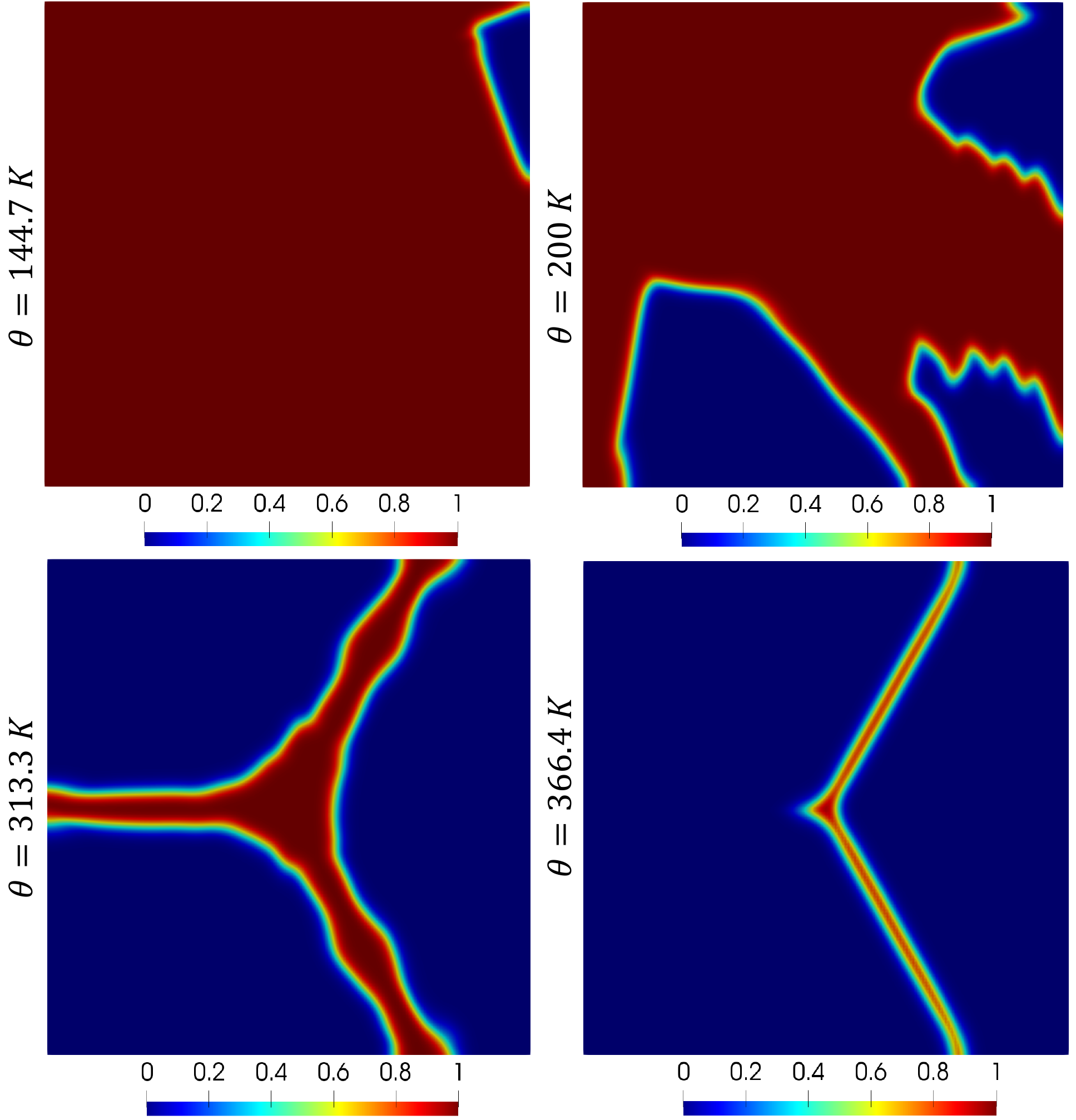}
	}\hspace{-1mm}
    \subfigure[$\eta_{eq}$ plots during reverse MT]{
    \includegraphics[width=2.8in, height=2.8in] {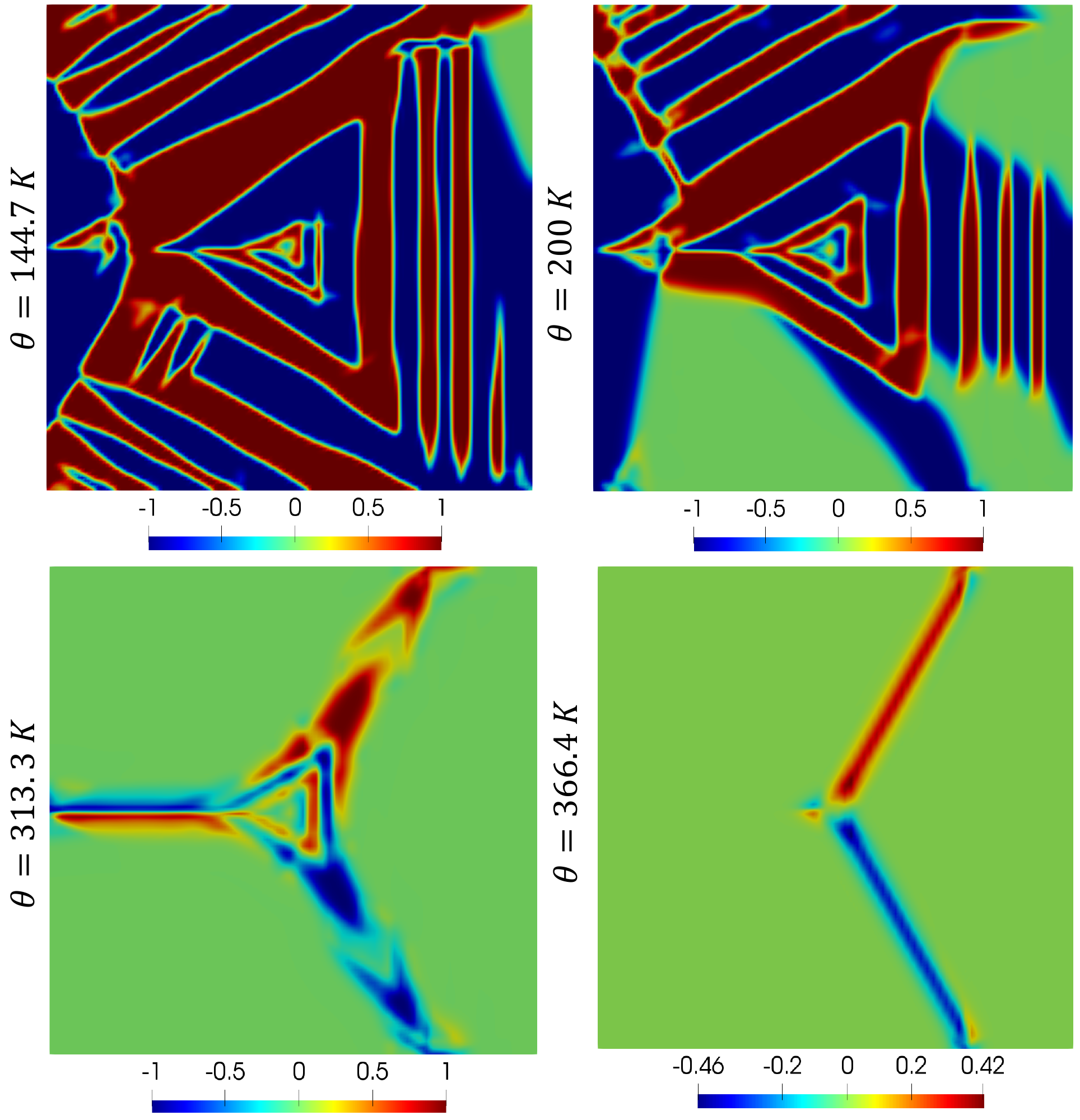}
}
\vspace{-7mm}
\caption{Microstructure evolution during (a,b) forward, and (c,d) reverse MTs in a $30$ nm $\times$ $30$ nm tricrystal having symmetric planar GBs with $\theta$ variation. We used $\Delta\gamma_{12}^{gb} = \Delta\gamma_{13}^{gb} =\Delta\gamma_{23}^{gb} =-0.5$ N/m, $\delta_{12}^{gb}=\delta_{13}^{gb}=\delta_{23}^{gb}=1$ nm, and $\rho_0K_{IJ}^{gb}=\rho_0K_{123}=800$ GPa. No strain applied.}
\label{triGBwidth1nm}
\end{figure}
\begin{figure}[t!]
\centering
\hspace{-8mm}
  \includegraphics[width=5.1in, height=3.4in] {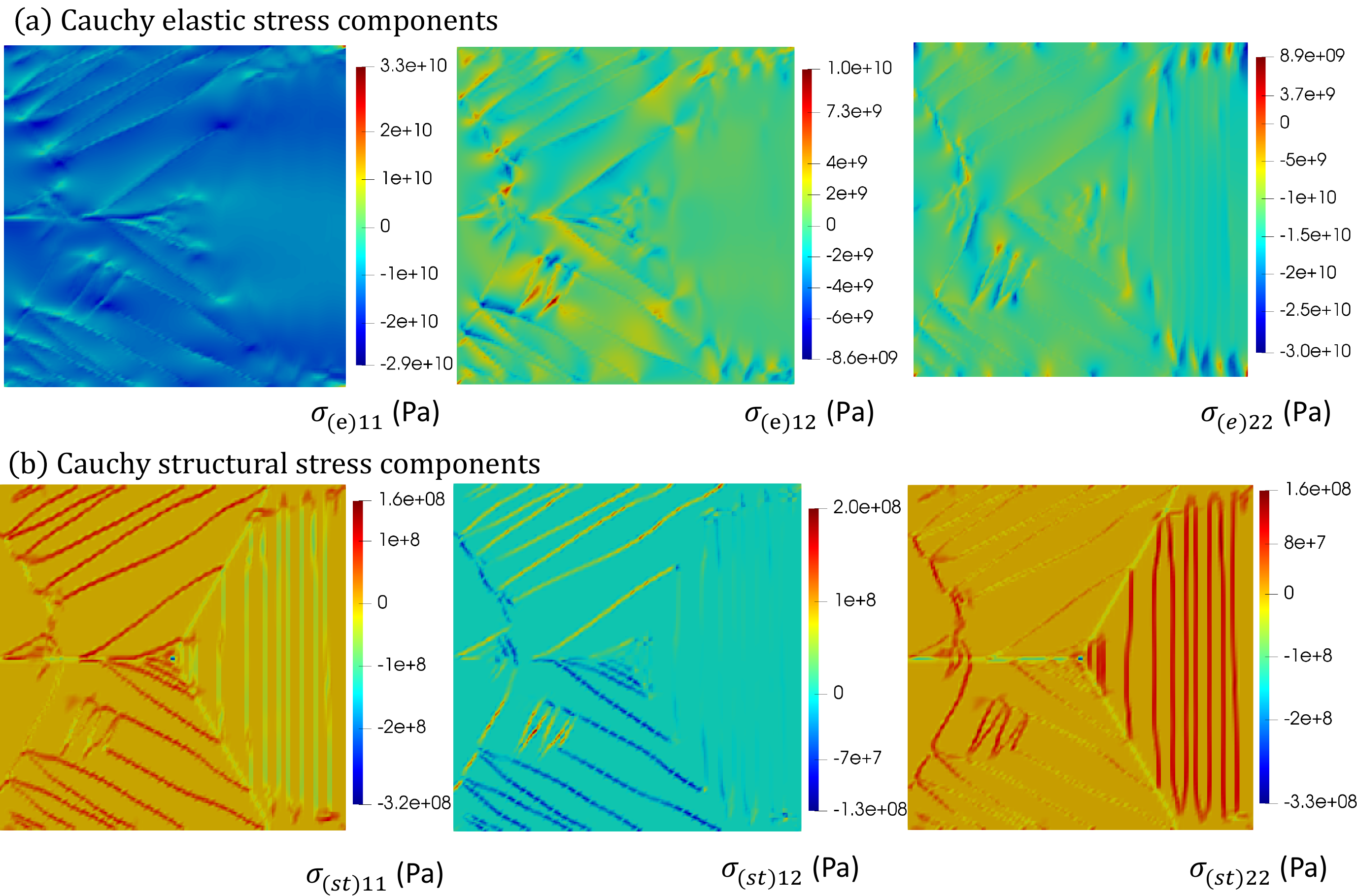}
	\caption{Plots for the (a) Cauchy elastic stresses $\sigma_{(e)ij}$, and (b) Cauchy structural stresses $\sigma_{(st)ij}$ at $\theta=85$ K within the martensitic tricrystal shown in Fig. \ref{triGBwidth1nm}(a,b) obtained by forward MT.}
\label{triGBwidth1nmStresses}
\end{figure}

\subsection{Results for tricrystal}
\label{tricryslt}
We now present the results for the tricrystals with three planar symmetric tilt GBs; a typical austenitic tricrystal is shown in Fig. \ref{xiplots}(c). The grain orientation angles are taken as $\vartheta_1=-15^\circ$, $\vartheta_2=-75^\circ$,  and $\vartheta_3=-45^\circ$. The misorientations of the three GBs as per the definition given by Eq. \eqref{misortnn} are $\vartheta_{12}=60^\circ$, $\vartheta_{23}=-30^\circ$, and $\vartheta_{13}=30^\circ$. However, noticing the cubic symmetry of the parent $\sf A$ phase in 2D (see Fig. \ref{unit_cells}), it can be easily verified that all the austenitic GBs are symmetric and equivalent to having $30^\circ$ misorientations. We thus consider identical austenitic GB energy  ($\gamma^A_{12}=\gamma^A_{13}=\gamma^A_{23}$), and  identical change in GB energy due to MT $\Delta\gamma^{gb}_{12}=\Delta\gamma^{gb}_{13}=\Delta\gamma^{gb}_{23}$. The width for the GBs is also taken identical: $\delta^{gb}_{12}=\delta^{gb}_{13}=\delta^{gb}_{23}$. We will now study the effect of the parameters, including the GB width, change in GB energy during MTs, triple junction energy, and sample size on nucleation of the phases and the subsequent microstructures evolution in detail. No external strain is applied to the surfaces. All other BCs and the ICs are already prescribed at the beginning of the section.

\begin{figure}[t!]
\centering
\hspace{-8mm}
\subfigure[$\eta_0$ plots during forward MT]{
  \includegraphics[width=3.0in, height=3.00in] {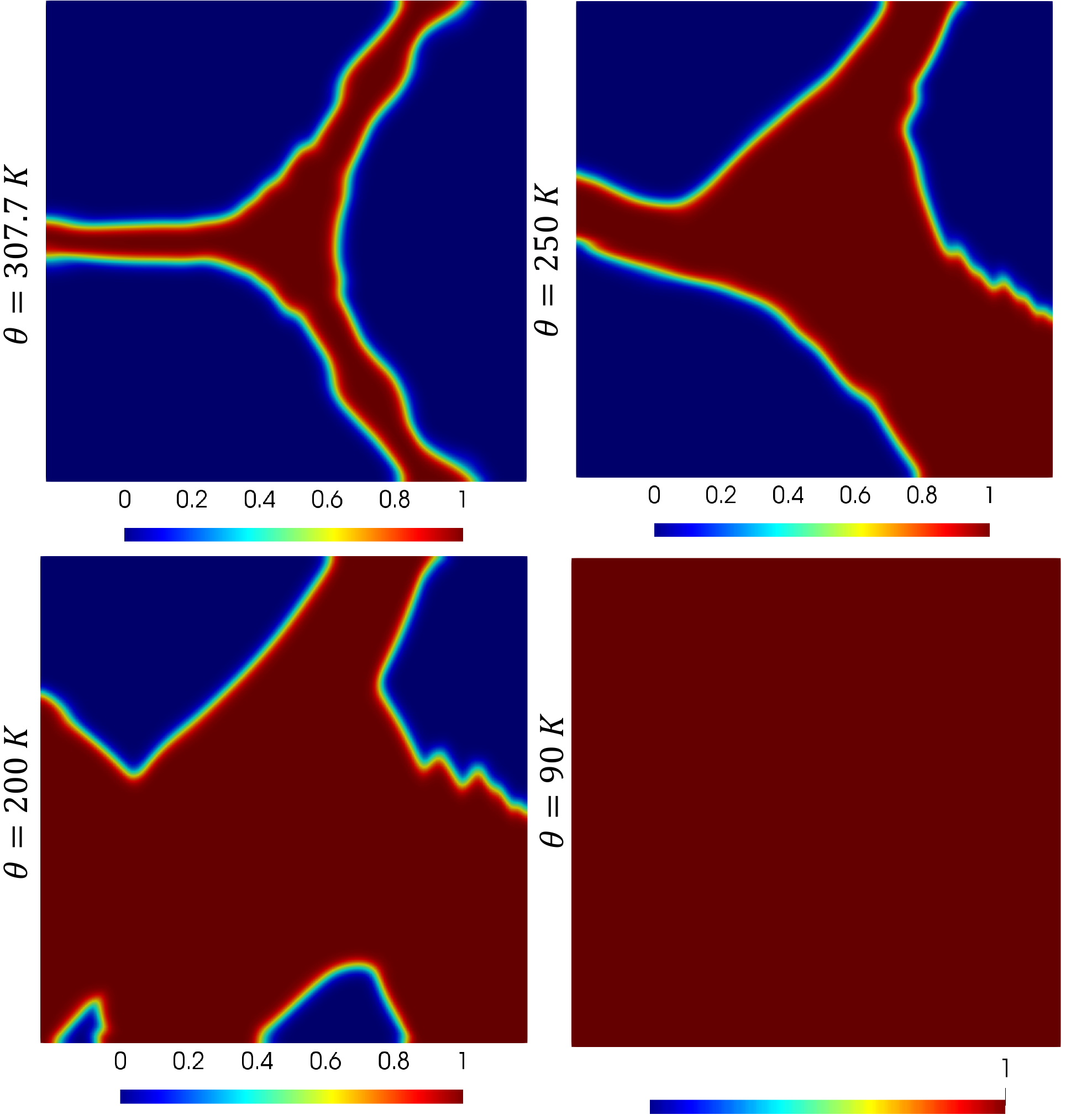}
	}\hspace{1mm}
    \subfigure[$\eta_{eq}$ plots during forward MT]{
    \includegraphics[width=3.0in, height=3.00in] {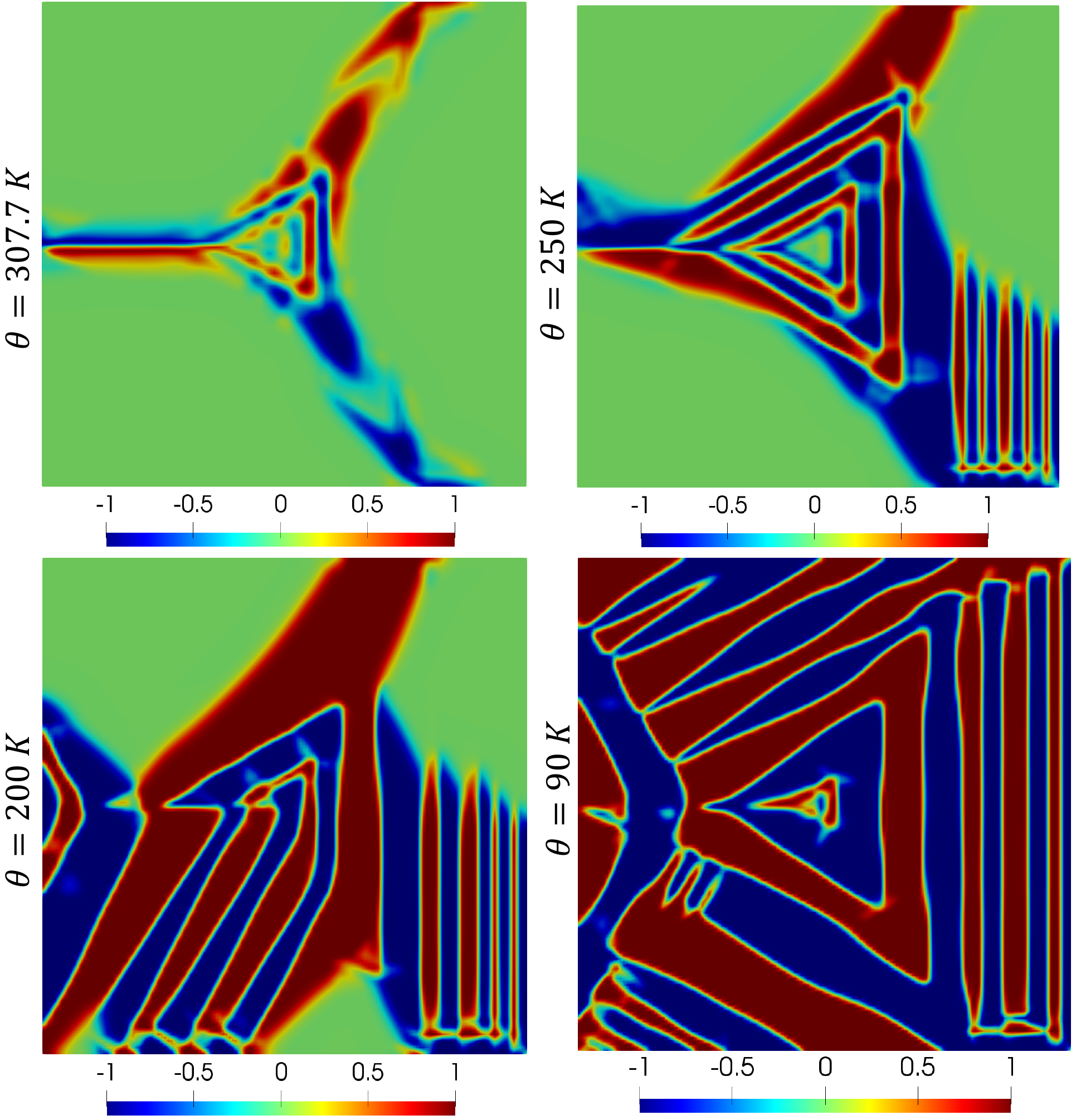}
}
\subfigure[$\eta_0$ plots during reverse MT]{
  \includegraphics[width=3.0in, height=3.00in] {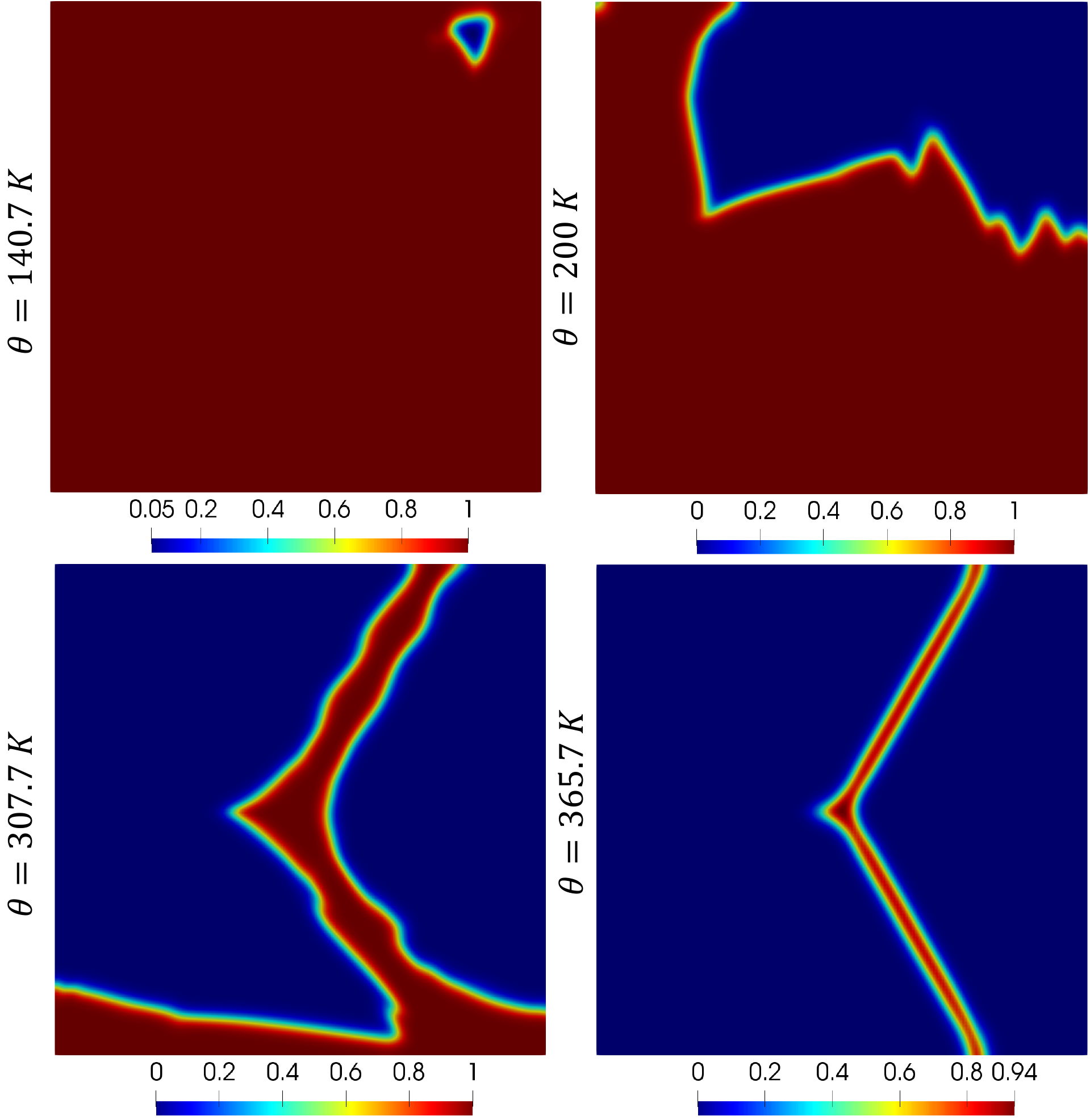}
	}\hspace{1mm}
    \subfigure[$\eta_{eq}$ plots during reverse MT]{
    \includegraphics[width=3.0in, height=3.00in] {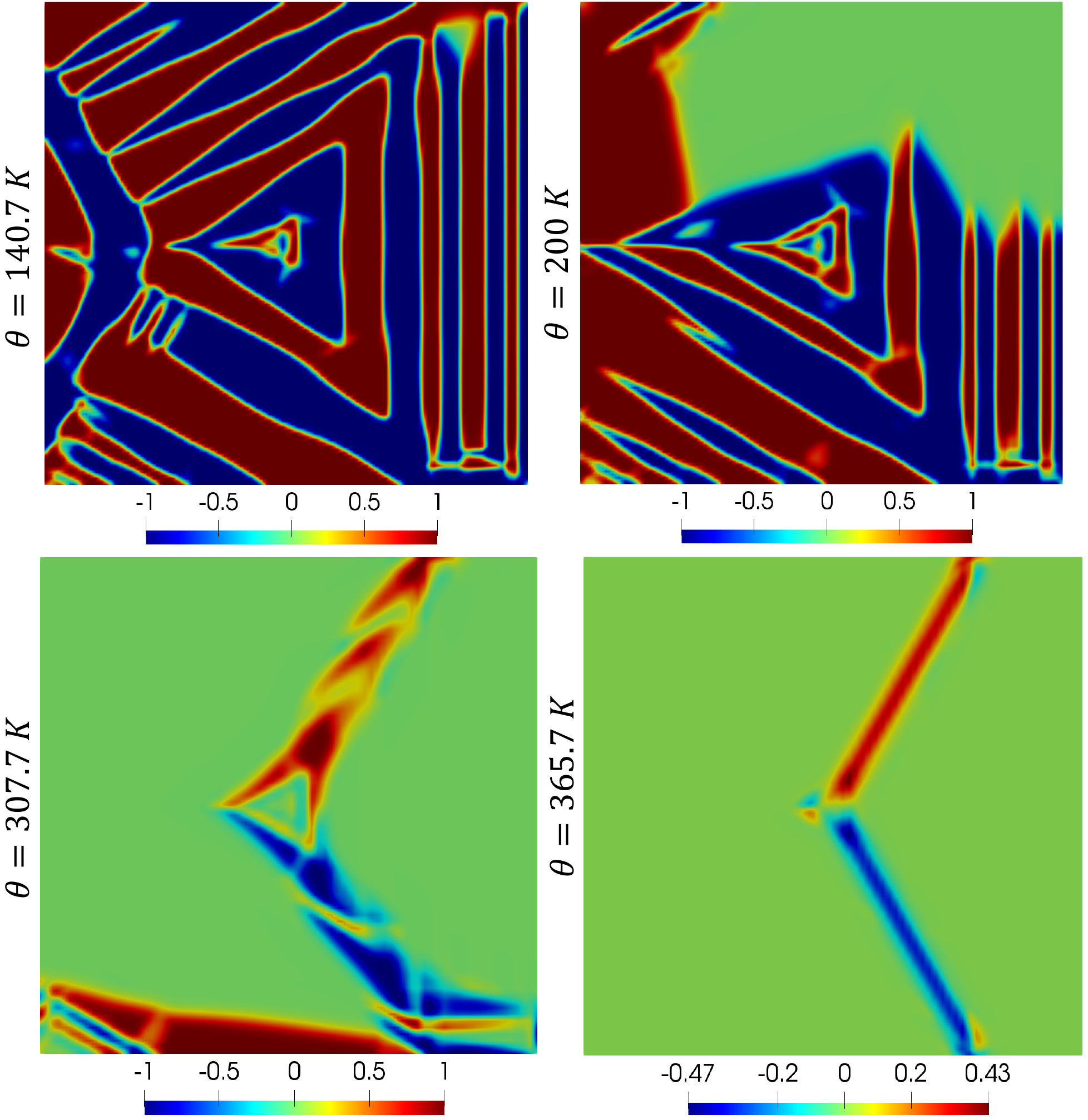}
}
\caption{Microstructure evolution during (a,b) forward, and (c,d) reverse MTs in a $30$ nm $\times$ $30$ nm tricrystal having symmetric planar GBs with $\theta$ variation. We used $\Delta\gamma_{12}^{gb} = \Delta\gamma_{13}^{gb} =\Delta\gamma_{23}^{gb} = -0.5$ N/m,  $\delta_{12}^{gb}=\delta_{13}^{gb}=\delta_{23}^{gb}=1.5$ nm, and $\rho_0K_{IJ}^{gb}=\rho_0K_{123}=800$ GPa. No strain applied.}
\label{triGBwidth1pt5nm}
\end{figure}
\begin{figure}[t!]
\centering
\hspace{-8mm}
  \includegraphics[width=5.1in, height=3.4in] {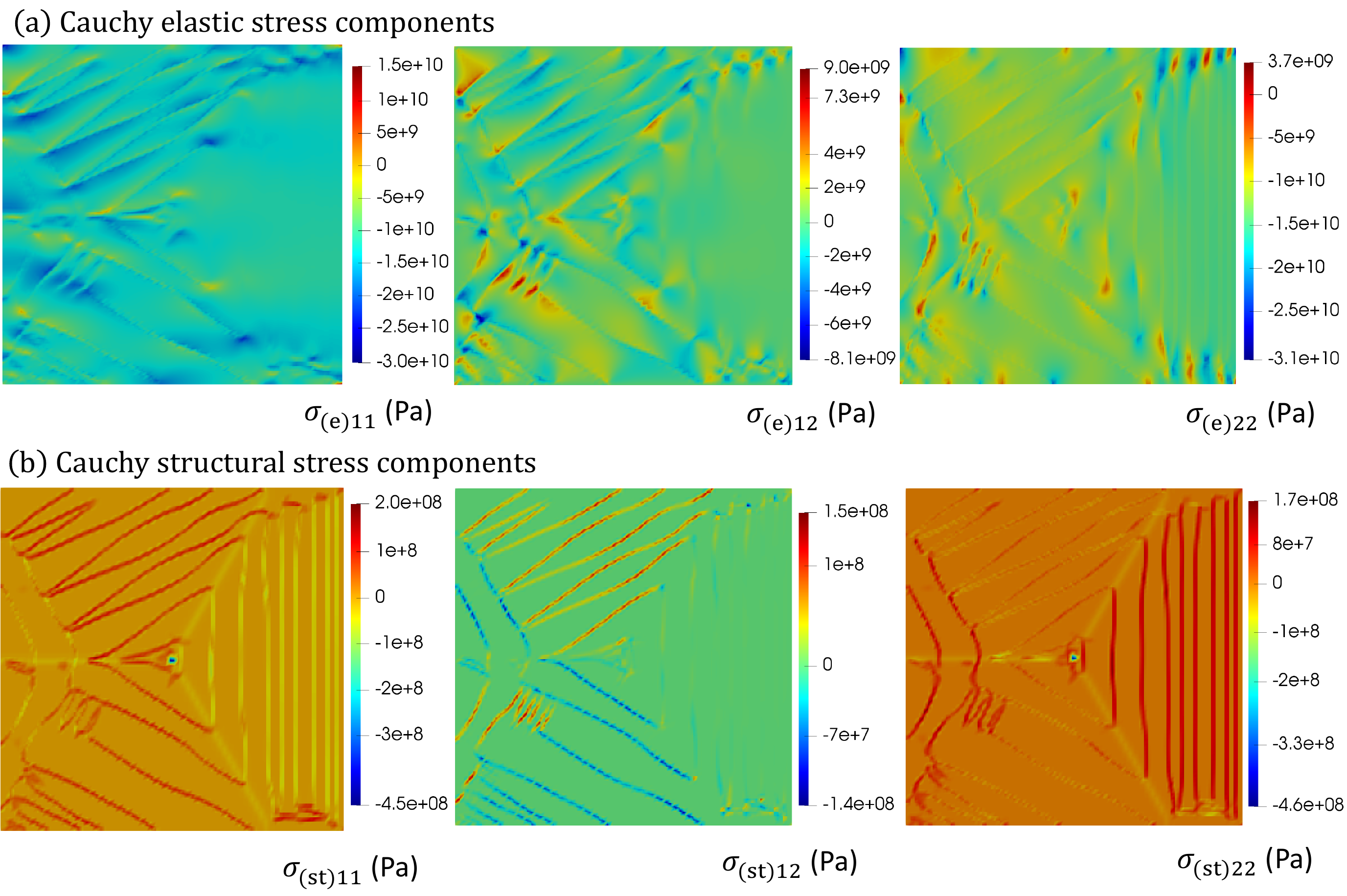}
	\caption{Plots for the (a) Cauchy elastic stresses $\sigma_{(e)ij}$, and (b) Cauchy structural stresses $\sigma_{(st)ij}$ at $\theta=90$ K within the martensitic tricrystal  shown in Fig. \ref{triGBwidth1pt5nm}(a,b) obtained by forward MT.}
\label{triGBwidth1pt5nmStresses}
\end{figure}

\subsubsection{Effect of GB width}
\label{TJGBwidth}
We consider two $30$ nm $\times$ $30$ nm-sized tricrystals where the stress-free austenitic GBs have theoretical widths of $\delta_{12}^{gb}=\delta_{23}^{gb}=\delta_{13}^{gb}=1$ nm and $\delta_{12}^{gb}=\delta_{23}^{gb}=\delta_{13}=1.5$ nm, respectively, but all other parameters are identical: $\gamma_{IJ}^A = 0.9$ N/m, $\Delta\gamma_{IJ}^{gb} = -0.5$ N/m,  and $\rho_0K_{12}=\rho_0K_{13}=\rho_0K_{23}=\rho_0K_{123}=800$ GPa (see Eq. \eqref{engiso5}).  The stress-free austenitic tricrystal shown in Fig. \ref{xiplots}(c) is obtained by solving Eqs. \eqref{bicrystgovt}$_{3,4,5}$ and neglecting mechanics (also see \cite{Basak-2021-PCCP}). The numerical width of the GBs  would differ from the theoretical widths away from the TJ region since a fraction of nonzero $\xi_H$ would always be present on the GB $\Gamma_{IJ}$ for $H\neq I,J$  (see \cite{Basak-2021-PCCP} for numerical examples). Following \cite{Steinbach-09}, we define the numerical width of a GB, say, $\Gamma_{12}$, as the distance between the points across the GB where $\xi_1=0.05$ (or $\xi_2=0.95$) and $\xi_1=0.95$ (or $\xi_2=0.05$).  For the tricrystals under consideration, we obtain the numerical widths of the GBs as $1.07$ nm and $1.58$ nm, respectively, showing a deviation of $7$\% and $5$\%  from the respective theoretical widths. The microstructure evolution during the forward and reverse MTs for these two samples is shown in Figs. \ref{triGBwidth1nm} and \ref{triGBwidth1pt5nm}, respectively.

\noindent{\bf Forward MT.} \hspace{3mm} Thick $\sf M$ layers containing both $\sf M_1$ and $\sf M_2$ have nucleated in the junction and GB regions at $\theta_n^f=313.3$ K and $307.7$ K in a jump-like manner within the tricrystals with $\delta_{IJ}^{gb}=1$ nm and $1.5$ nm, respectively. However, the samples were fully austenitic at $313.4$ K and $307.8$ K, respectively. The overall GB and junction region contain both variants, with an equal volume fraction. We also note that each of the grains nucleates both the variants near the GBs and TJ.  The $\sf M$ phase within the small TJ region contains incomplete $\sf M_1$ and $\sf M_2$. With a further reduction in $\theta$, the variant plates grow and progress inside the grains.  The two samples are completely martensitic at $85$ K and $90$ K, respectively. According to the crystallographic theory of martensite, the pair of variant plates for cubic to tetragonal MTs make $45^\circ$ or $135^\circ$ with the $\fg c_1$-axis of the cubic $\sf A$ unit cell in an isolated large single grain (see e.g. Chapter 5 of \cite{Bha04}). This implies that the plates in the isolated $G_1$ would have made $30^\circ$ or $60^\circ$ with the $\fg e_1$-axis, and they should have made $30^\circ$ or $160^\circ$ angles in the isolated $G_2$, and $90^\circ$ or $0^\circ$ in the isolated $G_3$  with the $\fg e_1$-axis. However, since the grains are constrained, the evolving microstructures in the samples are rather more complex and not so intuitive, especially near the GBs and TJ. The orientations of some of the variant plates in each grain away from the GB regions are close to the above mentioned crystallographic solution in the fully martensitic samples of Figs. \ref{triGBwidth1nm}(b) and \ref{triGBwidth1pt5nm}(b). However, the orientation of the plates deviates in the GB and TJ regions in trying to form a compatible $\sf M_1$-$\sf M_1$ and $\sf M_2$-$\sf M_2$ microstructure therein, as all the GBs are symmetric (see Appendix \ref{compatibilityBiTri}). Interestingly we note in Figs. \ref{triGBwidth1nm} and \ref{triGBwidth1pt5nm} that the variant plates obtained in the grain $G_2$ at $200$ K rearrange themselves perpendicularly in the completely martensitic samples at $85$ K and $90$ K, respectively. The fully $\sf M$ samples in Figs. \ref{triGBwidth1nm}(b) and \ref{triGBwidth1pt5nm}(b) are similar except for variations in the variant plate thickness in some regions within the grains and near the TJs. 

We can qualitatively compare the results for the tricrystal in Fig. \ref{triGBwidth1nm} with that of the bicrystal whown in Fig. \ref{BivaryGbmisor} where the theoretical GB width, the misorientation, and all other material parameters are identical. However, a quantitative comparison between the nucleation temperatures $\theta_n^f$ in these two samples is not possible, as the numerical GB widths in the tricrystals differ from the analytical ones (=$1$ nm). In contrast, the bicrystals were obtained using the analytical solution for $\xi_1$ (Eq. \eqref{analyticeta}), and the GB width therein is hence the theoretical value only  (=$1$ nm).
In contrast to the tricrystals, where thick $\sf M$ layers consisting of both the variants nucleated in the GB  and TJ regions, much thinner layers of pre-$\sf M_2$ nucleated on the GBs of the bicrystals at $\theta_n^f$.  The TJ size and energy (controlled by the parameter $\rho_0K_{123}$ and $\rho_0K_{IJ}^{gb}$; see Eq. \eqref{engiso5})  play an important role in determining the $\theta_n^f$ and the corresponding volume fraction of $\sf M$ at different temperatures in the tricrystals. 

We could not find out the high-resolution martensitic microstructures near the TJs at the nanoscale in the literature to directly compare our numerical results.  However, a qualitatively similar arrangement of $\sf M$ plates has been reported near the triple junctions of CuNiAl SMA  \cite{Schuh-13} and ceramics \cite{Arlt-90} at the micron scale. The Cauchy elastic and structural stress components in the fully $\sf M $ tricrystals obtained at $85$ K and $90$ K are shown in Figs. \ref{triGBwidth1nmStresses} and \ref{triGBwidth1pt5nmStresses}, respectively. The stresses are mainly concentrated across the GBs, TJ, and variant-variant interfaces, which is as per the experimental observations in polycrystalline SMAs (see, e.g. \cite{Schuh-13}).

\noindent{\bf Reverse MT.} \hspace{3mm} We now consider the fully $\sf M$ samples shown in Figs. \ref{triGBwidth1nm}(a,b) and \ref{triGBwidth1pt5nm}(a,b) considering the ICs mentioned above to study the reverse MTs. All the material parameters remain identical to that used during the forward MT. The $\sf A$ and $\sf PM$  nucleates in these two samples at $\theta_n^r=144.7$ K and $\theta_n^r=140.7$ K, respectively, as shown in Figs.  \ref{triGBwidth1nm}(c,d) and \ref{triGBwidth1pt5nm}(c,d). Both the nucleated regions are near the upper-right corners of the respective samples. The nucleated $\sf A$ region in the first sample is much larger than the $\sf PM$ region nucleated in the other sample. As $\theta$ of the samples are raised, the volume fraction of $\sf A$ of the sample increases as desired.  A fraction of the grains $G_2$ and $G_3$ in the first sample transform to $\sf A$ at  $200$ K, but $G_1$ still remains martensitic, as shown in Figs. \ref{triGBwidth1nm}(c,d). On the other hand, only parts of the $G_1$ and $G_3$ of the other sample transform to $\sf A$ at that $\theta$, where $G_2$ is still fully martensitic, as shown in Fig. \ref{triGBwidth1pt5nm}(c,d). On further increase in $\theta$, the $\sf M$ from the grains' interiors continuously transforms to $\sf A$, leaving a residual $\sf M$ on the GB and junction regions only, as shown in Figs.  \ref{triGBwidth1nm}(c,d) and \ref{triGBwidth1pt5nm}(c,d). When $\theta=\theta_n^f$, residual $\sf M$ is still observed in all the GBs and TJ  in Fig.  \ref{triGBwidth1nm}(c,d); the residual $\sf M$ is mainly observed on the two GBs, in the TJ, and on the bottom surface of the other sample shown in Fig. \ref{triGBwidth1pt5nm}(c,d). The residual $\sf M$ continuously transforms to $\sf PM$ on a further increase of $\theta$. The layers of $\sf PM$ in the GB and junction regions of the two samples are shown at $366.4$ K and $365.7$ K, respectively, which barrierlessly disappear on the increase in the respective sample temperatures by $0.1$ K to yield the original austenitic samples.


\begin{figure}[t!]
\centering
\hspace{-8mm}
\subfigure[$\eta_0$ plots during forward MT]{
  \includegraphics[width=3.0in, height=3.00in] {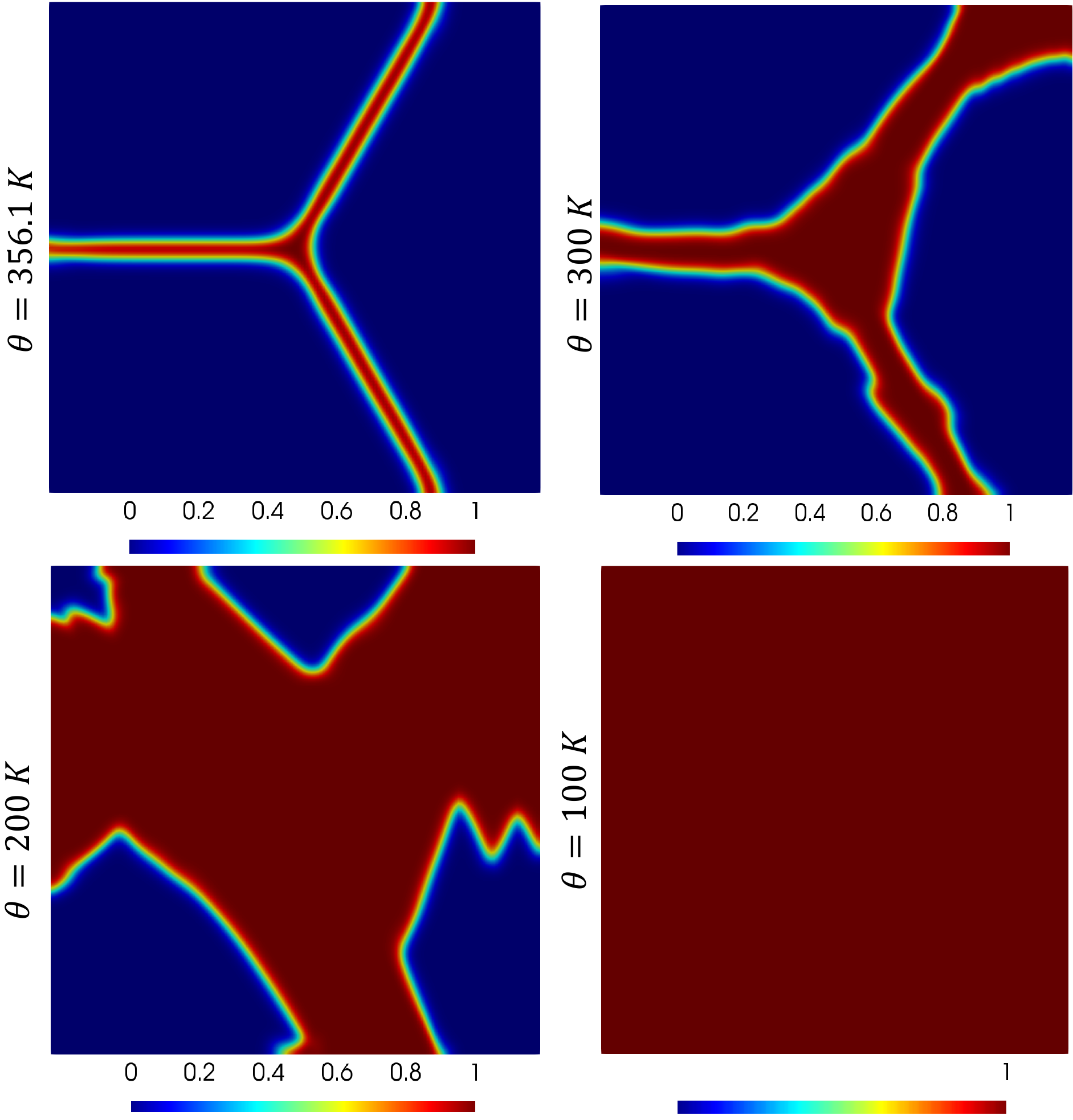}
	}\hspace{1mm}
    \subfigure[$\eta_{eq}$ plots during forward MT]{
    \includegraphics[width=3.0in, height=3.00in] {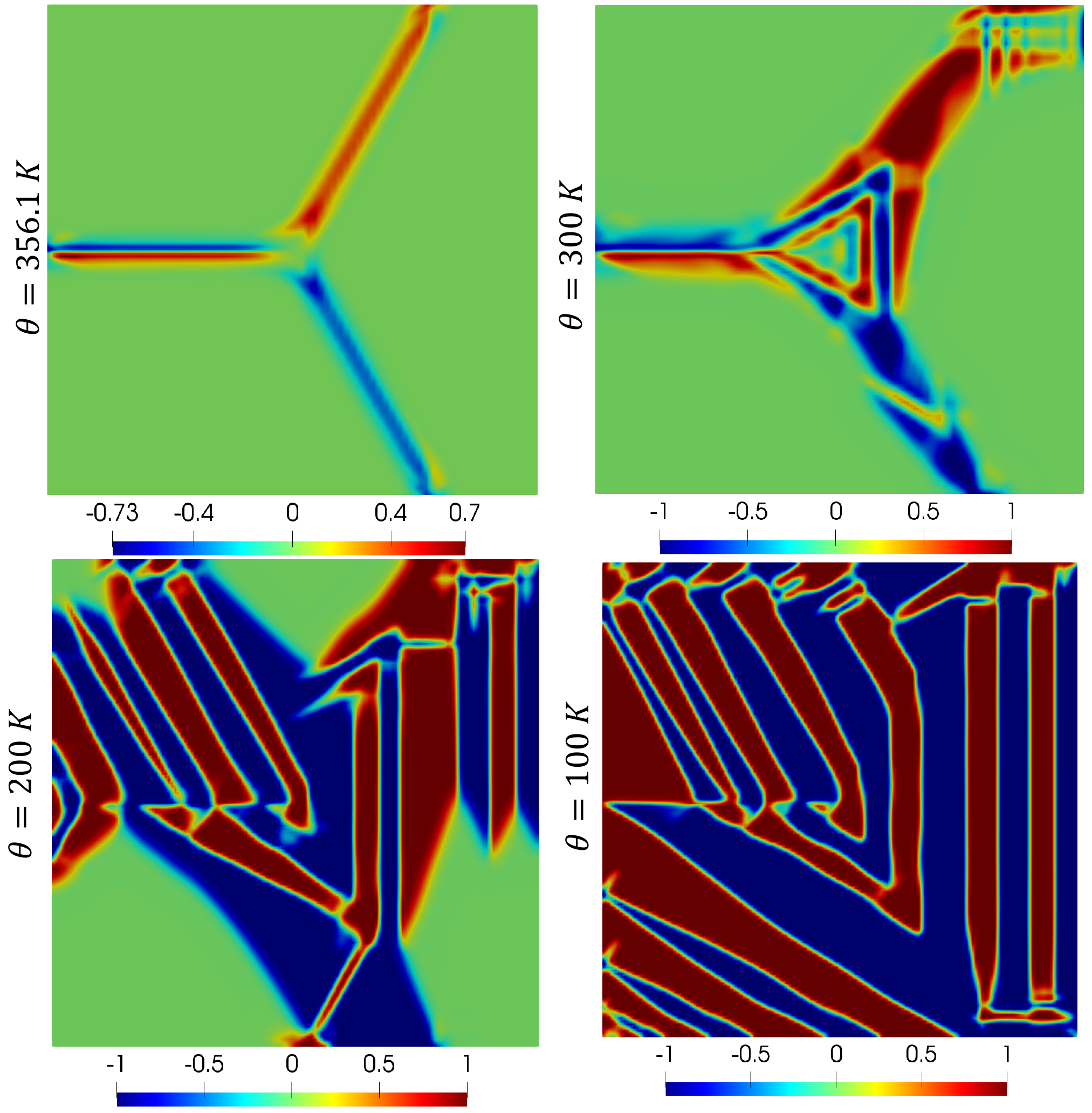}
}
\subfigure[$\eta_0$ plots during reverse MT]{
  \includegraphics[width=3.0in, height=3.00in] {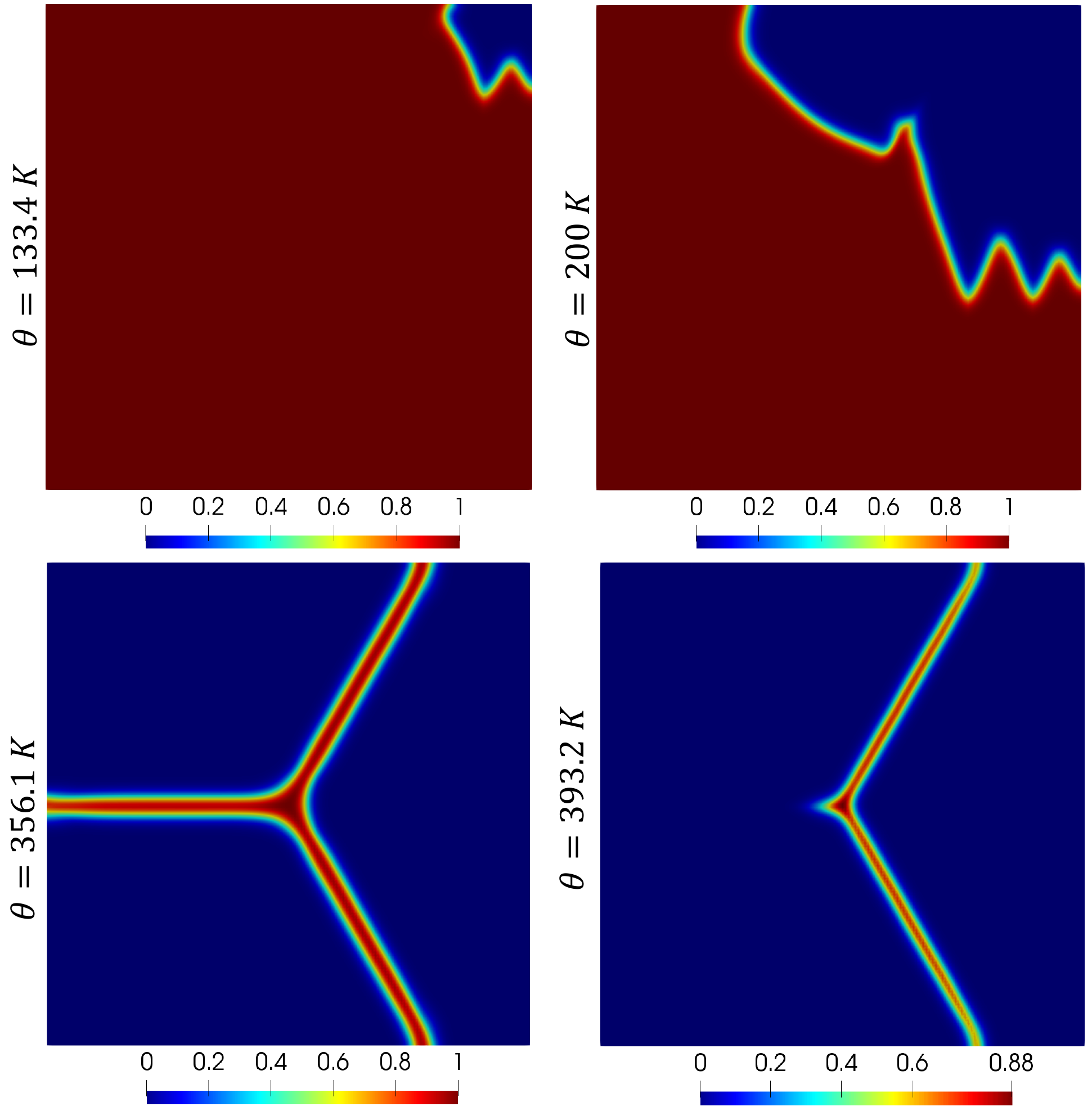}
	}\hspace{1mm}
    \subfigure[$\eta_{eq}$ plots during reverse MT]{
    \includegraphics[width=3.0in, height=3.00in] {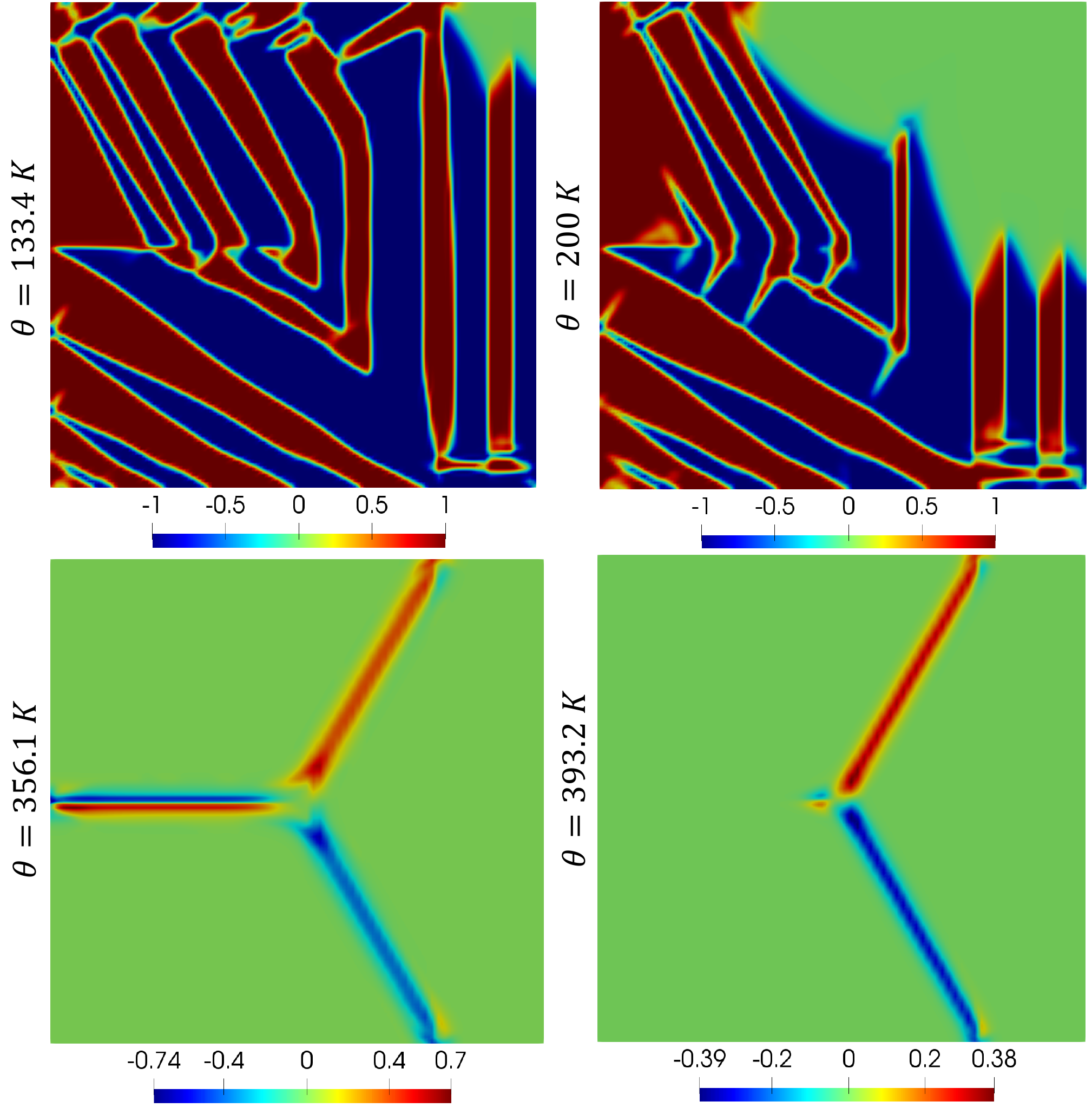}
}
\caption{Microstructure evolution during (a,b) forward, and (c,d) reverse MTs in a $30$ nm $\times$ $30$ nm tricrystal having symmetric planar GBs with $\theta$ variation. We used $\Delta\gamma_{12}^{gb} = \Delta\gamma_{13}^{gb} =\Delta\gamma_{23}^{gb} = -0.7$ N/m, $\delta_{12}^{gb}=\delta_{13}^{gb}=\delta_{23}^{gb}=1$ nm, and  $\rho_0K_{IJ}^{gb}=\rho_0K_{123}=800$ GPa. No strain applied.}
\label{triGBenergy}
\end{figure}

\subsubsection{Effect of change in GB energy $\Delta\gamma_{IJ}^{gb}$}
\label{TJGBenergy}
We now consider another $30$ nm $\times$ $30$ nm-sized tricrystal and obtain the results for $\Delta\gamma_{12}^{gb}=\Delta\gamma_{13}^{gb}=\Delta\gamma_{23}^{gb}=-0.7$ N/m as shown in Fig. \ref{triGBenergy}. All other parameters  are identical to that considered in sample shown in Fig. \ref{triGBwidth1nm}: $\delta^{gb}_{IJ}=1$ nm,  and $\rho_0K_{IJ}^{gb}=\rho_0K_{123}=800$ GPa. The numerical width of the dry GBs in both  the austenitic samples $V_0$ is  $1.07$ nm. The results for $\Delta\gamma_{IJ}^{gb}=-0.5$ N/m and $\Delta\gamma_{IJ}^{gb}=-0.7$ N/m are also compared. 

\noindent{\bf Forward MT.} \hspace{3mm} The $\sf M$ phase nucleates in a jump-like manner on the GBs and TJ region at $\theta_n^f=356.1$ K as shown in Fig. \ref{triGBenergy}(a,b), which is $42.8$ K higher than the case when $\Delta\gamma_{IJ}^{gb}=-0.5$ N/m. The volume fraction of nucleated $\sf M$ in the sample of Fig. \ref{triGBenergy}(a,b) is much less than that of Fig. \ref{triGBwidth1nm}(a,b). As $\theta$ is decreased, the volume fraction of $\sf M$ increases and progresses inside the grains. The microstructures at $300$ K, $200$ K, and $100$ K are shown in Fig. \ref{triGBenergy}(a,b). Although, there is not much difference in the volume fraction of the $\sf M$ phase at these temperatures between the samples in Figs. \ref{triGBwidth1nm}(a,b) and \ref{triGBenergy}(a,b), the evolution of the microstructure significantly differs.  The arrangement of the twin plates is also different in $G_1$ and $G_2$ between these samples. Orientation of the plates far from the GBs is close to the crystallographic solutions for a large isolated grain mentioned in Sec. \ref{TJGBwidth}. The arrangement of the variants in the junctions regions of the fully martensitic samples is  different.  

\noindent{\bf Reverse MT.} \hspace{3mm}  As $\theta$ is raised from $100$ K, an $\sf A$ region nucleates at $\theta_n^r=133.4$ K, as shown in  Fig. \ref{triGBenergy}(c,d), which is $11.3$ K lower than the corresponding $\theta_n^r$ for the sample with $\Delta\gamma_{IJ}^{gb}=-0.5$ N/m (see Fig. \ref{triGBwidth1nm}(c,d)). On the further rise of $\theta$, we observe a large area of $G_1$ and $G_3$ transform to $\sf A$, while $G_2$ is still fully martensitic at $200$ K as shown in Fig. \ref{triGBenergy}(c,d). At $\theta=\theta_n^f=356.1$ K, the thin layers of residual $\sf PM$ are observed in the GB, and junction regions similar to the $\sf PM$ appeared during the forward MT. The GBs and TJ  retained $\sf PM$ up to $\theta=393.2$ K, and the entire tricrystal transformed to austenite in a jump-like manner at $393.3$ K. 

\begin{figure}[t!]
\centering
\hspace{-8mm}
\subfigure[$\eta_0$ plots during forward MT]{
  \includegraphics[width=3.0in, height=3.00in] {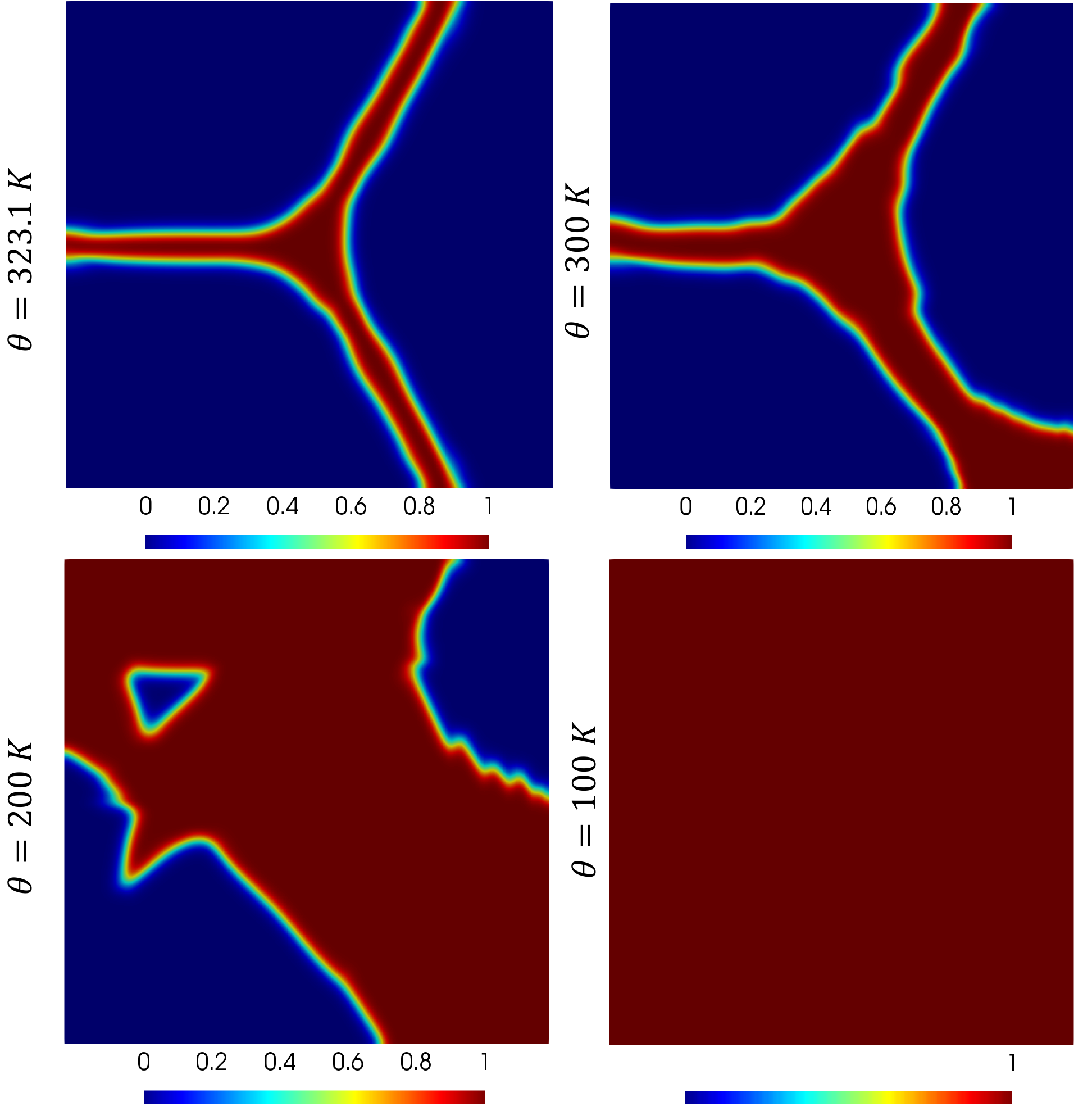}
	}\hspace{1mm}
    \subfigure[$\eta_{eq}$ plots during forward MT]{
    \includegraphics[width=3.0in, height=3.00in] {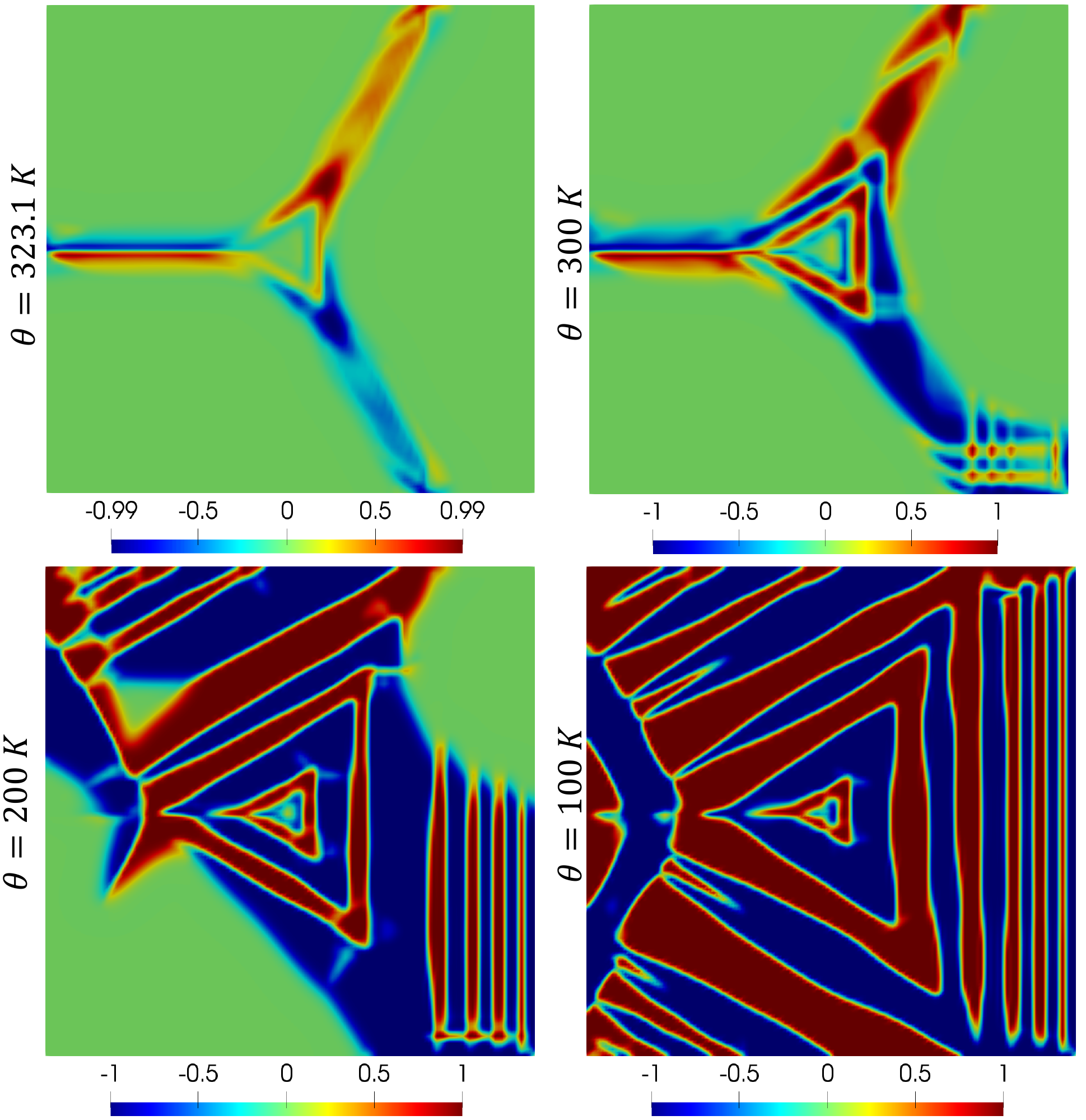}
}
\hspace{-8mm}
\subfigure[$\eta_0$ plots during reverse MT]{
  \includegraphics[width=3.0in, height=3.00in] {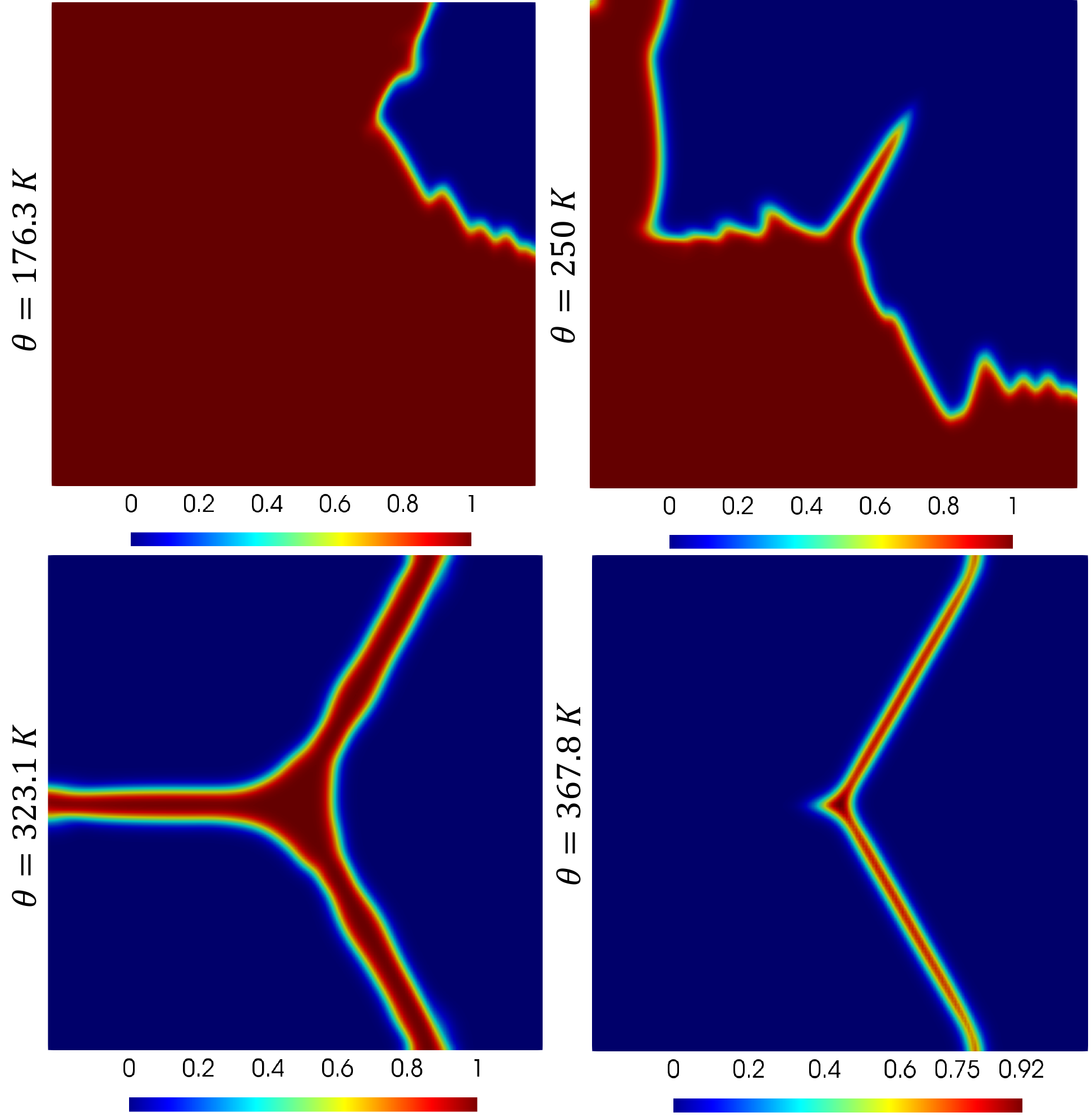}
	}\hspace{1mm}
    \subfigure[$\eta_{eq}$ plots during reverse MT]{
    \includegraphics[width=3.0in, height=3.00in] {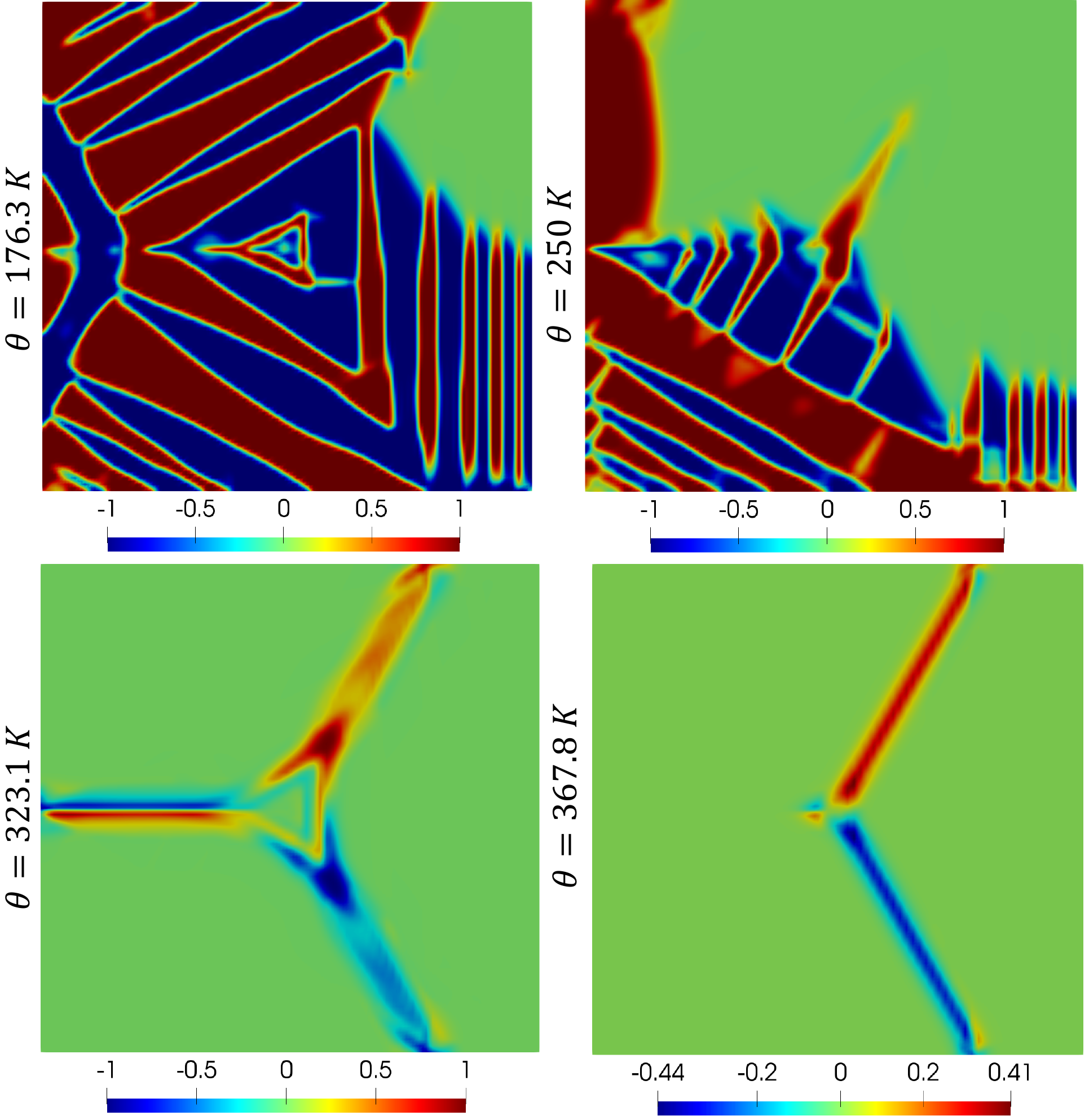}
}
\caption{Microstructure evolution during (a,b) forward, and (c,d) reverse MTs in a $30$ nm $\times$ $30$ nm tricrystal having symmetric planar GBs with $\theta$ variation. We used $\Delta\gamma_{12}^{gb} = \Delta\gamma_{13}^{gb} =\Delta\gamma_{23}^{gb} = -0.5$ N/m,  $\delta_{12}^{gb}=\delta_{13}^{gb}=\delta_{23}^{gb}=1$ nm, and $\rho_0K_{IJ}^{gb}=\rho_0K_{123}=1200$ GPa. No strain applied.}
\label{TJenrgyy}
\end{figure}

\subsubsection{Effect of junction energy}
\label{TJenergy}
We recall that in Eq. \eqref{MF2}, we introduced the  parameter $K_{HIJ}$, and by varying it, we can control the energy and size of the TJ region. We take an austenitic tricrystal of $30$ nm $\times$ $30$ nm size obtained by considering $\rho_0K_{IJ}^{gb}=\rho_0K_{123}=1200$ GPa, while all other parameters are identical to that used for Fig. \ref{triGBwidth1nm}: $\delta_{IJ}^{gb}=1$ nm, and $\Delta\gamma_{IJ}^{gb} = -0.5$ N/m.  The numerical width of all the GBs in the austenitic tricrystal is obtained as $1.06$ nm, which is closer to the analytical width than the case with $\rho_0K_{IJ}^{gb}=\rho_0K_{123}=800$ GPa (see Sec. \ref{TJGBwidth}). The microstructure evolution during  the forward and reverse MTs are shown in Fig. \ref{TJenrgyy} when  $\rho_0K_{IJ}^{gb}=\rho_0K_{123}=1200$ GPa, and we compare the results with Fig. \ref{triGBwidth1nm} for $\rho_0K_{IJ}^{gb}=\rho_0K_{123}=800$ GPa .

\noindent{\bf Forward MT.} \hspace{3mm}  The  $\sf M$ nucleates in a jump-like manner on the GBs and TJ at $\theta_n^f=323.1$ K, which is  $9.8$ K higher than the sample with lower junction energy ($\rho_0K_{IJ}^{gb}=\rho_0K_{123}=800$ GPa). We thus conclude that the combined effect of a slightly lower numerical width and the higher junction energy promotes $\sf M$ nucleation. With a further decrease in $\theta$, the volume fraction of $\sf M$ increases, and we  get a fully martensitic sample at $100$ K. In the wholly transformed samples shown in Figs. \ref{triGBwidth1nm}(a,b) and \ref{TJenrgyy}(a,b), the microstructures are qualitatively similar with some differences in the number of $\sf M$ plates. However, the microstructures at the intermediate temperatures in these samples were different. 

\noindent{\bf Reverse MT.} \hspace{3mm} As depicted in Fig. \ref{TJenrgyy}(c,d), a large $\sf A$ region nucleates within $G_3$ at the upper-right corner at $\theta_n^r=176.3$ K. At $250$ K, a significant volume fraction of $G_1$ and $G_3$ are transformed to $\sf A$, while $G_2$ is still fully martensitic; see Fig. \ref{TJenrgyy}(a,b). At $\theta=\theta_n^f=323.1$ K, residual $\sf M$ is observed only in the GB and junction regions, and the obtained microstructure is almost the same as that obtained during the forward MT, as shown in Fig. \ref{TJenrgyy}(a,b). A residual $\sf PM$ is observed on $\Gamma_{13}$, $\Gamma_{23}$, and the junction regions at $367.8$ K, which  barrierlessly transforms to $\sf A$ at $367.9$ K leaving behind the original austenitic tricrystal.

\subsubsection{Effect of sample size}
\label{TJgrainsenergy}
\begin{figure}[t!]
\centering
\hspace{-8mm}
\subfigure[$\eta_0$ plots during forward MT]{
  \includegraphics[width=3.05in, height=3.3in] {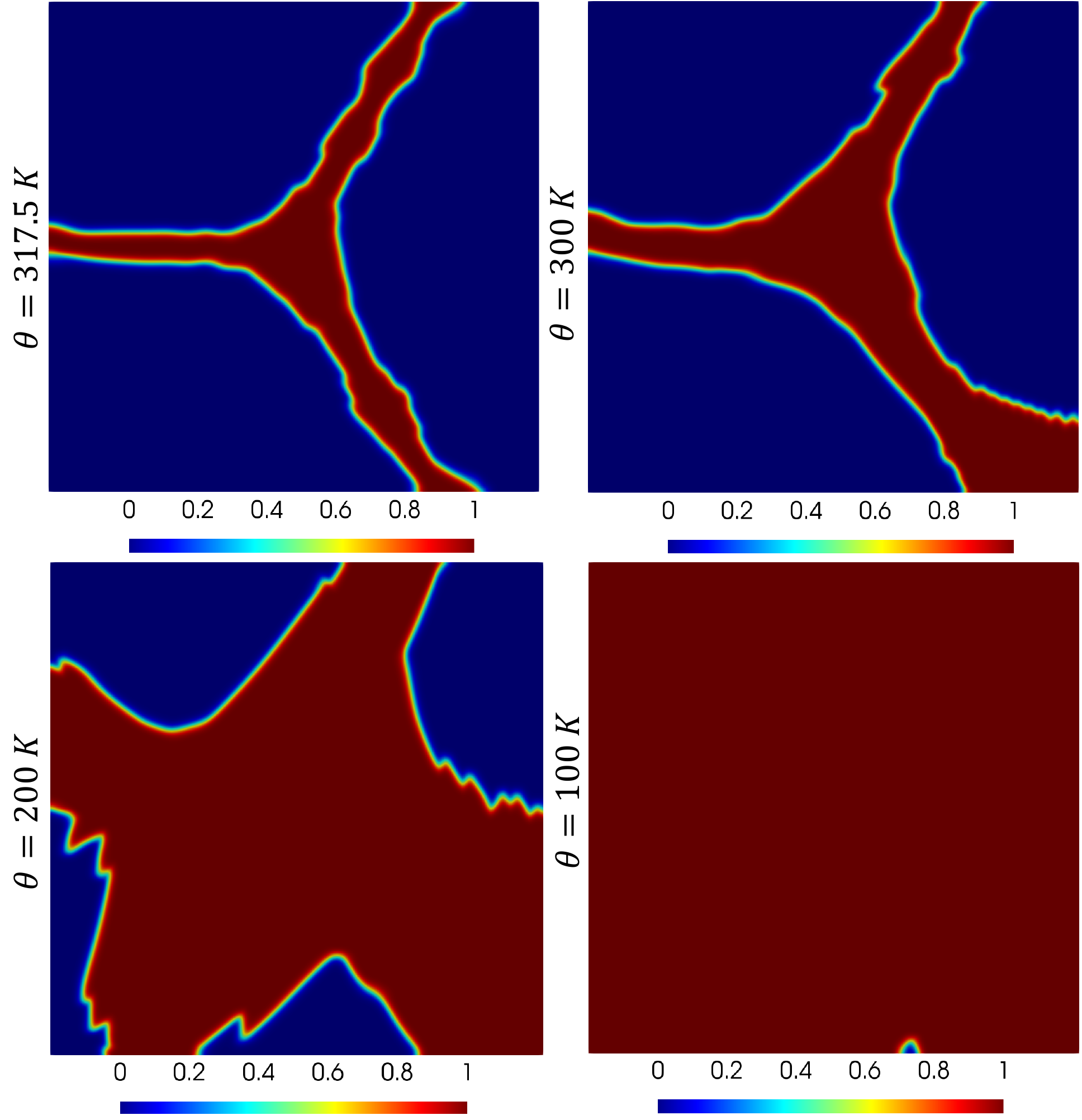}
	}\hspace{1mm}
    \subfigure[$\eta_{eq}$ plots during forward MT]{
    \includegraphics[width=3.05in, height=3.3in] {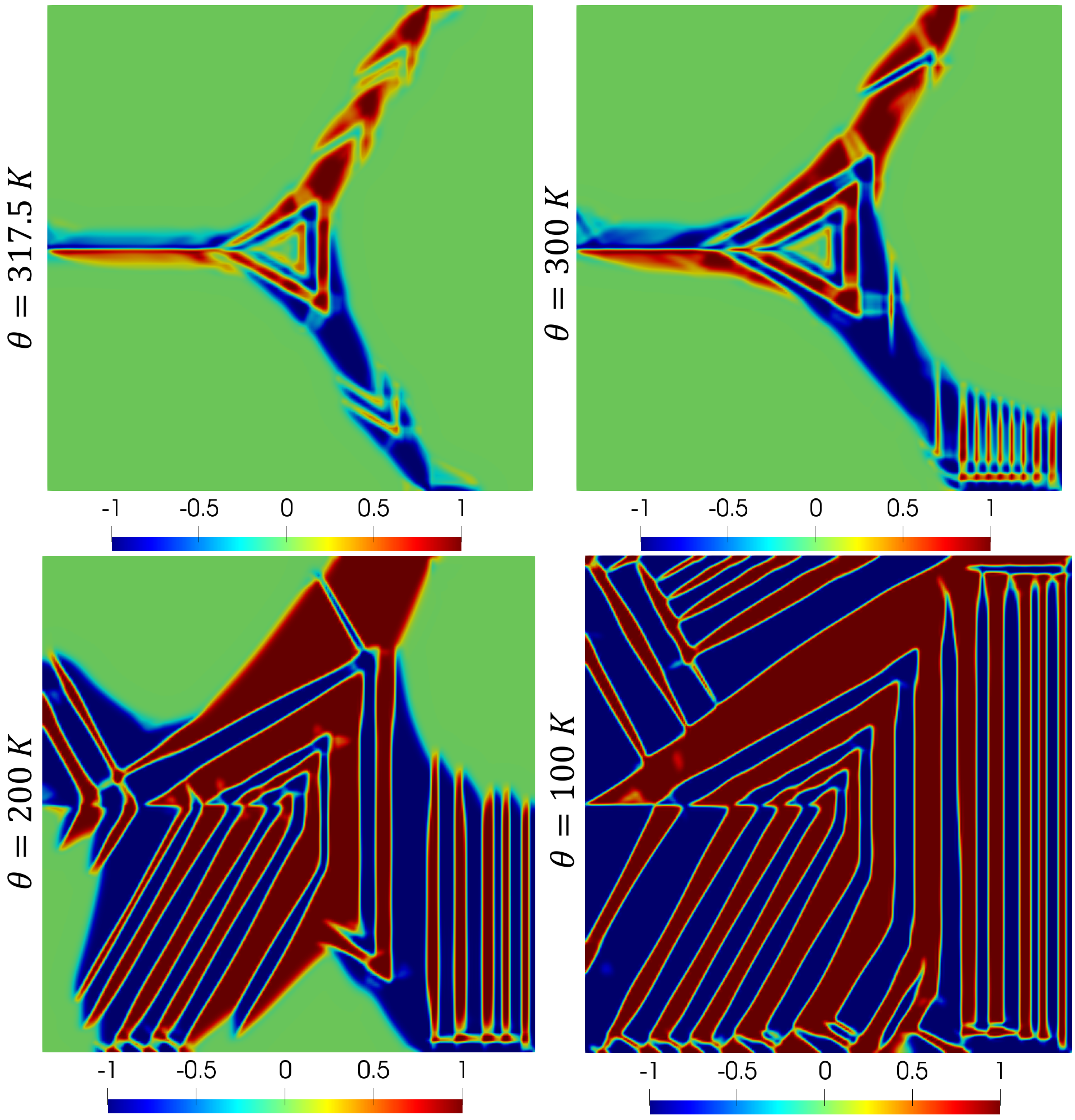}
}
\caption{Microstructure evolution during (a,b) forward, and (c,d) reverse MTs in a $50$ nm $\times$ $50$ nm tricrystal having symmetric planar GBs with $\theta$ variation. We used $\Delta\gamma_{12}^{gb} = \Delta\gamma_{13}^{gb} =\Delta\gamma_{23}^{gb} = -0.5$ N/m, $\delta_{12}^{gb}=\delta_{13}^{gb}=\delta_{23}^{gb}=1$ nm, and $\rho_0K_{IJ}^{gb}=\rho_0K_{123}=800$ GPa. No strain applied.}
\label{trigrainsize}
\end{figure}

To show the effect of the sample size, we consider an austenitic tricrystal of $50$ nm $\times 50$ nm size. All the parameters are identical to the example shown in Fig. \ref{triGBwidth1nm}. The results for the forward MT is shown in Fig. \ref{trigrainsize}. The numerical width of the GB is obtained as $1.07$ nm. The $\sf M$ phase nucleates at $\theta_n^f=317.5$ K in a jump-like manner.  Although the volume fraction of $\sf M$ at a given $\theta$ is almost the same in both the samples, the variant plates are longer and more regularly spaced in the larger sample. The sample is fully martensitic at $100$ K. There is a difference in the orientation of the plates in both the martensitic samples. In grain $G_1$ of the larger sample, two sets of variant plates oriented at $30^\circ$ and  $60^\circ$  with respect to the $\fg e_1$-axis are obtained. The variant plates are mainly oriented at $30^\circ$ with respect to $\fg e_1$-axis in $G_2$ except for a small region of the bottom surface. The plates are vertical in $G_3$. However, a pair of horizontal plates are seen near the top and bottom surfaces. 

\noindent{\bf Remarks:} We could not find the relevant experimental or atomistic results for quantitatively comparing the transformation temperatures influenced by GBs in the presence of TJs with our numerical results. However, the nucleation at the GBs and microstructure development about the GBs are in agreement with the experimental and atomistic results in bicrystals \cite{Ueda-2001Acta,Ueda-2004ISIJ,Qin-2018}, as discussed in Sec. \ref{sussub1} in details. Notably, the existing PF models initiate nucleation at the GBs by placing artificial nuclei or considering noise in the order parameters or displacement fields as discussed in details in Sec. \ref{introd}. on the other hand, the GB parameters $\Delta\gamma_{IJ}^{gb}$ and $\delta_{IJ}^{gb}$ in the bicrystals, and an additional TJ parameter $\rho_0K_{IJK}$ play the central role in determining the nucleation of $\sf M$ and the subsequent microstructure evolution, which justifies the novely of the present model.

\section{Concluding remarks}
\label{consls}
We have developed a thermodynamically consistent novel multiphase PF approach for studying GB-induced MTs in polycrystalline solids, where $N$ independent order parameters are used for describing $\sf A$ and $N(>1)$ martensitic variants, and another $M$ independent order parameters are used for describing the $M(>1)$ grains of the sample. The transformation stretch tensor for each of the grains is taken as a linear combination of the Bain tensors of all the variants (appropriately rotated about a reference frame) multiplied with nonlinear interpolation functions of the order parameters related to the phases. The transformation stretch tensor of the sample is then taken as a linear combination of the transformation strain tensor for each grain multiplied with a nonlinear interpolation function related to the grains. The free energy of the system is composed of the elastic energy, barrier energy between $\sf A$-$\sf M$ and also between all the variants, the thermal (or chemical) energy of the phases, the energy penalizing the deviation of the transformation paths between the variant from the prescribed ones, the energy of the interfaces between $\sf A$-$\sf M$ and the ones between the variants, and the isotropic energy (independent of GB orientation) of the GB region. The energy of the GB region is considered to be composed of the barrier energy between the grains, gradient energies of the GBs, and energies of the triple junctions, where the barrier energy coefficients and the gradient energy coefficients are considered to be functions of the order parameter $\eta_0$, which allows the GB energy to vary during the $\sf A\leftrightarrow\sf M$ transformations due to the structural changes. The total stresses are composed of the elastic and the structural parts within the $\sf A$-$\sf M$ interfaces, variant-variant boundaries, and the GBs. In order to obtain the correct form of the structural stresses, all the barrier energies and the gradient energies are multiplied with the determinant of the deformation gradient tensor.  The coupled system of mechanical equilibrium equations and the Ginzburg-Landau equations for all the order parameters related to the $\sf A$-$\sf M$ phases, the variants, and the grains are derived. The governing partial differential equations are solved using the finite element method and assuming the isotropic elastic response of the solid under plane stress condition. The results for the 2D bicrystals with a planar symmetric tilt GB, a planar asymmetric tilt GB, and a circular symmetric tilt GB are presented, where the strong effect of the stress-free austenitic GB width, change in GB energy due to the MTs, GB misorientation, applied strain, and sample size are studied  in details during both the forward ($\sf A \to \sf M$) and reverse ($\sf M \to \sf A$) transformations. A rich plot for the temperatures of barrierless (jump-like) transformations between $\sf A$, $\sf PM$, and $\sf M$ in the GB region within a bicrystal with symmetric planar tilt GB with a variation of the stress-free austenitic GB width was plotted, keeping all other parameters identical. All the transformation temperatures increase with a decrease in the GB width. The effect of the GB width on  temperature hysteresis is also studied. The role of the incompatibility of the transformation strains for each variant across the GB on microstructures is studied. The nucleation of $\sf PM$ or $\sf A$ from a fully martensitic sample during the reverse transformation was seen at the highest energetic sites of the sample. The GB parameters are shown to play an important role on the microstructures and austenite finish temperature.
 The effect of the GB width, change in GB energy due to MTs, TJ energy, and sample size on the transformation temperatures and subsequent microstructure evolution during the forward and reverse MTs is studied for 2D  tricrystals consisting of three planar symmetric tilt GBs meeting in the middle of the sample at $120^\circ$ dihedral angles. The compatibility of the twinned microstructures across the GBs and TJ is also studied. The elastic and structural stresses within the fully martensitic bicrystals and tricrystals are shown, which are observed to be concentrated mainly within the GB and TJ regions, and across the  variant-variant interfaces. Understanding of the stress distribution is important for the materials design and their failure analysis \cite{Schuh-13}.

The numerical results have been compared with the limited experimental and atomistic results available in the literature, and there is a good qualitative agreement between them.
This paper motivates the researchers to conduct controlled experiments and atomistic studies to better understand the role of the GB parameters on the nucleation of $\sf PM$ and $\sf M$ at the nanoscale. The present PF results also suggest that TJ energy and size play a significant role in the temperatures of transformations and microstructures, which is yet to be confirmed systematically, both experimentally and through atomistic simulations. Systematic studies are needed for quantification of the change in energies during to MTs within the GBs and TJ regions, which can be compared with the present model for the GB  energy (see Eqs. \eqref{MF2}and \eqref{propertiesf}).
 The present PF model can further study the GB-induced MTs in a more realistic polycrystalline sample with numerous grains.  The model can also be extended for studying the role of other heterogeneities, including point defects, dislocation pileups, external surface, and the role of diffusion in polycrystalline solids coupled to GB-induced MTs. 
 
\appendix
{\bf Appendix}
\section{Derivation for the dissipation inequality}
\label{psidot}
In this appendix, we derive the dissipation inequality \eqref{dissiIneq1}, which is finally used in Sec. \ref{dississipat} for deriving the total, elastic, and structural stresses and the Ginzburg-Landau equations. The general thermodynamically consistent framework at large strains developed for a single grain in \cite{Levitas-14a,Basak-Levitas-2018JMPS}  is extended for polycrystalline samples in this paper. 
We directly start with the following dissipation inequality, which is derived using the first and second laws of thermodynamics like that of \cite{Levitas-14a,Basak-Levitas-2018JMPS}:
\begin{equation}
 \rho_0{\mathcal D}=\fg P:\dot{\fg F}^T-\rho_0\dot\psi-\rho_0 s \dot\theta+\nabla_0\cdot(\fg q_0^\eta\dot\eta_0)
 +\sum_{i=1}^N\nabla_0\cdot(\fg q_i^\eta\dot\eta_i) +\sum_{I=1}^M\nabla_0\cdot(\fg q_I^\xi\dot\xi_I) \geq 0
 \quad \text{in }\Omega_0,
\label{dissiIneq10}
\end{equation}
where $s$ is the specific entropy, the vectors $\fg q_0^\eta$, $\fg q_i^\eta$, and $\fg q_I^\xi$ are the generalized forces which are conjugate to the gradient of the rate of the associated order parameters in $\Omega_0$, and all other symbols are defined in Sec. \ref{system_eqns}.  Note that $\fg q_0^\eta$, $\fg q_i^\eta$, and $\fg q_I^\xi$, which are introduced here for balancing some terms from the inequality \eqref{dissiIneq10}, are yet to be determined (see \cite{Levitas-14a} for details). In addition to the inequality \eqref{dissiIneq10}, we also get the Fourier's inequality $-\frac{1}{\theta}\fg h_0\cdot\nabla_0\theta \geq 0$ \cite{Levitas-14a}, where $\fg h_0$ is the heat flux vector. Based on this inequality, one can assume the Fourier's law  $\fg h_0 = -\fg K_\theta\cdot\nabla_0\theta$, where $\fg K_\theta$ is the heat conductivity tensor which is symmetric and positive semi-definite. Note that $\fg h_0=\fg 0$ for a sample with uniform temperature. 

 We now work on the inequality \eqref{dissiIneq10} and rewrite it in a suitable form.  The material time derivative of the free energy give by Eq. (\ref{MF0}) is obtained as
\begin{eqnarray}
 \dot\psi &=& \frac{J_t}{\rho_0}\frac{\partial\psi_e}{\partial\fg F_e}\cdot \fg F_t^{-T}:\dot{\fg F}^T
 -\frac{J_t}{\rho_0}\fg F_e^T\cdot\frac{\partial\psi_e}{\partial\fg F_e}\cdot\fg F_t^{-T}:\dot{\fg F}_t^T
  + \frac{J_t\psi_e}{\rho_0}\fg F_t^{-T}:\dot{\fg F}_t^T
 +J(\breve{\psi}^{\theta}+\psi^\nabla+\psi^\xi){\fg F}^{-T}:\dot{\fg F}^T+
 \nonumber\\  && \sum_{i=0}^N\left(J\frac{\partial(\breve{\psi}^{\theta}+\psi^\nabla)}{\partial\eta_i}+\frac{\partial(\tilde\psi^\theta+\psi^p)}
 {\partial\eta_i}+\frac{J_t}{\rho_0}\left.\frac{\partial\psi_e}{\partial\eta_i}\right|_{\fg F_e}\right)\dot\eta_i+J\frac{\partial{\psi^\xi}}{\partial\eta_0}\dot\eta_0 + \sum_{I=1}^M\left(J\frac{\partial{\psi}^{\xi}_{loc}}{\partial\xi_I}+\frac{J_t}{\rho_0}\left.\frac{\partial\psi_e}{\partial\xi_I}\right|_{\fg F_e}\right)\dot\xi_I
 +
 \nonumber\\
 &&
J\sum_{i=0}^N\frac{\partial\psi^\nabla}{\partial\nabla\eta_i}\cdot\dot{\overline{\nabla\eta_i}}  + 
  J\sum_{I=1}^M\frac{\partial\psi^\nabla}{\partial\nabla\xi_I}\cdot\dot{\overline{\nabla\xi_I}} + \frac{\partial\psi}{\partial\theta}\dot\theta  ,
\label{dotpsi1}
\end{eqnarray}
where we have used the following time derivatives $\dot{\overline{det\,\fg A}}=({det\,\fg A}) \fg A^{-T}:\dot{\fg A}^T$, and $\dot{\overline{\fg A^{-1}}}=-\fg A^{-1}\cdot\dot{\fg A}\cdot\fg A^{-1}$  for an arbitrary invertible second order tensor $\fg A(t)$. The following term from Eq. (\ref{dotpsi1}), for example, is rewritten as
\begin{equation}
\frac{\partial\psi^\nabla}{\partial\nabla\eta_i}\cdot\dot{\overline{\nabla\eta_i}}
 = \frac{\partial\psi^\nabla}{\partial\nabla\eta_i}\cdot\dot{\overline{(\fg F^{-T}\cdot{\nabla_0\eta_i})}}
 =\left( \fg F^{-1}\cdot\frac{\partial\psi^\nabla}{\partial\nabla\eta_i}\right)\cdot\dot{\overline{ \nabla_0\eta_i}}
 -\left(\nabla\eta_i\otimes\fg F^{-1}\cdot\frac{\partial\psi^\nabla}{\partial\nabla\eta_i} \right):\dot{\fg F}^T ,
\label{dotpsi2}
\end{equation}
where we have used the relation $ \nabla\eta_i=\fg F^{-T}\cdot\nabla_0\eta_i$ \cite{Levitas-14a}.
Using Eq. \eqref{dotpsi2} into Eq. \eqref{dotpsi1} and rearranging the terms,  $\dot\psi$ is rewritten as
\begin{eqnarray}
 \dot\psi &=& \left(\frac{J_t}{\rho_0}\frac{\partial\psi_e}{\partial\fg F_e}\cdot\fg F_t^{-T}
 +J(\breve{\psi}^{\theta}+\psi^\nabla+\psi^\xi){\fg F}^{-T} -J\sum_{i=0}^N\nabla\eta_i \otimes\fg F^{-1}\cdot\frac{\partial\psi^\nabla}{\partial\nabla\eta_i}
 -J\sum_{I=1}^M\nabla\xi_I \otimes\fg F^{-1}\cdot\frac{\partial\psi^\xi_\nabla}{\partial\nabla\xi_I} \right):\dot{\fg F}^T+  
     \nonumber\\ &&
\sum_{i=0}^N\left(J\frac{\partial(\breve{\psi}^{\theta}+\psi^\nabla)}{\partial\eta_i}+\frac{\partial({\tilde\psi}^\theta+\psi^p)} {\partial\eta_i}
    +\frac{J_t}{\rho_0}\left.\frac{\partial\psi_e}{\partial\eta_i}\right|_{\fg F_e}\right)\dot\eta_i +J\frac{\partial{\psi^\xi}}{\partial\eta_0}\dot\eta_0+
   \frac{J_t}{\rho_0}  \left(\psi_e\fg F_t^{-T} - \fg F_e^T\cdot\frac{\partial\psi_e}{\partial\fg F_e}\cdot\fg F_t^{-T}
 \right):\dot{\fg F}_t^T  + 
     \nonumber\\ &&
 \sum_{I=1}^M\left(J\frac{\partial{\psi}^{\xi}_{loc}}{\partial\xi_I}+ \frac{J_t}{\rho_0}\left.\frac{\partial\psi_e}{\partial\xi_I}\right|_{\fg F_e}\right)\dot\xi_I +
 \sum_{i=0}^N J\left(\fg F^{-1}\cdot\frac{\partial\psi^\nabla}{\partial\nabla\eta_i}\right)
 \cdot\dot{\overline{ \nabla_0\eta}}+
  \sum_{I=1}^MJ\left(\fg F^{-1}\cdot\frac{\partial\psi^\nabla}{\partial\nabla\xi_I}\right)\cdot\dot{\overline{\nabla_0\xi_I}} + \frac{\partial\psi}{\partial\theta}\dot\theta.
\label{dotpsi3y}
\end{eqnarray}
Since $\fg F_t$ is considered to be a function of all the order parameters $\eta_0,\eta_1,\ldots,\eta_N$ and $\xi_1,\ldots,\xi_M$ (see Eq. \eqref{utilde}), Eq. (\ref{dotpsi3y}) can further be rewritten as
\begin{eqnarray}
 \dot\psi &=& \left(\frac{J_t}{\rho_0}\frac{\partial\psi_e}{\partial\fg F_e}\cdot\fg F_t^{-T}
 +J(\breve{\psi}^{\theta}+\psi^\nabla+\psi^\xi){\fg F}^{-T} -J
\sum_{i=0}^N\nabla\eta_i \otimes\fg F^{-1}\cdot\frac{\partial\psi^\nabla}{\partial\nabla\eta_i} -J\sum_{I=1}^M\nabla\xi_I \otimes\fg F^{-1}\cdot\frac{\partial\psi^\xi_\nabla}{\partial\nabla\xi_I}\right):\dot{\fg F}^T
+    \nonumber\\ &&
 \sum_{i=0}^N\left[J\frac{\partial(\breve{\psi}^{\theta}+\psi^\nabla)}{\partial\eta_i}+    \frac{\partial({\tilde\psi}^\theta+\psi^p)} {\partial\eta_i}
    +\frac{J_t}{\rho_0}\left.\frac{\partial\psi_e}{\partial\eta_i}\right|_{\fg F_e}+
  \frac{J_t}{\rho_0}\left(\psi_e\fg F_t^{-T} -\fg F_e^T\cdot\frac{\partial\psi_e}{\partial\fg F_e}\cdot\fg F_t^{-T}
 \right):\frac{\partial{\fg F}_t^T}{\partial\eta_i}\right]\dot\eta_i +
     \nonumber\\ &&
J\frac{\partial{\psi^\xi}}{\partial\eta_0}\dot\eta_0+   \sum_{I=1}^M\left[J\frac{\partial{\psi}^{\xi}}{\partial\xi_I}+  \frac{J_t}{\rho_0}\left.\frac{\partial\psi_e}{\partial\xi_I}\right|_{\fg F_e} +
    \frac{J_t}{\rho_0} \left(\psi_e\fg F_t^{-T} - \fg F_e^T\cdot\frac{\partial\psi_e}{\partial\fg F_e}\cdot\fg F_t^{-T}
 \right):\frac{\partial{\fg F}_t^T}{\partial\xi_I}
     \right]\dot\xi_I +
  \nonumber\\ &&
   \sum_{i=0}^N J\left(\fg F^{-1}\cdot\frac{\partial\psi^\nabla}{\partial\nabla\eta_i}\right)
 \cdot\dot{\overline{ \nabla_0\eta}}+
  \sum_{I=1}^MJ\left(\fg F^{-1}\cdot \frac{\partial\psi^\nabla}{\partial\nabla\xi_I}\right)\cdot\dot{\overline{\nabla_0\xi_I}} + \frac{\partial\psi}{\partial\theta}\dot\theta.
\label{dotpsi3}
\end{eqnarray}
Using  Eq. \eqref{dotpsi3}  in the inequality \eqref{dissiIneq10}, and assuming that the dissipation rate is independent of $\dot\theta$,  $\dot{\overline{ \nabla_0\eta_l}}$  (for $l=0,\ldots,N$) and  $\dot{\overline{ \nabla_0\xi_I}}$  (for $I=0,\ldots,M$) \cite{Levitas-14a}, we get the specific entropy as  $s=-{\partial\psi}/{\partial\theta}$, and the generalized force vectors as
\begin{equation}
\fg q_{l}^\eta = \rho_0J\fg F^{-1}\cdot\frac{\partial\psi^\nabla}{\partial\nabla\eta_l} \quad \text{for } l=0,1,\ldots,N;  \quad \text{and}\quad
\fg q_{I}^\xi = \rho_0J\fg F^{-1}\cdot\frac{\partial\psi^\nabla}{\partial\nabla\xi_I} \quad \text{for} \quad I=1,\ldots,M.
\label{hl5f342}
\end{equation}
Using Eqs. \eqref{dotpsi3} and \eqref{hl5f342}, the dissipation inequality \eqref{dissiIneq10} is finally expressed in the form given by \eqref{dissiIneq1}, and it is used in Sec.  \ref{dississipat} for deriving the governing equations for the stresses and the Ginzburg-Landau equations.

\section{A list of the governing equations in the general form}
\label{Listequns2d}
\begin{itemize}
\item{{\bf Order parameters:} $\eta_0$ is the order parameter describing $\sf A\leftrightarrow\sf M$ transformations; $\eta_1,\eta_2,\ldots,\eta_N$ are the order parameters describing the variants ${\sf M}_1, {\sf M}_2, \ldots, {\sf M}_N$, respectively; $\xi_1,\xi_2,\ldots,\xi_M$ are the order parameters describing grains $G_1, G_2, \ldots, G_M$, respectively. The order parameters satisfy
\begin{equation}
\sum_{i=1}^N\eta_i=1; \quad \text{and}\quad  \sum_{I=1}^M\xi_I=1.
 \label{multdecom_B1cons}
\end{equation}
}
\item{ Kinematic relations:}
\begin{equation}
{\boldsymbol F} = \fg V_e\cdot \fg R\cdot \fg U_t;  \,\,
{\boldsymbol E} = \frac{1}{2}({\boldsymbol F}^T\cdot{\boldsymbol F}-{\boldsymbol I}); \,\,
 {\boldsymbol E}_e = \frac{1}{2}({\boldsymbol F}_e^T\cdot{\boldsymbol F}_e-{\boldsymbol I}); \,\,
 \fg b = \frac{1}{2}({\boldsymbol F}\cdot{\boldsymbol F}^T-{\boldsymbol I}); \,\,
 \fg b_e =\frac{1}{2}({\boldsymbol F}_e\cdot{\boldsymbol F}_e^T-{\boldsymbol I}).
 \label{multdecom_B1}
\end{equation}
Transformation deformation gradient:
\begin{eqnarray}
\text{ }\quad{\boldsymbol F}_t &=&{\boldsymbol U}_t =\fg I+\sum_{I=1}^M  \sum_{i=1}^N \fg\varepsilon_{ti}^{(I)} \phi_i(\eta_i)\varphi(a_\varepsilon,\eta_0) \phi_I^\xi(\xi_I) ; \nonumber\\
 \varphi(a_\varepsilon, \eta_0)
&=& a_\varepsilon \eta_0^2(1-\eta_0)^2+\eta_0^3(4-3 \eta_0); \quad \phi_i(\eta_i)=\eta_i^2(3-2\eta_i); \quad  \phi_I^\xi(\xi_I)=\xi_I^2(3-2\xi_I).
 \label{transstr_B1}
\end{eqnarray}

\item{ Helmholtz's free energy density:}
\begin{eqnarray}
 \psi &=& \frac{J_t}{\rho_0}\psi_e(\fg E_e,\tilde\eta, \tilde\xi,\theta)
+J\breve{\psi}^{\theta}(\tilde\eta,\theta) +\tilde{\psi}^\theta(\tilde\eta,\theta)+{\psi}^p(\tilde\eta)  + J\psi^\nabla(\eta_0,\tilde\eta^\nabla) +J\psi^\xi(\tilde\xi,\tilde\xi^\nabla,\eta_0), \, \text{where}\nonumber\\
\psi_e &=& \frac{1}{2} \fg E_e:\left({\boldsymbol{\mathbb C}}:\fg E_e\right); \quad{\boldsymbol{\mathbb C}}(\tilde\eta,\tilde\xi) = \sum_{I=1}^M  \left[(1-\varphi(a,\eta_0)) {\boldsymbol{\mathbb C}}_0^{(I)}
 + \varphi(a,\eta_0)\sum_{i=1}^N \phi_i (\eta_i)  {\boldsymbol{\mathbb C}}_i^{(I)}\right] \phi_I^\xi(\xi_I);  \nonumber\\
\breve\psi^\theta &=& [A_{0M}+(a_\theta-3)\Delta\psi^\theta(\theta)]\eta_0^2(1-\eta_0)^2+\varphi(a_b,\eta_0)\sum_{i=1}^{N-1}\sum_{j=i+1}^{N}{A}^\eta_{ij}\eta_i^2\eta_j^2 ; \nonumber\\
 \tilde{\psi}^\theta  &=&  \psi^\theta_0(\theta)  + \eta_0^2(3-2\eta_0) \Delta \psi^\theta (\theta); \nonumber\\
\psi^p &=& \sum_{i=1}^{N-1}\sum_{j=i+1}^N  K_{ij}^\eta( \eta_i+\eta_j-1)^2\eta_i^2\eta_j^2; \nonumber\\
 \psi^\nabla &=& \frac{1}{2\rho_0} \left[\beta_{0M}|\nabla\eta_0|^2- \tilde\varphi(a_\beta,a_0,\eta_0) \sum_{i=1}^{N-1}\sum_{j=i+1}^{N}\beta_{ij}\nabla\eta_i\cdot\nabla\eta_j \right];\nonumber\\
  \psi^\xi &=& \psi^\xi_{loc} + \psi^\xi_{\nabla},  \qquad\text{where} \quad
 \psi^\xi_\nabla = -\sum_{I=1}^{M-1}\sum_{J=I+1}^{M}  \frac{\beta_{IJ}^{gb}(\eta_0)}{2\rho_0} \nabla\xi_I\cdot\nabla\xi_J ,  \nonumber\\
 \psi^\xi_{loc}&=& \sum_{I=1}^{M-1}\sum_{J=I+1}^{M} \left[ A_{IJ}^{gb}(\eta_0)\xi_I^2\xi_J^2 +  K_{IJ}^{gb}( \xi_I+\xi_J-1)^2\xi_I^2\xi_J^2 \right]+ \sum_{H=1}^{M-2}\sum_{I=H+1}^{M-1} \sum_{J=I+1}^{M} K_{HIJ}\xi_H^2\xi_I^2\xi_J^2;
  \nonumber\\
 A_{IJ}^{gb}(\eta_0) &=& \frac{18\gamma_{IJ}^{gb}(\eta_0)}{\delta_{IJ}^{gb}}, \quad  \beta_{IJ}^{gb}(\eta_0) = \delta_{IJ}^{gb}\gamma_{IJ}^{gb}(\eta_0); \quad \gamma_{IJ}^{gb}(\eta_0) = \gamma_{IJ}^A+\Delta \gamma_{IJ}^{gb} \varphi(\eta_0,a); \nonumber\\
  \tilde\varphi(a_\beta, a_0,\eta_0)&=&{a_\beta }{\eta_0^2} -2[{a_\beta }-2(1-a_0)]{\eta_0^3}+ [{a_\beta }- 3(1-a_0)]{\eta_0^4} + a_0.
\label{MF0_B1b}
\end{eqnarray}

\item{{Total, elastic, and structural stresses:}
\begin{eqnarray}
{\boldsymbol P} &=& {\boldsymbol P}_e+{\boldsymbol P}_{st}, \quad \text{where }  
 {\boldsymbol P}_e =  J_t\frac{\partial\psi_e(\fg F_e)}{\partial\fg F_e}\cdot\fg F_t^{-T};
 \nonumber\\
  {\boldsymbol P}_{st} &=& J\rho_0(\breve{\psi}_{\theta}+\psi_\nabla+\psi_\xi){\fg F}^{-T}
  -J\left( \beta_{0M}\nabla\eta_0\otimes\nabla\eta_0 -
  \frac{\tilde\varphi(a_\beta,a_0,\eta_0)}{2}\sum_{i=1}^{N}\sum_{j=1,\neq i}^N\beta_{ij}^\eta\nabla\eta_i\otimes\nabla\eta_j  -  \right. \nonumber\\
  && \left. \sum_{I=1}^{M}\sum_{H=1,\neq I}^M\frac{\beta_{IH}^{gb}(\eta_0)}{2}\nabla\xi_I\otimes\nabla\xi_H \right )\cdot{\fg F}^{-T} ;
\label{stress_compospk}
\end{eqnarray}
\begin{eqnarray}
 \fg\sigma &=&\fg\sigma_e+\fg\sigma_{st}, \quad \text{where }    \fg\sigma_e=J^{-1}_e \frac{\partial\psi_e(\fg F_e)}{\partial {\boldsymbol F}_e}\cdot{\boldsymbol F}_e^T
=J_t\,{\boldsymbol F}_e\cdot \frac{\partial \psi_e(\fg E_e)}{\partial{\boldsymbol  E}_e }\cdot{\boldsymbol F}_e^T;  \nonumber\\
 \fg\sigma_{st} &=& J\rho(\breve{\psi}_{\theta}+\psi_\nabla+\psi_\xi){\fg I}
  - \beta_{0M}\nabla\eta_0\otimes\nabla\eta_0 +
  \frac{\tilde\varphi(a_\beta,a_0,\eta_0)}{2}\sum_{i=1}^{N}\sum_{j=1,\neq i}^N\beta_{ij}^\eta\nabla\eta_i\otimes\nabla\eta_j  +  \nonumber\\
  && \sum_{I=1}^{M}\sum_{H=1,\neq I}^M\frac{\beta_{IH}^{gb}(\eta_0)}{2}\nabla\xi_I\otimes\nabla\xi_H.
\label{stress_compos}
\end{eqnarray}
}
\item{Mechanical equilibrium equations (neglecting body forces and inertia):
\begin{equation}
\nabla_0\cdot\fg P = \fg 0  \quad\text{in }\Omega_0 \quad\text{or equivalently, }
\quad \nabla\cdot\fg\sigma=\fg 0 \quad\text{in }\Omega.
 \label{eql_eqs1}
\end{equation}
}
\item{Ginzburg-Landau equations}
\begin{equation}
 \dot\eta_0 = L_{0M} X_{0M}; \quad
 \dot\eta_i = \sum_{j=1 \,(\neq i)}^N L_{ij}^\eta (X_i^\eta-X_j^\eta);  \quad
 \dot\xi_I = \sum_{J=1\,(\neq I)}^M L_{IJ}^\xi (X_I^\xi-X_J^\xi) \quad\text{(no sum on }i,\,I\text{), where}
 \label{final_kinetic1Box1}
\end{equation}
\begin{eqnarray}
X_{0M} &=& \left(J\fg F^{-1}\cdot\fg \sigma_e\cdot\fg F-J_t\psi_e\fg I\right): \fg F_t^{-1}\cdot\frac{\partial\fg F_t}{\partial\eta_0}-
\frac{J_t}{2}\fg E_e:\left(\frac{\partial\fg{\mathbb C}}{\partial\eta_0}:\fg E_e\right)-J\rho_0\frac{\partial\varphi(a_b,\eta_0)}{\partial\eta_0}\sum_{i=1}^{N-1}\sum_{j=i+1}^NA_{ij}^\eta\eta_i^2\eta_j^2
\nonumber\\
&&
-\rho_0J [A(\theta)+(a_\theta-3)\Delta\psi^\theta(\theta)] (2\eta_0-6\eta_0^2+4\eta_0^3)
+\frac{J}{2}\frac{\partial\tilde\varphi(a_\beta,a_0,\eta_0)}
{\partial\eta_0}\sum_{i=1}^{N-1}\sum_{j=i+1}^N\beta_{ij}^\eta\nabla\eta_i\cdot\nabla\eta_j \nonumber\\
&&-\frac{\partial\varphi(\eta_0,a)}
{\partial\eta_0} \sum_{I=1}^{M-1}\sum_{H=I+1}^{M} J\rho_0\Delta\gamma_{IH}^{gb}  \left[ \frac{18 }{\delta_{IH}^{gb}}\xi_I^2\xi_H^2- \frac{\delta_{IH}^{gb}}{2\rho_0} \nabla\xi_I\cdot\nabla\xi_H \right]- \rho_0(6\eta_0-6\eta_0^2)\Delta\psi^\theta
+J\nabla\cdot\left( \beta_{0M}\nabla\eta_0\right);  
\label{forcessbox1}
\end{eqnarray}
\begin{eqnarray}
X_i^\eta &=& \left(J\fg F^{-1}\fg \sigma_e\fg F-J_t\psi_e\fg I\right):\fg F_t^{-1}\cdot\frac{\partial\fg F_t}{\partial\eta_i}-
\frac{J_t}{2}\fg E_e:\left(\frac{\partial\fg{\mathbb C}}{\partial\eta_i}:\fg E_e\right)- 2\rho_0J\sum_{j=1,j\neq i}^N A_{ij}^\eta\eta_i\eta_j^2\varphi(a_b,\eta_0)-\nonumber\\
&& 2 \sum_{j=1,j\neq i}^N \rho_0K_{ij}(\eta_i+\eta_j-1)(2\eta_i+\eta_j-1)\eta_i\eta_j^2
-\frac{J}{2}\nabla\cdot\left( \sum_{j=1, j\neq i}^N \tilde\varphi\beta_{ij}^\eta\nabla\eta_j\right), \quad\text{for } i=1,2, \ldots, N; 
\label{forcessbox15}
\end{eqnarray}
\begin{eqnarray}
X_I^\xi &=& \left(J\fg F^{-1}\fg \sigma_e\fg F-J_t\psi_e\fg I\right):\fg F_t^{-1}\cdot\frac{\partial\fg F_t}{\partial\xi_I}-
\frac{J_t}{2}\fg E_e:\left(\frac{\partial\fg{\mathbb C}}{\partial\xi_I}:\fg E_e\right)-
 \sum_{H=1, H\neq I}^{M} J \frac{36\gamma_{IH}^{gb}(\eta_0) }{(\delta_{IH}^{gb})^2}\xi_I\xi_H^2-\nonumber\\
&& \frac{J}{2}\nabla\cdot\left( \sum_{H=1, H\neq I}^M  \beta_{IH}^\xi\nabla\xi_H\right) \quad \text{for } I=1,2, \ldots, M.
\label{forcessbox16}
\end{eqnarray}

\item{Boundary conditions for the order parameters}
\begin{equation}
\nabla\eta_0\cdot\fg n=0; \quad \sum_{j=1}^N \beta^\eta_{ij}\nabla\eta_j\cdot\fg n=0   \quad\text{for } i=1, \ldots, N; \quad \sum_{H=1}^M\beta^\xi_{IH}\nabla\xi_H\cdot\fg n=0 
 \quad \text{for } I=1, \ldots, M  \quad\text{on }S.
 \label{neumbcs}
\end{equation}

\end{itemize}


\section{List of equations for  MTs in bicrystals and tricrystals with two variants under plane stress  isotropic elasticity}
\label{planestresseqs}
We list down the governing equations for the 2D bicrystals and tricrystals used for the simulations under plane stress condition and consider transformations between $\sf A$ and two variants ${\sf M}_1$ and ${\sf M}_2$. Furthermore, we assume isotropic elastic response and the Lam\'{e} constants are identical for $\sf A$ and both the variants, which are denoted by $\lambda$ and $\mu$. Using Eq. \eqref{constraint11}, we conclude that two independent order parameters $\eta_0$ and $\eta_1$ are sufficient to describe the phases. The order parameter $\xi_1$ describes the  grains in the bicrystals, and the grains in the tricrystals are described by the order parameters $\xi_1$, $\xi_2$, and $\xi_3$, which satisfy $\xi_1+\xi_2+\xi_3=1$ (see Eq. \eqref{constraint}).

\begin{itemize}
\item{Elastic strain and total strain tensor components (indices $a,b=1,2$)
\begin{eqnarray}
&& b_{ab}= \frac{1}{2}\left(\sum_{c=1}^2F_{ac}F_{bc}-\delta_{ab}\right); \quad  b_{(e)ab}= \frac{1}{2}\left(\sum_{c=1}^2F_{(e)ac}F_{(e)bc}-\delta_{ab}\right); \quad
b_{(e)33}=-\frac{\lambda' }{2\mu}(b_{(e)11}+b_{(e)22}); 
\nonumber\\
&& b_{33}=\frac{1}{2}(F_{33}^2-1);
\quad F_{(e)33} = V_{(e)33}=\sqrt{1+2 b_{(e)33}}; \quad F_{33}=V_{(e)33}U_{(t)33}, \quad \text{where }\lambda'=\frac{2\lambda\mu}{\lambda+2\mu}.
\label{stress_compost}
\end{eqnarray}
\begin{equation}
\text{Also, } b_{a3}=b_{(e)a3}=F_{a3}=F_{(e)a3}=V_{a3}=V_{(e)a3}=U_{(t)a3}=0 \quad \text{for all } a=1,2.
\label{stress_compostdf}
\end{equation}
\begin{equation}
J = (F_{11}F_{22}-F_{12}F_{21})F_{33}; \quad J_e = (F_{(e)11}F_{(e)22}-F_{(e)12}F_{(e)21})F_{(e)33}; \quad
J_t = (F_{(t)11}F_{(t)22}-F_{(t)12}F_{(t)21})F_{(t)33}.
 \label{jacobians}
\end{equation}
\item{Transformation stretch tensor components (indices $a,b=1,2$)}

{\bf Bicrystals:}
\begin{eqnarray}
F_{(t)ab} &=& \delta_{ab}+ \phi_1^\xi(\xi_1) \varepsilon_{(t)ab}^{(1)}+[1-\phi_1^\xi(\xi_1)] \varepsilon_{(t)ab}^{(2)}, \quad \text{using } \phi_1^\xi(\xi_1)+\phi_2^\xi(\xi_2)=1; \,\, \xi_1+\xi_2=1, \text{ where}
\nonumber\\
 \varepsilon_{(t)ab}^{(I)} &=& \left[\varepsilon_{(t2)ab}^{(I)} +
(\varepsilon_{(t1)ab}^{(I)}-\fg\varepsilon_{(t2)ab}^{(I)})\phi_1(\eta_1)\right] \varphi(a_\varepsilon,\eta_0) \quad \text{for } I=1,2.\nonumber\\
F_{(t)33}&=&\alpha; \,\, F_{(t)13}=F_{(t)23}=F_{(t)31}=F_{(t)32}=0.
 \label{utildebox1}
\end{eqnarray}

{\bf Tricrystals: }
\begin{eqnarray}
 F_{(t)ab} &=& \delta_{ab}+ \phi_1^\xi(\xi_1) \varepsilon_{(t)ab}^{(1)} + \phi_2^\xi(\xi_2) \varepsilon_{(t)ab}^{(2)} + \phi_3^\xi(\xi_3) \varepsilon_{(t)ab}^{(3)}, \,\,\,\text{where } \xi_1+\xi_2+\xi_3=1; \nonumber\\
 \varepsilon_{(t)ab}^{(I)} &=& \left[\varepsilon_{(t2)ab}^{(I)} +
(\varepsilon_{(t1)ab}^{(I)}-\fg\varepsilon_{(t2)ab}^{(I)})\phi_1(\eta_1)\right] \varphi(a_\varepsilon,\eta_0) \quad \text{for } I=1,2,3.
\nonumber\\
 F_{(t)33} &=& 1+ (\alpha-1) \left(\phi_1^\xi(\xi_1)+ \phi_2^\xi(\xi_2)+\phi_3^\xi(\xi_3)\right); \,\,\, F_{(t)13}=F_{(t)23}=F_{(t)31}=F_{(t)32}=0.
 \label{utildebox1tt}
\end{eqnarray}
}
\item{Free energy (indices $a,b,c,d=1,2$)
\begin{eqnarray}
&&\psi_e =\left[0.5\lambda'(b_{(e)11}+b_{(e)22})^2+\mu \sum_{a,b=1}^2 b_{(e)ab}b_{(e)ab}\right] \sum_{I=1}^M\phi^\xi_I,  \text{ using } 
\label{engiso}
\\
&&({\mathbb C}_0^{(I)})_{abcd}=({\mathbb C}_1^{(I)})_{abcd}=({\mathbb C}_2^{(I)})_{abcd}=\left[\lambda \delta_{ab}\delta_{cd}+\mu(\delta_{ad}\delta_{bc}+\delta_{ac}\delta_{bd})\right]\sum_{I=1}^M\phi^\xi_I;
\label{engisoff}
\\
&&\rho_0\breve\psi^\theta = \rho_0[A_{0M}(\theta)+(a_\theta-3)\Delta\psi^\theta(\theta)]\eta_0^2(1-\eta_0)^2+\rho_0A_{12}^\eta
\eta_1^2(1-\eta_1)^2\varphi(a_b,\eta_0);
\label{engiso1}
\\
&& \rho_0\tilde{\psi}^\theta  = \rho_0 \psi_{0}^{\theta}(\theta)  + \rho_0\Delta \psi^{\theta} (\theta) \eta_0^2(3-2\eta_0) \quad
\text{with }\rho_0\Delta\psi^\theta = -\Delta s(\theta-\theta_e);
\label{engiso2}
\\ 
&& \rho_0\psi^p = 0,   \quad \text{using Eq. } \eqref{constraint11};
\label{engiso3}
\\
 &&\rho_0\psi^\nabla = \frac{1}{2}\sum_{a=1}^2\left(\beta_{0M}\frac{\partial\eta_0}{\partial r_a}\frac{\partial\eta_0}
{\partial r_a}+\beta_{12}^\eta\frac{\partial\eta_1}{\partial r_a}\frac{\partial\eta_1}{\partial r_a}\tilde\varphi(a_\beta,a_0,\eta_0)\right);
\label{engiso4}
\\
&& \tilde\varphi(a_\beta, a_0,\eta_0)={a_\beta }{\eta_0^2} -2[{a_\beta }-2(1-a_0)]{\eta_0^3}+ [{a_\beta }
- 3(1-a_0)]{\eta_0^4} + a_0.
\label{MF0_B2}
\end{eqnarray}

\begin{eqnarray}
 \text{\bf Bicrystals:}\hspace{-5mm}
 &&\rho_0\psi_\xi = \rho_0A_{12}^{gb}(\eta_0)\xi_1^2(1-\xi_1)^2 + \frac{\beta_{12}^{gb}(\eta_0)}{2}\sum_{a=1}^2 \frac{\partial\xi_1}{\partial r_a} \frac{\partial\xi_1}{\partial r_a};
 \\\text{\bf Tricrystals:}\hspace{-5mm}
 &&\rho_0\psi_\xi = \rho_0A_{12}^{gb}(\eta_0)\xi_1^2\xi_2^2 +\rho_0A_{23}^{gb}(\eta_0)\xi_2^2\xi_3^2 +\rho_0A_{13}^{gb}(\eta_0)\xi_1^2\xi_3^2  - \frac{1}{2}\sum_{a=1}^2 \left[{\beta_{12}^{gb}(\eta_0)} \frac{\partial\xi_1}{\partial r_a} \frac{\partial\xi_2}{\partial r_a}+ \right. \nonumber\\
 && \left. {\beta_{23}^{gb}(\eta_0)} \frac{\partial\xi_2}{\partial r_a} \frac{\partial\xi_3}{\partial r_a}+{\beta_{13}^{gb}(\eta_0)} \frac{\partial\xi_1}{\partial r_a} \frac{\partial\xi_3}{\partial r_a} \right]+\rho_0(K_{12}^{gb}+K_{23}^{gb}+K_{13}^{gb}+K_{123})\xi_1^2\xi_2^2\xi_3^2; \quad
\label{engiso5}
\\
&& \rho_0A_{IJ}^{gb}(\eta_0) = \frac{18\gamma_{IJ}^{gb}(\eta_0)}{\delta_{IJ}^{gb}}, \,\,  \beta_{IJ}^{gb}(\eta_0) = \delta_{IJ}^{gb}\gamma_{IJ}^{gb}(\eta_0),  \, \text{and } \gamma_{IJ}^{gb} = \gamma_{IJ}^A + (\gamma_{IJ}^M- \gamma_{IJ}^A) \varphi(a,\eta_0).
\end{eqnarray}

}
 \item{Elastic and structural stresses (indices $a,b=1,2$)
 \begin{eqnarray}
 &&   \sigma_{ab}=\sigma_{(e)ab}+\sigma_{(st)ab}, \quad \text{and}\quad P_{ab}=P_{(e)ab}+P_{(st)ab}; \quad \text{where} \nonumber\\
&&\sigma_{(e)ab}=J_e^{-1} \sum_{c,d=1}^2  V_{(e)ac}V_{(e)cd}\left[\lambda'(b_{(e)11}+b_{(e)22}) \delta_{db}+ 2\mu b_{(e)db}
\right];\\
  \label{surfaceStress2vbx2}
&& P_{(e)ab}= J_t \sum_{c,d,f=1}^2  V_{(e)ac}V_{(e)cd}\left[\lambda'(b_{(e)11}+b_{(e)22})
\delta_{df}+ 2\mu b_{(e)df}\right]F^{-1}_{bf}.
 \label{surfaceStress2vbx1}
   \end{eqnarray}
   \begin{eqnarray}
  \text{\bf Bicrystals:}&&  \nonumber\\
 P_{({st})ab} &=&J\rho_0(\breve{\psi}^{\theta}+\psi^\nabla+\psi^\xi)  F^{-1}_{ba}
-J\sum_{c=1}^2 \left[\beta_{0M}\frac{\partial\eta_0}{\partial r_a}\frac{\partial\eta_0}{\partial r_c}
+\beta_{12}^\eta\frac{\partial\eta_1}{\partial r_a}\frac{\partial\eta_1}{\partial r_c}\tilde\varphi +\beta_{12}^{gb}(\eta_0)\frac{\partial\xi_1}{\partial r_a}\frac{\partial\xi_1}{\partial r_c}\right]F^{-1}_{bc};
 \label{surfaceStress2vbx3bi} \nonumber
  \\
 \sigma_{({st})ab}&=&J\rho(\breve{\psi}^{\theta}+\psi^\nabla+\psi^\xi) \delta_{ab}-\beta_{0M}\frac{\partial\eta_0}{\partial r_a}
\frac{\partial\eta_0}{\partial r_b}-\beta_{12}^\eta\frac{\partial\eta_1}{\partial r_a}\frac{\partial\eta_1}{\partial r_b}\tilde\varphi-\beta_{12}^{gb}(\eta_0)\frac{\partial\xi_1}{\partial r_a}\frac{\partial\xi_1}{\partial r_b}.
 \label{surfaceStress2vbxbi}
  \end{eqnarray}
  \begin{eqnarray}
 \text{\bf Tricrystals:} &&\nonumber\\
 P_{({st})ab} &=&J\rho_0(\breve{\psi}^{\theta}+\psi^\nabla+\psi^\xi)  F^{-1}_{ba}
-J\sum_{c=1}^2 \left[\beta_{0M}\frac{\partial\eta_0}{\partial r_a}\frac{\partial\eta_0}{\partial r_c}
+\beta_{12}^\eta\frac{\partial\eta_1}{\partial r_a}\frac{\partial\eta_1}{\partial r_c}\tilde\varphi-\frac{\beta_{12}^{gb}(\eta_0)}{2}\left(\frac{\partial\xi_1}{\partial r_a}\frac{\partial\xi_2}{\partial r_c}+\right. \right. \nonumber\\
&& \left. \left. \frac{\partial\xi_2}{\partial r_a}\frac{\partial\xi_1}{\partial r_c}\right) - \frac{\beta_{13}^{gb}(\eta_0)}{2}\left(\frac{\partial\xi_1}{\partial r_a}\frac{\partial\xi_3}{\partial r_c}+\frac{\partial\xi_3}{\partial r_a}\frac{\partial\xi_1}{\partial r_c}\right)
-\frac{\beta_{23}^{gb}(\eta_0)}{2}\left(\frac{\partial\xi_2}{\partial r_a}\frac{\partial\xi_3}{\partial r_c}+\frac{\partial\xi_3}{\partial r_a}\frac{\partial\xi_2}{\partial r_c}\right)
\right]F^{-1}_{bc}; \nonumber
 \label{surfaceStress2vbx3tr}
 \end{eqnarray}
 \begin{eqnarray}
\sigma_{({st})ab}&=&J\rho(\breve{\psi}^{\theta}+\psi^\nabla+\psi^\xi) \delta_{ab}-\beta_{0M}\frac{\partial\eta_0}{\partial r_a}
\frac{\partial\eta_0}{\partial r_b}-\beta_{12}^\eta\frac{\partial\eta_1}{\partial r_a}\frac{\partial\eta_1}{\partial r_b}\tilde\varphi+
 \frac{\beta_{12}^{gb}(\eta_0)}{2}\left(\frac{\partial\xi_1}{\partial r_a}\frac{\partial\xi_2}{\partial r_c}+\frac{\partial\xi_2}{\partial r_a}\frac{\partial\xi_1}{\partial r_c}\right) +\nonumber\\
&& \frac{\beta_{13}^{gb}(\eta_0)}{2}\left(\frac{\partial\xi_1}{\partial r_a}\frac{\partial\xi_3}{\partial r_c}+\frac{\partial\xi_3}{\partial r_a}\frac{\partial\xi_1}{\partial r_c}\right)
+\frac{\beta_{23}^{gb}(\eta_0)}{2}\left(\frac{\partial\xi_2}{\partial r_a}\frac{\partial\xi_3}{\partial r_c}+\frac{\partial\xi_3}{\partial r_a}\frac{\partial\xi_2}{\partial r_c}\right).
 \label{surfaceStress2vbxtr}
 \end{eqnarray}
 \begin{equation}
  \sigma_{c3}=\sigma_{(e)c3}=\sigma_{({st})c3}=P_{c3}=P_{3c}=P_{(e)c3}=P_{(e)3c}=P_{({st})c3}=P_{({st})3c}=0 \,\, \text{for } c=1,2,3.
 \label{trivialStress}
 \end{equation}
 
}

\item{Mechanical equilibrium equation neglecting body forces and inertia ($a=1,2$)
\begin{equation}
\sum_{b=1}^2\frac{\partial P_{ab}}{\partial r_{0b}}=0 \,\,\text{in }\Omega_0, \quad\text{or equivalently, }
\quad \sum_{b=1}^2 \frac{\partial \sigma_{ab}}{\partial r_b}=0 \,\,\text{in }\Omega.
 \label{eql_eqs}
\end{equation}
}
\item{Ginzburg-Landau equations}

{\bf Bicrystals:} 
\begin{equation}
\dot\eta_0 = L_{0M}^\eta X_{0M};    \quad \dot\eta_1 = L_{12}^\eta X_{12}^\eta;  \quad \dot\xi_1 = L_{12}^\xi X_{12}^\xi; \quad \text{ where}
\label{bicrystgov}
\end{equation}
\begin{eqnarray}
X_{0M} &=& \sum_{a,b,c,d,f=1}^2\left(J{F}^{-1}_{ac} \sigma_{(e)cf} F_{fb}-J_t\psi_e\delta_{ab}\right)
{ F}_{(t)bd}^{-1}\frac{\partial{F}_{(t)da}}{\partial\eta_0}-
\rho_0\Delta\psi_\theta(6\eta_0-6\eta_0^2)-\rho_0A_{12}^\eta J\frac{\partial\varphi(a_b,\eta_0)}{\partial\eta_0}\times
\nonumber\\&&
 \eta_1^2(1-\eta_1)^2- \rho_0 J [A_{0M}+(a_\theta-3)\Delta\psi^\theta(\theta)] (2\eta_0-6\eta_0^2+4\eta_0^3)-
 \frac{J \beta_{12}^\eta}{2}\frac{\partial\tilde\varphi(a_\beta,a_0,\eta_0)}{\partial\eta_0} \sum_{a=1}^2\frac{\partial\eta_1}{\partial r_a}\frac{\partial\eta_1}{\partial r_a}-
\nonumber\\
&&
 \Delta\gamma_{12}^{gb}\frac{\partial\varphi}{\partial\eta_0} \left[18\xi_1^2(1-\xi_1)^2+0.5\sum_{a=1}^2\frac{\partial\xi_1}{\partial r_a}\frac{\partial\xi_1}{\partial r_a}\right]
+ \sum_{a=1}^2  \frac{\partial}{\partial r_a} \left(\beta_{0M} \frac{\partial\eta_0}{\partial r_a}\right);
  \label{kineqs20a}
\end{eqnarray}
\begin{eqnarray}
 X_{12}^\eta &=& X_1^\eta-X_2^\eta= \sum_{a,b,c,d,f=1}^2\left(J{F}^{-1}_{ac} \sigma_{(e)cf} F_{fb}-J_t\psi_e\delta_{ab}\right)
{ F}_{(t)bd}^{-1}\frac{\partial{F}_{(t)da}}{\partial\eta_1}-
  \nonumber\\ 
&&\rho_0J A_{12}^\eta \, (2\eta_1-6\eta_1^2+4\eta_1^3)\varphi(a_b,\eta_0)+\sum_{a=1}^2  \frac{\partial}{\partial r_a} \left[\tilde\varphi(a_\beta,a_0,\eta_0)\,\beta_{12}^\eta \frac{\partial\eta_1}{\partial r_a} \right];
\quad \text{and}
\label{kineqs20b}
\end{eqnarray}
\begin{eqnarray}
X_{12}^\xi &=& X_1^\xi-X_2^\xi=  \sum_{a,b,c,d,f=1}^2\left(J{F}^{-1}_{ac} \sigma_{(e)cf} F_{fb}-J_t\psi_e\delta_{ab}\right)
{ F}_{(t)bd}^{-1}\frac{\partial{F}_{(t)da}}{\partial\xi_1}-A_{12}^{gb}\,(2\xi_1-6\xi_1^2+4\xi_1^3)+
\nonumber\\
&& \sum_{a=1}^2  \frac{\partial}{\partial r_a} \left[\beta_{12}^{gb} \frac{\partial\xi_1}{\partial r_a} \right].
 \label{kineqs20}
\end{eqnarray}

{\bf Tricrystals:} 
\begin{equation}
\dot\eta_0 = L_{0M}^\eta X_{0M};    \quad \dot\eta_1 = L_{12}^\eta X_{12}^\eta;  \quad  \dot\xi_1 = L_{12}^\xi X_{12}^\xi+L_{13}^\xi X_{13}^\xi; \quad  \dot\xi_2 = L_{12}^\xi X_{21}^\xi+L_{23}^\xi X_{23}^\xi;  \quad \dot\xi_3=-\dot\xi_1-\dot\xi_2 ; 
\label{bicrystgovt}
\end{equation}
where $X_{12}^\xi=X_1^\xi-X_2^\xi$; \quad $X_{13}^\xi=X_1^\xi-X_3^\xi$; \quad $X_{23}^\xi=X_2^\xi-X_3^\xi$; \quad $X_{21}^\xi =X_2^\xi-X_1^\xi$;
\begin{eqnarray}
X_{0M} &=&\sum_{a,b,c,d,f=1}^2\left(J{F}^{-1}_{ac} \sigma_{(e)cf} F_{fb}-J_t\psi_e\delta_{ab}\right)
{ F}_{(t)bd}^{-1}\frac{\partial{F}_{(t)da}}{\partial\eta_0}-
\rho_0\Delta\psi^\theta(6\eta_0-6\eta_0^2)
-\rho_0A_{12}^\eta J\frac{\partial\varphi(a_b,\eta_0)}{\partial\eta_0}\times
\nonumber\\&&
  \eta_1^2(1-\eta_1)^2-\rho_0 J [A_{0M}+(a_\theta-3)\Delta\psi_\theta(\theta)] (2\eta_0-6\eta_0^2+4\eta_0^3)-
\frac{J \beta_{12}^\eta}{2}\frac{\partial\tilde\varphi(a_\beta,a_0,\eta_0)}{\partial\eta_0}\sum_{a=1}^2\frac{\partial\eta_1}{\partial r_a}\frac{\partial\eta_1}{\partial r_a}-
 \nonumber\\
&&\frac{\partial\varphi(\eta_0,a)}
{\partial\eta_0} \sum_{I=1}^{2}\sum_{J=I+1}^{3} J\Delta\gamma_{IJ}^{gb}  \left[ \frac{18 }{\delta_{IJ}^{gb}}\xi_I^2\xi_J^2- \frac{\delta_{IJ}^{gb}}{2} \sum_{a=1}^2\frac{\partial\xi_I}{\partial r_a}\frac{\partial\xi_J}{\partial r_a} \right]
+ \sum_{a=1}^2  \frac{\partial}{\partial r_a} \left(\beta_{0M} \frac{\partial\eta_0}{\partial r_a}\right);
\label{kineqs20tri1}
\end{eqnarray}
\begin{eqnarray}
 X_{12}^\eta &=& X_1^\eta -X_2^\eta=\sum_{a,b,c,d,f=1}^2\left(J{F}^{-1}_{ac} \sigma_{(e)cf} F_{fb}-J_t\psi_e\delta_{ab}\right)
{ F}_{(t)bd}^{-1}\frac{\partial{F}_{(t)da}}{\partial\eta_1}
-\nonumber\\
&& \rho_0J A_{12}^\eta (2\eta_1-6\eta_1^2+4\eta_1^3)\varphi(a_b,\eta_0)
+ \sum_{a=1}^2  \frac{\partial}{\partial r_a} \left[\tilde\varphi(a_\beta,a_0,\eta_0)\,\beta_{12}^\eta \frac{\partial\eta_1}{\partial r_a} \right];
\label{kineqs20tri2}
\end{eqnarray}
\begin{eqnarray}
X_1^\xi &=&\sum_{a,b,c,d,f=1}^2\left(J{F}^{-1}_{ac} \sigma_{(e)cf} F_{fb}-J_t\psi_e\delta_{ab}\right)
{ F}_{(t)bd}^{-1}\frac{\partial{F}_{(t)da}}{\partial\xi_1}-J_t\psi^e \frac{\partial\phi^\xi_1}{\partial\xi_1}
- \sum_{H=2}^{3} J \frac{36 \gamma_{1H}^{gb}}{(\delta_{1H}^{gb})^2}\xi_1\xi_H^2-\nonumber\\
&& 2\rho_0(K_{12}^{gb}+K_{13}^{gb}+K_{23}^{gb}+K_{123})\xi_1\xi_2^2\xi_3^2-\frac{J}{2} \sum_{H=2}^3\sum_{a=1}^{2}\frac{\partial}{\partial r_a}\left(   \beta_{1H}^\xi \frac{\partial \xi_H}{\partial r_a}\right) ;
\label{kineqs20tri3}
\end{eqnarray}
\begin{eqnarray}
X_2^\xi &=&\sum_{a,b,c,d,f=1}^2\left(J{F}^{-1}_{ac} \sigma_{(e)cf} F_{fb}-J_t\psi_e\delta_{ab}\right)
{ F}_{(t)bd}^{-1}\frac{\partial{F}_{(t)da}}{\partial\xi_2}-J_t\psi^e \frac{\partial\phi^\xi_2}{\partial\xi_2}
- \sum_{H=1,\neq 2}^{3} J \frac{36 \gamma_{2H}^{gb}}{(\delta_{2H}^{gb})^2}\xi_2\xi_H^2-\nonumber\\
&& 2\rho_0(K_{12}^{gb}+K_{13}^{gb}+K_{23}^{gb}+K_{123})\xi_1^2\xi_2\xi_3^2-\frac{J}{2} \sum_{H=1,\neq 2}^3\sum_{a=1}^{2}\frac{\partial}{\partial r_a}\left(   \beta_{2H}^\xi \frac{\partial \xi_H}{\partial r_a}\right) ;
\label{kineqs20tri4}
\end{eqnarray}
\begin{eqnarray}
X_3^\xi &=&\sum_{a,b,c,d,f=1}^2\left(J{F}^{-1}_{ac} \sigma_{(e)cf} F_{fb}-J_t\psi_e\delta_{ab}\right)
{ F}_{(t)bd}^{-1}\frac{\partial{F}_{(t)da}}{\partial\xi_3}-J_t\psi^e \frac{\partial\phi^\xi_3}{\partial\xi_3}
- \sum_{H=1}^{2} J \frac{36 \gamma_{3H}^{gb}}{(\delta_{3H}^{gb})^2}\xi_3\xi_H^2-\nonumber\\
&& 2\rho_0(K_{12}^{gb}+K_{13}^{gb}+K_{23}^{gb}+K_{123})\xi_1^3\xi_2^2\xi_3-\frac{J}{2} \sum_{H=1}^2\sum_{a=1}^{2}\frac{\partial}{\partial r_a}\left(   \beta_{3H}^\xi \frac{\partial \xi_H}{\partial r_a}\right).
 \label{kineqs20tri}
\end{eqnarray}

\item{Boundary conditions for the order parameters
\begin{equation}
\sum_{a=1}^2\frac{\partial\eta_0}{\partial r_{0a}}n_{0a}=0; \quad \sum_{a=1}^2\frac{\partial\eta_1}{\partial r_{0a}}n_{0a}=0; 
\quad \sum_{a=1}^2\frac{\partial\xi_I}{\partial r_{0a}}n_{0a}=0 \,\, \text{for all }I=1,\ldots,M \quad\text{on }S_0,
 \label{neumbcs1}
\end{equation}
which are obtained using Eq. \eqref{multdecom_B1cons} in Eq. \eqref{neumbcs}, and assuming $\beta^\eta_{12}=\beta^\eta_{13}=\beta^\eta_{23}$ and $\beta^{gb}_{12}=\beta^{gb}_{13}=\beta^{gb}_{23}$.
}

\end{itemize}

\section{Compatibility of transformation strains across grain boundary}
\label{compatibilityBiTri}
To understand the martensitic microstructures across the symmetric and asymmetric GBs, we briefly describe the strain compatibility conditions across the GBs. Although a large strain-based theory is considered in this paper, we consider the compatibility condition in terms of the linearized strains across a sharp GB following Chapter 13 of \cite{Bha04}, which is sufficient for understanding the results in this paper. The Bain strains for the grain $G_1$ in the $\{\fg e_1,\fg e_2\}$ basis (see e.g. Fig. \ref{xiplots}(a)) are
\begin{equation}
\fg\varepsilon_{t1}^{(1)}=  \begin{bmatrix}
  \varepsilon_\chi\cos^2\vartheta_1+\varepsilon_\alpha\sin^2\vartheta_1 & (\varepsilon_\alpha-\varepsilon_\chi)\cos\vartheta_1\sin\vartheta_1  \\
    (\varepsilon_\alpha-\varepsilon_\chi)\cos\vartheta_1\sin\vartheta_1  &  \varepsilon_\alpha\cos^2\vartheta_1+\varepsilon_\chi\sin^2\vartheta_1  \\
\end{bmatrix}, 
\label{Bainstrainsrot1}
\end{equation}
\begin{equation}
\fg\varepsilon_{t2}^{(1)} =  \begin{bmatrix}
  \varepsilon_\alpha\cos^2\vartheta_1+\varepsilon_\chi\sin^2\vartheta_1 & (\varepsilon_\chi-\varepsilon_\alpha)\cos\vartheta_1\sin\vartheta_1  \\
    (\varepsilon_\chi-\varepsilon_\alpha)\cos\vartheta_1\sin\vartheta_1  &  \varepsilon_\chi\cos^2\vartheta_1+\varepsilon_\alpha\sin^2\vartheta_1  \\
\end{bmatrix},
\label{Bainstrainsrot2}
\end{equation}
where $\varepsilon_\alpha=\alpha-1$ and $\varepsilon_\chi=\chi-1$. The Bain strains $\fg\varepsilon_{t1}^{(2)}$ and $\fg\varepsilon_{t2}^{(2)}$ for the grain $G_2$ is simply obtained by replacing $\vartheta_1$ with $\vartheta_2$ in Eqs. \eqref{Bainstrainsrot1} and \eqref{Bainstrainsrot2}, respectively. The variant, say $\sf M_1$, would be compatible across the planar GB in the bicrystal considered in Fig. \ref{xiplots}(a) if it satisfies (Chapter 13 of \cite{Bha04})
\begin{equation}
\fg\varepsilon_{t1}^{(2)}-\fg\varepsilon_{t1}^{(1)} =  0.5(\fg s \otimes \fg e_2 +\fg e_2 \otimes \fg s),
\label{Bainstrainsrot3}
\end{equation}
where $\fg s$ is a vector related to the deformation which is to be determined, and we note that the GB normal is parallel to $\fg e_2$-axis. Operating $\fg e_1$ and $\fg e_2$ vectors in Eq. \eqref{Bainstrainsrot3} one can easily show that the compatibility of $\sf M_1$ across the GB requires to satisfy
\begin{equation}
(\fg\varepsilon_{t1}^{(2)})_{11}=(\fg\varepsilon_{t1}^{(1)})_{11}, \quad
(\fg\varepsilon_{t1}^{(2)})_{12}-(\fg\varepsilon_{t1}^{(1)})_{12} =  0.5\fg s \cdot \fg e_1, \quad \text{and} \quad 
(\fg\varepsilon_{t1}^{(2)})_{22}-(\fg\varepsilon_{t1}^{(1)})_{22} =  \fg s \cdot \fg e_2.
\label{Bainstrainsrot4}
\end{equation}
Using Eq. \eqref{Bainstrainsrot1} and the expression for $\fg\varepsilon_{t1}^{(2)}$ in Eq. \eqref{Bainstrainsrot4}$_{2,3}$, one can determine $\fg s$ in terms of $\varepsilon_\alpha$, $\varepsilon_\chi$, $\vartheta_1$, and $\vartheta_2$. Furthermore, the condition Eq. \eqref{Bainstrainsrot4}$_1$ requires
\begin{equation}
 \varepsilon_\chi\cos^2\vartheta_1+\varepsilon_\alpha\sin^2\vartheta_1  =  \varepsilon_\chi\cos^2\vartheta_2+\varepsilon_\alpha\sin^2\vartheta_2.
\label{Bainstrainsrot5}
\end{equation}
For the arbitrary  $ \varepsilon_\alpha$ and $ \varepsilon_\chi$, and $ \varepsilon_\alpha \neq \varepsilon_\chi$, the condition given by Eq. \eqref{Bainstrainsrot5} requires that $\vartheta_1=- \vartheta_2$ for a non-trivial GB, i.e. the GB must be symmetric. An asymmetric GB thus does not generally satisfy the strain compatibility equation \eqref{Bainstrainsrot3}. Like this we can show that the other variant $\sf M_2$ also satisfies the strain compatibility across the symmetric GBs, but not across the asymmetric ones.

{\bf Acknowledgements}

 The support from SERB (Grant number SRG/2020/001194) and the helpful comments and suggestions from Prof. Valery Levitas (Iowa State University) are gratefully acknowledged.

\end{document}